\documentclass[acmsmall,screen]{acmart}

\usepackage[]{collab}
\usepackage{xcolor}
\usepackage{colortbl}
\usepackage{tcolorbox}
\usepackage{mdframed}
\usepackage{listings}
\usepackage{makecell}
\usepackage{multirow}
\usepackage{array}

\usepackage{amsmath,amssymb,amsfonts}
\usepackage{algorithmic}
\usepackage{graphicx}
\usepackage{fbox}
\usepackage{textcomp}
\usepackage{minibox}
\usepackage{balance}
\usepackage{enumitem}
\usepackage{xspace}
\usepackage[T1]{fontenc}
\usepackage{booktabs}
\usepackage{threeparttable}
\usepackage{url}
\usepackage{soul}
\usepackage{ulem}
\usepackage{balance}
\usepackage{subfigure}
\usepackage{diagbox}

\def\BibTeX{{\rm B\kern-.05em{\sc i\kern-.025em b}\kern-.08em
    T\kern-.1667em\lower.7ex\hbox{E}\kern-.125emX}}

\newcommand\benchx{AI-NativeBench\xspace}
\newcommand\bench{AI-NativeBench}

\definecolor{ballblue}{rgb}{0.13, 0.67, 0.8}
\definecolor{grey}{rgb}{0.9, 0.9, 0.9}
\definecolor{googlered}{rgb}{0.914, 0.262, 0.207}
\definecolor{dandelion}{rgb}{0.95, 0.65, 0.0}
\definecolor{citecolor}{RGB}{106, 34, 107}

\definecolor{ballblue}{rgb}{0.13, 0.67, 0.8}
\definecolor{jcpink}{RGB}{255, 0, 96}
\collabAuthor{gb}{orange}{Guangba}
\collabAuthor{jc}{jcpink}{JC}
\collabAuthor{zh}{green}{Zhihan Jiang}
\collabAuthor{jj}{purple}{Junjie}
\collabAuthor{zr}{teal}{Zirui}
\collabAuthor{yl}{red}{Yulun}
\collabAuthor{yc}{blue}{Yichen}
\collabAuthor{ry}{dandelion}{Renyi}
\collabAuthor{yd}{orange}{Yuedong}

\definecolor{mygreen}{HTML}{AFCFA5}
\newcounter{summary}

\definecolor{mygreen}{HTML}{AFCFA5}
\newcounter{implication}

\definecolor{myyellow}{HTML}{FFF2CC}
\newcounter{finding}
\newcommand{\finding}[1]{\refstepcounter{finding}
	\begin{mdframed}[linecolor=gray!25,roundcorner=12pt,backgroundcolor=mygreen!30,linewidth=3pt,innerleftmargin=2pt, leftmargin=0cm,rightmargin=0cm,topline=false,bottomline=false,rightline=false,leftline=false,skipabove=3pt]
		\textbf{Finding \arabic{finding}.} #1
	\end{mdframed}
}

\usepackage{booktabs}       
\usepackage[table]{xcolor}  
\usepackage{pifont}         
\usepackage{bbding}         
\usepackage{graphicx}       
\usepackage{amsmath}        
\usepackage{subcaption}
\usepackage{longtable}
\usepackage{multirow}
\usepackage{placeins}
\usepackage{rotating}
\usepackage{siunitx}

\definecolor{MyHighlight}{gray}{0.92}
\definecolor{GreenCheck}{HTML}{009B50}
\definecolor{RedCross}{HTML}{D1191C}
\definecolor{OrangeWarn}{HTML}{FF8800}

\newcommand{\cmark}{\textcolor{GreenCheck}{\ding{51}}}
\newcommand{\xmark}{\textcolor{RedCross}{\ding{55}}}

\AtBeginDocument{%
  \providecommand\BibTeX{{%
    Bib\TeX}}}

\setcopyright{acmlicensed}
\copyrightyear{2025}
\acmYear{2025}
\acmDOI{XXXXXXX.XXXXXXX}

\acmJournal{TOSEM}

\begin{document}

\title{\bench: An Open-Source White-Box Agentic Benchmark Suite for AI-Native Systems}

\author{Zirui Wang}

\email{wangzr39@mail2.sysu.edu.cn}
\orcid{0009-0004-5773-8716}
\authornote{Equal Contribution.}
\affiliation{%
  \institution{Sun Yat-sen University}
  \city{Guangzhou City}
  \country{China}
}
\author{Guangba Yu}
\orcid{0000-0001-6195-9088}
\authornotemark[1]
\email{guangbayu@cuhk.edu.hk}
\affiliation{%
  \institution{The Chinese University of HongKong}
  \city{Hong Kong SAR}
  \country{China}
}

\author{Michael R. Lyu}
\orcid{0000-0002-3666-5798}
\email{lyu@cse.cuhk.edu.hk}
\affiliation{%
  \institution{The Chinese University of HongKong}
  \city{Hong Kong SAR}
  \country{China}
}








\renewcommand{\shortauthors}{Zirui Wang, Guangba Yu et al.}

\begin{abstract}
The transition from Cloud-Native to AI-Native architectures is fundamentally reshaping software engineering, replacing deterministic microservices with probabilistic agentic services. However, this shift renders traditional black-box evaluation paradigms insufficient: existing benchmarks measure raw model capabilities while remaining blind to system-level execution dynamics. To bridge this gap, we introduce AI-NativeBench, the first application-centric and white-box AI-Native benchmark suite grounded in Model Context Protocol (MCP) and Agent-to-Agent (A2A) standards. By treating agentic spans as first-class citizens within distributed traces, our methodology enables granular analysis of engineering characteristics beyond simple capabilities. 

Leveraging this benchmark across 21 system variants, we uncover critical engineering realities invisible to traditional metrics: a parameter paradox where lightweight models often surpass flagships in protocol adherence, a pervasive inference dominance that renders protocol overhead secondary, and an expensive failure pattern where self-healing mechanisms paradoxically act as cost multipliers on unviable workflows. This work provides the first systematic evidence to guide the transition from measuring model capability to engineering reliable AI-Native systems. To facilitate reproducibility and further research, we have open-sourced the benchmark and dataset.
\end{abstract}




\maketitle

\section{Introduction}
\label{sec:intro}

The integration of Large Language Models (LLMs) into the fabric of modern software represents a transformative shift with profound societal implications. AI-Native applications are rapidly transitioning from passive conversational interfaces to active workforce participants, fundamentally reshaping critical sectors ranging from automated software engineering~\cite{Codex,Cursor,Windsurf} to scientific discovery~\cite{AIScientist,AIScientist1,AIScientist2} and financial auditing~\cite{finance1,finance2,finance3}. As these systems are increasingly entrusted with high-stakes decision-making, their reliability becomes a matter of significant economic and social safety. This societal reliance is driving a fundamental architectural evolution in distributed computing: the transition from \textbf{Cloud-Native}~\cite{Microscaler,MicroRank,Nezha} to \textbf{AI-Native}~\cite{AINative,AgenticService} systems.

In the traditional Cloud-Native paradigm, the atomic unit of computation is the deterministic microservice, governed by rigid human-authored logic. In contrast, the AI-Native paradigm introduces the agentic service~\cite{AgenticService}, defined as autonomous entities driven by probabilistic LLMs (e.g., GPT-4, DeepSeek) that possess inherent capabilities for planning, reasoning, and tool execution. To facilitate scalable collaboration within this new ecosystem, ad-hoc integrations are being replaced by standardized interoperability layers, most notably the Model Context Protocol (MCP)~\cite{mcp} for tool abstraction and the Agent-to-Agent (A2A) protocol~\cite{a2a} for inter-service orchestration. However, this architectural shift introduces a distinctive engineering paradox. Unlike traditional systems that behave deterministically, AI-Native systems inherit infrastructure constraints while simultaneously introducing stochastic cognitive behaviors (such as hallucinations and reasoning loops) that create a layer of opacity which traditional engineering disciplines struggle to quantify.

Despite the rapid adoption of these architectures, the evaluation landscape remains rooted in a black-box paradigm. Predominant benchmarks, such as GAIA~\cite{GAIA}, AgentBench~\cite{AgentBench}, and WebShop~\cite{WebShop}, treat agentic systems as opaque entities, evaluating them based on final outcomes and coarse-grained trajectories to measure raw model capabilities. These suites measure success through binary pass rates or surface-level action sequences (e.g., API call history), effectively ignoring the complex internal execution dynamics. While this approach captures what actions were taken, it provides zero visibility into the system's internal states or architectural bottlenecks. It reveals that a step failed, but remains blind to the underlying cause. Engineers are left unable to determine whether a failure stemmed from a logic error in the framework, a serialization mismatch in the MCP layer, or a stochastic hallucination by the model. Consequently, the community lacks a benchmark that evaluates system-level engineering characteristics.

Departing from the traditional black-box paradigm, we introduce \benchx\footnote{https://github.com/AINativeOps/AINativeBench}, the first application-centric and white-box benchmark suite explicitly designed for AI-Native systems. Unlike prior works that treat agents as isolated inference endpoints, we evaluate them as complex distributed systems. Our suite comprises eight distinct applications spanning three domains, ranging from single-agent utilities to heterogeneous clusters of five agents (\S~\ref{sec:bench}). Crucially, we employ a trace-first methodology. By natively integrating distributed tracing via OpenTelemetry~\cite{Opentelemetry}, we treat agentic spans as first-class citizens, enabling precise attribution of latency and errors. This design allows researchers to dissect performance trade-offs across three architectural variations, comprising monolithic deployments, service-oriented MCP, and distributed A2A.

\textbf{Research Questions and Key Insights.}
To operationalize this evaluation framework and assess the engineering viability of AI-Native systems, we conduct an extensive empirical study involving seven LLMs and 21 application variants. To facilitate reproducibility and future research, we have open-sourced the complete dataset on Hugging Face~\footnote{https://huggingface.co/datasets/AINativeOps/AINativeBench}. We structure our analysis around three core questions that determine the viability of AI-Native systems:
\begin{itemize}[leftmargin=*]
    \item \textbf{RQ1: Behavioral Correctness.} \textit{How do varying model capabilities (e.g., parameter scale, reasoning mode) impact the behavioral correctness of AI-Native applications?} (\S~\ref{sec:rq1})
    Our analysis reveals a parameter paradox: lightweight models (e.g., \texttt{GPT-4o-mini}) often act as more compliant executors for following rigid protocols than flagship models. Furthermore, we identify a content-process divergence in reasoning models. While they generate deeper insights, they tend to internalize execution, bypassing the necessary tools and breaking the application's structural integrity.
    
    \item \textbf{RQ2: Performance Overhead.} \textit{What is the anatomy of latency in distributed AI-Native architectures?} (\S~\ref{sec:rq2})
    We identify a pervasive inference dominance, where LLM computation consumes between 86.9\% and 99.9\% of execution time. While faster models (e.g., GPT-5) reveal a measurable infrastructure footprint, reasoning models (e.g., DeepSeek-R1) exhibit a near-total monopoly on latency. This distribution renders protocol overheads statistically secondary in most high-capability workflows. Consequently, system latency is largely dictated by straggler agents on the critical path rather than communication transport.
    
    \item \textbf{RQ3: Token Economics.} \textit{What is the true economic cost of autonomy in AI-Native systems? }  (\S~\ref{sec:rq3})
    Our results demonstrate that current systems fail to adhere to the fail-fast principle. Instead, they exhibit an expensive failure pattern where failed workflows consume significantly more resources than successful ones. This occurs because agents exhaust their retry budgets on doomed workflow. We further find that while distributed architectures (A2A) mask latency, they incur a reliability tax due to the context redundancy required for synchronization.

\end{itemize}

\textbf{Contributions.}
In summary, this paper makes the following contributions:
\begin{itemize}[leftmargin=*]
    \item \textbf{The First White-Box AI-Native Benchmark Suite:} We present \benchx, an open-source suite comprising eight realistic applications across three domains. Unlike prior task-centric benchmarks, it fully implements emerging industry standards, specifically the MCP and A2A protocol, to support controlled architectural comparisons ranging from monolithic agents to heterogeneous multi-agent applications.
    
    \item \textbf{A Trace-First Evaluation Methodology:} We propose a novel evaluation framework that treats agentic spans as first-class citizens. By integrating distributed tracing via OpenTelemetry with semantic analysis for AI-Native systems, we enable granular fault localization that connects high-level behavioral anomalies (e.g., content-process divergence) to low-level system bottlenecks.
    
    \item \textbf{Comprehensive Empirical Study and Guidelines:} We conduct a large-scale study evaluating seven LLMs across 21 system variants. Our analysis uncovers critical engineering trade-offs, including the parameter paradox and inference dominance, providing evidence-based design guidelines for building cost-effective and reliable AI-Native systems.
\end{itemize}

The remainder of this paper is organized as follows: Section~\ref{sec:background} provides the background on the architectural shift from Cloud-Native to AI-Native systems and introduces the core protocols (MCP and A2A). Section~\ref{sec:bench} details the design principles and architectural diversity of the \bench suite. Section~\ref{sec:method} presents our trace-first evaluation methodology and metric definitions. Section~\ref{sec:evaluation} reports the extensive empirical results and answers our three research questions regarding correctness, performance, and economics. Section~\ref{sec:discussion} discusses the broader implications for system design and addresses threats to validity. Finally, Section~\ref{sec:conclusion} concludes the paper.

\section{Background and Related Work}\label{sec:background}

\subsection{From Cloud-Native to AI-Native}\label{sec:arc}

\begin{figure}[t]
	\centering
	\includegraphics[width=0.95\textwidth]{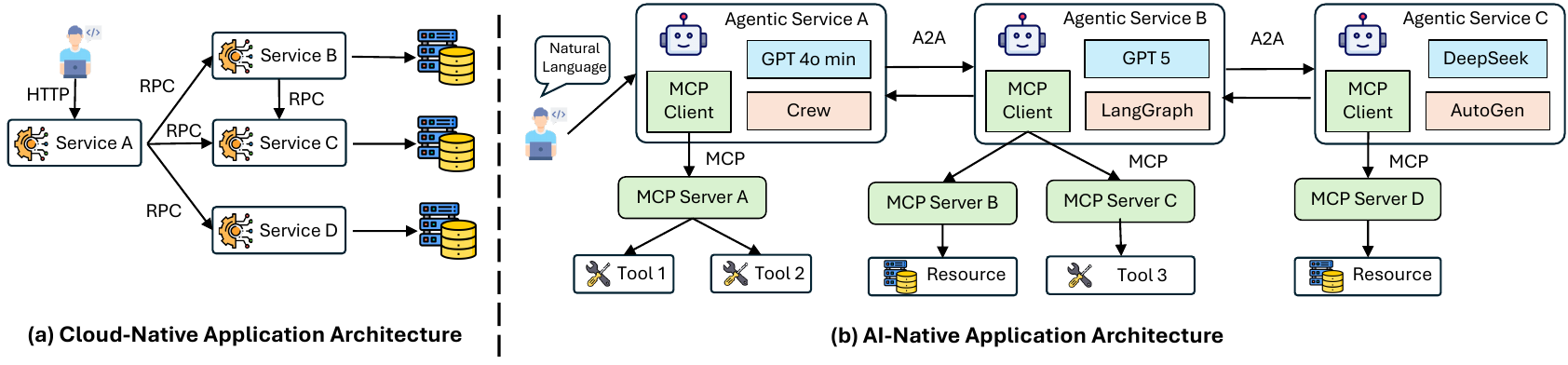}
	\vspace{-0.1in}
	\caption{Comparison of Cloud-Native and AI-Native application architecture.}
 	\label{fig:arch}
 	\vspace{-0.1in}
\end{figure}

\textbf{Cloud-Native System.} Cloud-Native is the established paradigm for modern distributed systems, defined by the Cloud Native Computing Foundation (CNCF)~\cite{CNCF} as the practice of building and running scalable and resilient applications. As illustrated in Figure~\ref{fig:arch}(a), the fundamental unit of a Cloud-Native architecture is the microservice. In this architecture, monolithic applications are decomposed into a set of small loosely-coupled services. These services are driven by deterministic, human-coded logic and communicate through established protocols like RPC or REST APIs, orchestrated by platforms such as Kubernetes. The primary focus of this architecture is agility, scalability, and system resilience.

\textbf{AI-Native System.} AI-Native represents the subsequent architectural paradigm, redesigning applications around Large Language Models (LLMs) as core coordinators~\cite{AINative}. As shown in Figure~\ref{fig:arch}(b), the basic building block evolves from the ``microservice'' to the ``agentic service.'' These services are not driven by fixed business logic but by autonomous AI agents (e.g., built with frameworks like AutoGen~\cite{autogen}, Crew~\cite{crew}, or LangGraph~\cite{langgraph}) that possess planning, reasoning, and decision-making capabilities, often powered by diverse LLMs (e.g., GPT-4, DeepSeek)~\cite{AgenticService}.

This evolution from ``coded service'' to ``autonomous agents'' necessitates a new standardized communication infrastructure; ad-hoc point-to-point integrations are unscalable and lack interoperability. Consequently, two core protocols are emerging as standards for building interoperable multi-agent systems:
\begin{itemize}[leftmargin=*,label=\textbullet]
    \item Model Context Protocol (MCP): This protocol standardizes the communication between an agentic service and its callable tools (e.g. APIs, databases, or functions)~\cite{mcp}, as shown by the interactions between the MCP Client and MCP Server in Fig.~\ref{fig:arch}(b).
    \item Agent-to-Agent (A2A) Protocol: This protocol standardizes the communication between agentic services~\cite{a2a}, enabling them to discover capabilities, negotiate, coordinate, and delegate tasks (e.g., the interactions between Agentic Service A and B in Fig.~\ref{fig:arch}(b)).
\end{itemize}

Therefore, an AI-Native application is a novel class of distributed system where end-to-end functionality is an emergent property of these dynamic, standardized interactions. This architectural shift introduces new, complex failure modes and performance bottlenecks at the protocol level (both A2A and MCP), which cannot be diagnosed by existing black-box or traditional microservice-oriented benchmarks.

\subsection{AI-Native Application Observability}\label{sec:obs}
The primary challenge in diagnosing AI-Native systems is their dual-failure nature. Unlike Cloud-Native applications, whose failures are primarily deterministic and technical (e.g., HTTP 500 errors), AI-Native applications are far more complex. They inherit all the deterministic failure modes of their underlying distributed infrastructure (e.g., protocol timeouts) while simultaneously introducing a new class of stochastic failures rooted in AI decision-making (e.g., incorrect tool-use, faulty planning, or hallucinated outputs) that are invisible to traditional APM tools~\cite{ChaosSurvey,zhu2025llmagentsfaillearn}. This behavioral blindness means that a successful observability strategy must be able to diagnose both types of failure, often co-located within the same component. Currently, three primary observability paradigms (i.e., metrics, trajectories, and traces) are used to try to address this challenge.

\begin{figure}[t]
	\centering
	\includegraphics[width=0.95\textwidth]{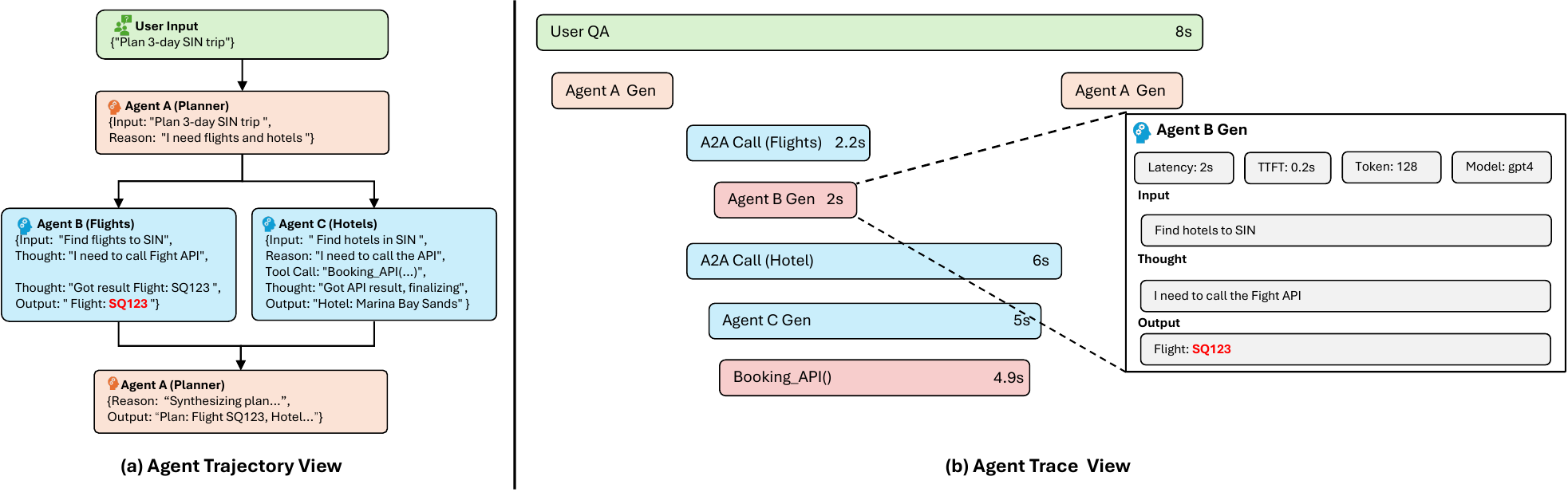}
	\vspace{-0.1in}
	\caption{Comparison betweem agent trajectory and trace view.}
 	\label{fig:obs-comparison}
 	\vspace{-0.1in}
\end{figure}

\textbf{Metrics.} This foundational layer provides a quantitative, aggregated view. It expands traditional Cloud-Native metrics (e.g., P99 latency)~\cite{TS-InvarNet} to include novel performance metrics (e.g., Time to First Token, Time Per Output Toekn) and AI semantic metrics (e.g., token costs and task success rates). However, while metrics reveal what happened (e.g., task success rates degrade), they are insufficient to explain why.

\textbf{Trajectory (Model Capability).} This ``gray-box'' paradigm, used by AI agent benchmarks~\cite{AgentBench,he2025trajectbench}, is designed to address the stochastic/decision-making part of the problem. As Fig.~\ref{fig:obs-comparison}(a) visualizes, this logical flowchart excels at capturing high-level semantic data (\texttt{Input}, \texttt{Thought}, \texttt{Output}). This view is effective at revealing AI decision-making (control-flow) failures. For example, a reviewer can clearly contrast \texttt{Agent B} and \texttt{Agent C}. While \texttt{Agent C} correctly executes its plan (\texttt{<code>Thought</code> → <code>Tool Call</code> → <code>Output</code>}), \texttt{Agent B} exhibits a critical decision failure: its \texttt{Thought} correctly identifies the need to call the API, but it fails to execute the \texttt{Tool Call} step, instead fabricating a hallucinated \texttt{Output: "Flight: SQ123"}. However, while the trajectory does capture this semantic failure, it is \textbf{performance-blind} and thus fails to capture performance issues. The view does not represent execution time, thus completely hiding the fact that \texttt{Agent C}'s \texttt{Booking\_API()} call is the system's primary performance bottleneck.

\textbf{Cloud-Native Trace (System Performance).} This ``white-box'' standard from Cloud-Native systems~\cite{Opentelemetry} is designed to address the deterministic part of the problem. The left part of Fig.~\ref{fig:obs-comparison}(b) shows this Gantt-like view. This pure trace excels at pinpointing technical bottlenecks. For example, it immediately reveals that the \texttt{A2A Call (Hotel)} (6s) and its child Span, \texttt{Hotel\_API()} (4.9s), are the dominant performance bottleneck, far slower than the \texttt{A2A Call (Flights)} (2.2s). However, this purely trace is semantically-blind. It has no native understanding of an AI's decision-making process. Therefore, it cannot tell that \texttt{Agent B} skipped its Tool Call step and its \texttt{Output} is a fabrication.

\textbf{AI Native Trace (System Capability and Performance).} Given that neither the trajectory (for decision failures) nor the pure trace (for performance failures) is sufficient alone, the necessary observability model is a unified, white-box trace~\cite{langfuse}. Fig.~\ref{fig:obs-comparison}(b) as a whole illustrates this fused model. This view enriches the technical trace (left part) by fusing it with deep semantic context (as conceptualized by the pop-out box on the right). This unified paradigm is the only one that can simultaneously reveal both distinct failures in their respective components: 
\begin{enumerate}[leftmargin=*,label=\textbullet]
\item \textbf{The Deterministic Failure:} The Gantt chart (left part) in Fig.~\ref{fig:obs-comparison}(b) immediately identifies the performance bottleneck: the \texttt{A2A Call (Hotel)} (6s) at \texttt{Agent C}, rooted in the \texttt{Hotel\_API()} Span (4.9s). This could be due to a network issue or abnormal tool performance.
\item \textbf{The Stochastic/Decision Failure:} A full unified trace would provide a semantic pop-out for every agent. An operator could then inspect \texttt{Agent B}'s trace, finding the ``white-box'' context for its failure: the skipped \texttt{Tool Call} step (which is absent in its trace) and the fabricated \texttt{Output: "Flight: SQ123"}. 
\end{enumerate} 
Therefore, we conclude that to properly evaluate AI-Native systems, a benchmark should adopt this fused trace to simultaneously measure performance overheads and semantic behavior.

\subsection{Related Work}
\textbf{AI Task Benchmark.}
The evaluation of LLM-driven AI agents has recently emerged as a central research area~\cite{KDDSurvey}. Predominant benchmark efforts, however, have concentrated on ``black-box'' evaluation, primarily assessing the upper bounds of agent capabilities. This paradigm broadly encompasses static, single-turn capability assessments like HELM~\cite{HELM}, as well as dynamic, interactive task evaluations. The latter category has seen significant milestones, including: (1) Interactive Environments such as Mind2Web~\cite{Mind2Web} , MineDojo~\cite{MINEDOJO}, ALFWorld~\cite{ALFWorld2021ICLR}, and WebShop~\cite{WebShop}; (2) Tool Use and Reasoning Benchmarks, spurred by works like Toolformer~\cite{Toolformer} and standardized by suites like ToolBench~\cite{ToolLLM}; and (3) Aggregate and Specialized Benchmarks that either combine multiple tasks (e.g., AgentBench~\cite{AgentBench}) or focus on specific domains like software engineering (SWE-bench~\cite{SWE-bench}) and machine learning (MLAgentBench~\cite{MLAgentBench}).

Currently, protocols such as A2A and MCP are recognized as essential to build scalable, interoperable AI-Native applications. However, research into evaluating these protocols remains nascent. Critically, emerging benchmarks such as MCP-Bench~\cite{MCP-Bench} and MCP-AgentBench~\cite{MCP-AgentBench} focus exclusively on using MCP as a standardized tool-calling interface for a single agent (i.e., Agent-to-Server or Agent-to-Tool). They do not address the multi-agent, agent-to-agent (A2A) communication vital for complex AI-Native applications.

The fundamental limitation of the existing work lies in its inherently task-centric outcome-based evaluation paradigm. By focusing on an agent's ability to complete a discrete task, these benchmarks typically assess validity by examining only the final output (e.g., task success) or externally observable action sequences (e.g., UI clicks). While valuable for gauging reasoning and planning capabilities, this black-box, task-centric approach is insufficient for diagnosing performance bottlenecks or failure modes within end-to-end AI-Native applications. This highlights a critical need for ``white-box'' trace-based methodologies that provide the necessary observability for system-level performance debugging and reliability analysis.

\begin{table*}[t!]
  \centering
  \footnotesize 
  \caption{
    Comparison of agentic benchmarks. 
  }
  \vspace{-0.1in}
  \label{tab:comparison}
  \resizebox{0.95\textwidth}{!}{
  \begin{tabular}{l c c c c c c} 
    \toprule
    \textbf{Benchmark} & \textbf{Focus} & \textbf{Paradigm} & \textbf{Goal} & \textbf{\shortstack{Multi- \\ Agent}} & \textbf{\shortstack{A2A \\ Support}} & \textbf{\shortstack{MCP \\ Support}} \\
    \midrule
    
    \rowcolor{MyHighlight}
    \multicolumn{7}{l}{\textit{Existent AI Task Benchmark}} \\
    \addlinespace[0.3em] 
    HELM~\cite{HELM}                  & Task-centric & Black-box (Outcome)    & Capability & \xmark & \xmark & \xmark \\
    WebShop~\cite{WebShop}            & Task-centric & Gray-box (Trajectory) & Capability & \xmark & \xmark & \xmark \\
    GAIA~\cite{GAIA}                  & Task-centric & Gray-box (Trajectory)    & Capability & \xmark & \xmark & \xmark \\
    AgentBench~\cite{AgentBench}      & Task-centric & Black-box (Outcome) & Capability & \xmark & \xmark   & \xmark \\
    MultiAgentBench~\cite{MultiAgentBench} & Task-centric & Gray-box (Trajectory) & Capability & \cmark &  \xmark   & \xmark \\
    MCP-Bench~\cite{MCP-Bench}        & Task-centric & Gray-box (Trajectory) & Capability & \xmark & \xmark & \cmark \\
    MCP-AgentBench~\cite{MCP-AgentBench}    & Task-centric & Black-box (Trajectory) & Capability & \xmark & \xmark & \cmark \\
    \addlinespace[0.5em]

    \rowcolor{MyHighlight}
    \multicolumn{7}{l}{\textit{Our AI-Native Application  Benchmark}} \\
    \addlinespace[0.3em]
    \textbf{AI-NativeBench}     & \textbf{App-centric} & \textbf{White-box (Trace)} & \makecell{\textbf{Performance \&} \\ \textbf{Reliability}} & \cmark & \cmark & \cmark \\
    \bottomrule
  \end{tabular}}
  \vspace{-0.2in}
\end{table*}

Consequently, as Table~\ref{tab:comparison} clearly illustrates, there is a significant gap in the current research landscape. There is a lack of a benchmark that (1) is grounded in a realistic AI-Native application architecture, such as those built with industry-adopted frameworks like AutoGen~\cite{autogen}, CrewAI~\cite{crew} and LangGraph~\cite{langgraph}, (2) explicitly incorporates emerging A2A and MCP protocols, and (3) facilitates quantitative white-box analysis of how system components and protocols impact end-to-end performance and reliability. Our research directly addresses this gap. We propose AI-NativeBench to shift the evaluation paradigm from assessing task-centric capability (``What can it do?'') to enabling application-centric diagnostics (``Why did it fail and why is it slow?'').

\textbf{Cloud-Native Benchmark.} The field of cloud-native applications, built on microservice architectures, has a mature body of research on performance benchmarking~\cite{Microscaler,MicroRank,Nezha,Deathstar}. To analyze these complex distributed systems, the SE community converged on white-box observability as the standard methodology. This approach relies on distributed traces (e.g., via Opentelemetry~\cite{Opentelemetry}) for system-level performance debugging and failure diagnosis~\cite{TraStrainer,Microsketch}.

A key insight of our work is that AI-Native applications, with their interacting agents and protocols, represent a new class of complex distributed systems. We argue that the mature ``white-box'' principles from cloud-native engineering are essential for this nascent domain. Our proposed trace-based methodology thus represents a critical adaptation and extension of these proven SE principles, tailored to the unique diagnostic challenges of multi-agent AI applications.

\section{AI-Native System Benchmark Suite}\label{sec:bench}

To bridge the growing gap between AI capability assessment and software engineering diagnostics, we introduce \benchx, an open-source, white-box agentic benchmark suite explicitly designed for AI-Native systems. To the best of our knowledge, \benchx is the first benchmark that adopts an \textbf{application-centric} perspective, grounding evaluation in realistic end-to-end multi-agent applications rather than isolated AI tasks. In contrast to prior benchmarks that treat agentic systems as black boxes and report only task-level success rates, \benchx emphasizes diagnosability by exposing internal execution behavior across diverse system configurations.

By natively integrating distributed tracing with modern agentic protocols (MCP and A2A), \benchx enables evaluation beyond \emph{what} an agent achieves (capability) toward a systematic analysis of \emph{how} the system behaves (performance and reliability) during end-to-end execution. This trace-first design allows researchers to attribute failures and overheads to specific architectural choices, communication protocols, and agent interactions.

As summarized in Table~\ref{tab:benchmark-suite}, \benchx comprises eight distinct AI-Native applications spanning three domains: Communication \& Collaboration, Software \& Data Engineering, and Content Generation. The suite intentionally covers a broad spectrum of scale and complexity, ranging from lightweight utilities (e.g., \texttt{Markdown Validator}, $\sim$1.2k LoC) to large engineering-intensive systems (e.g., \texttt{Landing Page Generator}, $\sim$5.1k LoC). Across these applications, systems orchestrate between one and five autonomous agents and interact with up to nine distinct tools.

Crucially, these applications are instantiated under multiple architectural variants, enabling controlled comparisons across increasing levels of system heterogeneity, ranging from monolithic single-framework deployments to distributed protocol-driven compositions. Several benchmarks are further augmented with Ground Truth (GT), enabling objective correctness assessment alongside performance and reliability diagnostics. The following subsections detail the workflow, architecture, and evaluation characteristics of each application.

Collectively, the applications in \benchx enable systematic investigation of (i) model selection under fixed system architectures, (ii) architectural and protocol trade-offs under controlled workloads, and (iii) cost--performance trade-offs in realistic AI-Native deployments.

\definecolor{GreenCheck}{HTML}{009B50}
\definecolor{RedCross}{HTML}{D1191C}
\definecolor{MyHighlight}{gray}{0.92}
\begin{table*}[t!]
  \centering
  \caption{Overview of the \benchx suite.}
  \vspace{-0.1in}
  \label{tab:benchmark-suite}
  \resizebox{0.9\textwidth}{!}{
  \begin{tabular}{l c c c c c c c c}
    \toprule
    & \multicolumn{4}{c}{\textbf{Scale}} 
    & \multicolumn{4}{c}{\textbf{Architectural Versions Tested}} \\
    \cmidrule(lr){2-5} \cmidrule(lr){6-9}
    \textbf{Application} 
      & \textbf{\# Agents} 
      & \textbf{\# Tools} 
      & \textbf{\# Resources} 
      & \textbf{LoC} 
      & \textbf{\shortstack{Pure CrewAI \\ Framework}} 
      & \textbf{+ MCP} 
      & \textbf{+ A2A} 
      & \textbf{\shortstack{+ H-A2A}} \\
    \midrule
    
    \rowcolor{MyHighlight}
    \multicolumn{9}{l}{\textit{Communication \& Collaboration}} \\
    \addlinespace[0.3em]
    \texttt{Email Responder} 
      & 2  
      & 3  
      & 1   
      & 1543   
      & \cmark & \cmark & \xmark & \xmark \\
    \texttt{Recruitment Assistant} 
      & 3 
      & 5 
      & 2   
      & 4319   
      & \xmark & \cmark & \cmark & \cmark \\
    \addlinespace[0.5em]

    \rowcolor{MyHighlight}
    \multicolumn{9}{l}{\textit{Software \& Data Engineering}} \\
    \addlinespace[0.3em]
    \texttt{Markdown Validator} (w/ GT)     
      & 1 
      & 1  
      & 1   
      & 1193   
      & \cmark & \cmark & \xmark & \xmark \\
    \texttt{Game Builder}            
      & 3 
      & 1 
      & 1   
      & 1314   
      & \cmark & \cmark & \xmark & \xmark \\
    \texttt{SQL Assistant} (w/ GT)    
      & 4 
      & 8 
      & 1   
      & 3969   
      & \xmark & \cmark & \cmark & \cmark \\
    \texttt{Landing Page Generator} (w/ GT)   
      & 3 
      & 4 
      & 3   
      & 5129   
      & \xmark & \cmark & \cmark & \cmark \\
    \addlinespace[0.5em]
    
    \rowcolor{MyHighlight}
    \multicolumn{9}{l}{\textit{Content Generation}} \\
    \addlinespace[0.3em]
    \texttt{Book Writer} 
      & 5 
      & 9 
      & 2   
      & 5068   
      & \xmark & \cmark & \cmark & \cmark \\
    \texttt{Social Media Manager}    
      & 3 
      & 7 
      & 0   
      & 3868   
      & \xmark & \cmark & \cmark & \cmark \\
    \bottomrule
  \end{tabular}
  } 
  \vspace{-0.2in}
\end{table*}

\subsection{Design Principles} 
\benchx is guided by three core design principles that collectively support holistic evaluation of modern AI-Native systems.


\textbf{Representativeness.}
The benchmark suite is designed to reflect the reality of AI-Native applications currently deployed in production rather than theoretical complexity. While we utilize widely adopted open-source frameworks (e.g., AutoGen~\cite{autogen}, CrewAI~\cite{crew}, LangGraph~\cite{langgraph}) to cover diverse domains like Software Engineering and Content Generation, we strictly adhere to industrial fidelity regarding topology scale. Recent empirical studies~\cite{pan2025measuringagentsproduction} on production-grade agents reveal a conservative landscape: 68\% of deployed systems restrict execution trajectories to fewer than 10 steps, with only 6.7\% permitting unbounded execution. Furthermore, research on agent scaling suggests that indiscriminately increasing agent count often degrades system intelligence due to coordination overhead~\cite{kim2025sciencescalingagentsystems}. Consequently, \benchx avoids artificial complexity. Our topologies range from single-agent tasks to focused multi-agent collaborations, capped at the complexity of our \texttt{Book Writer} application (5 agents, 9 tools). This design choice mirrors the practical constraints and bounded autonomy observed in high-value industrial systems.

\textbf{Trace-First (White-Box) Design.}
Trace-first design constitutes the core philosophical foundation of \benchx. We argue that black-box, capability-only evaluation is insufficient for understanding the behavior of complex agentic systems. Accordingly, all agent services and protocol interactions (MCP and A2A) are instrumented by default using OpenTelemetry. This instrumentation captures and unifies two complementary observability streams introduced in \S~\ref{sec:obs}: the semantic trajectory (e.g., \texttt{Input}, \texttt{Thought}, \texttt{Output}) and the technical execution trace (e.g., latency, errors). The resulting \emph{Unified Trace} enables systematic diagnosis of both deterministic failures (e.g., performance regressions) and stochastic failures arising from AI decision-making.

\textbf{Architectural Heterogeneity \& Modularity.}
Rather than a single monolithic system, \benchx is a configurable suite of applications with modular and reconfigurable architectures. This design mirrors the modularity and heterogeneity principles of microservice benchmarks. As detailed in Table~\ref{tab:benchmark-suite}, researchers can vary not only agents and models, but also communication protocols and underlying agent frameworks. This flexibility enables controlled empirical studies addressing key software engineering questions, such as the performance overhead of migrating from native framework calls to MCP-based tool invocation, or the reliability and latency trade-offs introduced by heterogeneous, multi-framework A2A communication.
Table~\ref{tab:benchmark-suite} summarizes the supported architectural variants for each application:
\begin{itemize}[leftmargin=*,label=\textbullet]
\item \textbf{Pure CrewAI Framework}: agents and tools are implemented  within a CrewAI framework.
\item \textbf{+ MCP}: tools are refactored into independent services and invoked via the standardized MCP.
\item \textbf{+ A2A}: inter-agent communication is standardized using the Agent-to-Agent (A2A) protocol.
\item \textbf{+ Heterogeneous A2A (H-A2A)}: applications are composed of agents built using multiple frameworks (e.g., AutoGen and CrewAI) communicating via A2A.
\end{itemize}

\newlength{\colheight}
\setlength{\colheight}{0.3\textheight} 
\setlength{\abovecaptionskip}{4pt}
\setlength{\belowcaptionskip}{0pt}
\begin{figure}[t]
  \centering

  \begin{minipage}[c]{0.35\linewidth}
    \centering

    \includegraphics[width=\linewidth,
                     height=0.6\colheight,
                     keepaspectratio]{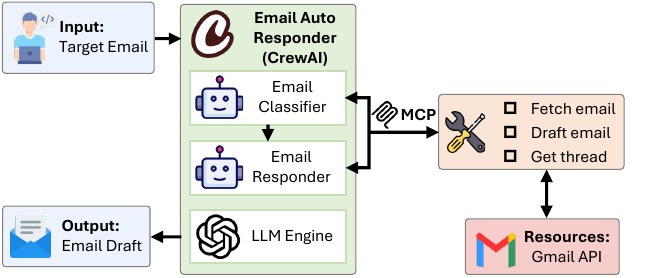}
    \caption{Email Responder(MCP).}
    \label{fig:email}

    \vspace{4mm}

    \includegraphics[width=\linewidth,
                     height=0.4\colheight,
                     keepaspectratio]{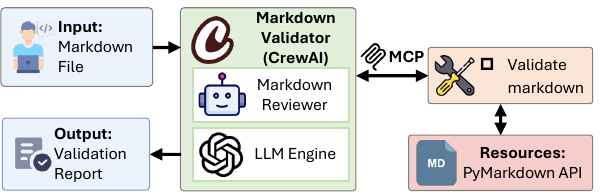}
    \caption{Markdown Validator(MCP).}
    \label{fig:markdown}
  \end{minipage}
  \hfill
  \begin{minipage}[c]{0.64\linewidth}
    \centering
    \includegraphics[width=\linewidth,
                     height=\colheight,
                     keepaspectratio]{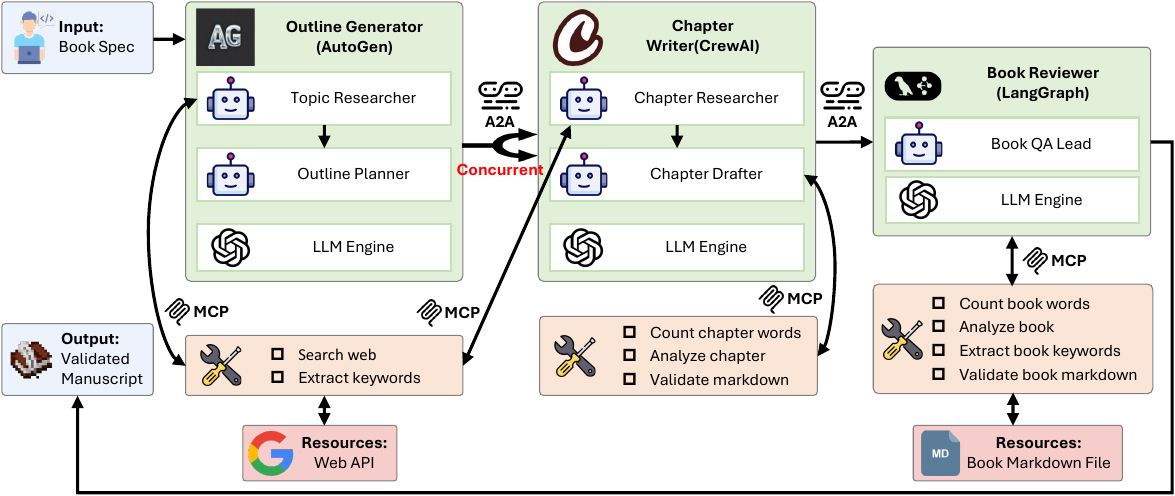}
    \caption{Book Writer(H-A2A).}
    \label{fig:book}
  \end{minipage}
  \vspace{-0.1in}
\end{figure}

\subsection{Communication \& Collaboration Task} 
Applications in this category model human-centric business workflows characterized by unstructured inputs and social context. Beyond task completion, these benchmarks evaluates an agentic system’s ability to coordinate roles, manage external resources, and maintain reliability under distributed execution. We include two representative applications that progress from standardized single-framework deployments to highly heterogeneous, multi-framework systems.

\subsubsection{Email Responder} 
\texttt{Email Responder} application simulates a two-agent workflow (Analyze $\rightarrow$ Respond) for intelligent email processing. As shown in Table~\ref{tab:benchmark-suite}, the application supports both a \textit{Pure CrewAI} variant and a \textit{+ MCP} variant, enabling direct comparison between monolithic and service-oriented architectures. Figure~\ref{fig:email} illustrates the \textit{+ MCP} version, in which tool access is decoupled from agent logic. Instead of directly invoking the \texttt{Gmail API}, \texttt{Email Classifier} and \texttt{Email Responder} interact with external services via standardized MCP calls. This design allows precise measurement of protocol-induced overheads relative to native framework execution.

\subsubsection{Recruitment Assistant} 
\begin{figure*}[t]
    \centering
    \includegraphics[width=1\linewidth]{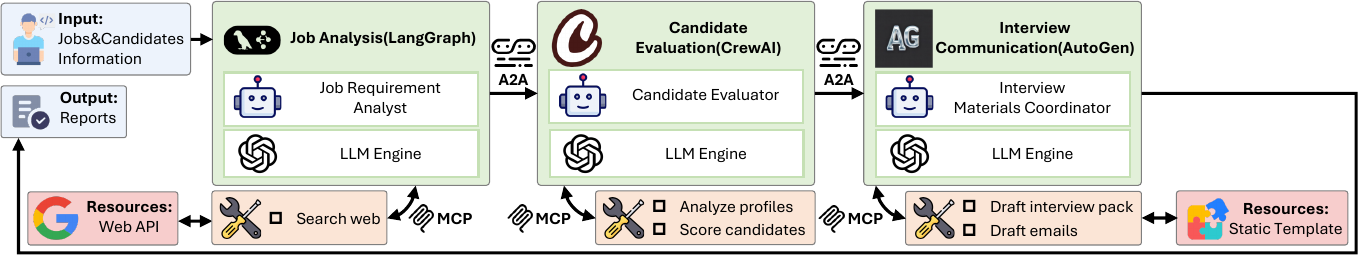}
    \caption{Recruitment Assistant(H-A2A).}
    \vspace{-0.2in}
    \label{fig:job}
\end{figure*}

\texttt{Recruitment Assistant} models a three-stage pipeline (Job Analysis $\rightarrow$ Candidate Evaluation $\rightarrow$ Interview Preparation) and focuses on distributed agent coordination. Supporting both \textit{+ A2A} and \textit{+ H-A2A} variant, this benchmark evaluates communication across heterogeneous agent frameworks. Figure~\ref{fig:job} depicts the most complex instantiation, where a LangGraph-based \texttt{Job Requirement Analyst}, a CrewAI-based \texttt{Candidate Evaluator}, and an AutoGen-based \texttt{Interview Materials Coordinator} interact via A2A while accessing external tools through MCP. This setting provides a rigorous testbed for analyzing end-to-end latency, failure propagation, and reliability in highly distributed agentic systems.

\subsection{Software \& Data Engineering}

Benchmarks in this domain automate technical workflows that demand high precision, strict schema adherence, and iterative self-correction. Unlike communication tasks, these applications emphasize correctness, reproducibility, and controlled interaction with execution environments. We include four benchmarks that progressively increase system complexity across code validation, software construction, and database interaction.

\subsubsection{Markdown Validator}
\texttt{Markdown Validator}, visualized in Fig.~\ref{fig:markdown}, represents the foundational \textit{+ MCP} architecture. Implemented as a single-agent system within CrewAI, it employs a \texttt{Markdown Reviewer} to perform automated syntax validation. The agent does not parse files directly. Instead, as shown in Fig.~\ref{fig:markdown}, it delegates the scanning logic to an independent MCP service wrapping the \texttt{PyMarkdown API} via the \texttt{Validate markdown} tool. This setup allows researchers to evaluate the performance impact of externalizing computational tasks (linting) from the agent's reasoning loop.

\subsubsection{Game Builder}
\texttt{Game Builder} simulates an iterative software development lifecycle through a three-agent team: a \texttt{Software Engineer}, a \texttt{QA Engineer}, and a \texttt{QA Lead}. As illustrated in Fig.~\ref{fig:game}, agents access a sandboxed local Python environment via MCP tools such as \texttt{Validate code}, enabling compilation and execution of generated artifacts. This Code $\rightarrow$ Review $\rightarrow$ Evaluate pipeline systematically exercises  multi-agent collaboration under strict execution constraints.

\subsubsection{Landing Page Generator}
\begin{figure*}
  \begin{minipage}[t]{0.35\textwidth}
    \centering
    \raisebox{\dimexpr-\height+\ht\strutbox}{\includegraphics[width=1\textwidth]{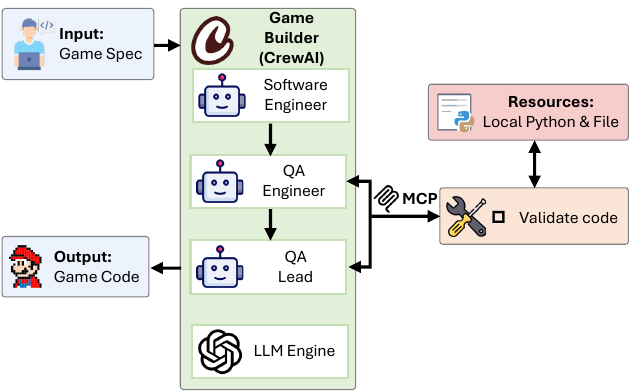}}
	\vspace{-0.08in}
	\caption{Game Builder(MCP).}
 	\label{fig:game}
  \end{minipage}
  \quad
  \begin{minipage}[t]{0.60\textwidth}
    \centering
    \raisebox{\dimexpr-\height+\ht\strutbox}{\includegraphics[width=1\textwidth]{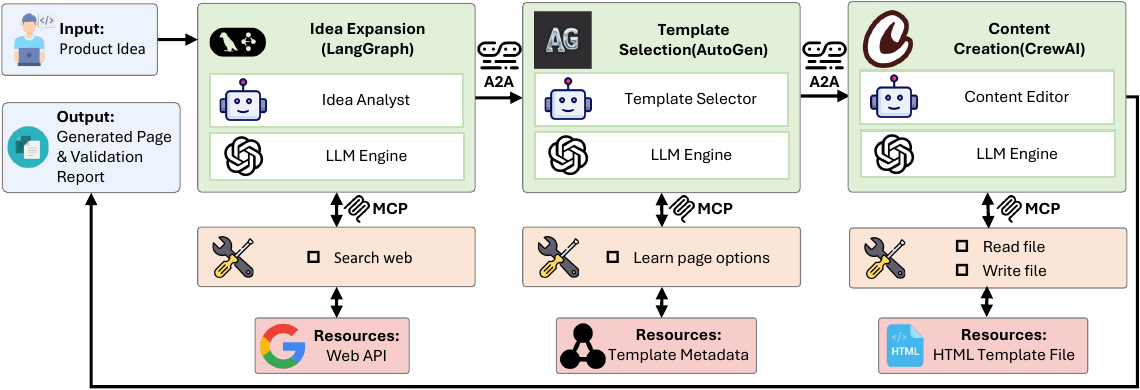}}
	\caption{Landing Page Generator(H-A2A).}
 	\label{fig:page}
  \end{minipage}
  \vspace{-0.1in}
\end{figure*}

\texttt{Landing Page Generator} introduces a creative workflow of designing landing pages. As illustrated in Fig.~\ref{fig:page}, this benchmark serves as a prime example of the \textit{+ H-A2A} configuration by orchestrating a sequential pipeline across three distinct platforms. A LangGraph-based \texttt{Idea Analyst} first expands the initial concept using \texttt{Web API} resources. It then passes the context via the A2A protocol to an AutoGen-based \texttt{Template Selector}, which chooses design patterns from \texttt{Template Metadata}. Finally, the task is handed off to a CrewAI-based \texttt{Content Editor} to synthesize the final HTML file. This setup demonstrates how specialized agents can be composed into a cohesive system using standardized protocols.

\setlength{\abovecaptionskip}{7pt}
\setlength{\belowcaptionskip}{0pt}

\subsubsection{SQL Assistant} 
\begin{figure*}[t]
    \centering
    \includegraphics[width=1\textwidth]{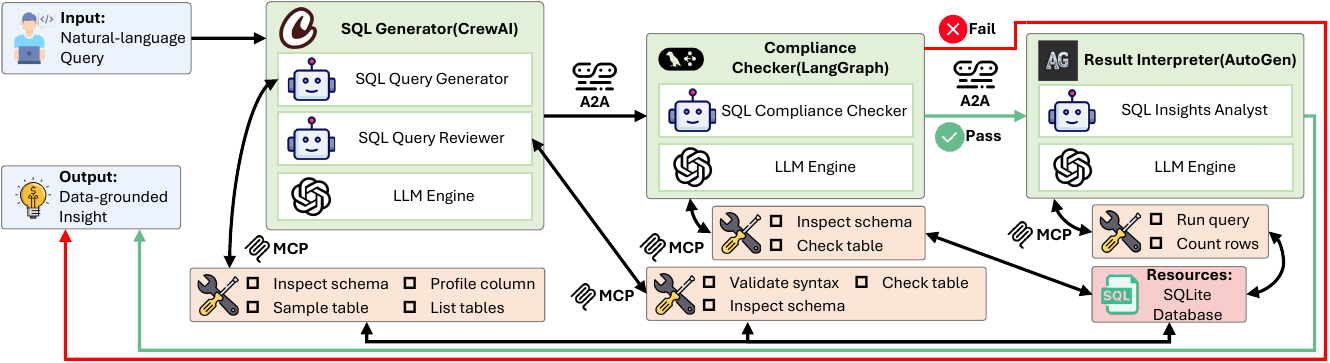}
    \caption{SQL Assistant(H-A2A).}
    \vspace{-0.2in}
    \label{fig:SQL}
\end{figure*}

\texttt{SQL Assistant} is the most sophisticated benchmark in this category, modeling an end-to-end Text-to-SQL workflow with built-in governance and retry mechanisms. As depicted in Fig.~\ref{fig:SQL}, the architecture coordinates a 4-agent system across three frameworks. A CrewAI generation crew (consisting of a \texttt{SQL Generator} and \texttt{Reviewer}) first drafts queries based on the schema. These queries are sent via A2A to a LangGraph-based \texttt{SQL Compliance Checker}, which acts as a gatekeeper. A critical feature of this benchmark is the feedback loop: if a query fails compliance checks (the ``Fail'' path), it is routed back to the CrewAI team for revision; only compliant queries (``Pass'') are forwarded via A2A to the AutoGen-based \texttt{SQL Insights Analyst} for execution against the SQLite database. This workflow tests the system's ability to handle complex control flows, recover from errors, and interact with data sources via MCP tools.

\subsection{Content Generation}
Applications in this category focus on the autonomous creation of unstructured content. Unlike engineering tasks that target a single correct solution, content generation requires agents to exhibit creativity, maintain stylistic consistency over long contexts, and perform iterative refinement. We selected two benchmarks that test these capabilities at different scales: from generating short, highly stylized social media posts to authoring entire books through hierarchical planning and parallel execution.

\subsubsection{Book Writer} \texttt{Book Writer} is the most computationally intensive benchmark in our suite, simulating a hierarchical, 5-agent editorial team designed to evaluate long-context coherency and parallel execution. As illustrated in Fig.~\ref{fig:book}, the workflow is orchestrated across three functional stages. First, a research crew (consisting of a \texttt{Topic Researcher} and \texttt{Outline Planner}) establishes the book's global structure. Next, the system enters a parallel execution phase where a dedicated crew (a \texttt{Chapter Researcher} and \texttt{Chapter Drafter}) processes multiple chapters concurrently using a map-reduce style pattern. Finally, a \texttt{Book QA Lead} consolidates the manuscript for quality assurance. These agents leverage MCP tools to interact with a shared \texttt{Book Markdown File}, rigorously testing the architectural limits of agent orchestration and concurrency.

\subsubsection{Social Media Manager} 
\begin{figure*}[t]
    \centering
    \includegraphics[width=1\textwidth]{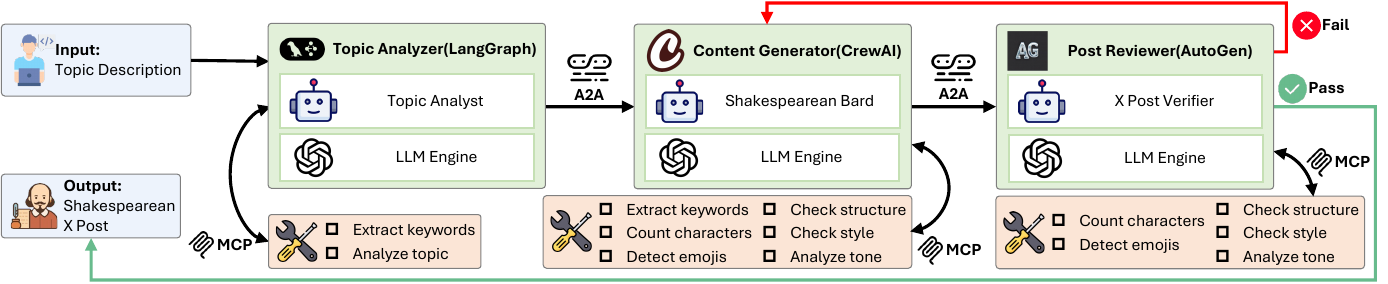}
    \caption{Social Media Manager(H-A2A).}
    \label{fig:post}
    \vspace{-0.2in}
\end{figure*}

\texttt{Social Media Manager}, visualized in Fig.~\ref{fig:post}, serves as a compact yet rigorous testbed for iterative self-correction within a \textit{+ H-A2A} architecture. The workflow orchestrates a feedback-driven loop between three specialized agents: a \texttt{Topic Analyst} first extracts key insights from raw input; a \texttt{Shakespearean Bard} then generates the initial creative content; and finally, an \texttt{X Post Verifier} acts as a critic to assess style and structure. Crucially, this benchmark implements a conditional retry mechanism: if the \texttt{Verifier} rejects a post (e.g., due to incorrect tone), the feedback is routed back to the \texttt{Bard} for regeneration. This Analyze → Generate → Review → Retry cycle evaluates the system's ability to interpret qualitative feedback and converge on a desired output format across heterogeneous agent frameworks.

\section{Evaluation Methodology}\label{sec:method}

To strictly assess the engineering characteristics of AI-Native systems, \benchx departs from traditional capability-based benchmarks. We adopt an application-centric methodology that treats the agent not merely as a model inference endpoint but as a complex distributed system. Crucially, we prioritize white-box diagnostics over black-box scoring. Although black-box metrics like LLM-as-a-judge scores~\cite{gu2025surveyllmasajudge} reveal what the outcome was, they often fail to explain why a system performed that way. Using our unified OpenTelemetry instrumentation, we inspect the internal execution states to provide a granular diagnosis of the system's behavior.

Guided by this philosophy, we pose three core research questions (RQs) to drive our analysis:
\begin{itemize}[leftmargin=*,label=\textbullet]
    \item \textbf{RQ1 (Dissecting Behavioral Correctness):}  \textit{How do varying model capabilities (e.g., parameter scale, reasoning mode) impact the behavioral correctness of agentic workflows?} We dissect whether escalating model intelligence translates to stricter workflow adherence or paradoxically induces an outcome-process divergence where agents bypass protocols to achieve correct results.
    \item \textbf{RQ2 (Dissecting Performance Overhead):}  \textit{What is the anatomy of latency in distributed AI-Native architectures?} We investigate whether the primary performance bottleneck stems from the infrastructure overhead of modular protocols, the inference dominance of heavy reasoning models, or the stochastic interaction dynamics between heterogeneous agents.
    \item \textbf{RQ3 (Dissecting Token Usage):}  \textit{What is the true economic cost of autonomy in AI-Native systems?} We dissect the token tax driven by architectural redundancy and self-healing loops to determine whether agents adhere to cost-efficient fail-fast principles or exhibit expensive failure patterns where resources are exhausted on doomed trajectories.
\end{itemize}

\subsection{RQ1: Behavioral Correctness}

\begin{figure*}[t]
  \centering
  \includegraphics[width=\textwidth]{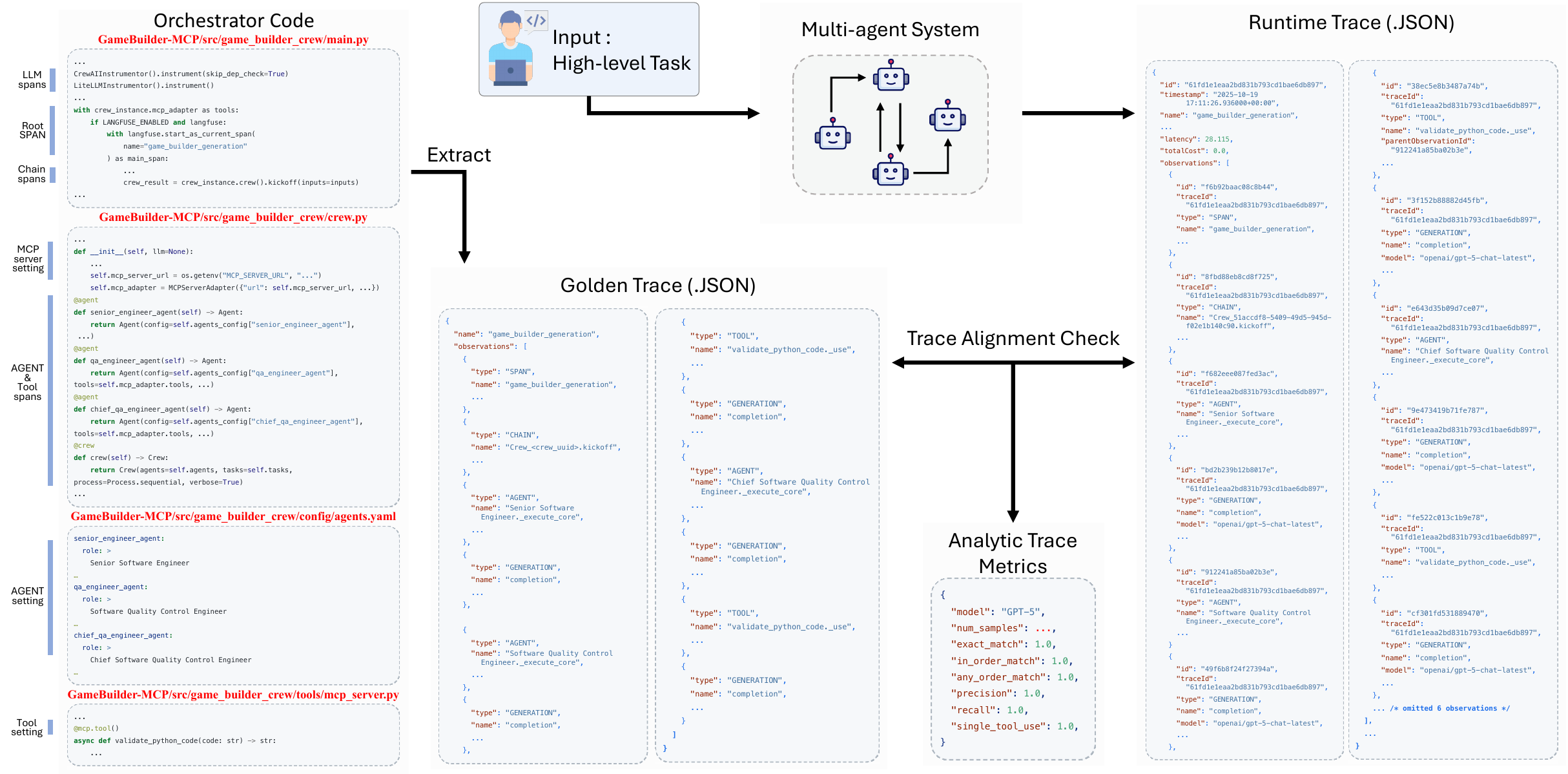}
  
  \caption{The trace-first evaluation pipeline of AI-NativeBench.}
  \label{fig:eval_workflow}
\end{figure*}

To answer RQ1, we move beyond simple binary success metrics to perform a multi-dimensional assessment of execution quality. To operationalize this, we implemented the trace-first evaluation pipeline illustrated in Figure \ref{fig:eval_workflow}, which captures granular execution data to evaluate agents across three distinct layers.

\textbf{1. Outcome.} We adopt a hybrid evaluation strategy for different applications.

\begin{itemize}[leftmargin=*,label=\textbullet]
    \item \textbf{Outcome Score (Deterministic Applications):} For applications with verifiable ground truth (e.g., \texttt{SQL Assistant}, \texttt{Game Builder}), we evaluate the functional correctness of the final output. The score measures the percentage of verified assertions (e.g., correct code blocks, database entries) satisfied by the agent's deliverable:
    \begin{equation}
        Score = 
        \begin{cases} 
        0 & \text{if task interrupted or assertion failed} \\
        \frac{N_{verified}}{N_{total}} \times 100 & \text{if task completed}
        \end{cases}.
    \end{equation}
    \textit{Metric Rationale:} This metric penalizes fragility (giving 0 for crashes) while differentiating between agents that achieve the core logic versus those that handle edge cases perfectly.

    \item \textbf{Pass Rate (Open-Ended Applications):}
    For applications without a single canonical answer (e.g., \texttt{Social Media Manager}), we utilize a binary metric focused on operational survivability. A run is considered a ``Pass'' if and only if the agent autonomously traverses the entire workflow and produces a final deliverable that satisfies basic structural constraints (e.g., non-empty output, correct file format) without runtime exceptions. 
    \begin{equation}
        R_{pass} = \frac{N_{completed\_runs}}{N_{total\_runs}}.
    \end{equation}
    \textit{Metric Rationale:} We explicitly prioritize engineering robustness over subjective content quality. While metrics like LLM-as-a-judge evaluate the \textit{intrinsic capability} of the underlying model (e.g., creativity, tone), they act as confounding variables when assessing architectural reliability. In the context of Software Engineering, the primary challenge for autonomous agents is execution stability, defined as the ability to complete multi-step protocols without diverging or crashing. Thus, pass rate serves as the definitive measure of architectural validity.
\end{itemize}

\textbf{2. Smoothness.} We assess the friction of execution by monitoring the agent's reliance on self-healing mechanisms.
\begin{itemize}[leftmargin=*,label=\textbullet]
    \item \textbf{Retry Rate:}  Defined as the percentage of runs that triggered the self-correction mechanism (e.g., due to format violations, or logical checks) at least once during execution.
    \begin{equation}
        R_{retry} = \frac{N_{runs\_with\_retries}}{N_{total\_runs}}.
    \end{equation}
    \textit{Metric Rationale:} A high retry rate indicates that the agent struggles to produce correct actions in a single shot, leading to higher latency and token costs.
\end{itemize}

\textbf{3. Process Alignment.} To evaluate the consistency of the execution process, we compare the agent's actual execution trace ($T_{agent}$) with a reference golden trace ($T_{gold}$).

As illustrated in the ``Extract'' phase of Figure \ref{fig:eval_workflow}, $T_{gold}$ is generated via static analysis of the orchestration configuration (e.g., \texttt{agents.yaml}) and task requirements, rather than derived from stochastic runtime execution. This trace defines the reference path by specifying the required sequence of \texttt{AGENT} spans and \texttt{TOOL} signatures, and can be viewed as an ideal plan that is both complete and non-redundant: it covers the required tool usage to complete the task while avoiding unnecessary repeated tool calls or extra LLM-driven loops. This exclusion of non-deterministic content, such as internal reasoning, establishes a deterministic structural baseline for evaluation. When a subset of required tools has no mandatory ordering constraint, $T_{gold}$ is not unique; in such cases, different permutations are treated as equivalent and the ordering is instantiated to match the observed valid execution, since the relative order should not be a penalty source.

Based on this baseline, we calculate the following white-box metrics:

\begin{itemize}[leftmargin=*,label=\textbullet]
    \item \textbf{Exact Match vs. Any-Order Match:} 
    \textit{Exact Match} requires strict sequential identity ($T_{agent} \equiv T_{gold}$), testing for adherence to application logic. \textit{Any-Order Match} treats traces as sets ($Set(T_{gold}) \subseteq Set(T_{agent})$), testing for goal completeness regardless of sequence.
    \\
    \textit{Metric Rationale:} The gap between these two metrics reveals the "Alignment Gap" where agents achieve goals via non-standard or risky paths.

    \item \textbf{Precision \& Recall (Tool Usage):} 
We further compute \emph{precision} and \emph{recall} over tool calls by treating the golden and runtime trace as sets of relevant actions, measuring respectively the correctness and completeness of tool usage.
Let $S_{agent}$ and $S_{gold}$ be the sets of unique tool signatures (defined as the tuple of tool name and critical arguments) appearing in the agent's trace and the golden trace, respectively.
\begin{equation}
    Precision = \frac{|S_{agent} \cap S_{gold}|}{|S_{agent}|}, \quad
    Recall = \frac{|S_{agent} \cap S_{gold}|}{|S_{gold}|}.
\end{equation}
\textit{Metric Rationale:} Precision (Correctness): Penalizes \textit{hallucinations}. A low precision indicates the agent invoked unnecessary or incorrect tools (noise) that were not part of the standard solution ($|S_{agent}| > |S_{agent} \cap S_{gold}|$). Recall (Completeness): Penalizes \textit{omissions}. A low recall indicates the agent failed to invoke required tools needed to solve the task ($|S_{gold}| > |S_{agent} \cap S_{gold}|$).
\end{itemize}

\subsection{RQ2: Performance Overhead}

To answer RQ2, we adopt a system-centric perspective to perform a granular latency breakdown. We measure \emph{end-to-end latency} ($L_{E2E}$) as the time from task initiation to completion.

To pinpoint bottlenecks, we perform decomposition at three levels:

\textbf{1. Agent-Level Breakdown.}
We verify per-agent latency contributions from the distributed traces to identify which agents dominate the overall response time. By partitioning $L_{E2E}$ by the active agent role (e.g., \texttt{Software Engineer} vs. \texttt{QA Lead}), we can distinguish whether latency is driven by a computationally intensive agent or an inefficient coordination step.

\textbf{2. Infrastructure-Level Breakdown ($L_{non-LLM}$).}
To strictly separate the ``cost of intelligence'' from the ``cost of infrastructure,'' we isolate \emph{non-LLM latency}, defined as the portion of latency not attributable to token generation. We decompose this overhead into four distinct components:
\begin{equation}
    L_{non-LLM} = L_{server} + L_{A2A} + L_{framework} + L_{tool},
\end{equation}
where:
\begin{itemize}[label=\textbullet]
    \item \textbf{$L_{server}$:} Baseline runtime overhead (backend/hosting).
    \item \textbf{$L_{A2A}$:} Protocol-specific serialization and network delays.
    \item \textbf{$L_{framework}$:} Logic overhead within the agent library (e.g., CrewAI orchestration).
    \item \textbf{$L_{tool}$:} Execution time of external tools.
\end{itemize}

\textbf{3. Protocol Comparison Across Architectures.}
We compare three architectural settings (i.e., MCP, A2A, and H-A2A) under the same task and model. For each setting, we report the distribution of $L_{E2E}$. To quantify the impact of introducing a protocol, we compute paired latency differences whenever runs share the same input. We then attribute observed differences using per-agent latency deltas from the traces.

\subsection{RQ3: Token Economics}

To answer RQ3, we evaluate the economic efficiency using token usage metrics conditioned on operational outcomes. As LLMs operate under a strict pay-per-token model, evaluating the \emph{performance-per-cost} ratio is as critical as evaluating accuracy~\cite{chen2023frugalgpt}. We analyze token consumption through two specific economic lenses.
\begin{itemize}[leftmargin=*,label=\textbullet]
    \item \textbf{Cost of Reliability (Retry Penalty):} We compare the token distribution of \emph{Success (with retry)} against \emph{Success (no retry)}. This quantifies the retry penalty, defined as the additional resources the system must expend to self-heal from intermediate errors. 
    \item \textbf{Failure Efficiency:} We analyze the token footprint of failed runs to diagnose the system's resilience pattern. A well-architected system should exhibit ``fail-fast'' behavior~\cite{nygard2018release}, identifying unrecoverable errors early. In contrast, inefficient systems exhibit expensive failures, where agents exhaust the maximum context window or retry budget before eventually failing, leading to significant economic waste without value generation.
\end{itemize}

This analysis provides the empirical data needed to guide decisions on budget-aware circuit breakers, helping engineers balance the trade-off between aggressive error-recovery mechanisms and strict operational budgets.

\subsection{Evaluation Setup}\label{sec:setup}

\subsubsection{Testbed Environment} All experiments are conducted on the dedicated Huawei Cloud ECS instances. The instance is equipped with 16 vCPUs (Intel Xeon Gold 6161 @ 2.20 GHz) and 64 GiB of memory. The host operating system is Ubuntu 24.04.2 LTS running the Linux kernel 6.8.x. To strictly isolate the benchmark workload, we enforce a single-job policy, ensuring that no other co-located workloads interfere with the CPU or memory bandwidth during execution.

\subsubsection{Implementation \& Software Stack} 
All applications in \benchx are implemented in Python 3.10, sharing a unified and version-locked dependency stack to ensure consistency. We employ three widely-used agent frameworks, including CrewAI~\cite{crew}, LangGraph~\cite{langgraph}, and AutoGen~\cite{autogen}. Protocol-level interoperability is enforced via the MCP~\cite{mcp} for tool isolation and A2A SDK~\cite{a2a} for inter-agent communication. Consistent with our white-box design principle, we utilize a comprehensive observability stack comprising OpenLit~\cite{openlit}, Langfuse~\cite{langfuse}, and OpenInference~\cite{openinference}. This pipeline automatically instruments CrewAI, LangChain, and LiteLLM calls, exporting unified traces that capture both heavy-weight LLM generations and light-weight tool invocations.


\subsubsection{Models Evaluated} 

We evaluate \benchx with seven LLMs that jointly cover proprietary and open-source, general-purpose and reasoning-enhanced backends: GPT-5, GPT-4o-mini, DeepSeek-V3.1, DeepSeek-R1, Gemini-2.5-flash, Gemini-2.5-flash-nothinking, and Qwen3-235b. Each model is accessed through its provider's API and is treated as a black-box text generation service behind a unified interface. Unless otherwise specified, all experiments use the same decoding configuration with \texttt{TEMPERATURE=0.85} and \texttt{TOP\_P=0.90}. For each application, we use identical system prompts and task instructions across all models and orchestration architectures.


\subsubsection{Workload}

To ensure \benchx reflects real-world complexity, we strictly curate workloads from diverse authentic data sources. For applications marked as ``w/ GT'' in Table~\ref{tab:benchmark-suite}, ground-truth outputs are derived directly from source datasets to enable precise evaluation. Given the computational cost of multi-agent execution, we curate balanced datasets (typically $N=60$) to provide statistical representativeness within a feasible budget.

\textbf{Communication \& Collaboration.} Workloads in this domain mimic professional workflows. \texttt{Email Responder} utilizes email scenarios sampled and reformatted from the Phishing Email dataset~\cite{phishingEmailDataset}. \texttt{Recruitment Assistant} constructs recruitment cases using job-candidate pairs derived from the Resume Score Details dataset~\cite{resumeScoreDataset}.

\textbf{Software \& Data Engineering.} These workloads focus on technical adherence. \texttt{Markdown Validator} operates on complex documents recombined from the Markdown-it demo suite~\cite{markdownItDemo}. \texttt{Game Builder} uses natural language briefs synthesizing logic from 25 classic games (e.g., \emph{Tetris}). For data tasks, \texttt{SQL Assistant} samples NL-to-SQL pairs from the ZSQL dataset~\cite{zsqlDataset}, while \texttt{Landing Page Generator} employs synthetic product briefs modeled after real-world app descriptions from public corpora (e.g., Google Play)~\cite{hou2024bridging, googlePlayDataset}.

\textbf{Content Generation.} Tasks here demand creativity. \texttt{Book Writer} leverages narrative prompts from the Reasoning Engaging Story dataset~\cite{reasoningStoryDataset} to trigger hierarchical generation. \texttt{Social Media Manager} utilizes technical topics from the Tech Keywords dataset~\cite{techKeywordsDataset} to seed the iterative refinement of stylized content.


\section{Detailed Distributed Trace Analysis for \benchx}\label{sec:evaluation}



\subsection{RQ1: Dissecting Behavioral Correctness}\label{sec:rq1}

\begin{figure}[t!]
    \centering
    \includegraphics[width=\textwidth]{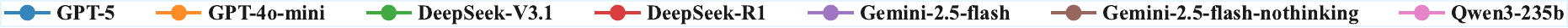}

    \begin{minipage}{0.45\linewidth}
        \centering
        \includegraphics[width=\linewidth]{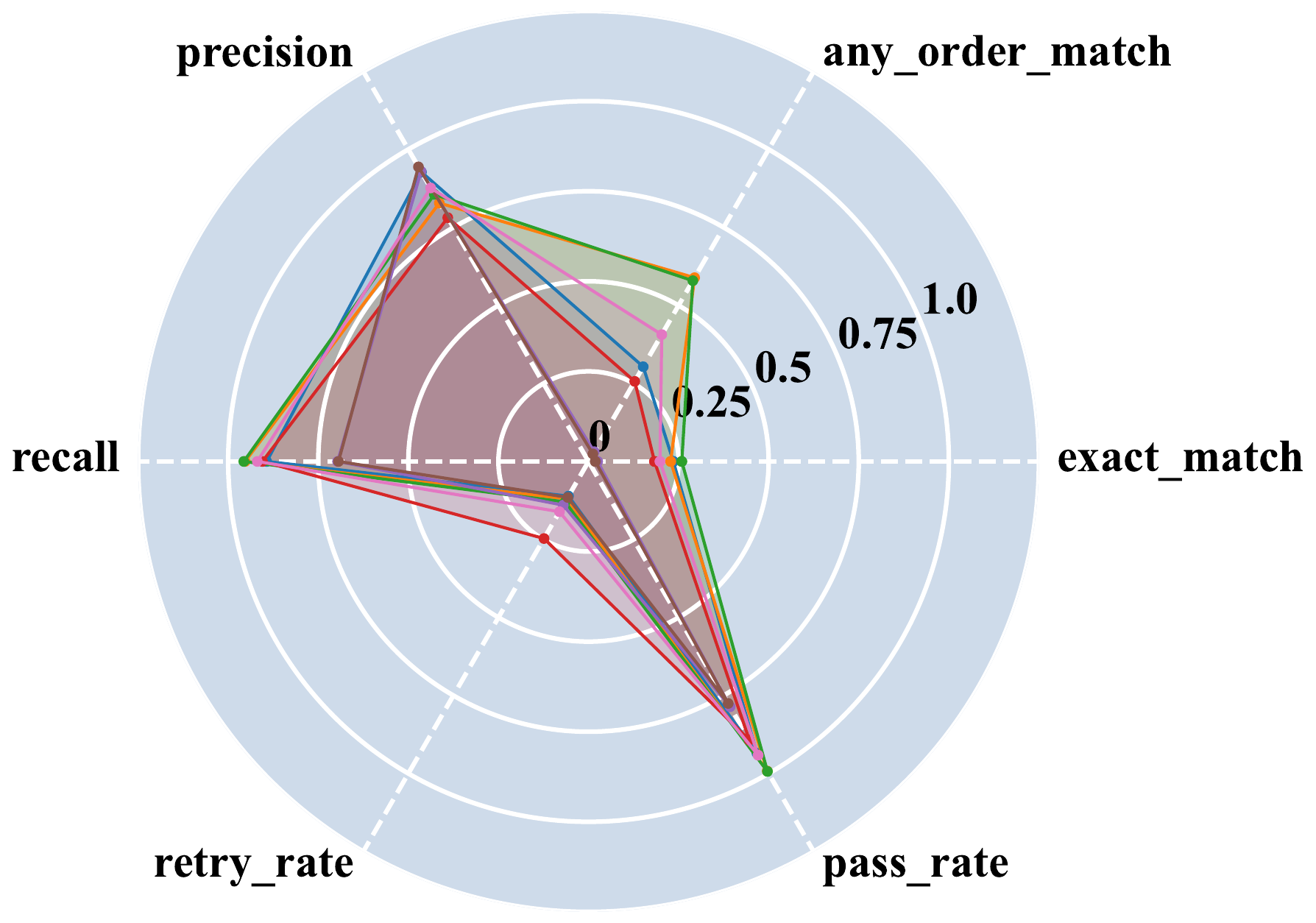}
        \caption{Comparison of overall behavioral correctness across all projects.}
        \label{fig:overall_average}
    \end{minipage}\hfill
    \begin{minipage}{0.45\linewidth}
        \centering
        \includegraphics[width=\linewidth]{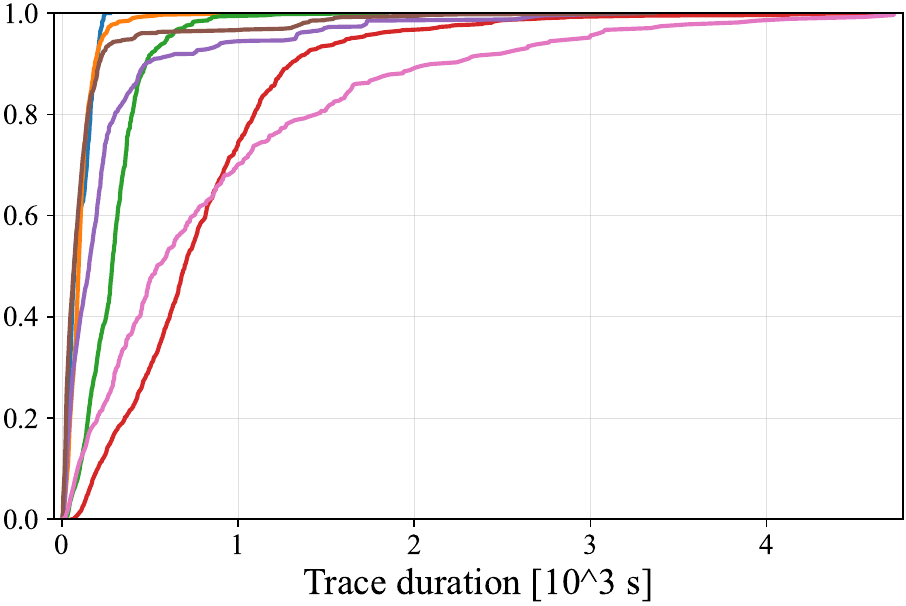}
        \caption{Overall end-to-end latency distributions across all projects.}
        \label{fig:overall_ecdf}
    \end{minipage}
    \vspace{-0.2in}
\end{figure}

\begin{figure}[t!]
    \centering
    \includegraphics[width=\textwidth]{fig/Failure_Breakdown/Model_Legend_horizontal.pdf}
    \begin{minipage}{0.1\textwidth} \hfill \end{minipage}
    \begin{minipage}{0.05\textwidth}
        \rotatebox{90}{\scriptsize \textbf{Email Responder}}
    \end{minipage}%
    \begin{minipage}{0.35\textwidth}
        \includegraphics[width=\linewidth]{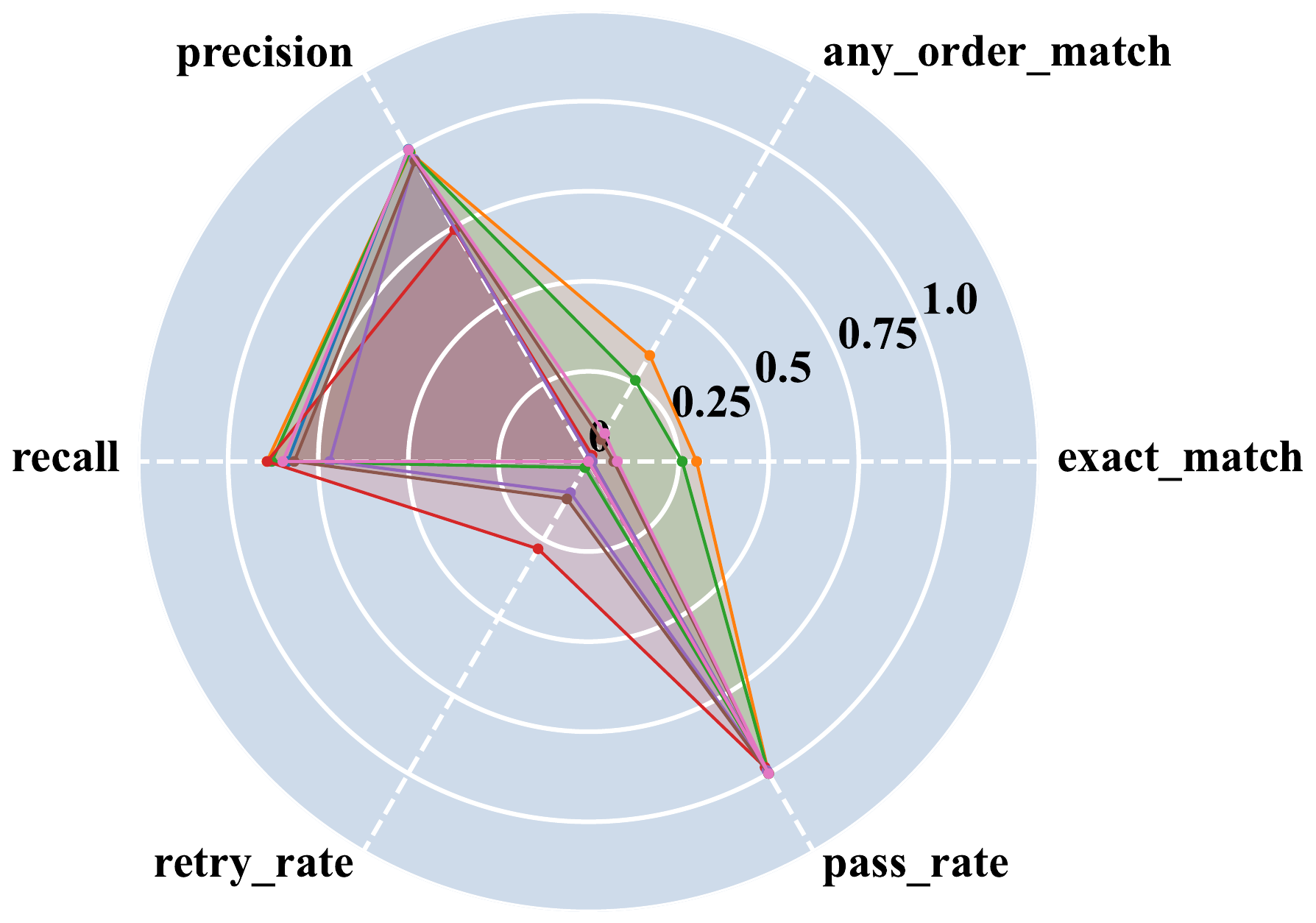}
    \end{minipage}%
    \hfill
    \begin{minipage}{0.35\textwidth}
        \includegraphics[width=\linewidth]{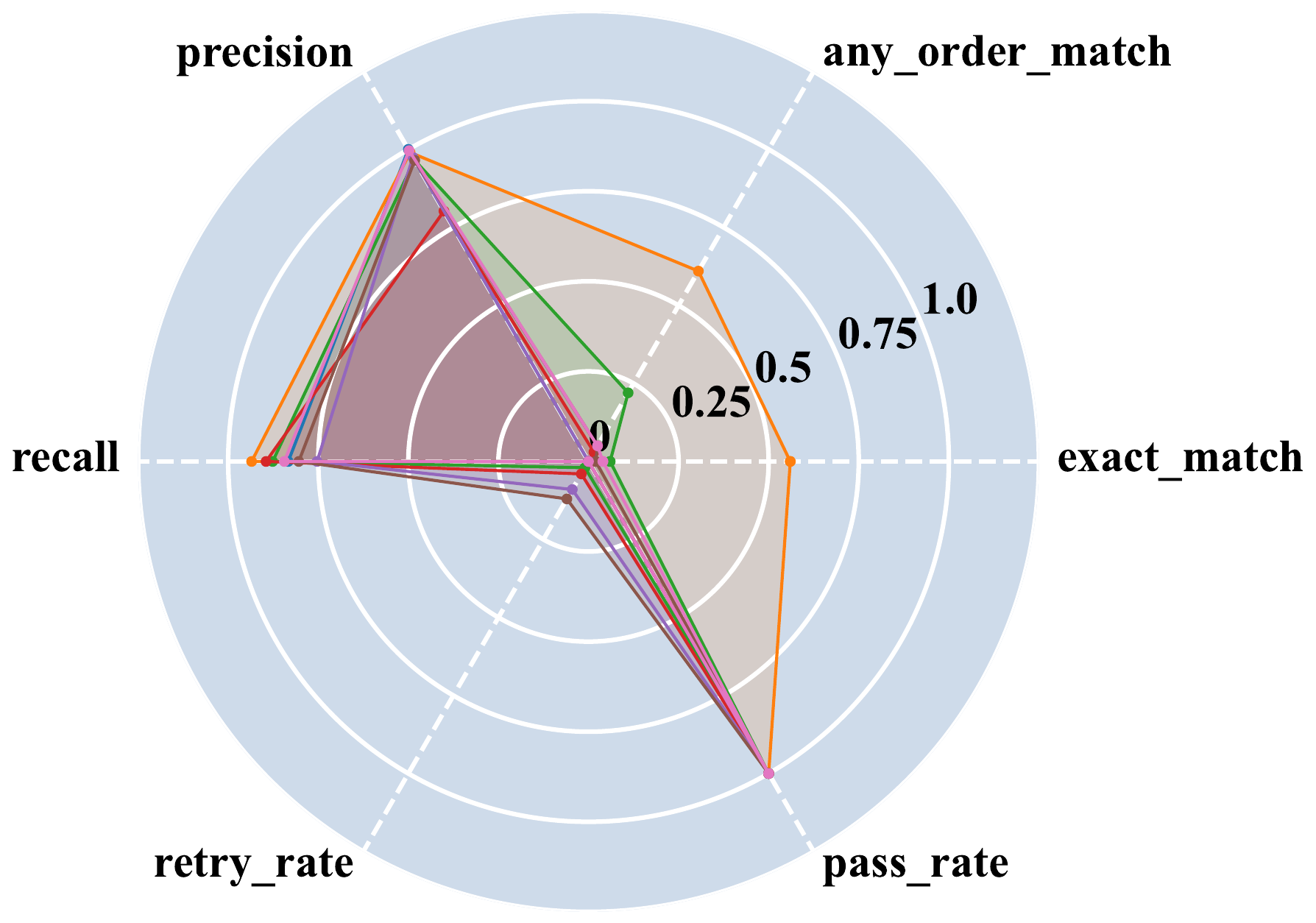}
    \end{minipage}
    \begin{minipage}{0.1\textwidth} \hfill \end{minipage}
    \vspace{0.4em}

    \begin{minipage}{0.1\textwidth} \hfill \end{minipage}
    \begin{minipage}{0.05\textwidth}
        \rotatebox{90}{\scriptsize \textbf{Game Builder}}
    \end{minipage}%
    \begin{minipage}{0.35\textwidth}
        \includegraphics[width=\linewidth]{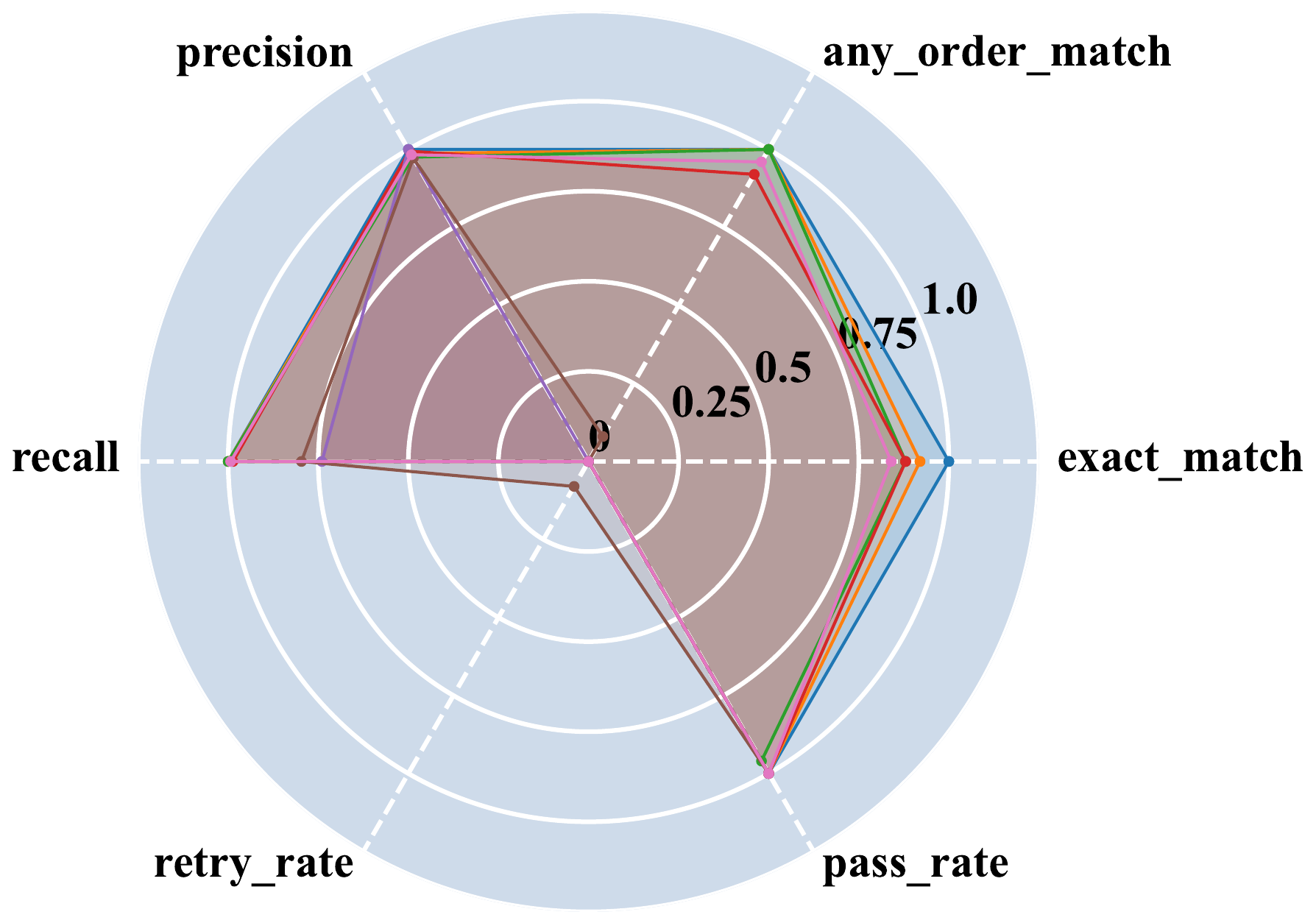}
    \end{minipage}%
    \hfill
    \begin{minipage}{0.35\textwidth}
        \includegraphics[width=\linewidth]{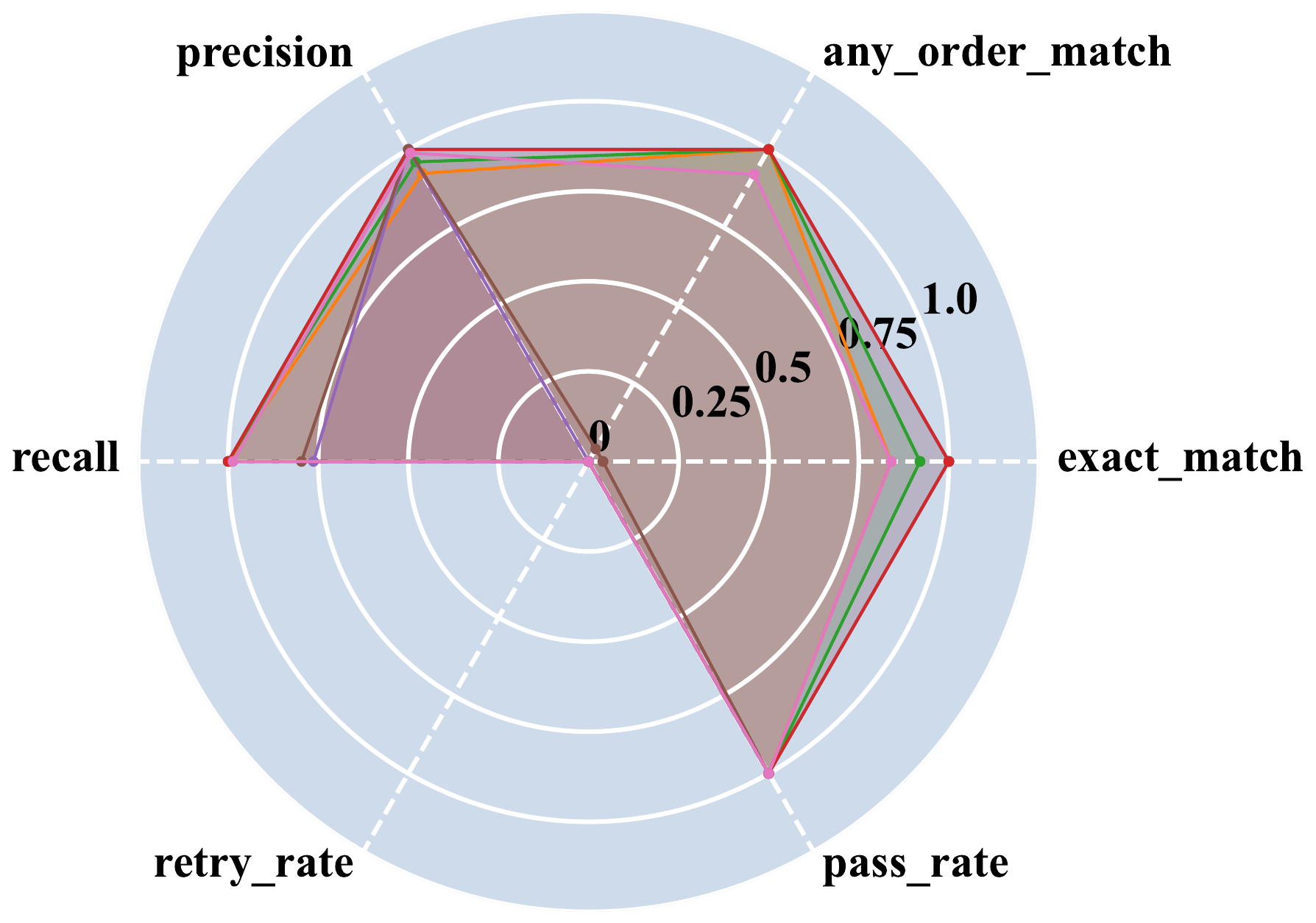}
    \end{minipage}
    \begin{minipage}{0.1\textwidth} \hfill \end{minipage}
    \vspace{0.4em}

    \begin{minipage}{0.1\textwidth} \hfill \end{minipage}
    \begin{minipage}{0.05\textwidth}
        \rotatebox{90}{\scriptsize \textbf{Markdown Val.(GT)}}
    \end{minipage}%
    \begin{minipage}{0.35\textwidth}
        \includegraphics[width=\linewidth]{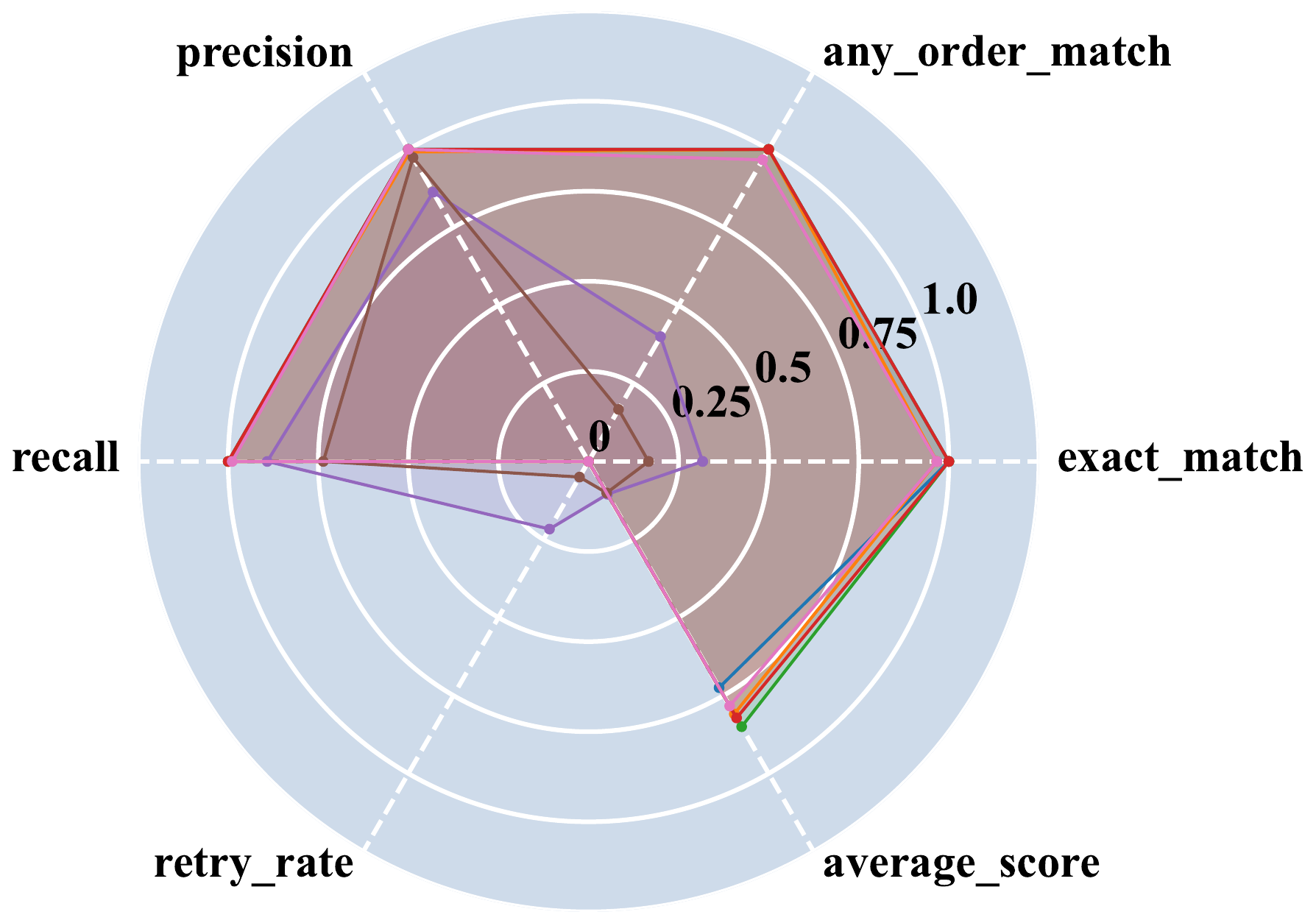}
    \end{minipage}%
    \hfill
    \begin{minipage}{0.35\textwidth}
        \includegraphics[width=\linewidth]{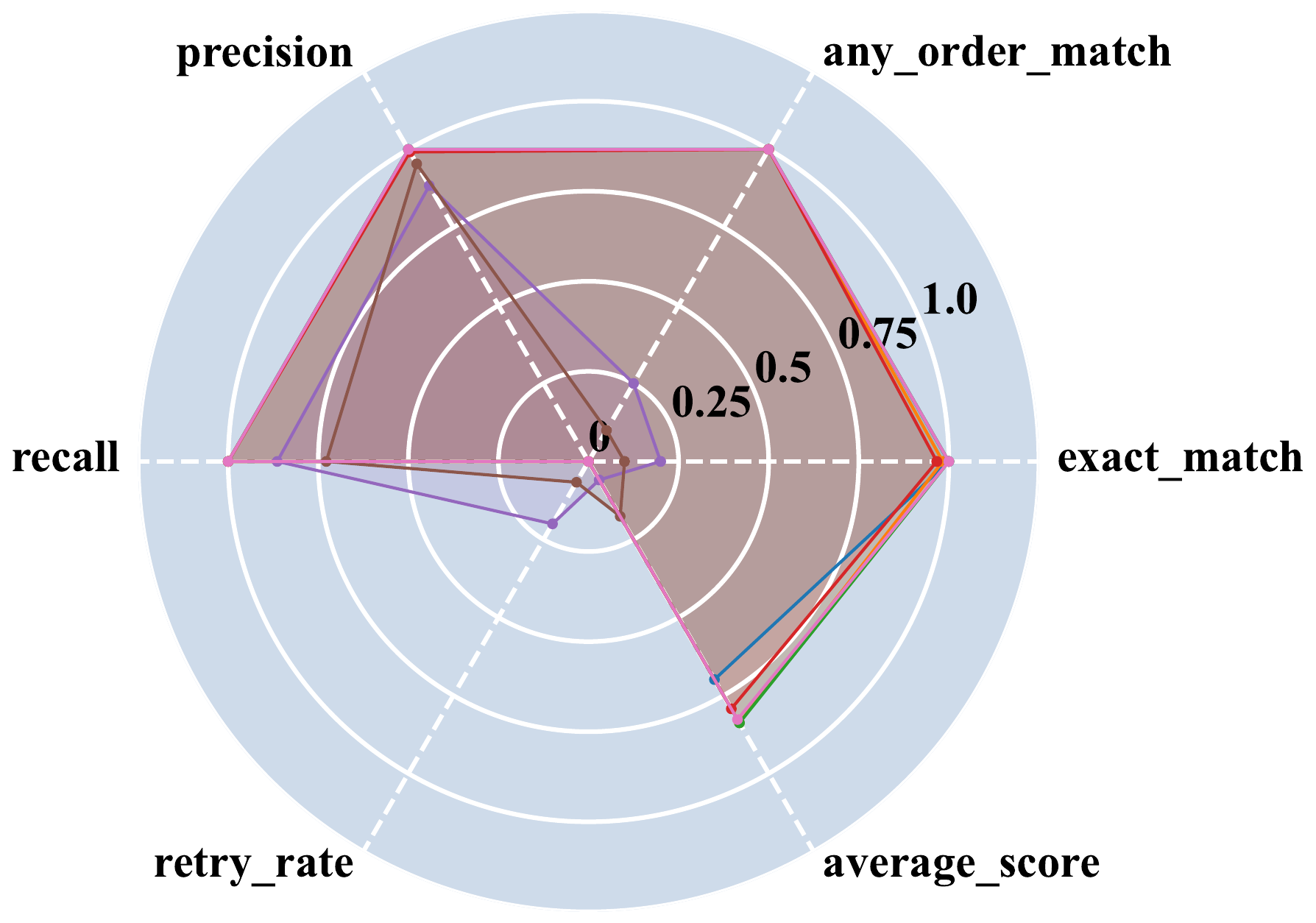}
    \end{minipage}
    \begin{minipage}{0.1\textwidth} \hfill \end{minipage}

    \vspace{0.3em}
    \begin{minipage}{0.15\textwidth} \hfill \end{minipage}
    \begin{minipage}{0.35\textwidth} \centering \textbf{\small{Pure CrewAI}} \end{minipage}
    \hfill
    \begin{minipage}{0.35\textwidth} \centering \textbf{\small{MCP}} \end{minipage}
    \begin{minipage}{0.1\textwidth} \hfill \end{minipage}
    \caption{Comparison of behavioral correctness for applications with two variants (Pure CrewAI and MCP).}
    \vspace{-0.2in}
    \label{fig:radar_2_cols}
\end{figure}

\begin{figure}[t!] 
    \centering
    \includegraphics[width=\textwidth]{fig/Failure_Breakdown/Model_Legend_horizontal.pdf}

    \begin{minipage}{0.03\textwidth}
        \rotatebox{90}{\scriptsize \textbf{SQL Asst.(GT)}}
    \end{minipage}%
    \begin{minipage}{0.32\textwidth}
        \includegraphics[width=\linewidth]{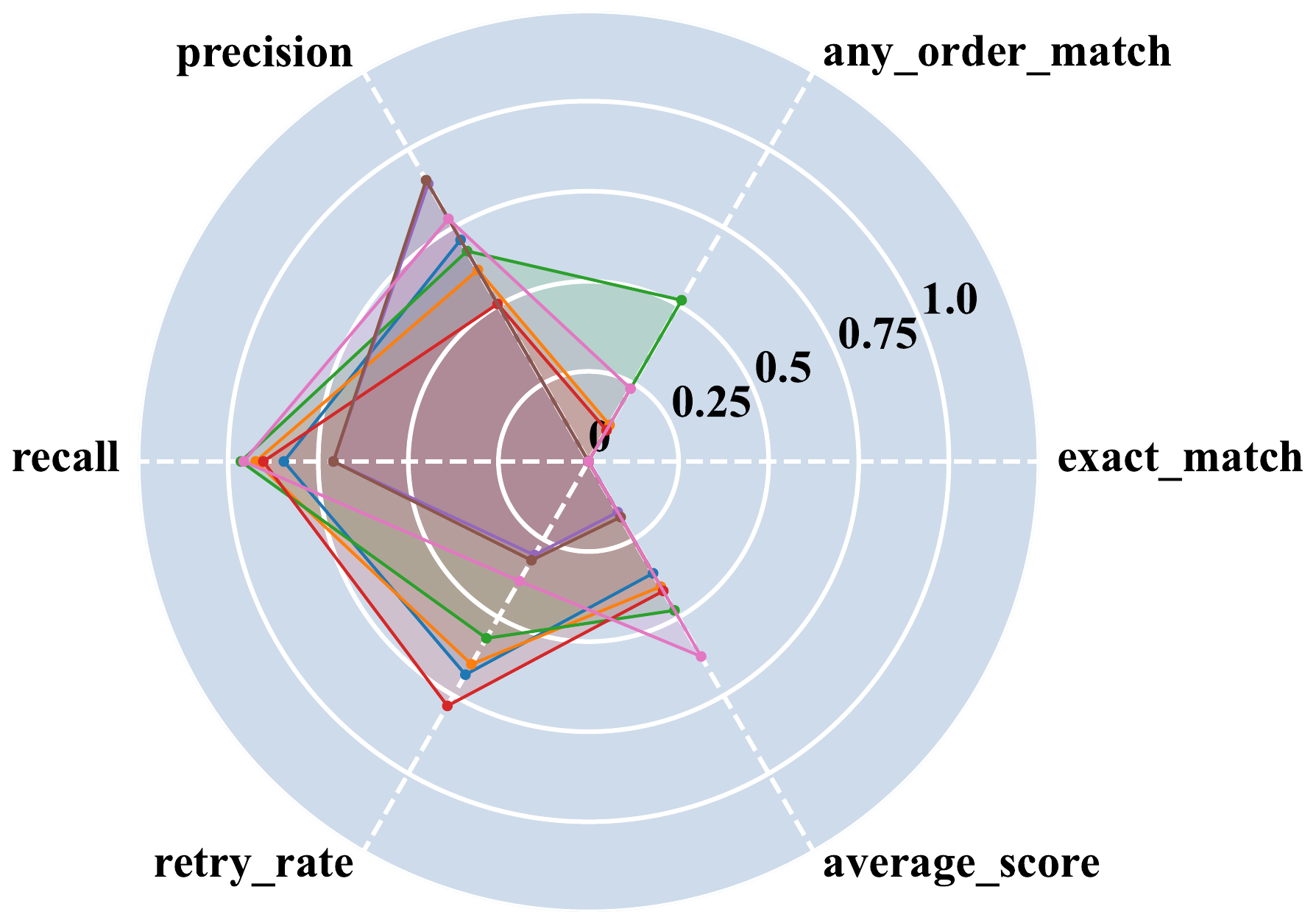}
    \end{minipage}%
    \begin{minipage}{0.32\textwidth}
        \includegraphics[width=\linewidth]{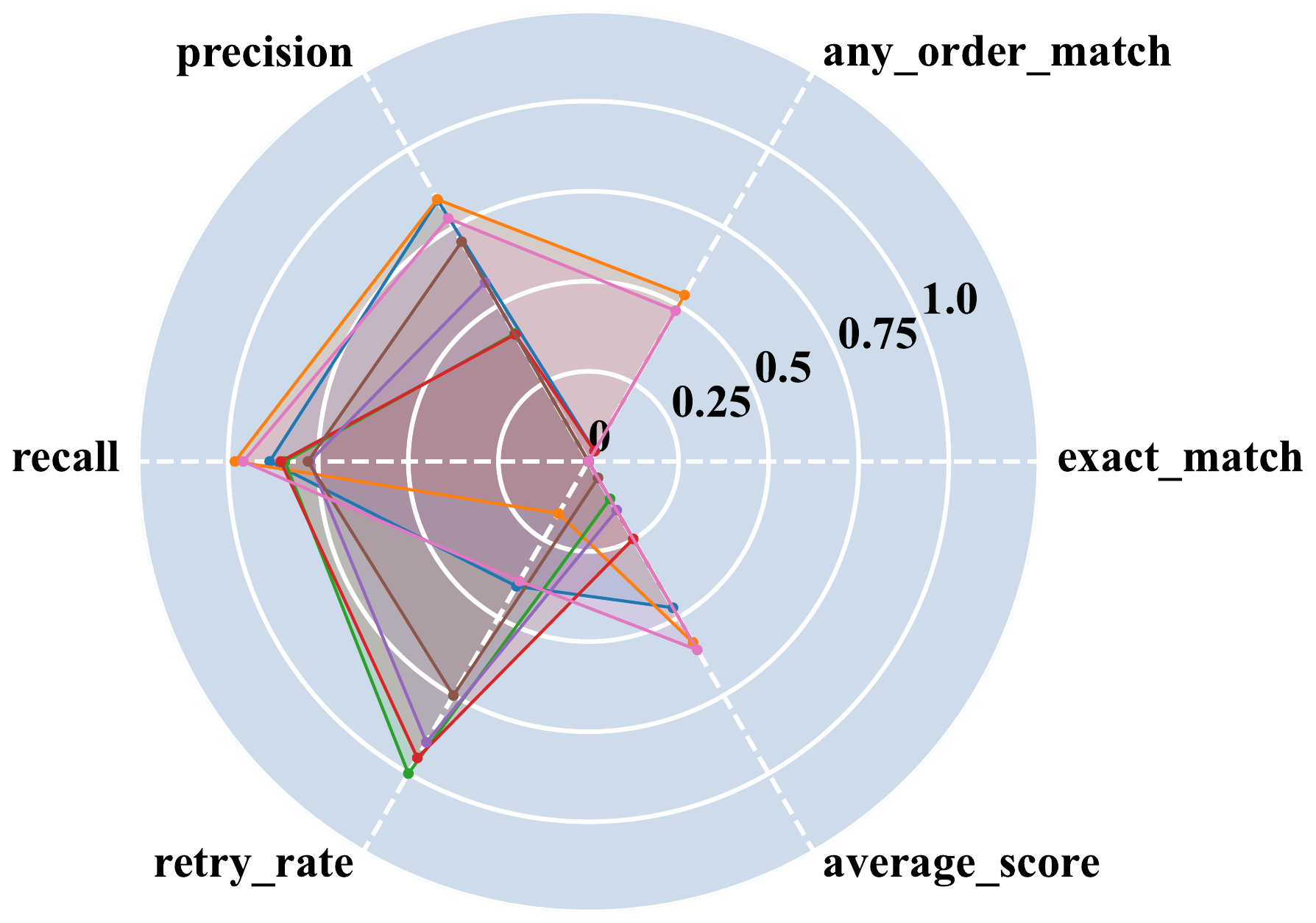}
    \end{minipage}%
    \begin{minipage}{0.32\textwidth}
        \includegraphics[width=\linewidth]{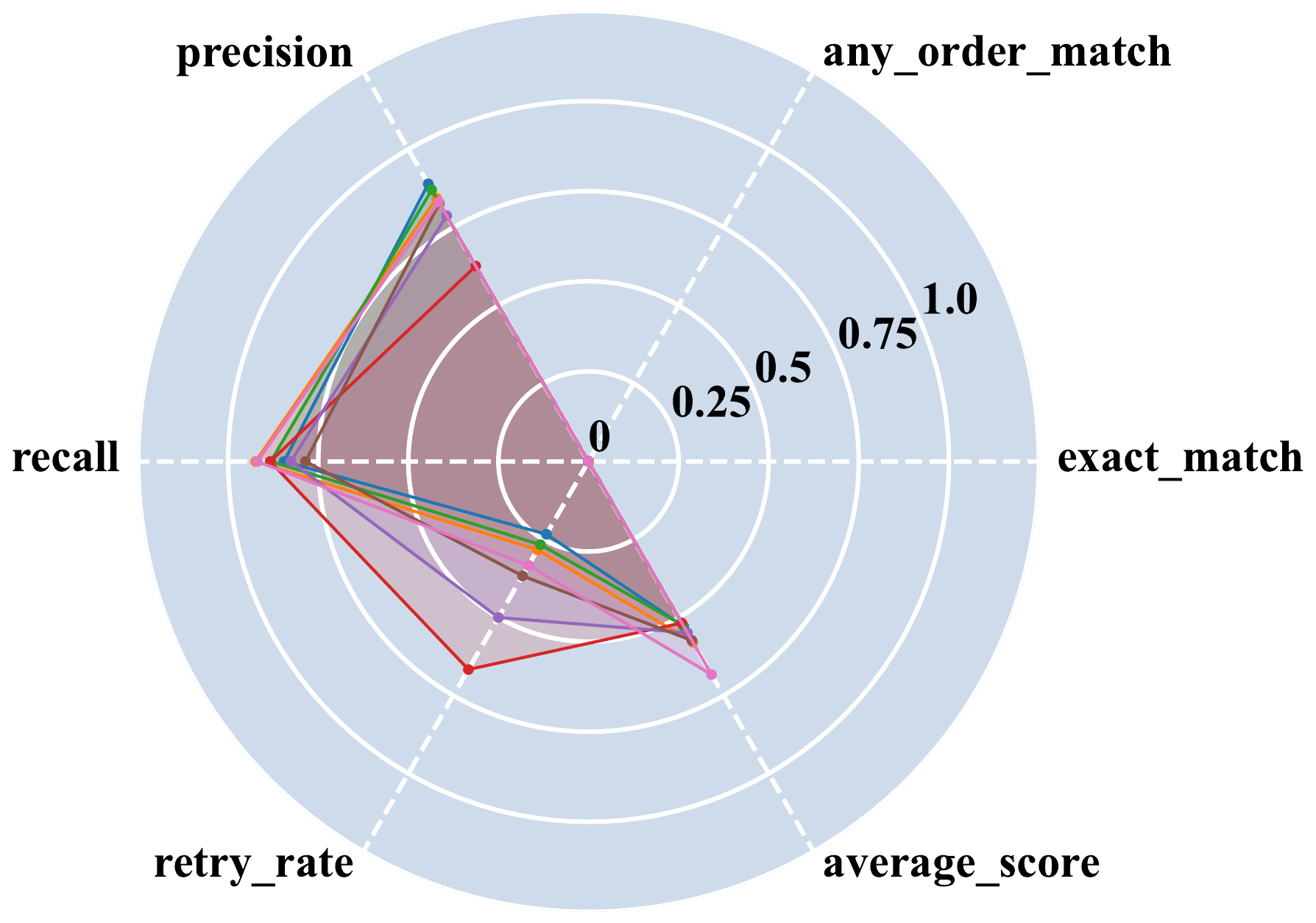}
    \end{minipage}
    \vspace{0.4em}

    \begin{minipage}{0.03\textwidth}
        \rotatebox{90}{\scriptsize \textbf{Recruitment}}
    \end{minipage}%
    \begin{minipage}{0.32\textwidth}
        \includegraphics[width=\linewidth]{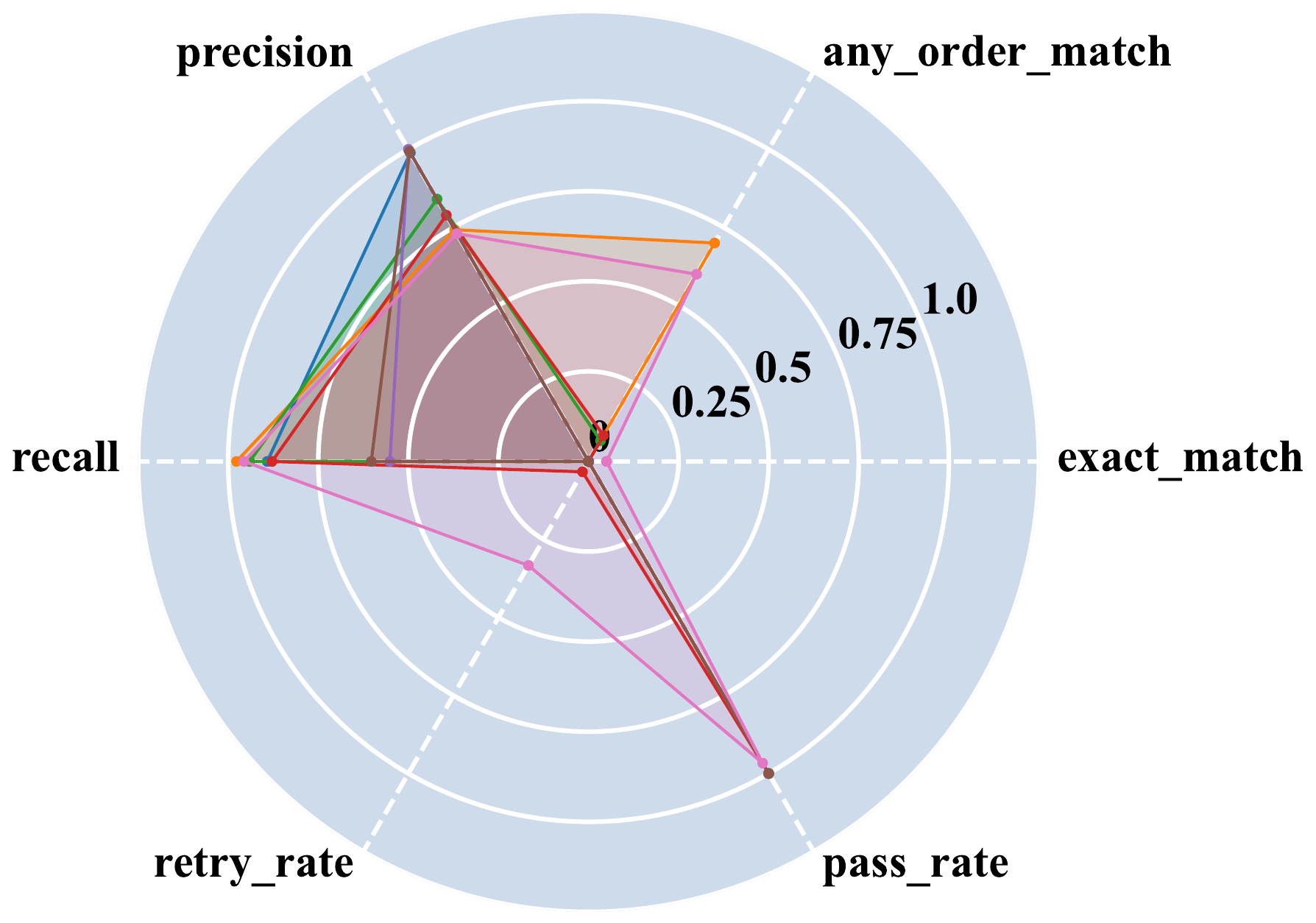}
    \end{minipage}%
    \begin{minipage}{0.32\textwidth}
        \includegraphics[width=\linewidth]{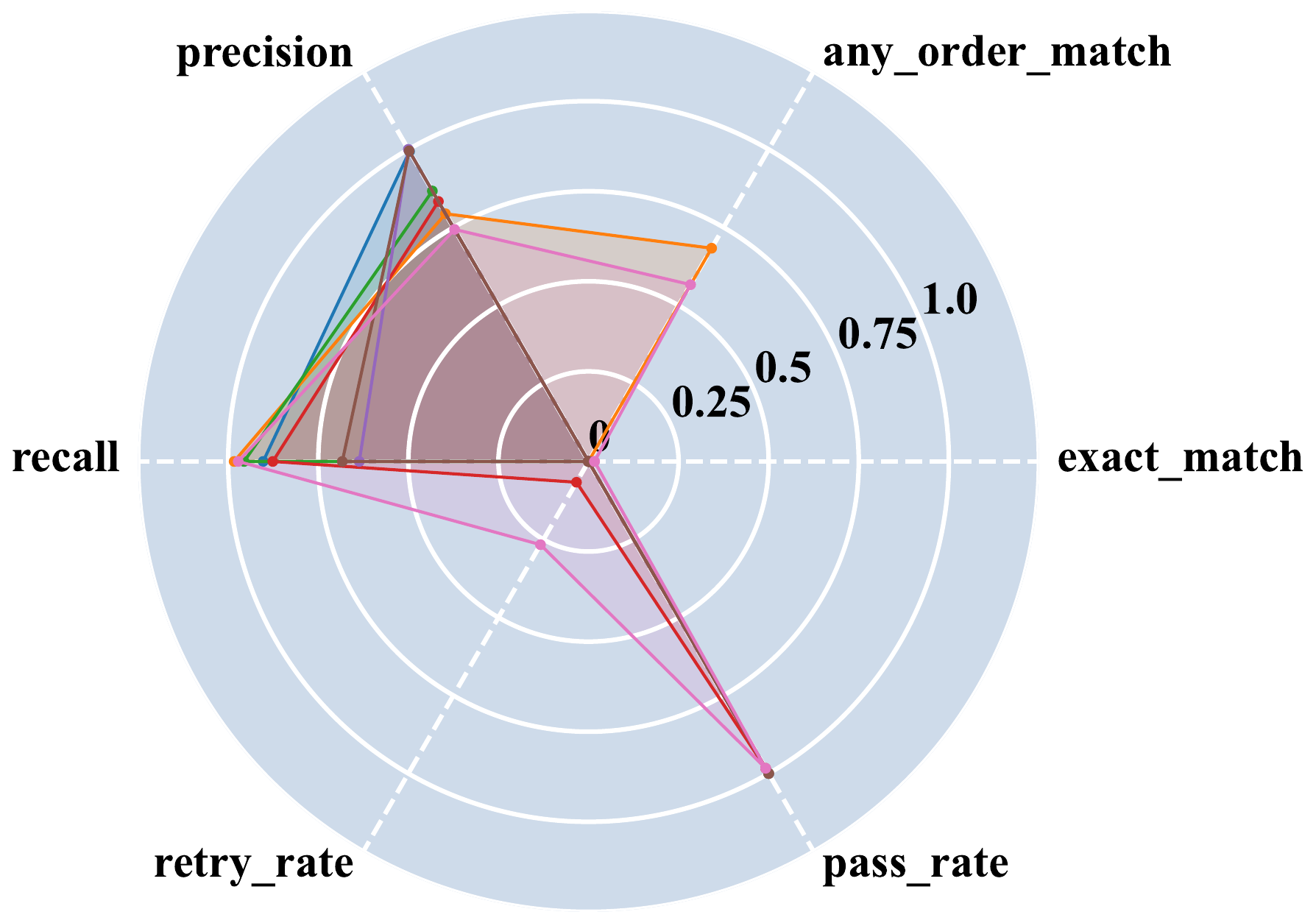}
    \end{minipage}%
    \begin{minipage}{0.32\textwidth}
        \includegraphics[width=\linewidth]{fig/Failure_Breakdown/intelligent_recruitment_platform-A2A_mix_radar.pdf}
    \end{minipage}
    \vspace{0.4em}
    
    \begin{minipage}{0.03\textwidth}
        \rotatebox{90}{\scriptsize \textbf{Landing Pg.(GT)}}
    \end{minipage}%
    \begin{minipage}{0.32\textwidth}
        \includegraphics[width=\linewidth]{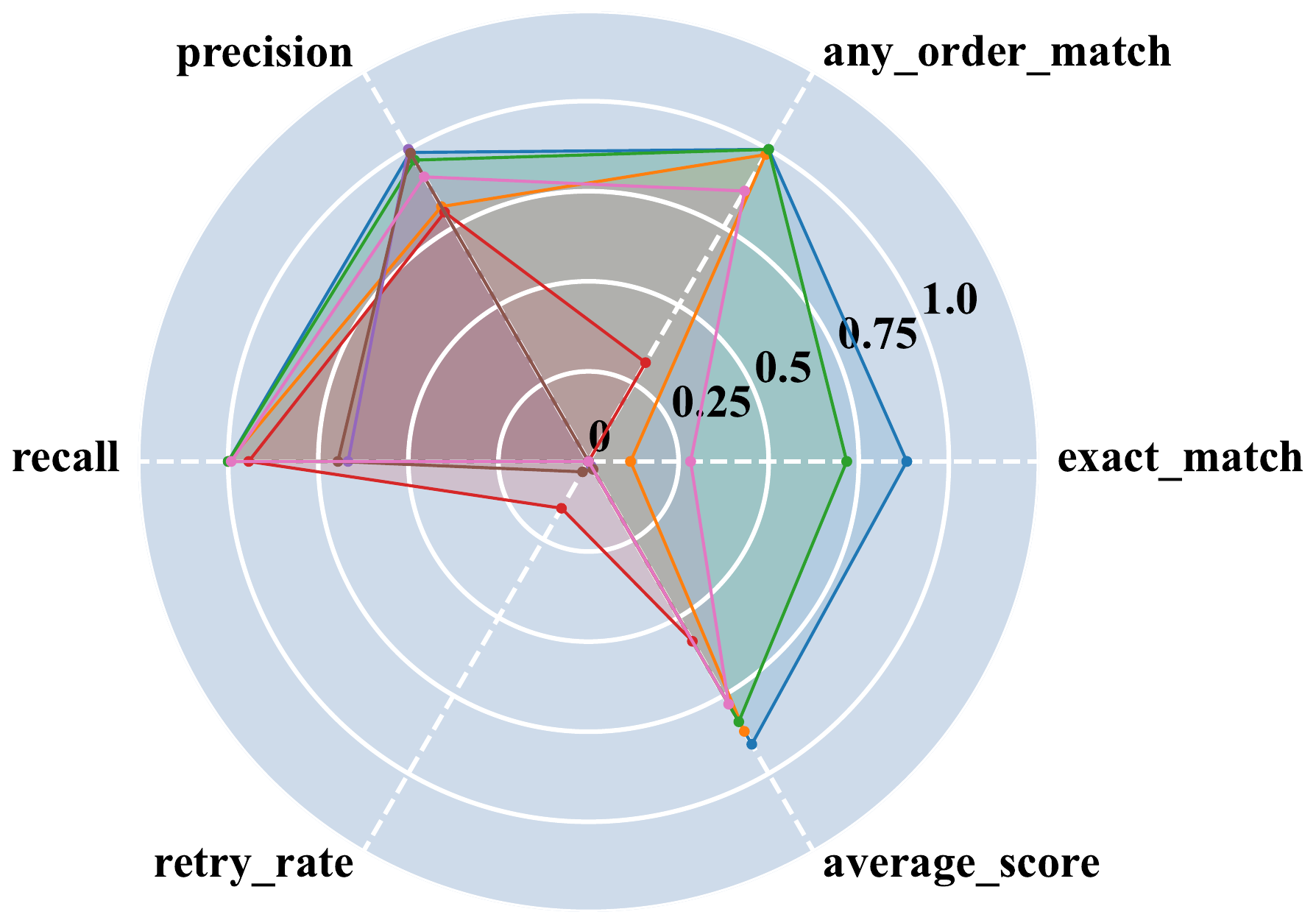}
    \end{minipage}%
    \begin{minipage}{0.32\textwidth}
        \includegraphics[width=\linewidth]{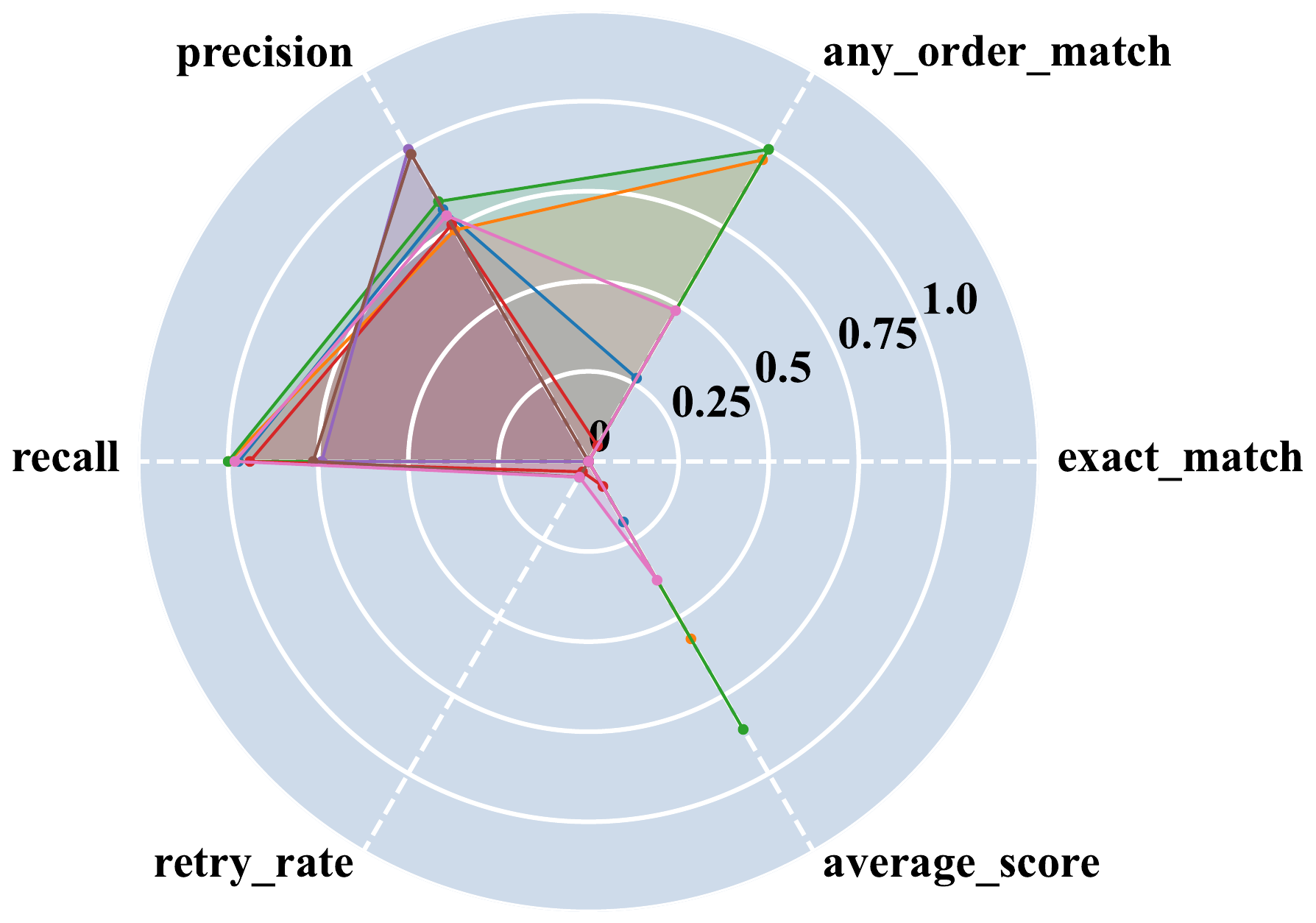}
    \end{minipage}%
    \begin{minipage}{0.32\textwidth}
        \includegraphics[width=\linewidth]{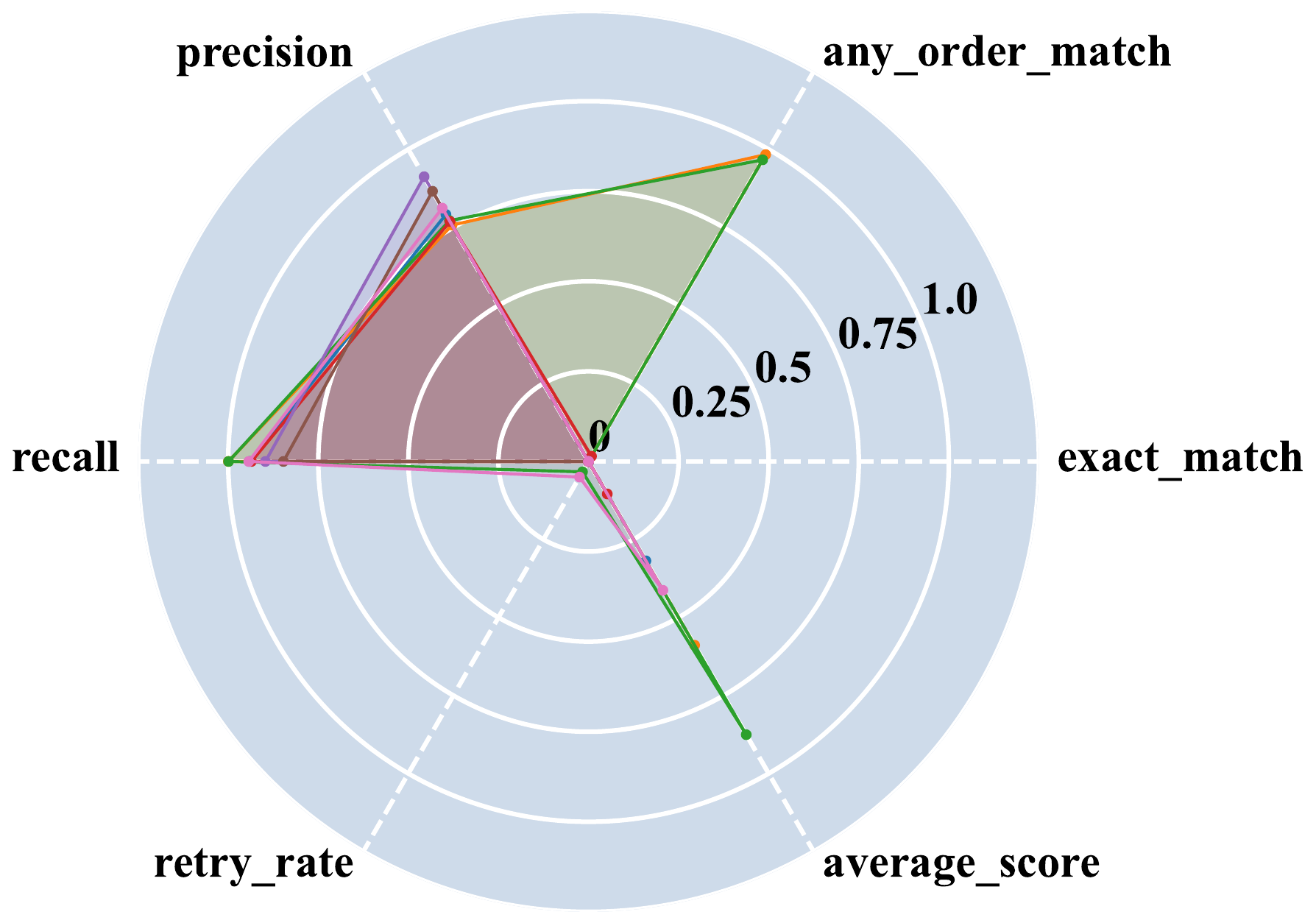}
    \end{minipage}
    \vspace{0.4em}

    \begin{minipage}{0.03\textwidth}
        \rotatebox{90}{\scriptsize \textbf{Social M. M.}}
    \end{minipage}%
    \begin{minipage}{0.32\textwidth}
        \includegraphics[width=\linewidth]{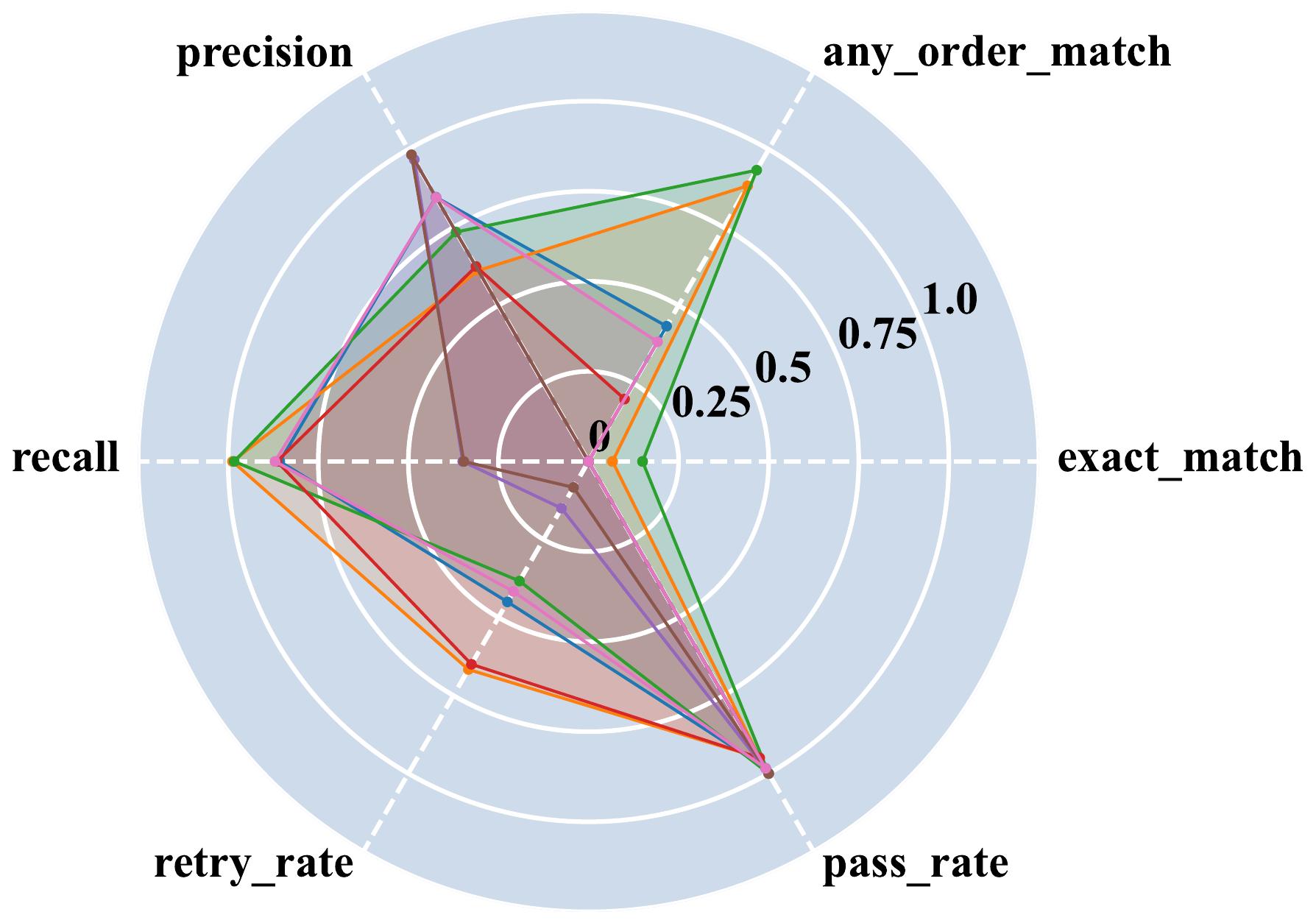}
    \end{minipage}%
    \begin{minipage}{0.32\textwidth}
        \includegraphics[width=\linewidth]{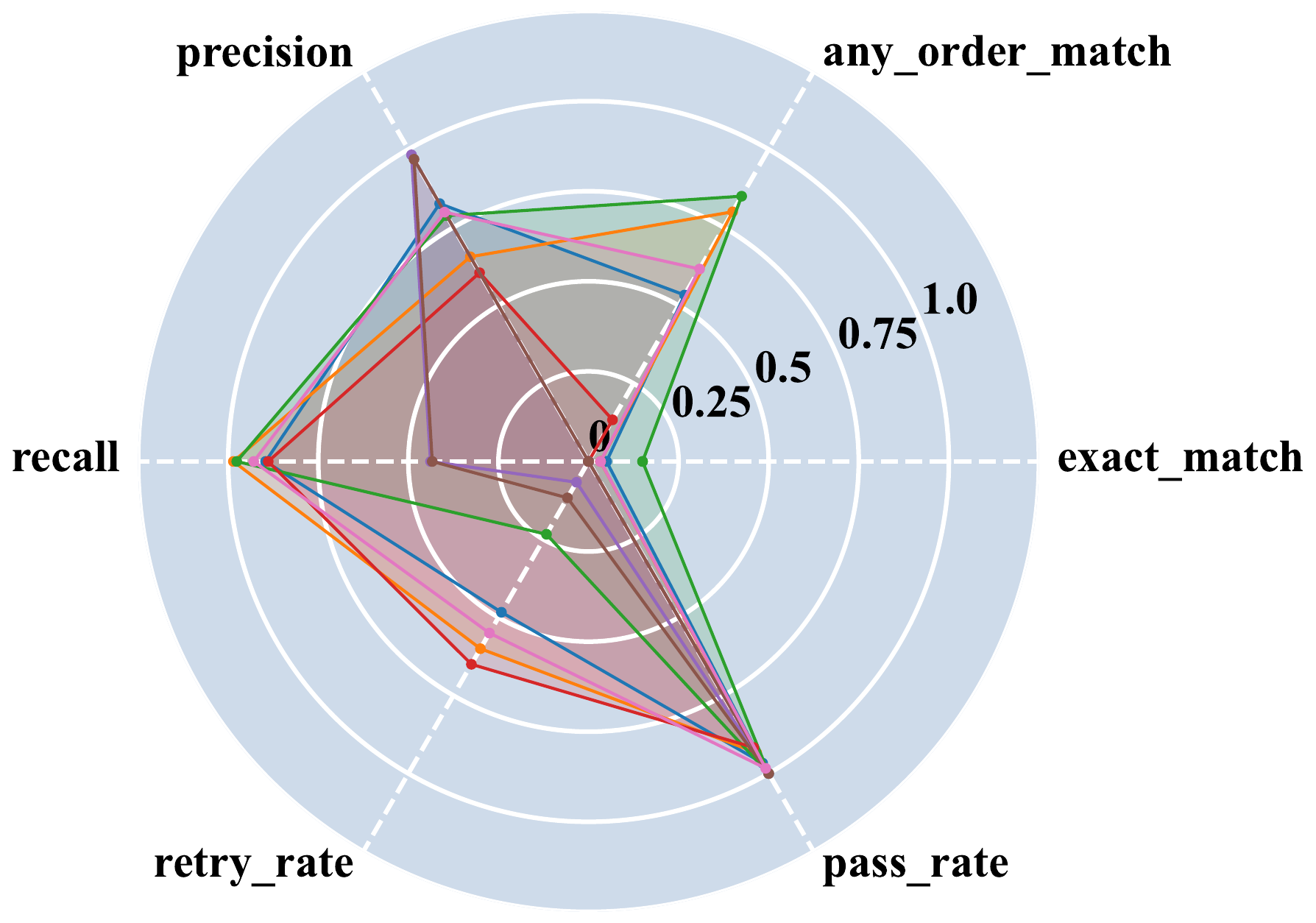}
    \end{minipage}%
    \begin{minipage}{0.32\textwidth}
        \includegraphics[width=\linewidth]{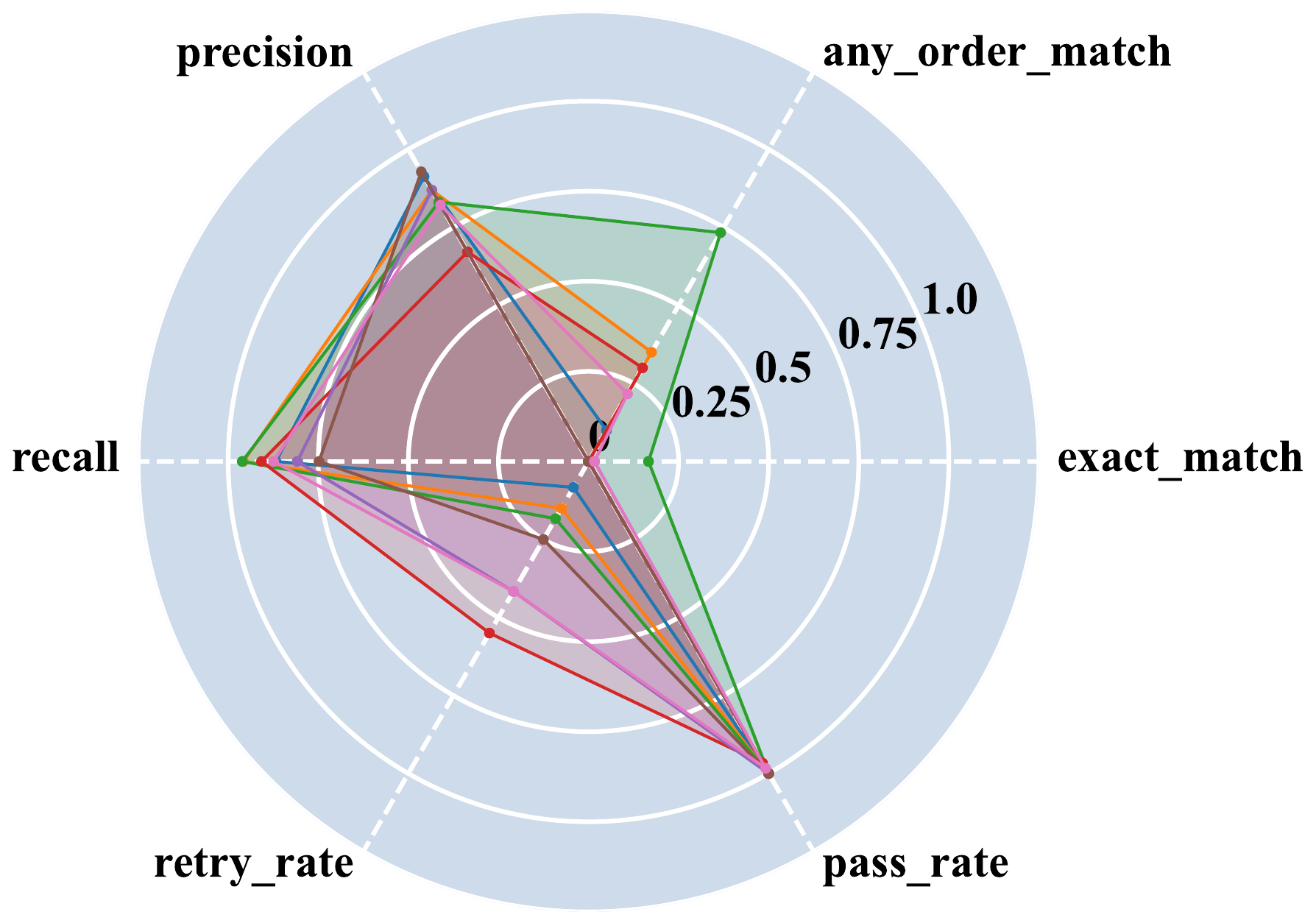}
    \end{minipage}
    \vspace{0.4em}

    \begin{minipage}{0.03\textwidth}
        \rotatebox{90}{\scriptsize \textbf{Book Writer}}
    \end{minipage}%
    \begin{minipage}{0.32\textwidth}
        \includegraphics[width=\linewidth]{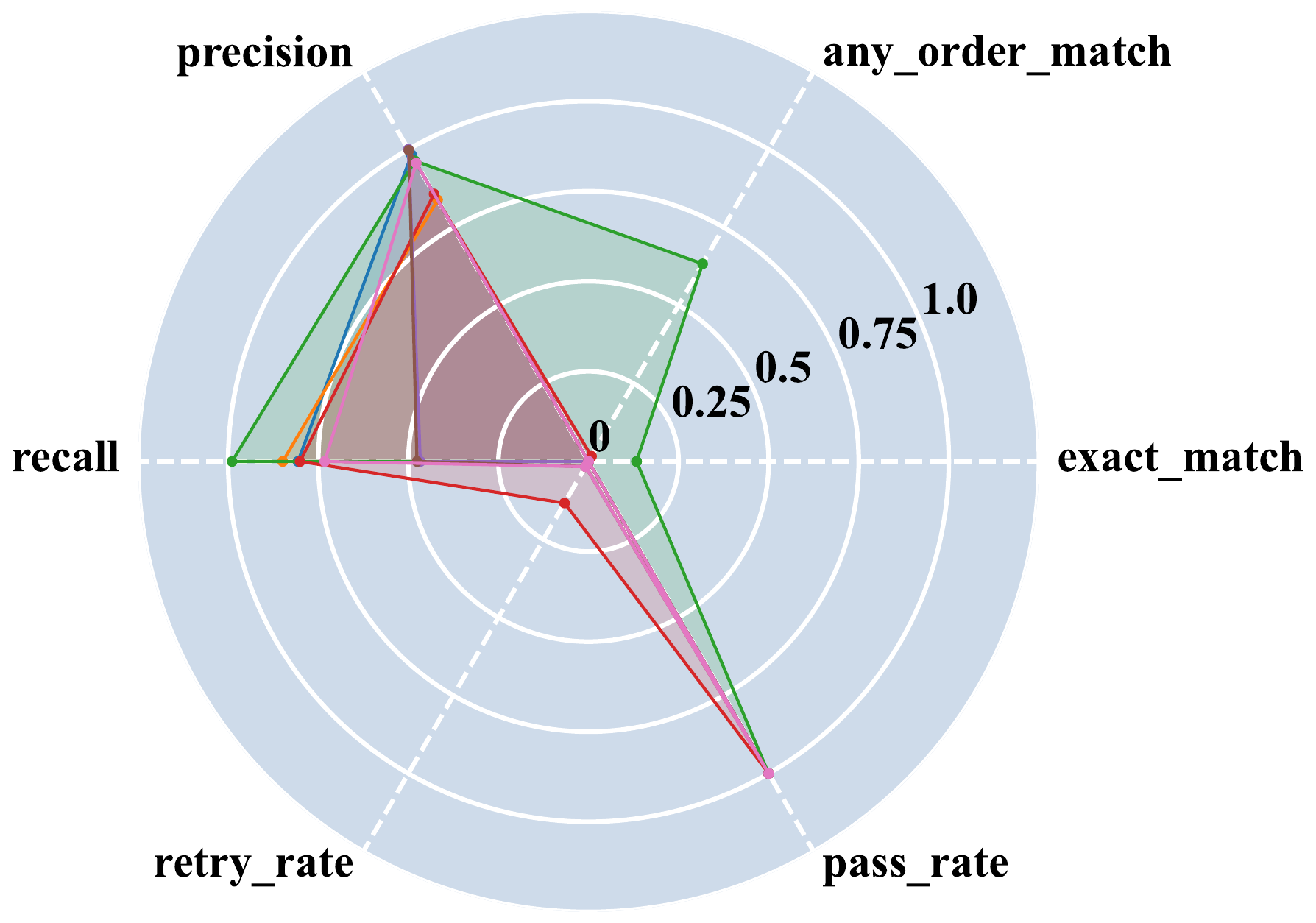}
    \end{minipage}%
    \begin{minipage}{0.32\textwidth}
        \includegraphics[width=\linewidth]{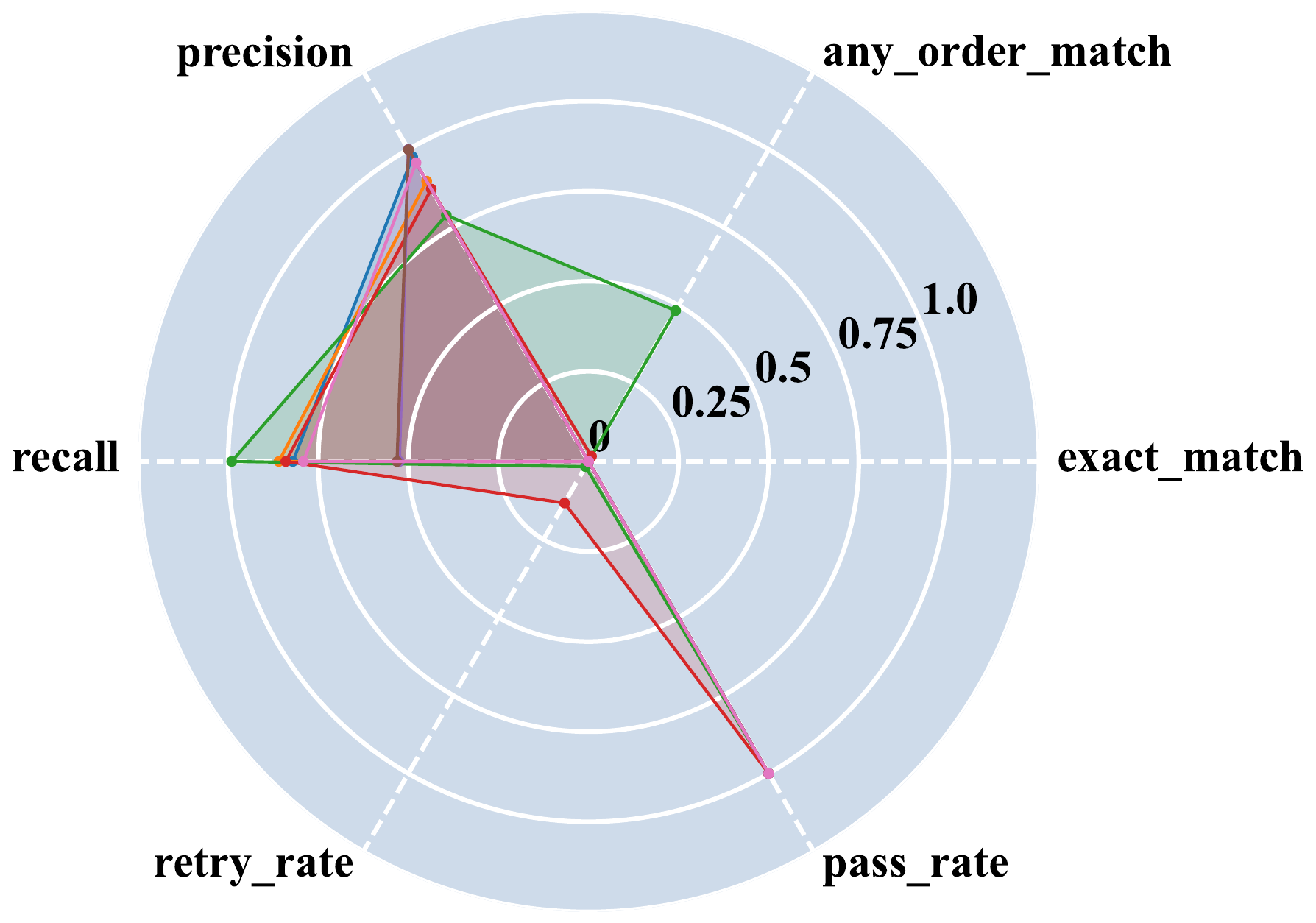}
    \end{minipage}%
    \begin{minipage}{0.32\textwidth}
        \includegraphics[width=\linewidth]{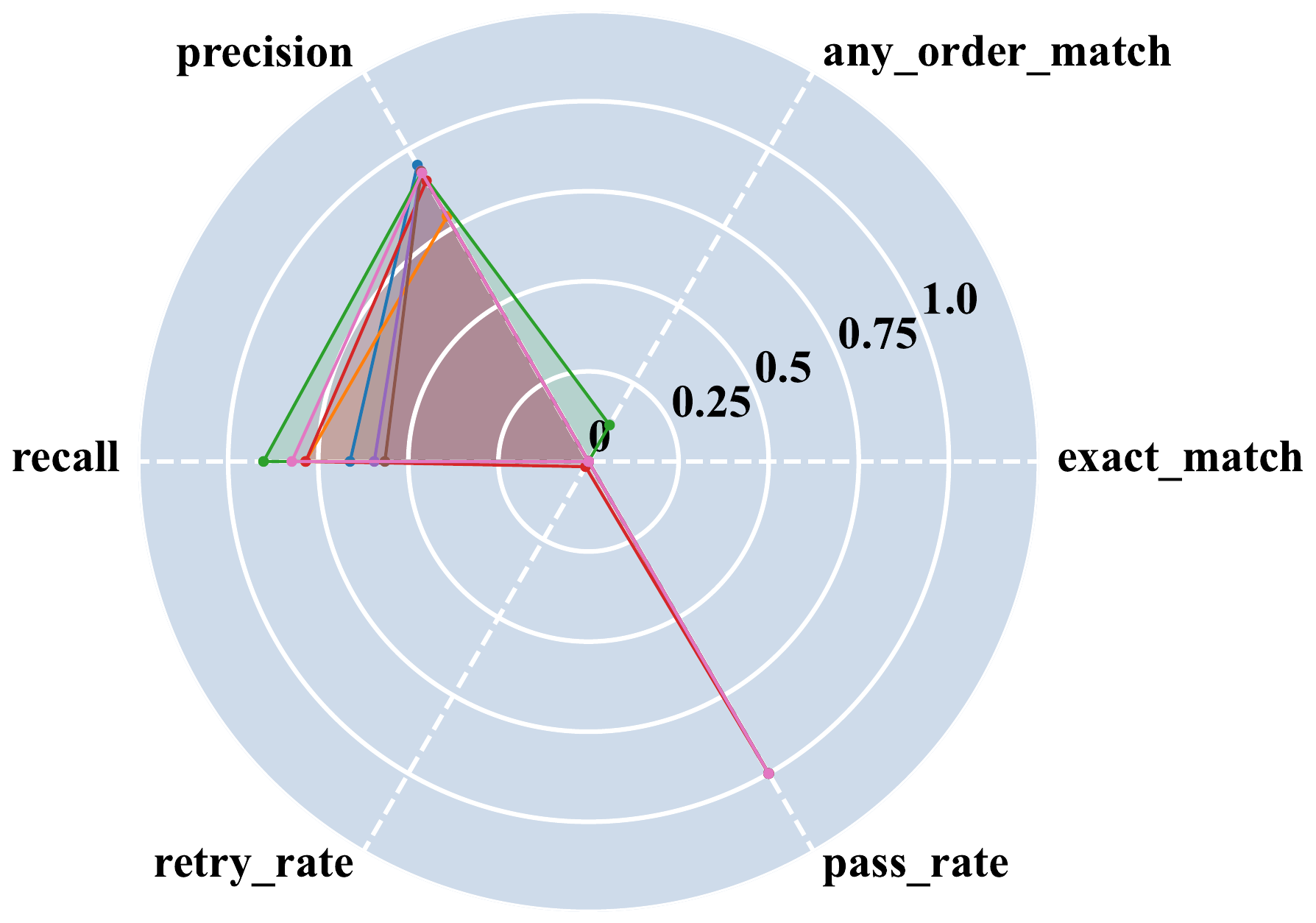}
    \end{minipage}

    \vspace{0.3em}
    \begin{minipage}{0.03\textwidth} \hfill \end{minipage}%
    \begin{minipage}{0.32\textwidth} \centering \textbf{MCP} \end{minipage}
    \begin{minipage}{0.32\textwidth} \centering \textbf{A2A} \end{minipage}
    \begin{minipage}{0.32\textwidth} \centering \textbf{H-A2A} \end{minipage}

    \caption{Comparison of behavioral correctness for applications with three variants (MCP, A2A, H-A2A).}
    \vspace{-0.2in}
    \label{fig:radar_3_cols}
\end{figure}

To address RQ1, we evaluate the structural integrity of agent workflows by contrasting final task success against procedural adherence. Fig.~\ref{fig:overall_average} presents the aggregated performance metrics across all experimental configurations, while Fig.~\ref{fig:radar_2_cols} and Fig.~\ref{fig:radar_3_cols} visualize the multidimensional trade-offs for specific applications. The data reveals that superior model capabilities do not linearly translate to engineering reliability. Instead, we observe a pervasive tension between functional completion and process fidelity. We dissect this phenomenon through three key dimensions: the counter-intuitive impact of parameter scale, the dual-edged nature of reasoning modes, and the structural constraints of distributed architectures.

\subsubsection{Impact of Parameter Paradox}

A striking phenomenon emerges, termed the ``Parameter Paradox.'' Flagship models do not always exhibit better workflow adherence and output quality than their lightweight counterparts. Aggregated metrics in Fig.~\ref{fig:overall_average} provide evidence. Both the flagship \texttt{GPT-5} and the lightweight \texttt{GPT-4o-mini} achieve near-perfect task completion (\textit{pass\_rate} $\approx 1.00$ vs.\ $0.99$). However, \texttt{GPT-4o-mini} consistently demonstrates superior process and ground truth alignment. Specifically, \texttt{GPT-4o-mini} achieves higher \textit{any\_order\_match} ($0.64$ vs.\ $0.35$) and better \textit{average\_score} ($0.67$ vs.\ $0.55$) in applications with GT labels. This substantial gap indicates that the lightweight model captures the requisite set of actions in complex workflows more consistently.

In contrast, \texttt{GPT-5} behaves as a ``high-precision specialist.'' It exhibits superior tool-calling \textit{precision} ($0.94$ vs.\ $0.85$) but lower \textit{recall} ($0.90$ vs.\ $0.95$). \texttt{GPT-5} rarely hallucinates tool parameters due to its high precision. However, it is more conservative and prone to omitting required actions. This behavior may be acceptable in general QA tasks. Yet, it poses risks to auditability and verification in software engineering contexts. Step-by-step transparency is critical for fault localization.

The performance divergence depends on context. It reveals a clear trade-off determined by the clarity and scope of the task:
\begin{itemize}[leftmargin=*,label=\textbullet]
    \item \textbf{Complex Coordination:} Fig.~\ref{fig:radar_3_cols} illustrates workflows requiring extensive multi-step coordination. In \texttt{Landing Page Generator} (Overall), \texttt{GPT-4o-mini} dominates alignment metrics (\textit{any\_order\_match} $0.98$ vs.\ $0.42$) and outcome quality (\textit{average\_score} $0.67$ vs.\ $0.47$). Similarly, in \texttt{Recruitment Assistant} (Overall), \texttt{GPT-5} fails to align with the golden traces entirely (\textit{any\_order\_match} $0.00$). In contrast, \texttt{GPT-4o-mini} achieves a respectable $0.70$.
    \item \textbf{Structured Deterministic Tasks:} Flagship models excel in structured tasks with rigid logic and compact scope (e.g., \texttt{Game Builder} (Overall), Fig.~\ref{fig:radar_2_cols}). They demonstrate perfect execution (\textit{exact\_match} $1.00$) due to superior instruction-following precision. Their higher \textit{precision} ($1.00$ vs.\ $0.96$) is ideal for scenarios where false positives outweigh the risk of omission.
\end{itemize}

\finding{\textbf{Parameter Paradox:} Model scale lacks direct correlation with process fidelity. Lightweight models exhibit consistent execution in complex workflows, while larger models yield outcome efficiency in structured tasks but reduced adherence in verification-intensive workflows.}\label{fnd:parameter}

\subsubsection{Impact of Reasoning Mode}
This section investigates whether enabling ``reasoning mode'' (i.e., chain-of-thought capabilities) universally enhances agentic collaboration. Contrary to expectations, stronger reasoning does not guarantee superior planning. Instead, a complex trade-off emerges. Reasoning capabilities often improve internal decision quality but compromise protocol adherence and execution stability.

\begin{figure}[t]
    \centering
    \begin{minipage}[t]{0.48\linewidth}
        \centering
        \includegraphics[width=\linewidth]{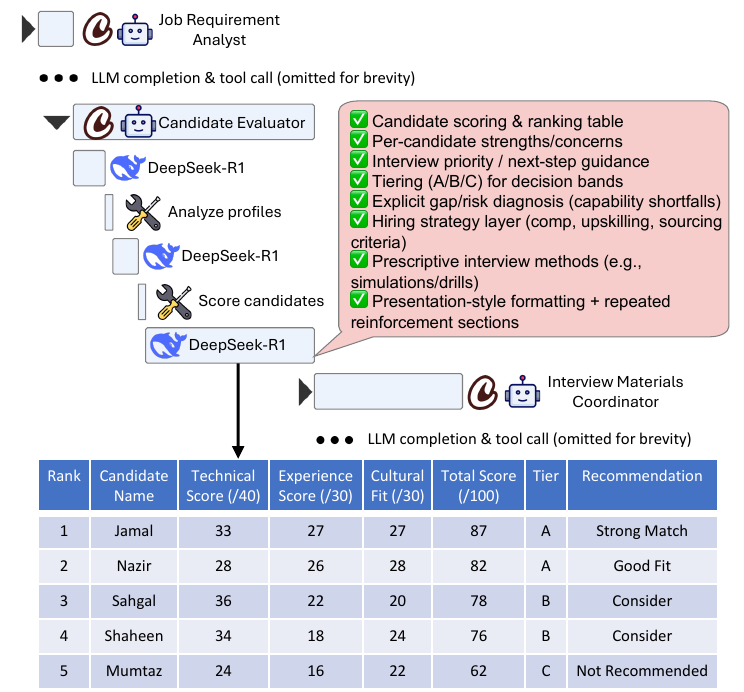}
        \centerline{\small (a) DeepSeek-R1 (Reasoning)}
    \end{minipage}%
    \hfill
    \begin{minipage}[t]{0.48\linewidth}
        \centering
        \includegraphics[width=\linewidth]{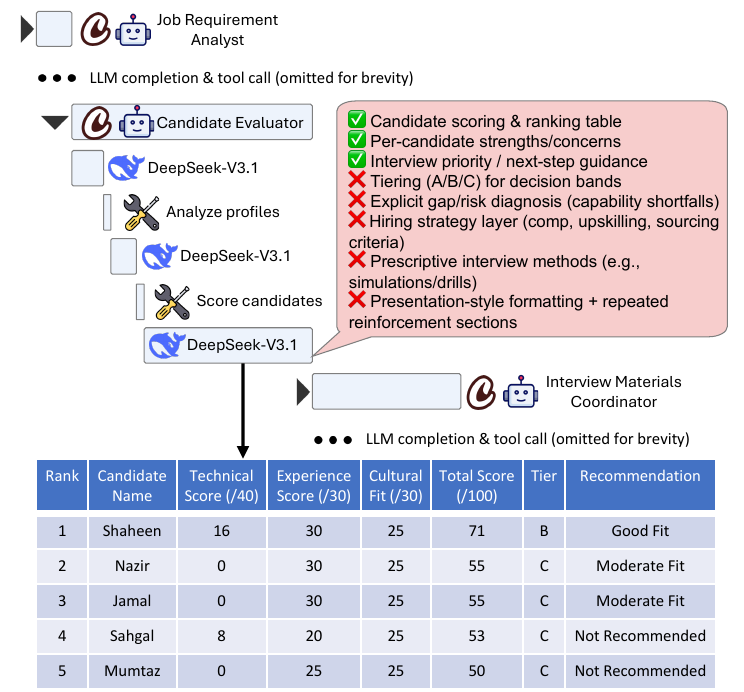}
        \centerline{\small (b) DeepSeek-V3.1 (Standard)}
    \end{minipage}%
    \caption{Case study on \texttt{Recruitment Assistant}: qualitative divergence in decision content. Comparison of the \textit{Candidate Evaluator} agent outputs.}
    \label{fig:recruitment_case}
    \vspace{-0.2in}
\end{figure}

\textbf{1. DeepSeek Family: The Content-Process Divergence.}
Aggregating results over all projects (see Fig.~\ref{fig:overall_average}), we observe a significant performance inversion within the \texttt{DeepSeek} family. The standard model, \texttt{DeepSeek-V3.1}, consistently outperforms its reasoning-enhanced counterpart, \texttt{DeepSeek-R1}, across critical process-oriented dimensions. Specifically, \texttt{DeepSeek-R1} exhibits a significant degradation in workflow adherence (\textit{any\_order\_match} $0.31$ vs.\ $0.65$) and tool-use correctness (\textit{precision} $0.80$ vs.\ $0.87$), leading to substantially higher execution instability (\textit{retry\_rate} $0.22$ vs.\ $0.11$). This trade-off manifests in two distinct behavioral patterns:
\begin{itemize}[leftmargin=*,label=\textbullet]
    \item \textbf{Process Degradation via Protocol Bypass:}  \texttt{DeepSeek-R1} tends to ``internalize'' interactions within its chain-of-thought rather than executing required protocols in workflows requiring strict coordination. For instance, in \texttt{Book Writer} (Overall), Fig.~\ref{fig:radar_3_cols} shows that \texttt{DeepSeek-R1} suffers a near-total loss of alignment (\textit{any\_order\_match} $\approx 0.01$ vs.\ $0.58$), effectively bypassing necessary communication steps. Similarly, in \texttt{Landing Page Generator} (Overall), this inability to sustain sequential hand-offs results in a collapse of the \textit{average\_score} ($0.25$ vs.\ $0.86$), confirming that reasoning capabilities can disrupt the operational flow of agentic systems.
    
    \item \textbf{Content Superiority in Decision-Making:} Despite trajectory degradation, reasoning capabilities significantly enhance decision quality. As illustrated in the \texttt{Recruitment Assistant} case (Fig.~\ref{fig:recruitment_case}), \texttt{DeepSeek-R1} demonstrates greater robustness and granularity than \texttt{DeepSeek-V3.1}. Quantitatively, in terms of candidate evaluation scores, \texttt{DeepSeek-R1} avoids extreme valuations (e.g., giving a candidate a $0$) and repetitive assessments (e.g., assigning identical ``Cultural Fit'' ratings of $25$ across candidates), instead producing a stratified distribution of total valuations ($62$--$87$) with clear separation. \texttt{DeepSeek-R1} transcends \texttt{DeepSeek-V3.1}'s partial coverage to provide multi-dimensional analysis, explicitly including gap diagnosis and prescriptive strategies. This indicates that although \texttt{DeepSeek-R1} faces challenges with workflow logistics, its reasoning chain supports comprehensive and persuasive decision-making for complex cognitive tasks. 
    
\end{itemize}

\textbf{2. Gemini Family: The Adherence-Stability Trade-off.}
 The \texttt{Gemini-2.5-flash} family, in contrast to \texttt{DeepSeek} family, demonstrates that reasoning can improve process adherence, but often at the expense of operational efficiency. The \texttt{Markdown Validator} panel in Fig.~\ref{fig:radar_2_cols} provides a example of this dichotomy, where ``thinking'' mode improves the process but degrades the outcome:
\begin{itemize}[leftmargin=*,label=\textbullet]
    \item \textbf{Enhanced Adherence \& Discovery:} Enabling reasoning capabilities significantly boosts the agent's ability to follow the standard workflow and identify issues. In \texttt{Markdown Validator} (Overall), the \textit{exact\_match} score doubles ($0.13 \rightarrow 0.26$) and \textit{any\_order\_match} increases substantially ($0.13 \rightarrow 0.33$). Furthermore, \textit{recall} improves ($0.73 \rightarrow 0.88$), indicating the agent successfully catches more validation errors.
    \item \textbf{Operational Friction via Over-Analysis:} Still within \texttt{Markdown Validator} (Overall), this heightened sensitivity introduces a penalty on reliability. The \textit{precision} drops ($0.96 \rightarrow 0.87$), implying more false positives due to over-analysis. Crucially, the system struggles to stabilize this complex reasoning, causing \textit{retry\_rate} to spike ($0.06 \rightarrow 0.21$). Consequently, despite better adherence, the final \textit{average\_score} actually declines ($0.14 \rightarrow 0.08$), illustrating a case where the cost of self-correction outweighs the benefits of deeper reasoning.
\end{itemize}

\finding{\textbf{Content-Process Divergence:} Reasoning capabilities introduce complex trade-offs rather than uniform improvements. Enhanced decision depth often correlates with degraded protocol adherence, as models internalize execution steps or exhibit instability due to over-analysis.}\label{fnd:reasoning}


\subsubsection{Impact of Architectural Decoupling}

This section investigates the impact of architectural evolution from monolithic local tools to distributed protocols. The results reveal a clear distinction. Tool decoupling (MCP) stabilizes performance. In contrast, agent decoupling (A2A/H-A2A) progressively separates intent from execution.

\textbf{1. MCP: Modularity via Protocol Enforcement.}
Refactoring local tools into independent services via MCP maintains behavioral consistency. It also yields significant reliability dividends. Fig.~\ref{fig:radar_2_cols} shows overlapping performance envelopes for \texttt{Markdown Validator} (Overall), \texttt{Game Builder} (Overall), and \texttt{Email Responder} (Overall). The transition imposes negligible latency or accuracy penalties on the LLM's reasoning loop. Both architectures achieve near-identical \textit{pass\_rate}, with MCP variant slightly higher ($1.00$ vs.\ $0.99$). This confirms that the protocol's serialization layer is effectively transparent to the model. In addition, the rigid schema of MCP simplifies the model's task. By enforcing a strict contract, MCP reduces the model's burden to hallucinate tool formats. 
This reduces the aggregate \textit{retry\_rate} ($0.04\!\rightarrow\!0.03$). This effect is amplified in reasoning models. For example, in \texttt{Email Responder} (Overall), \texttt{DeepSeek-R1}'s runtime \textit{retry\_rate} drops sharply ($0.28\!\rightarrow\!0.04$).  This validates that standardized interfaces serve as effective structural guardrails.

\begin{figure}[t]
    \centering
    \begin{minipage}[t]{0.35\linewidth}
        \centering
        \includegraphics[width=\linewidth]{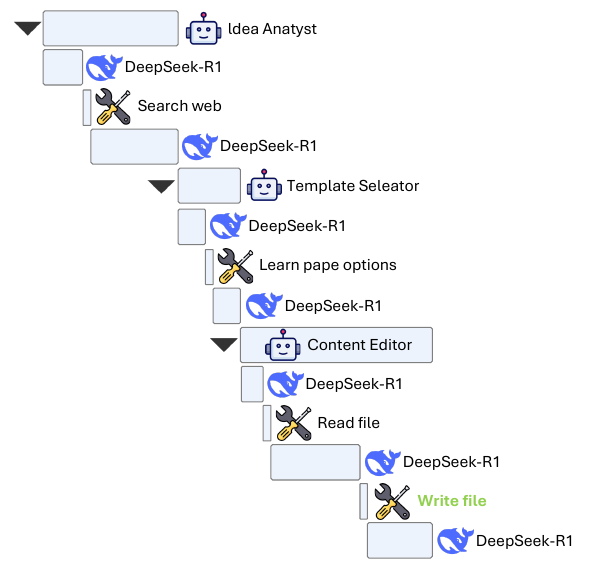}
        \centerline{\small (a)  Golden Trace }
    \end{minipage}%
    \hfill
    \begin{minipage}[t]{0.315\linewidth}
        \centering
        \includegraphics[width=\linewidth]{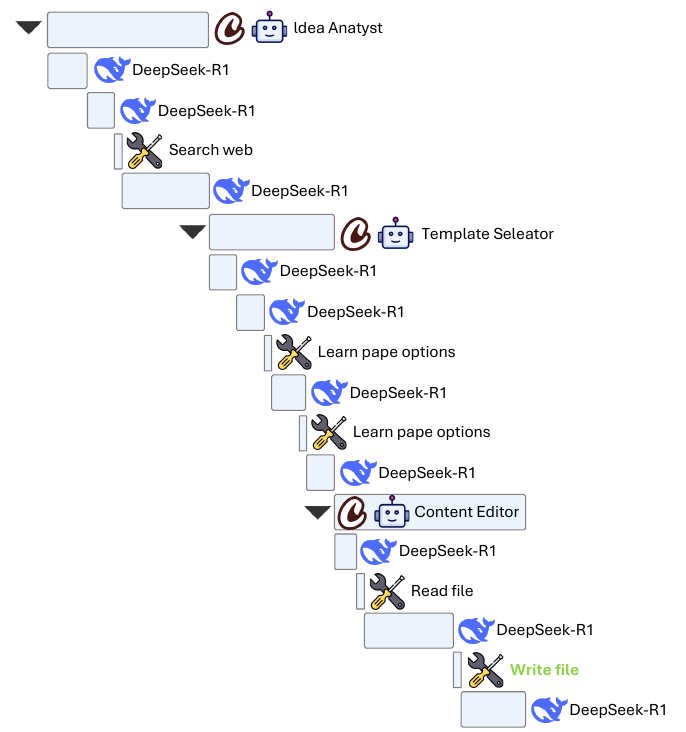}
        \centerline{\small (b) MCP  Runtime Trace }
    \end{minipage}%
    \hfill
    \begin{minipage}[t]{0.32\linewidth}
        \centering
        \includegraphics[width=\linewidth]{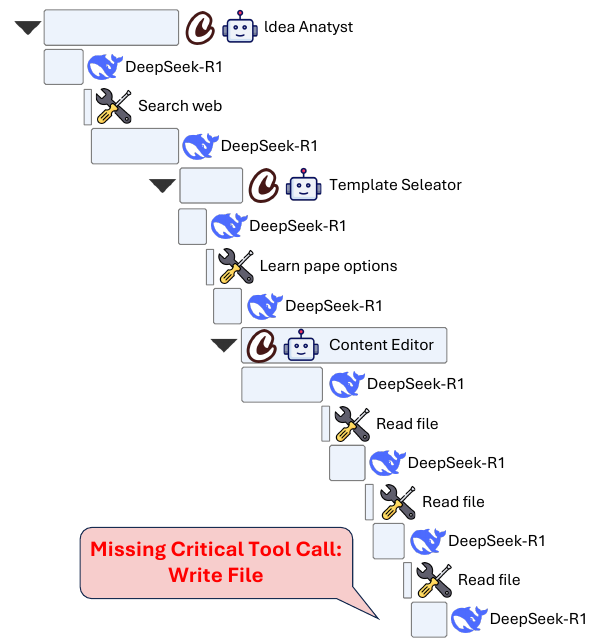}
        \centerline{\small (c) A2A Runtime Trace}
    \end{minipage}%
    
    \caption{Comparison of traces on \texttt{Landing Page Generator} under different architectures.}
    \label{fig:landing_page_case}
    \vspace{-0.2in}
\end{figure}

\textbf{2. A2A: The Selection-Adherence Gap.}
Unlike the stability of MCP, adopting the A2A protocol disrupts semantic continuity. Aggregated across all distributed applications (Fig.~\ref{fig:radar_3_cols}), switching to A2A triggers a regression in behavioral correctness (\textit{any\_order\_match} $0.29\!\rightarrow\!0.25$). This occurs despite a stable \textit{pass\_rate}.

This degradation is not a failure of discovery because \textit{recall} increases ($0.81\!\rightarrow\!0.84$). This reveals a ``Selection-Adherence Gap.'' Distributed agents successfully identify what to use (high \textit{recall}) but fail to execute when to use it (low alignment). The coordination overhead fragments the context, causing agents to deviate from optimal parameterization and sequencing despite correct tool selection.

The impact of this coordination overhead depends on the nature of the workload: 
(1) For noise-sensitive tasks like \texttt{SQL Assistant}, A2A's context isolation acts as a filter. \texttt{GPT-4o-mini} achieves a massive boost in trajectory alignment (\textit{any\_order\_match} $0.12\!\rightarrow\!0.53$) and improved runtime stability (\textit{retry\_rate} $0.65\!\rightarrow\!0.17$). This proves that compartmentalization benefits lightweight models by reducing context pollution. (2) Conversely, for dependency-heavy workflows like \texttt{Landing Page Generator}, isolation acts as a barrier. \texttt{DeepSeek-R1}'s performance collapses (\textit{average\_score} $0.58\!\rightarrow\!0.08$). We attribute this to context fragmentation severing the execution chain. As depicted in the case study (Fig.~\ref{fig:landing_page_case}), compared to the golden trace's complete execution sequence, the MCP runtime trace maintains full execution, while the A2A trace terminates prematurely. It successfully performs the \texttt{read} operation (validating retrieval capability) but consistently fails to execute the subsequent  \texttt{write} operation required to persist the output. This specific failure pattern is prevalent across the majority of test cases, serving as the primary driver for the precipitous drop in overall scores. This confirms that preventing downstream agents from accessing the full upstream reasoning chain destroys performance in coupled tasks.

\begin{figure}[t]
    \centering
    \begin{minipage}[t]{0.285\linewidth}
        \centering
        \includegraphics[width=\linewidth]{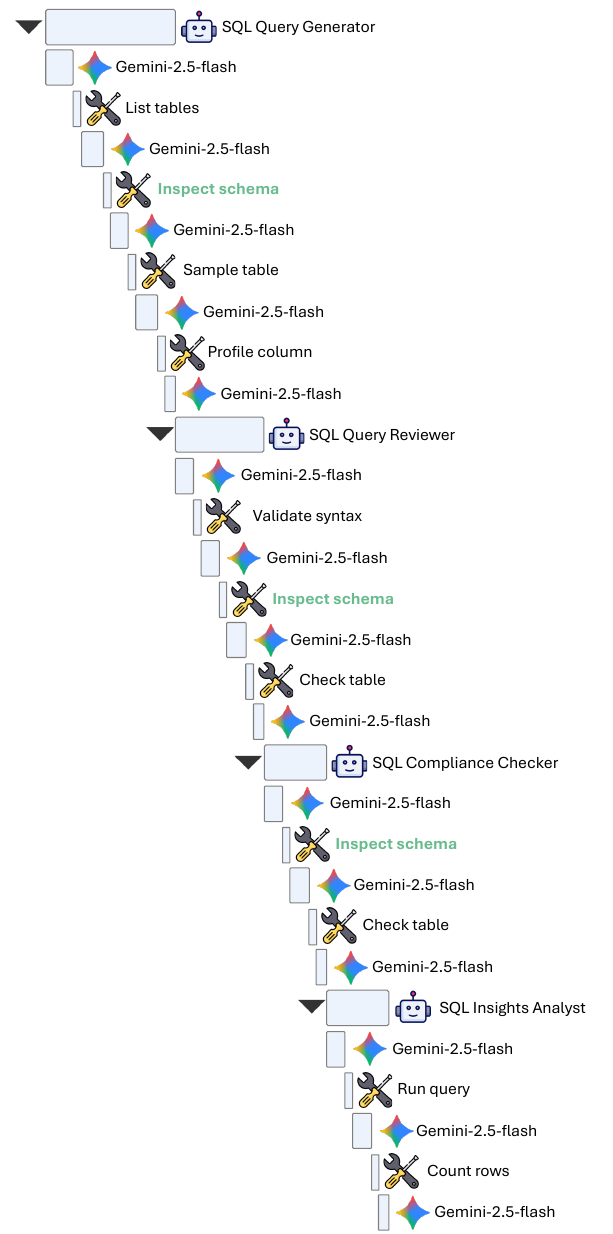}
        \centerline{\small (a) Golden Trace}
    \end{minipage}%
    \hfill
    \begin{minipage}[t]{0.33\linewidth}
        \centering
        \includegraphics[width=\linewidth]{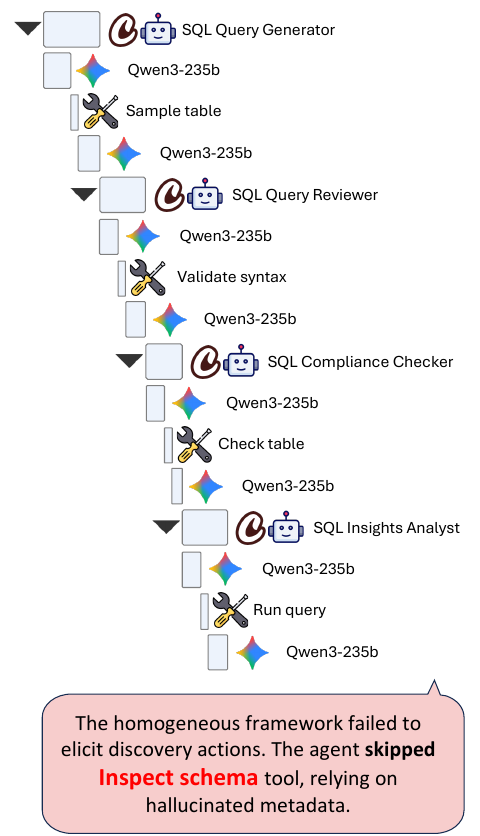}
        \centerline{\small (b) A2A Runtime Trace}
    \end{minipage}%
    \hfill
    \begin{minipage}[t]{0.34\linewidth}
        \centering
        \includegraphics[width=\linewidth]{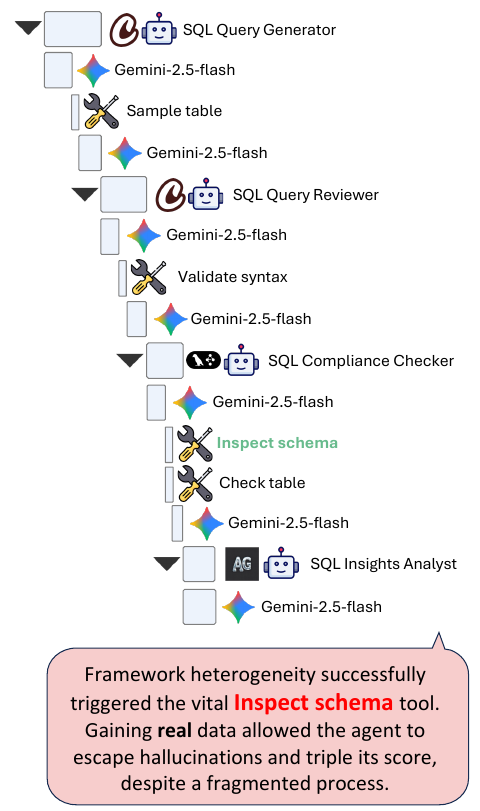}
        \centerline{\small (c) H-A2A  Runtime Trace}
    \end{minipage}%
    
    \caption{Comparison of traces on \texttt{SQL Assistant} under different architectures.}
    \label{fig:sql_case_study}
    \vspace{-0.2in}
\end{figure}

\textbf{3. H-A2A: The Outcome-Process Divergence.}
The transition to Heterogeneous A2A (H-A2A) represents the most extreme shift. It prioritizes functional completion over procedural correctness. As confirmed in the \texttt{Book Writer} radar charts (Fig.~\ref{fig:radar_3_cols}), the alignment axes collapse to near-zero. However, outcome axes remain saturated.

Quantitative analysis corroborates this. H-A2A boosts \textit{average\_score} ($0.31\!\rightarrow\!0.45$) but drives \textit{any\_order\_match} to negligible levels ($0.25\!\rightarrow\!0.19$). This creates an ``Outcome-Process Divergence.'' Agents achieve the final results not by following the engineering specification. Instead, they rely on opaque brute-force interactions that are mathematically misaligned with the intended protocol. This renders the system functionally effective but rigorous to maintain. The impact of heterogeneity is polarized by model capability:

(1) \texttt{Gemini-2.5-flash} presents a paradigmatic case of this divergence in \texttt{SQL Assistant} in Fig.~\ref{fig:radar_3_cols}. The architectural shift to H-A2A triggers a comprehensive improvement in outcome-oriented metrics: \textit{average\_score} almost triples ($0.16\!\rightarrow\!0.55$), \textit{precision} rises ($0.57\!\rightarrow\!0.81$), and operational instability is nearly halved (\textit{retry\_rate} $0.90\!\rightarrow\!0.50$). Yet, this success masks a critical paradox: trace analysis confirms that \textit{any\_order\_match} remains exactly $0.00$. As visualized in the case study (Fig.~\ref{fig:sql_case_study}), the framework transition serendipitously triggered critical data-retrieval actions—absent in the A2A execution—that allowed the agent to escape pure hallucination. However, the agent still failed to follow the standard execution process, achieving the correct result via a non-compliant trace rather than adherence to the engineering specification. This proves that H-A2A can unlock capabilities (improving outcomes) without remedying the fragmentation of the reasoning trace.

\begin{figure}[t]
    \centering
    \begin{minipage}[t]{0.37\linewidth}
        \centering
        \includegraphics[width=\linewidth]{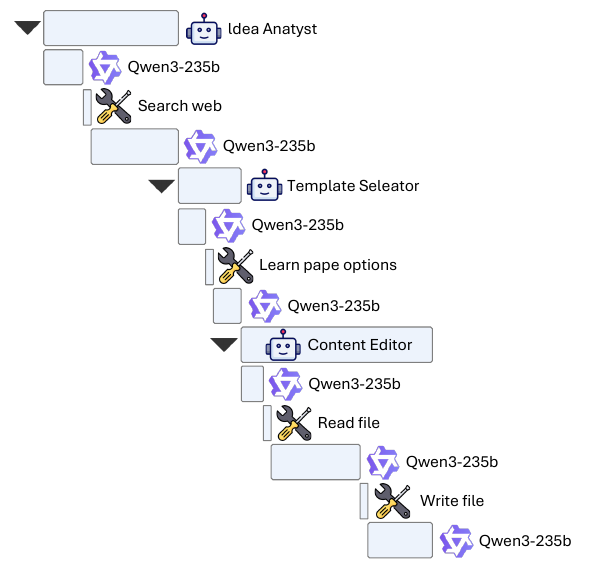}
        \centerline{\small (a)  Golden trace }
    \end{minipage}%
    \hfill
    \begin{minipage}[t]{0.29\linewidth}
        \centering
        \includegraphics[width=\linewidth]{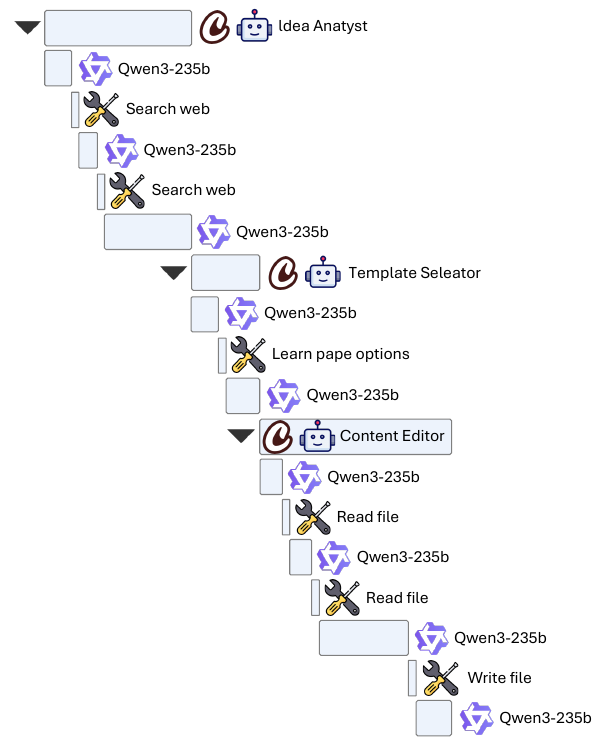}
        \centerline{\small (b) A2A Runtime Trace}
    \end{minipage}%
    \hfill
    \begin{minipage}[t]{0.255\linewidth}
        \centering
        \includegraphics[width=\linewidth]{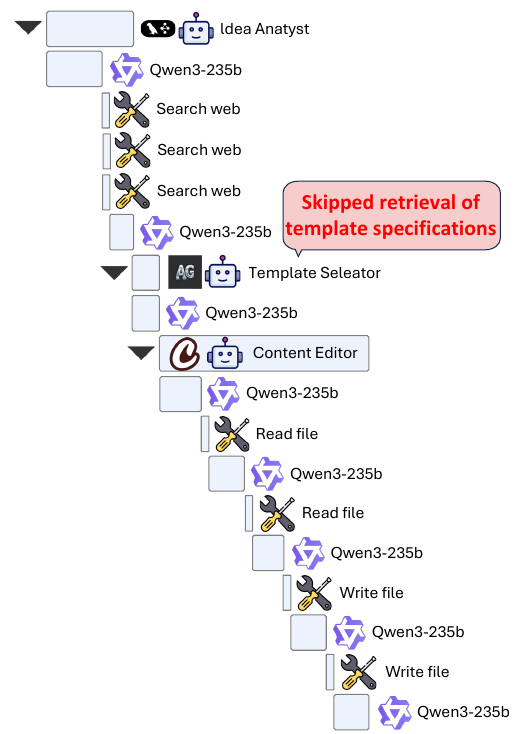}
        \centerline{\small (c) H-A2A Runtime Trace}
    \end{minipage}%
    
    \caption{Comparison of traces on \texttt{Landing Page Generator} under different architectures.}
    \label{fig:case_study_qwen}
    \vspace{-0.2in}
\end{figure}

(2) \texttt{Qwen3-235b} show a divergence between outcomes and processes in heterogeneous environments. Overall, the alignment metric for \texttt{Qwen3-235b} decreases (\textit{any\_order\_match} $0.43 \!\rightarrow\! 0.10$) while outcome scores remain stable. This pattern is evident in the \textit{Landing Page Generator} in Fig.~\ref{fig:radar_3_cols}. Switching to H-A2A results in an alignment score of zero (\textit{any\_order\_match} $0.48 \rightarrow 0.00$), whereas the average score increases ($0.38 \rightarrow 0.41$). These observations suggest that heterogeneity affects established execution paths. Fig.~\ref{fig:case_study_qwen} illustrates that while homogeneous agents are more likely to cover essential actions, the context fragmentation in H-A2A leads the agent to omit the step of retrieving available template specifications. The model uses parametric knowledge to generate the final artifact directly. Consequently, the trace differs from the golden trace, but remains functionally correct, prioritizing task completion over following a specific execution path.

\finding{
\textbf{Architectural Opacity}: The transition to heterogeneous distributed architectures (H-A2A) creates an outcome-process divergence. While improving functional completion rates via brute-force interactions, these architectures compromise procedural visibility and maintainability compared to monolithic protocol enforcement.}\label{fnd:architecture}




\subsection{RQ2: Dissecting Performance Overhead}\label{sec:rq2}
\label{sec:perf-breakdown}
This section dissects the system's performance profile by analyzing end-to-end application latency distributions (Fig.~\ref{fig:overall_ecdf}), computational breakdown (Fig.~\ref{fig:llm_share_heatmap}), and agent-level bottlenecks (Fig.~\ref{fig:agent_time_share_2_cols} and Fig.~\ref{fig:agent_time_share_3_cols}). The analysis yields critical insights regarding the interplay between model capability, architectural choice, and system optimization.

\subsubsection{Application-Level Breakdown}
We dissect the performance overhead at the application level to characterize the system's temporal behavior. Our analysis highlights two observations: latency varies widely across models, and inference time dominates end-to-end latency.

\textbf{1. The Tail Latency Trap.}
The end-to-end latency distribution (Fig.~\ref{fig:overall_ecdf}) reveals a severe ``long-tail'' phenomenon driven by model selection. Quantitative results across all $21$ projects show substantial cross-model divergence: \texttt{GPT-5} serves as the lowest-latency baseline in most scenarios, with a global mean of $45.52$~s. More critically, a clear gap emerges within the same model family: the reasoning-enhanced \texttt{DeepSeek-R1} averages $589.60$~s, versus $191.58$~s for the non-reasoning \texttt{DeepSeek-V3.1}, i.e., an approximately $3\times$ slowdown. This extreme variance indicates that reasoning models should be employed with caution in latency-sensitive agent systems. Specifically, the stochastic nature of the thinking process undermines the deterministic timing required for real-time interaction.

\begin{figure}[t!]
    \centering
    \includegraphics[width=\textwidth]{fig/Failure_Breakdown/Model_Legend_horizontal.pdf}

    \begin{minipage}{0.1\textwidth} \hfill \end{minipage}
    \begin{minipage}{0.05\textwidth}
        \rotatebox{90}{\scriptsize \textbf{Email Responder}}
    \end{minipage}%
    \begin{minipage}{0.35\textwidth}
        \includegraphics[width=\linewidth]{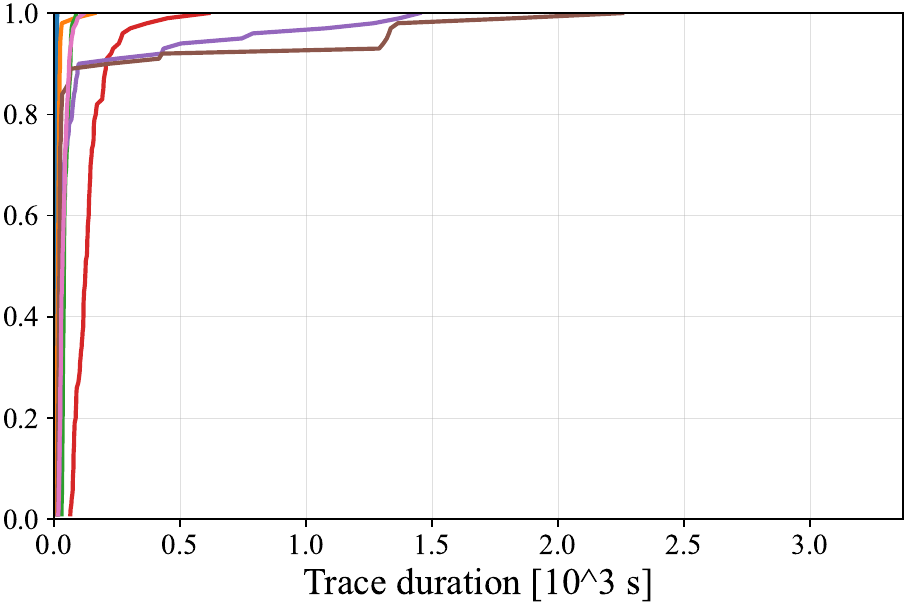}
    \end{minipage}%
    \hfill
    \begin{minipage}{0.35\textwidth}
        \includegraphics[width=\linewidth]{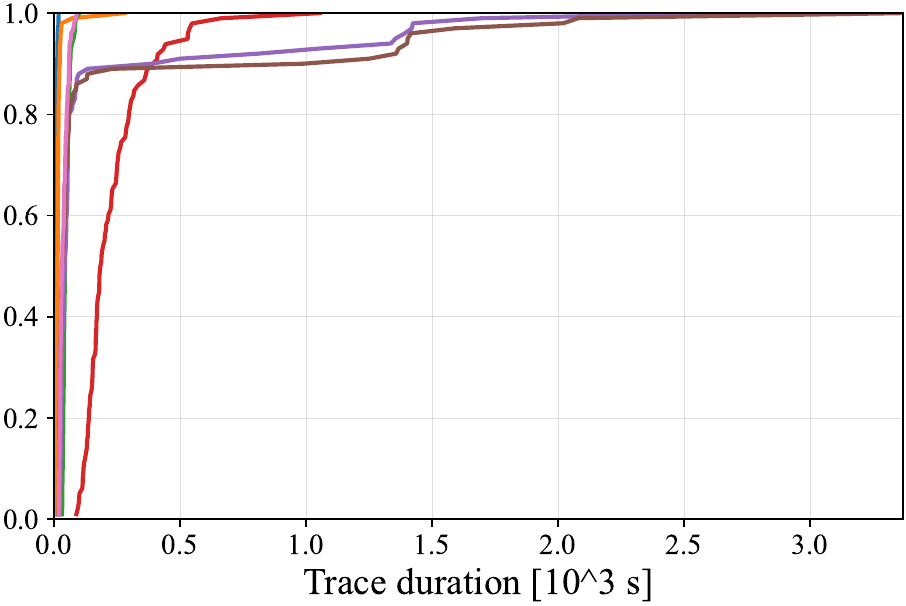}
    \end{minipage}
    \begin{minipage}{0.1\textwidth} \hfill \end{minipage}
    \vspace{0.4em}

    \begin{minipage}{0.1\textwidth} \hfill \end{minipage}
    \begin{minipage}{0.05\textwidth}
        \rotatebox{90}{\scriptsize \textbf{Game Builder}}
    \end{minipage}%
    \begin{minipage}{0.35\textwidth}
        \includegraphics[width=\linewidth]{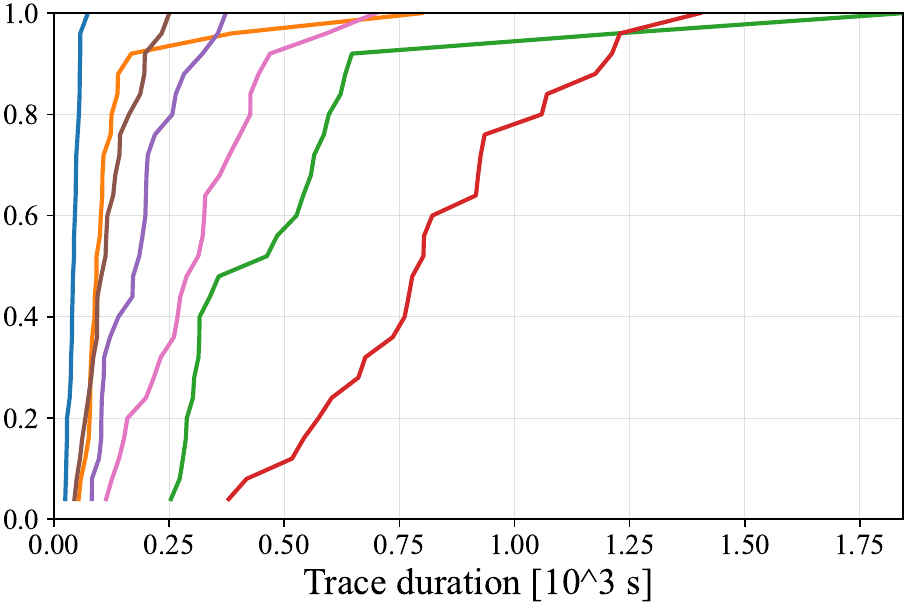}
    \end{minipage}%
    \hfill
    \begin{minipage}{0.35\textwidth}
        \includegraphics[width=\linewidth]{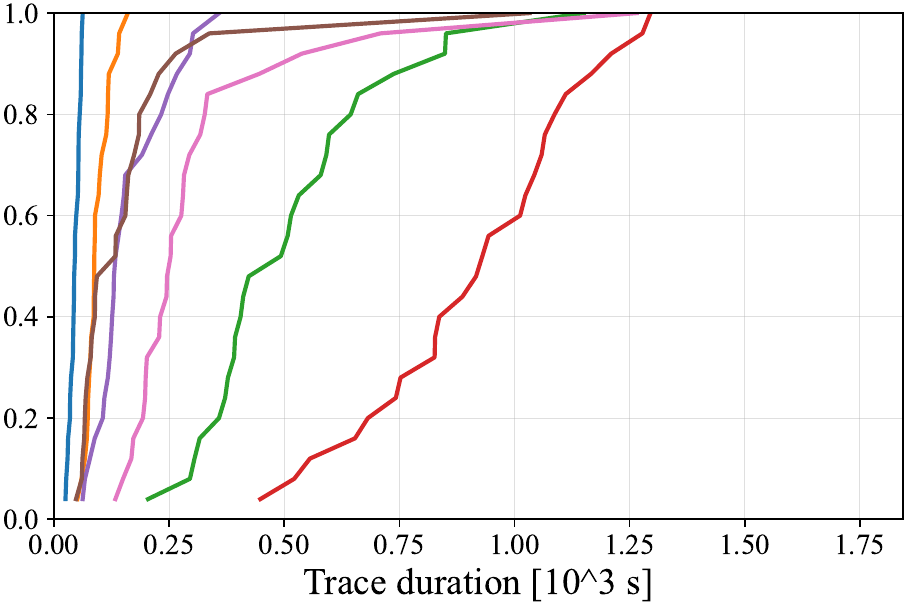}
    \end{minipage}
    \begin{minipage}{0.1\textwidth} \hfill \end{minipage}
    \vspace{0.4em}

    \begin{minipage}{0.1\textwidth} \hfill \end{minipage}
    \begin{minipage}{0.05\textwidth}
        \rotatebox{90}{\scriptsize \textbf{Markdown Val.}}
    \end{minipage}%
    \begin{minipage}{0.35\textwidth}
        \includegraphics[width=\linewidth]{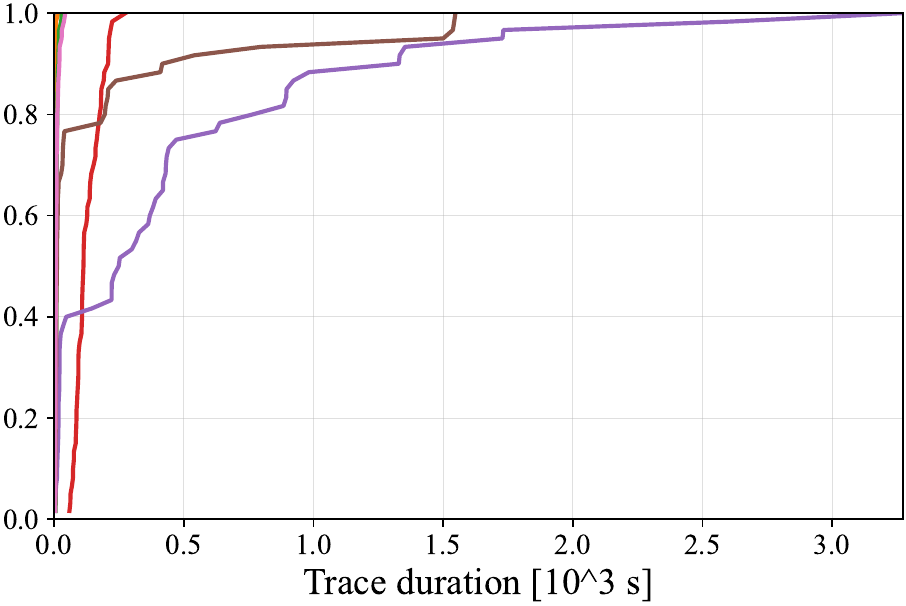}
    \end{minipage}%
    \hfill
    \begin{minipage}{0.35\textwidth}
        \includegraphics[width=\linewidth]{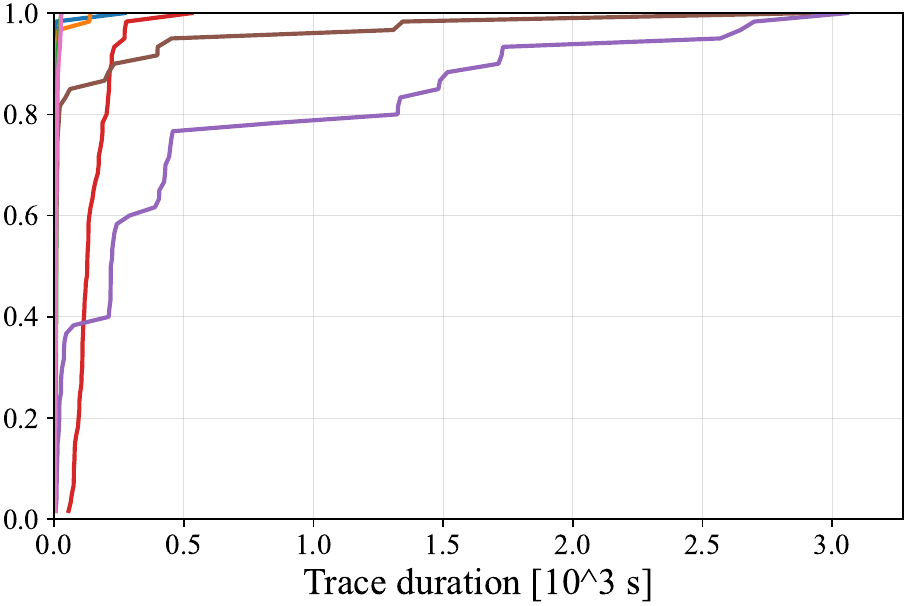}
    \end{minipage}
    \begin{minipage}{0.1\textwidth} \hfill \end{minipage}

    \vspace{0.3em}
    \begin{minipage}{0.15\textwidth} \hfill \end{minipage}
    \begin{minipage}{0.35\textwidth} \centering \textbf{\small{Pure CrewAI}} \end{minipage}
    \hfill
    \begin{minipage}{0.35\textwidth} \centering \textbf{\small{MCP}} \end{minipage}
    \begin{minipage}{0.1\textwidth} \hfill \end{minipage}
    \caption{End-to-end latency distributions for applications with two variants.}
    \label{fig:ecdf_2_cols}
    \vspace{-0.2in}

\end{figure}

\begin{figure*}[t!] 
    \centering
    \includegraphics[width=\textwidth]{fig/Failure_Breakdown/Model_Legend_horizontal.pdf}

    \begin{minipage}{0.03\textwidth}
        \rotatebox{90}{\scriptsize \textbf{SQL Asst.}}
    \end{minipage}%
    \begin{minipage}{0.32\textwidth}
        \includegraphics[width=\linewidth]{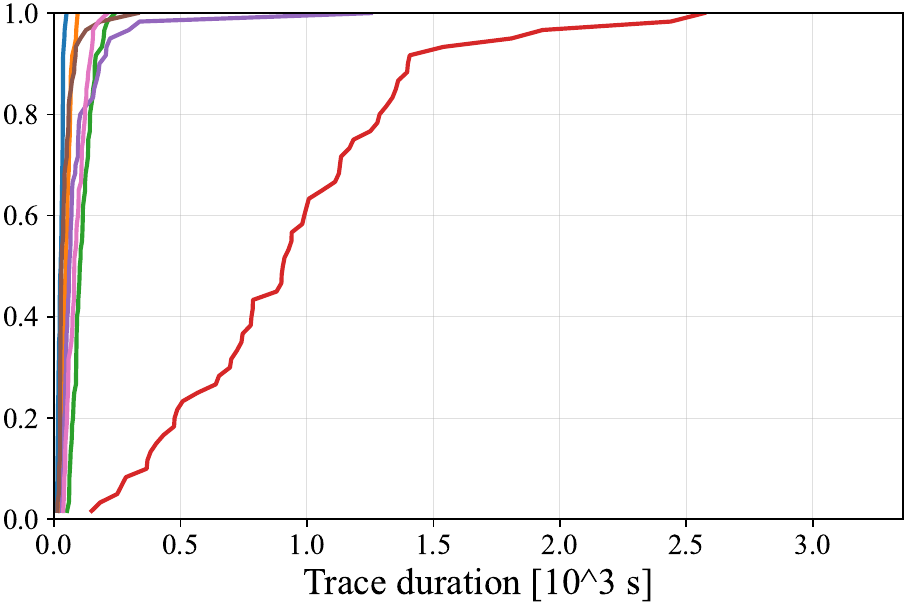}
    \end{minipage}%
    \begin{minipage}{0.32\textwidth}
        \includegraphics[width=\linewidth]{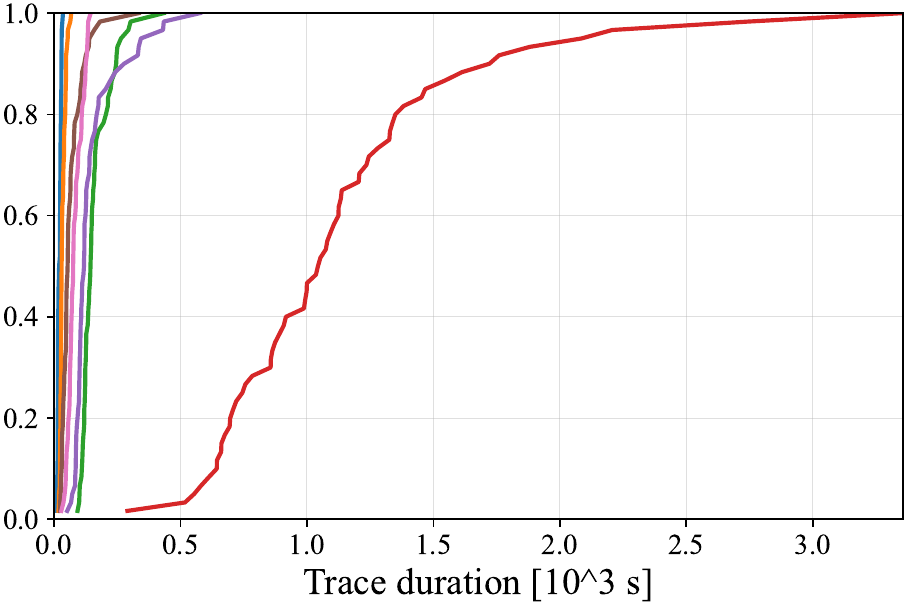}
    \end{minipage}%
    \begin{minipage}{0.32\textwidth}
        \includegraphics[width=\linewidth]{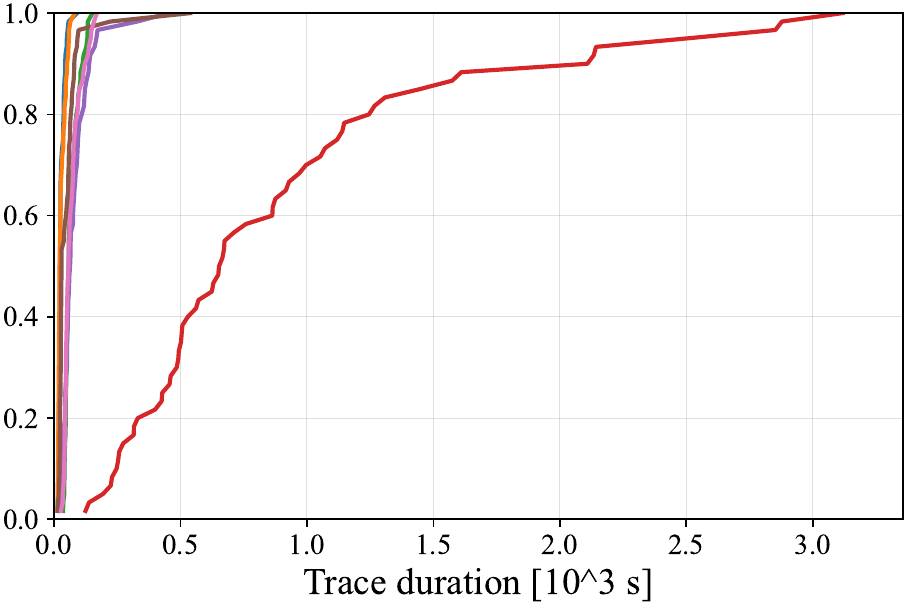}
    \end{minipage}
    \vspace{0.4em}

    \begin{minipage}{0.03\textwidth}
        \rotatebox{90}{\scriptsize \textbf{Recruitment}}
    \end{minipage}%
    \begin{minipage}{0.32\textwidth}
        \includegraphics[width=\linewidth]{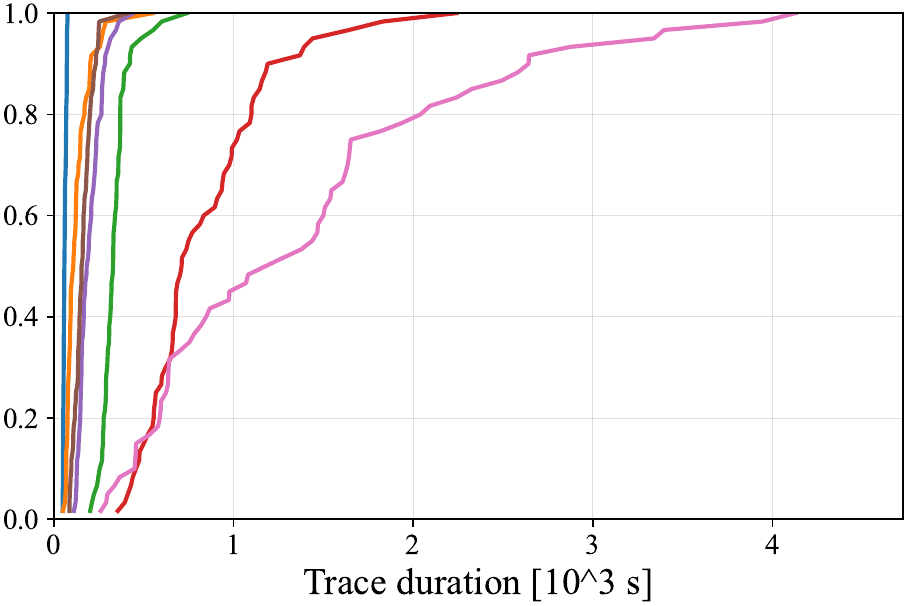}
    \end{minipage}%
    \begin{minipage}{0.32\textwidth}
        \includegraphics[width=\linewidth]{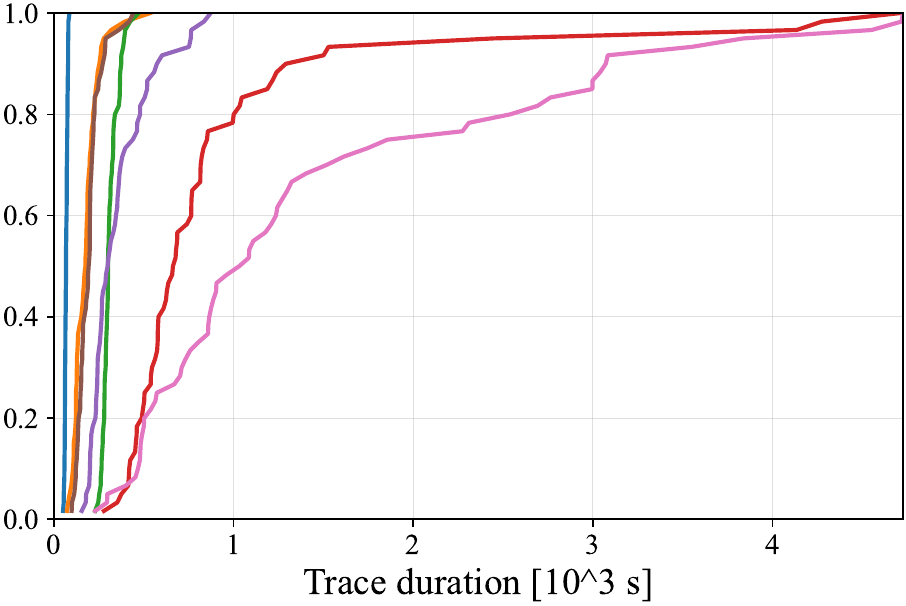}
    \end{minipage}%
    \begin{minipage}{0.32\textwidth}
        \includegraphics[width=\linewidth]{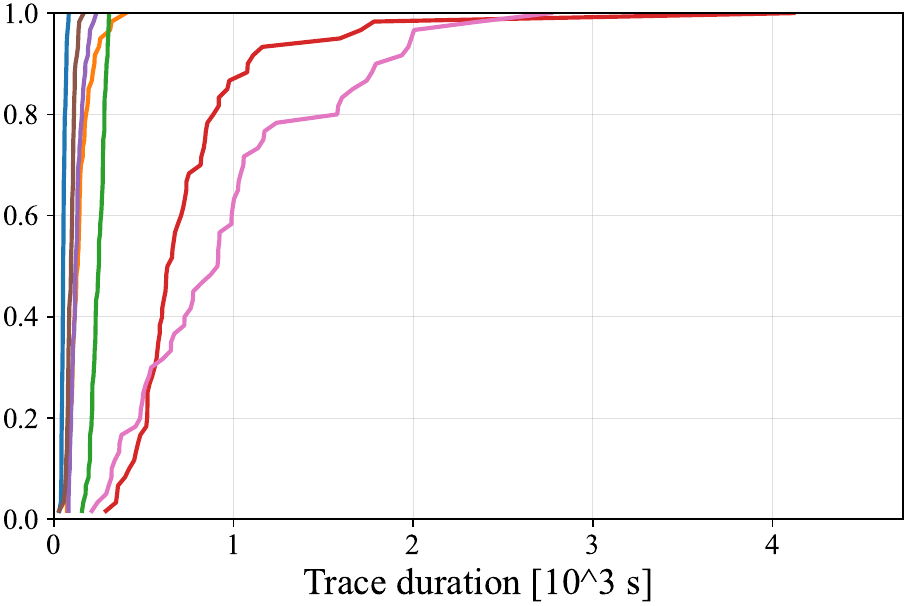}
    \end{minipage}
    \vspace{0.4em}
    
    \begin{minipage}{0.03\textwidth}
        \rotatebox{90}{\scriptsize \textbf{Landing Pg.}}
    \end{minipage}%
    \begin{minipage}{0.32\textwidth}
        \includegraphics[width=\linewidth]{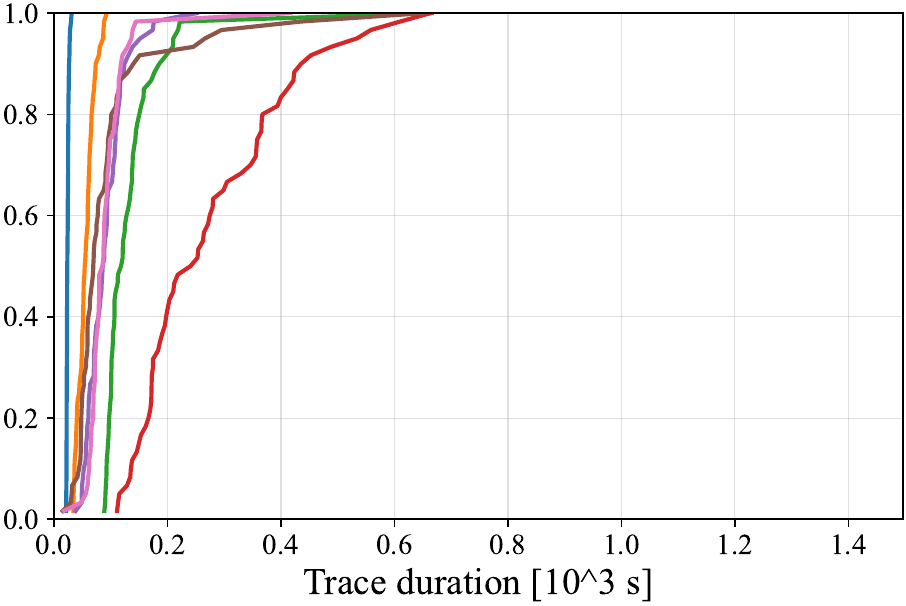}
    \end{minipage}%
    \begin{minipage}{0.32\textwidth}
        \includegraphics[width=\linewidth]{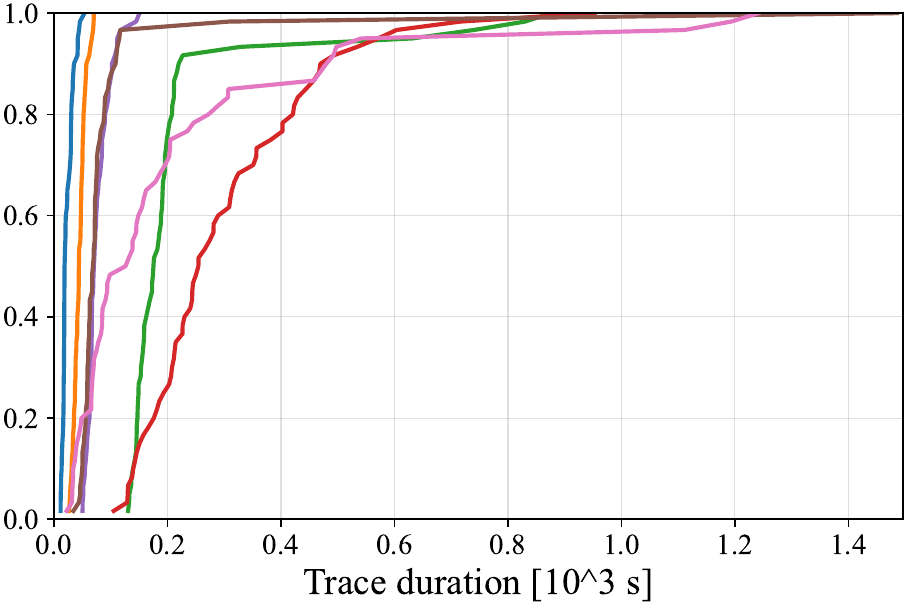}
    \end{minipage}%
    \begin{minipage}{0.32\textwidth}
        \includegraphics[width=\linewidth]{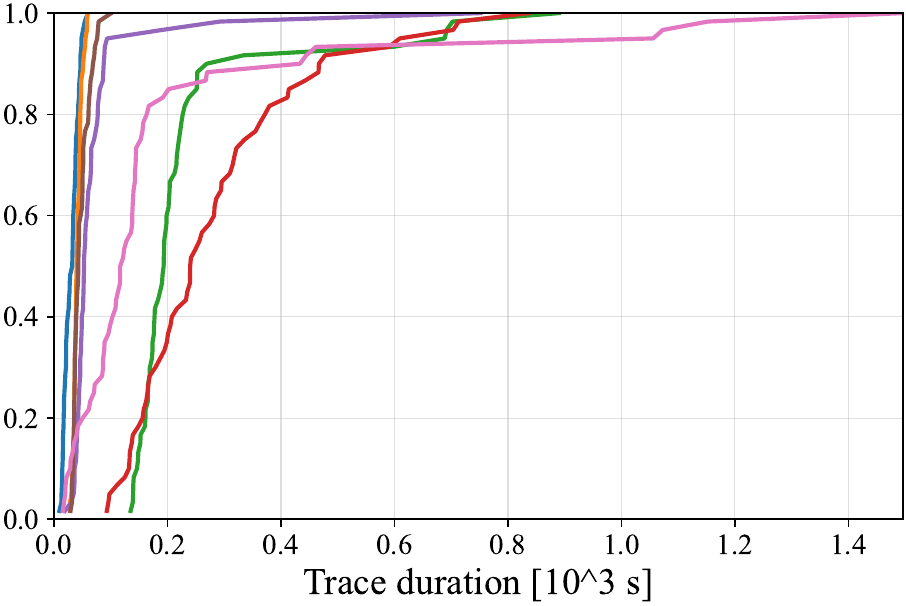}
    \end{minipage}
    \vspace{0.4em}

    \begin{minipage}{0.03\textwidth}
        \rotatebox{90}{\scriptsize \textbf{Social M. M.}}
    \end{minipage}%
    \begin{minipage}{0.32\textwidth}
        \includegraphics[width=\linewidth]{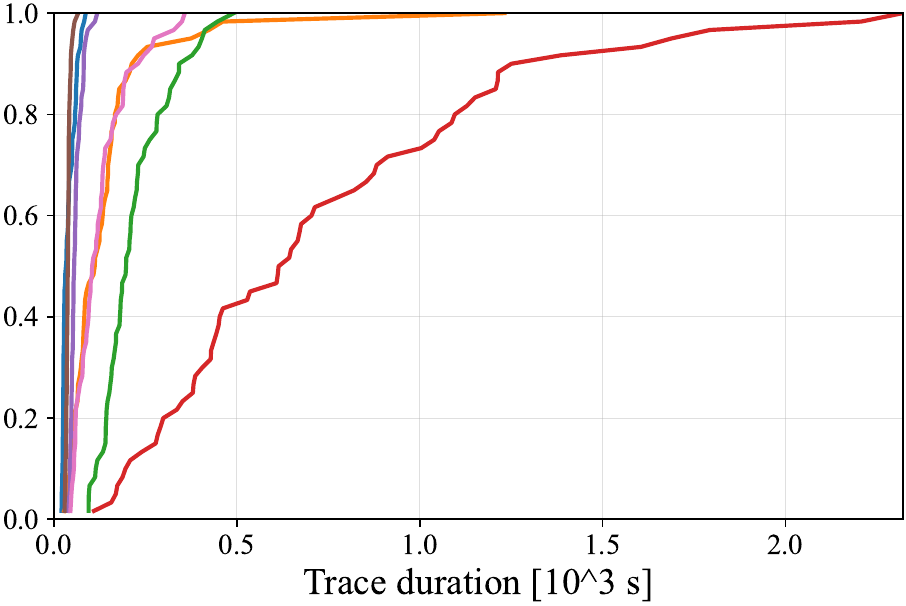}
    \end{minipage}%
    \begin{minipage}{0.32\textwidth}
        \includegraphics[width=\linewidth]{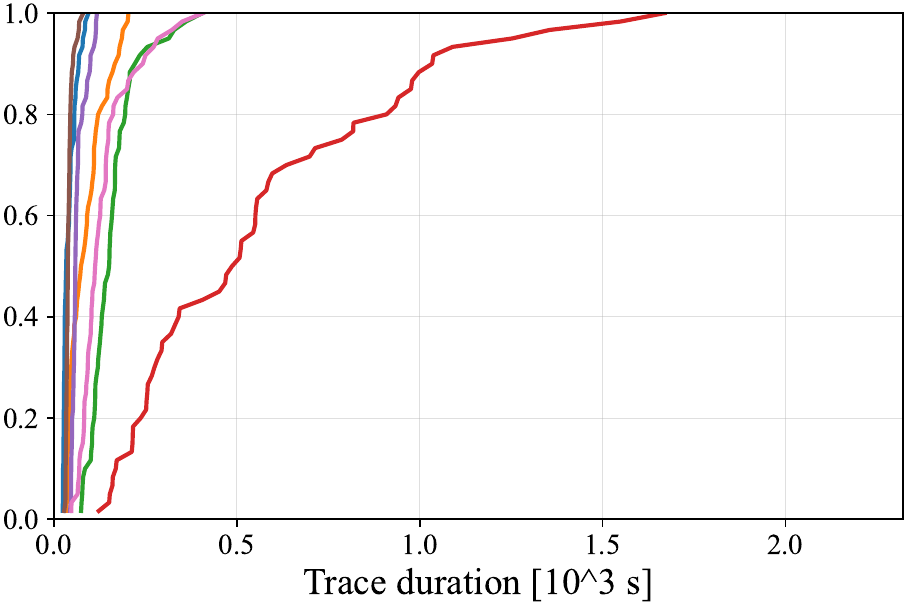}
    \end{minipage}%
    \begin{minipage}{0.32\textwidth}
        \includegraphics[width=\linewidth]{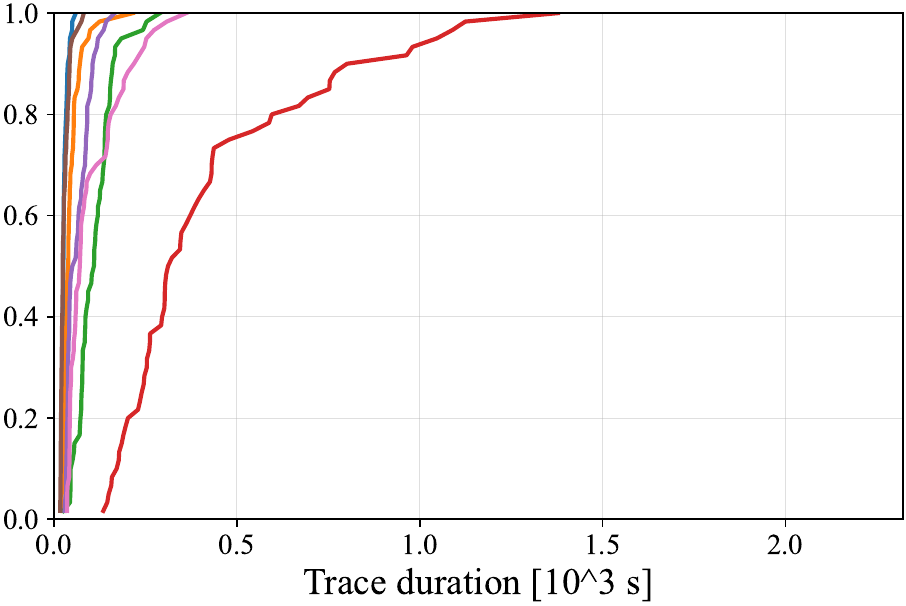}
    \end{minipage}
    \vspace{0.4em}

    \begin{minipage}{0.03\textwidth}
        \rotatebox{90}{\scriptsize \textbf{Book Writer}}
    \end{minipage}%
    \begin{minipage}{0.32\textwidth}
        \includegraphics[width=\linewidth]{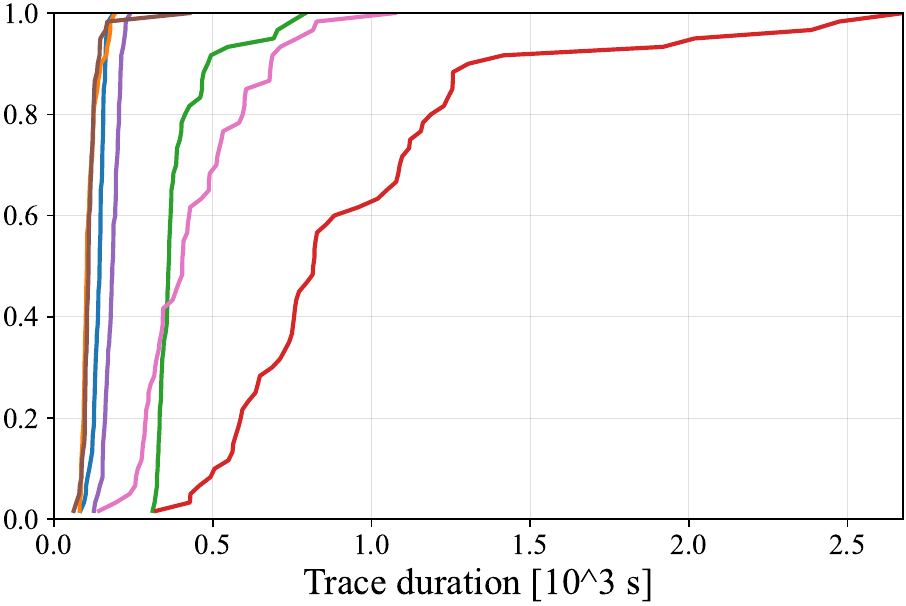}
    \end{minipage}%
    \begin{minipage}{0.32\textwidth}
        \includegraphics[width=\linewidth]{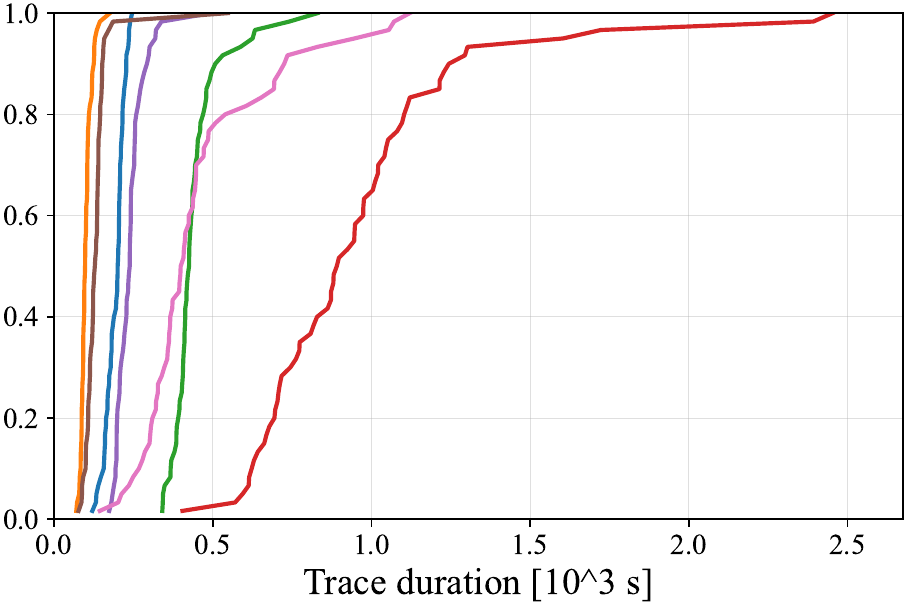}
    \end{minipage}%
    \begin{minipage}{0.32\textwidth}
        \includegraphics[width=\linewidth]{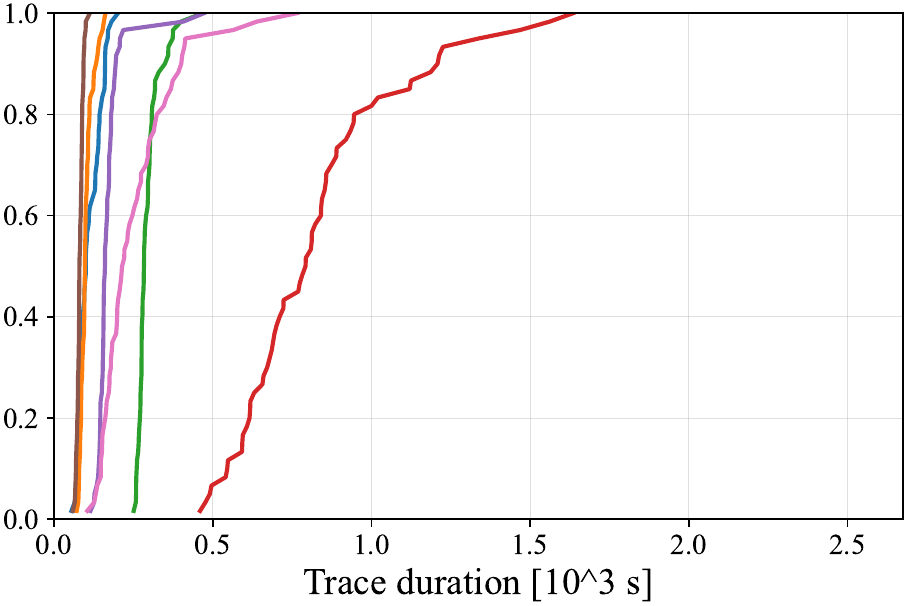}
    \end{minipage}

    \vspace{0.3em}
    \begin{minipage}{0.03\textwidth} \hfill \end{minipage}%
    \begin{minipage}{0.32\textwidth} \centering \textbf{MCP} \end{minipage}
    \begin{minipage}{0.32\textwidth} \centering \textbf{A2A} \end{minipage}
    \begin{minipage}{0.32\textwidth} \centering \textbf{H-A2A} \end{minipage}

    \caption{End-to-end latency distributions for applications with three variants. }
    \label{fig:ecdf_3_cols}
    \vspace{-0.2in}
\end{figure*}

This pattern is also visible in Fig.~\ref{fig:ecdf_2_cols} and Fig.~\ref{fig:ecdf_3_cols}. The choice of a reasoning model acts as the primary driver of latency degradation, overshadowing task complexity or architecture. This impact is most visible in the complex applications such as \texttt{SQL Assistant} and \texttt{Book Writer}. In \texttt{SQL Assistant}, \texttt{DeepSeek-R1} introduces a staggering delay: under the A2A variant, its mean latency increases by over $+5000\%$ ($1130.69$~s vs.\ $21.10$~s) relative to \texttt{GPT-5}. Similarly, across the \texttt{Book Writer} variants, \texttt{DeepSeek-R1} consistently creates a bottleneck, slowing the A2A workflow by approximately $9.5\times$ relative to \texttt{GPT-4o-mini}. Critically, this slowdown persists even in lightweight tasks: in \texttt{Email Responder} (Pure CrewAI), \texttt{DeepSeek-R1} is approximately $24\times$ slower than \texttt{GPT-5} ($233.49$~s vs.\ $9.41$~s). This validates that the latency penalty is intrinsic to the reasoning model and far exceeds the impact of network or architectural overhead.

\finding{\textbf{Latency Volatility:} AI-Native systems exhibit long-tail latency distributions influenced by model selection. Reasoning models introduce inherent latency overheads, reducing real-time predictability across varying task complexities.}\label{fnd:tail_latency}


\begin{figure*}[t]
  \centering
  \includegraphics[width=\textwidth]{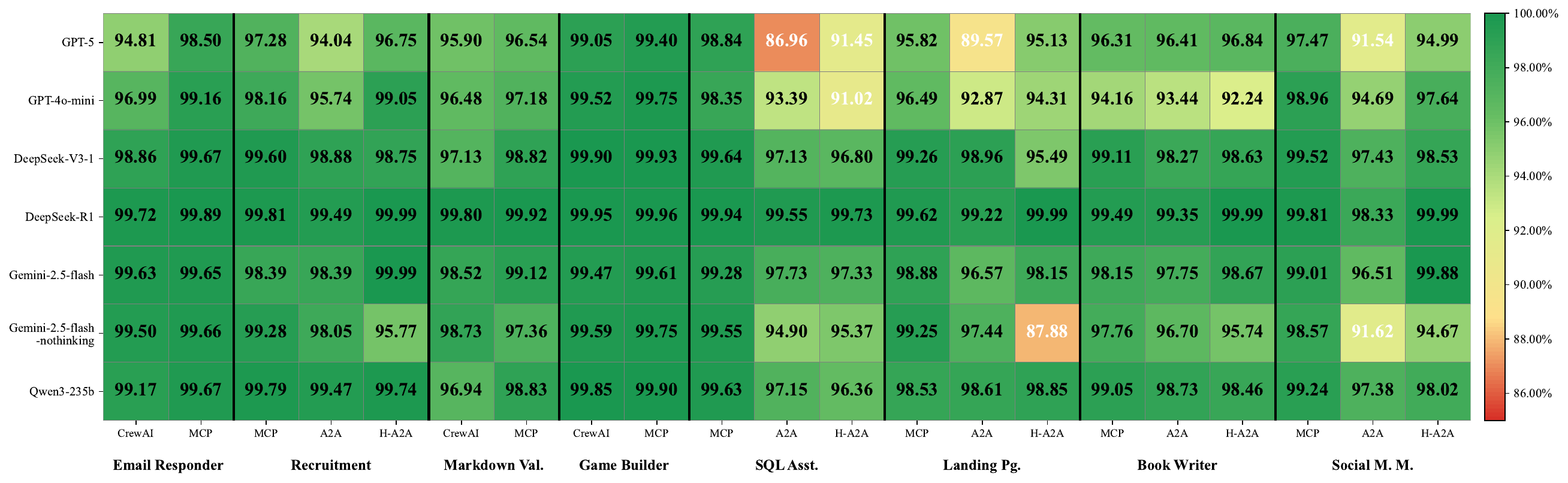}
  \vspace{-0.2in}
  \caption{End-to-end latency spent on LLM computation for seven models (rows) across application (columns).}
  \label{fig:llm_share_heatmap}
  \vspace{-0.2in}
\end{figure*}

\textbf{2. The Inference Dominance.}
To pinpoint the root cause of the observed latency, the end-to-end duration is decomposed into LLM inference time versus non-LLM overhead (e.g., network I/O, serialization, and framework logic). Fig.~\ref{fig:llm_share_heatmap} presents the ratio of LLM computation time across all experimental configurations.

Fig.~\ref{fig:llm_share_heatmap} reveals a pervasive inference dominance. Across the vast majority of model-task combinations, LLM computation overwhelms all other factors. It accounts for $86.9\%$ to $99.9\%$ of the total execution time. This dominance reaches its peak with reasoning models. \texttt{DeepSeek-R1} consistently consumes over $99\%$ of the latency budget. For instance, it reaches $99.99\%$ in the H-A2A variants of \texttt{Recruitment Assistant} and \texttt{Landing Page Generator}. This leaves less than $0.1\%$ for the entire distributed infrastructure. This breakdown fundamentally alters the cost-benefit analysis of distributed agents. Contrary to concerns that multi-agent protocols introduce unacceptable delays, the results indicate that protocol overhead is statistically negligible relative to inference time. Consequently, the benefits derived from enhanced maintainability, fault isolation, and independent scalability outweigh the minimal latency cost. Complex decoupled topologies (A2A/H-A2A) can be adopted without fear of performance regression.


\finding{\textbf{Inference Dominance:} LLM inference constitutes 86.9\% to 99.9\% of end-to-end latency. This dominance renders protocol overhead statistically secondary, particularly for reasoning-intensive models, facilitating the adoption of complex architectures.}\label{fnd:latency_dominance}


\subsubsection{Agent-Level Breakdown}

\begin{figure*}[t!]

    \begin{minipage}{0.1\textwidth} \hfill \end{minipage}
    \begin{minipage}{0.05\textwidth}
        \rotatebox{90}{\scriptsize \textbf{Email Responder}}
    \end{minipage}%
    \begin{minipage}{0.35\textwidth}
        \includegraphics[width=\linewidth]{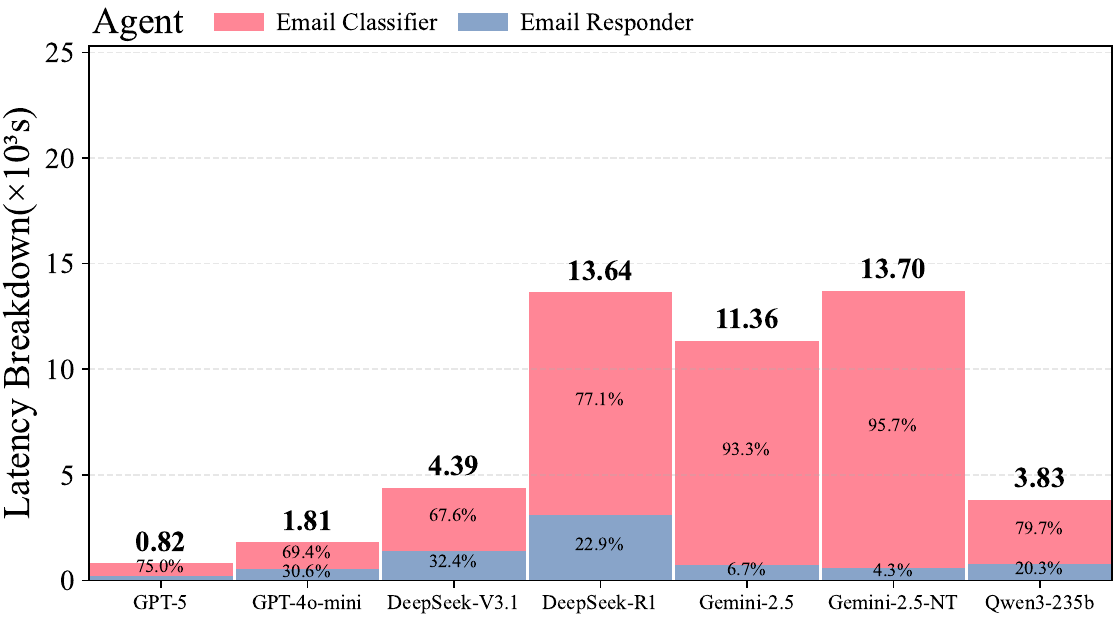}
    \end{minipage}%
    \hfill
    \begin{minipage}{0.35\textwidth}
        \includegraphics[width=\linewidth]{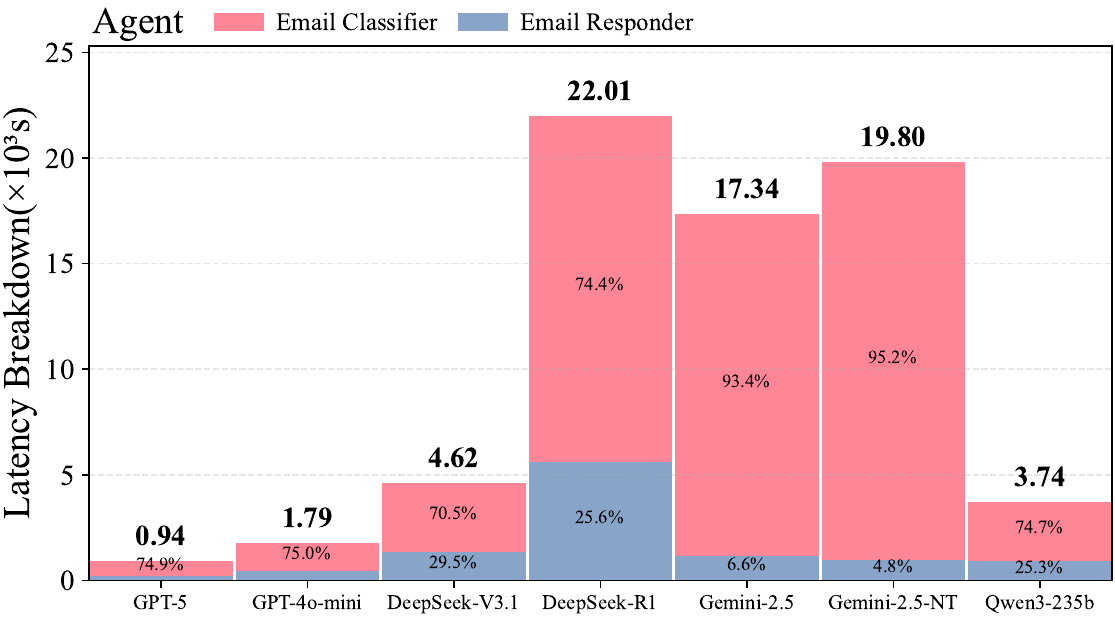}
    \end{minipage}
    \begin{minipage}{0.1\textwidth} \hfill \end{minipage}
    \vspace{0.4em}

    \begin{minipage}{0.1\textwidth} \hfill \end{minipage}
    \begin{minipage}{0.05\textwidth}
        \rotatebox{90}{\scriptsize \textbf{Game Builder}}
    \end{minipage}%
    \begin{minipage}{0.35\textwidth}
        \includegraphics[width=\linewidth]{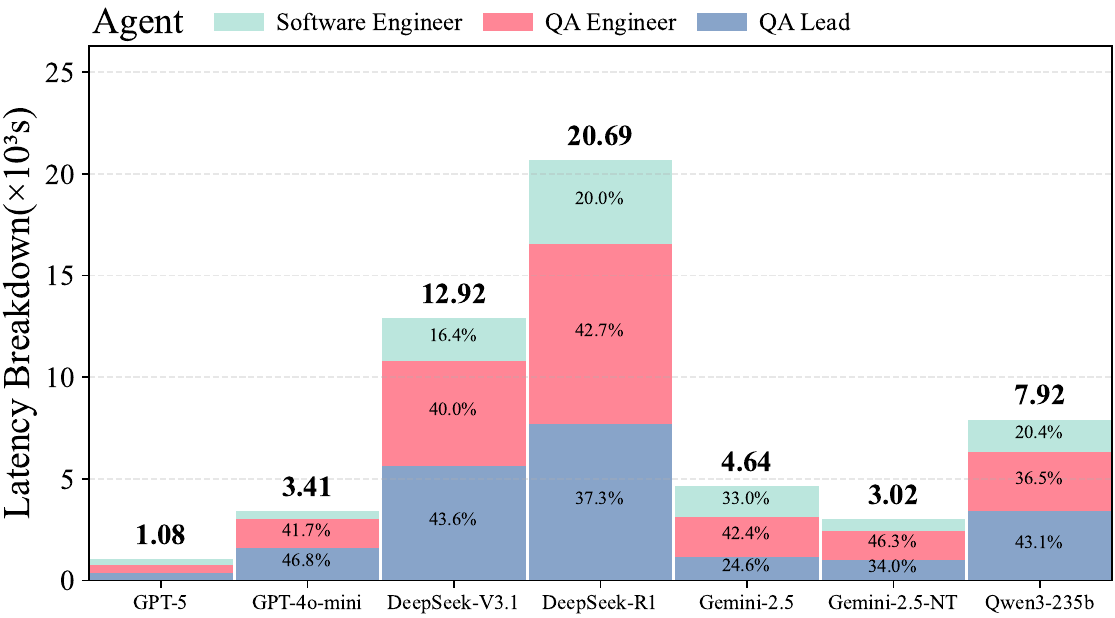}
    \end{minipage}%
    \hfill
    \begin{minipage}{0.35\textwidth}
        \includegraphics[width=\linewidth]{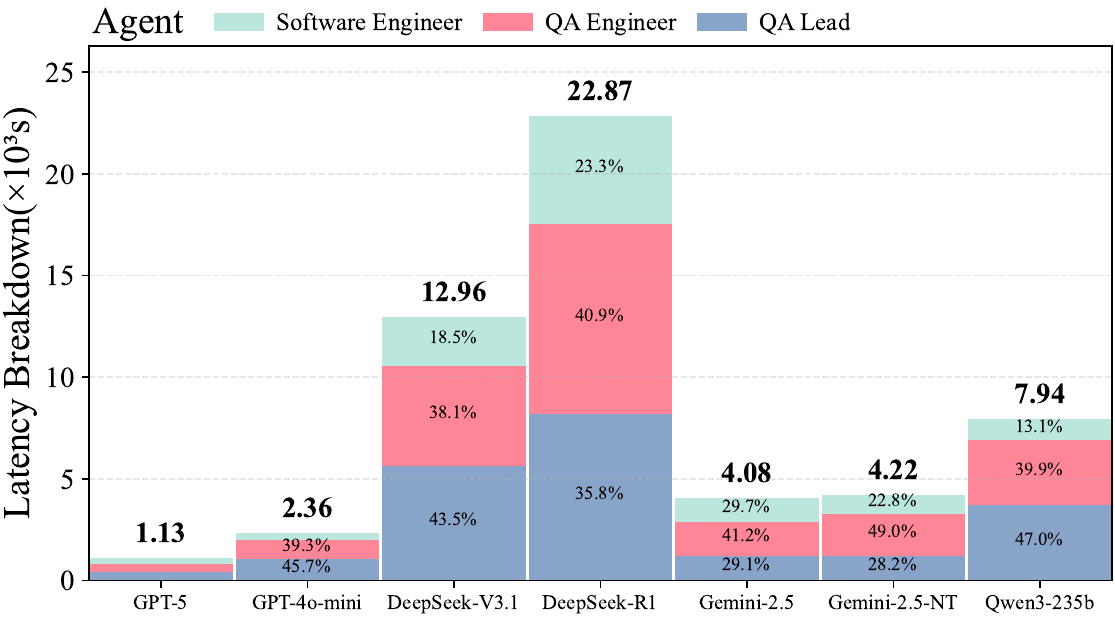}
    \end{minipage}
    \begin{minipage}{0.1\textwidth} \hfill \end{minipage}
    \vspace{0.4em}

    \begin{minipage}{0.1\textwidth} \hfill \end{minipage}
    \begin{minipage}{0.05\textwidth}
        \rotatebox{90}{\scriptsize \textbf{Markdown Val.}}
    \end{minipage}%
    \begin{minipage}{0.35\textwidth}
        \includegraphics[width=\linewidth]{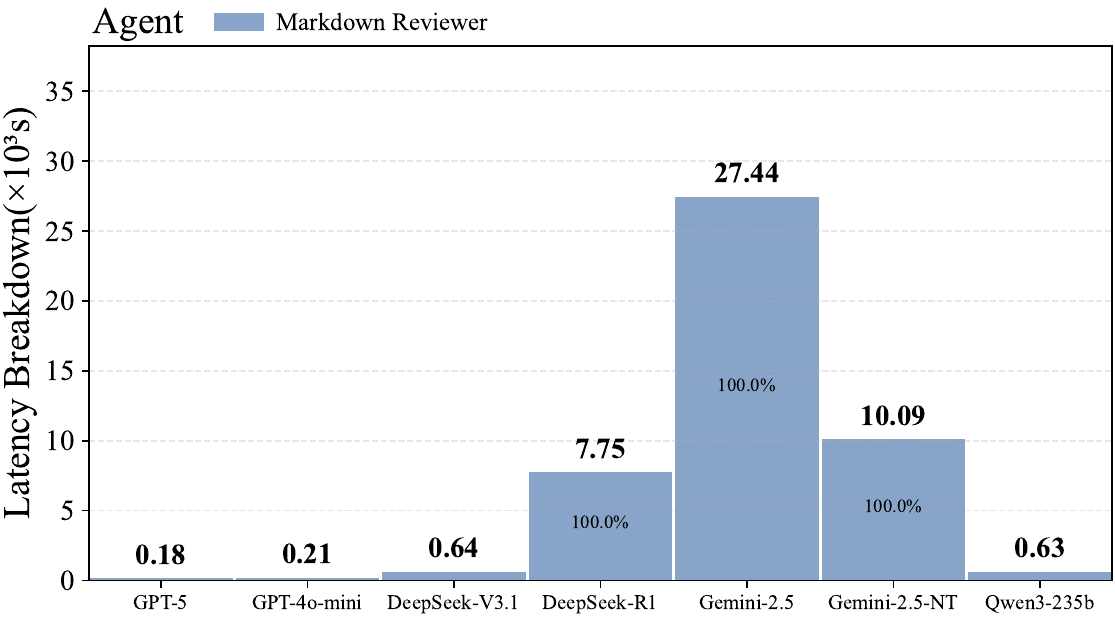}
    \end{minipage}%
    \hfill
    \begin{minipage}{0.35\textwidth}
        \includegraphics[width=\linewidth]{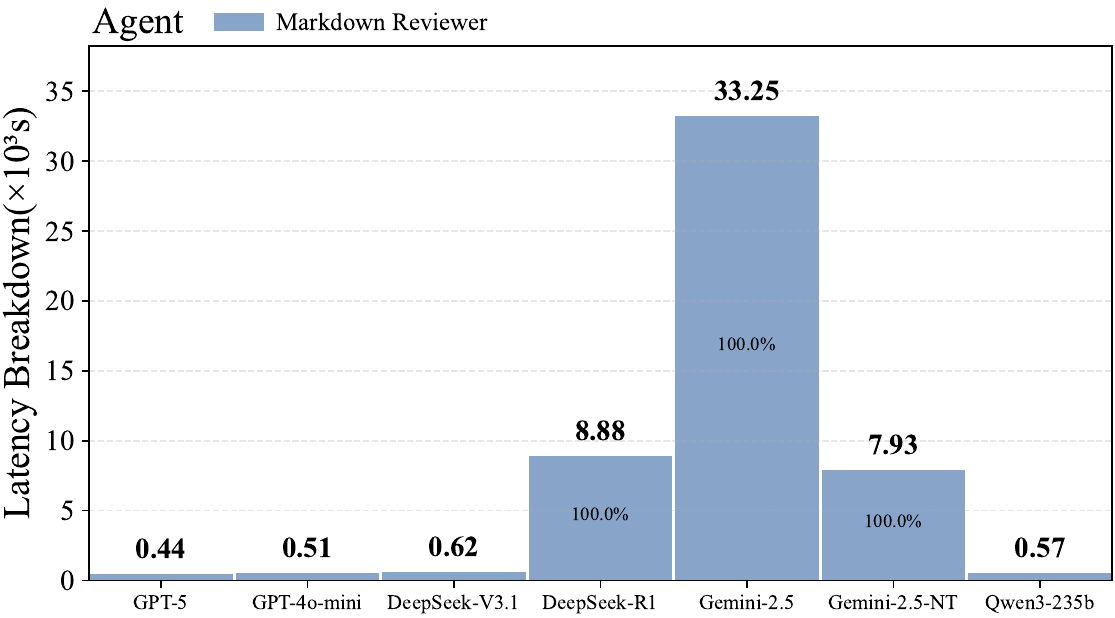}
    \end{minipage}
    \begin{minipage}{0.1\textwidth} \hfill \end{minipage}

    \vspace{0.3em}
    \begin{minipage}{0.15\textwidth} \hfill \end{minipage}
    \begin{minipage}{0.35\textwidth} \centering \textbf{\small{Pure CrewAI}} \end{minipage}
    \hfill
    \begin{minipage}{0.35\textwidth} \centering \textbf{\small{MCP}} \end{minipage}
    \begin{minipage}{0.1\textwidth} \hfill \end{minipage}
    \caption{Comparison of agent latency breakdown for applications with two variants.}
    \label{fig:agent_time_share_2_cols}
    \vspace{-0.2in}
\end{figure*}

\begin{figure*}[t]
    \begin{minipage}{0.03\textwidth}
        \rotatebox{90}{\scriptsize \textbf{SQL Asst.}}
    \end{minipage}%
    \begin{minipage}{0.32\textwidth}
        \includegraphics[width=\linewidth]{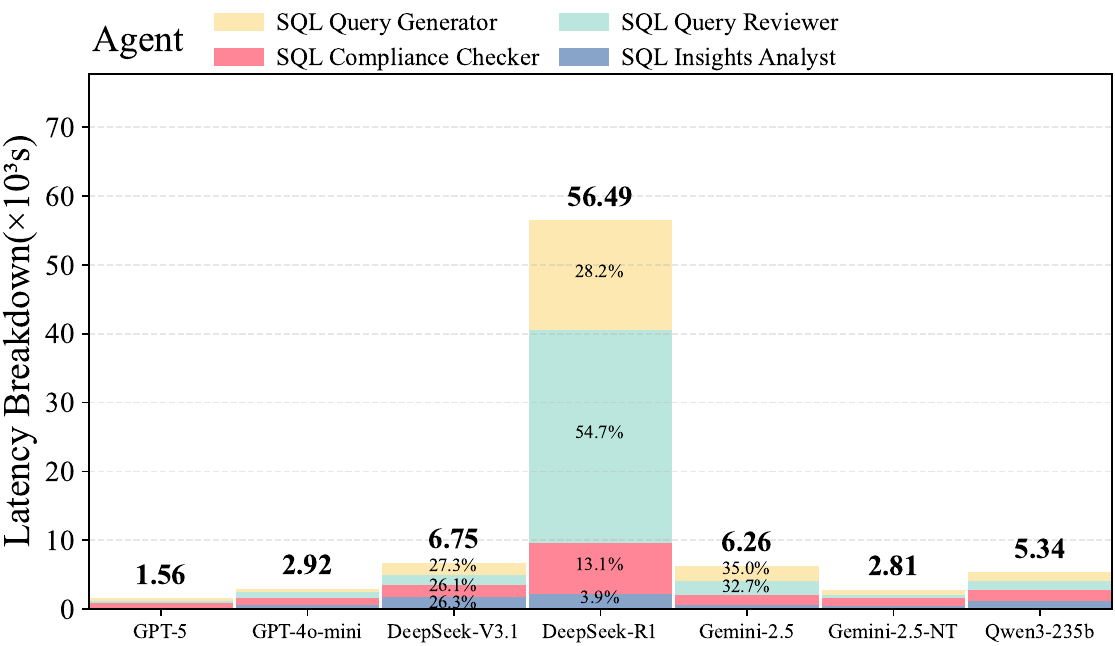}
    \end{minipage}%
    \begin{minipage}{0.32\textwidth}
        \includegraphics[width=\linewidth]{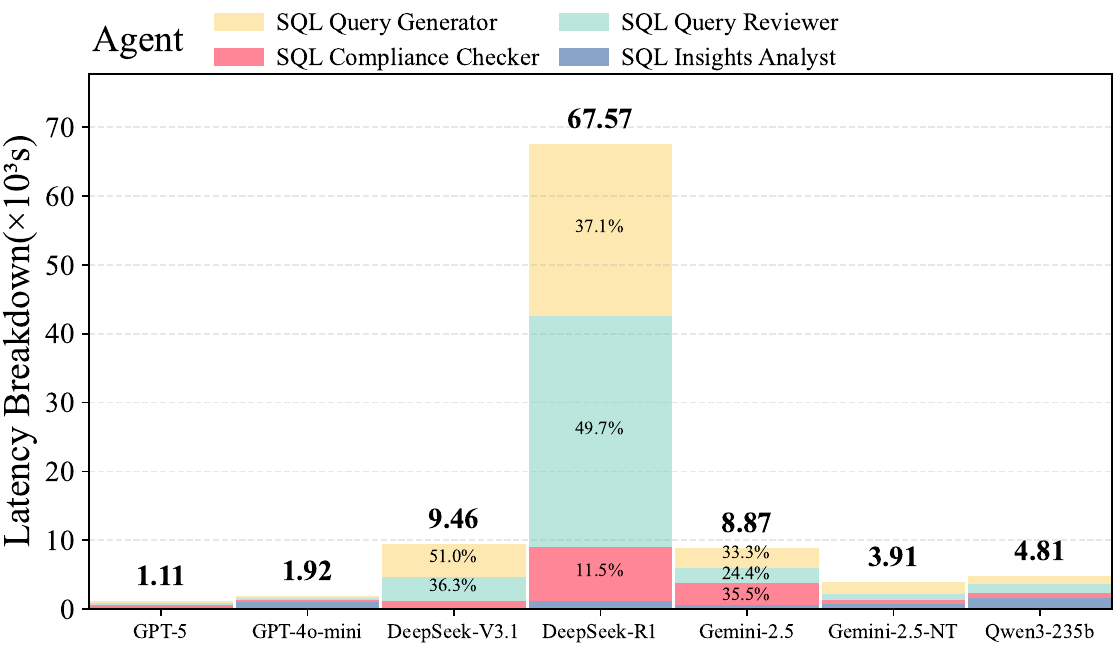}
    \end{minipage}%
    \begin{minipage}{0.32\textwidth}
        \includegraphics[width=\linewidth]{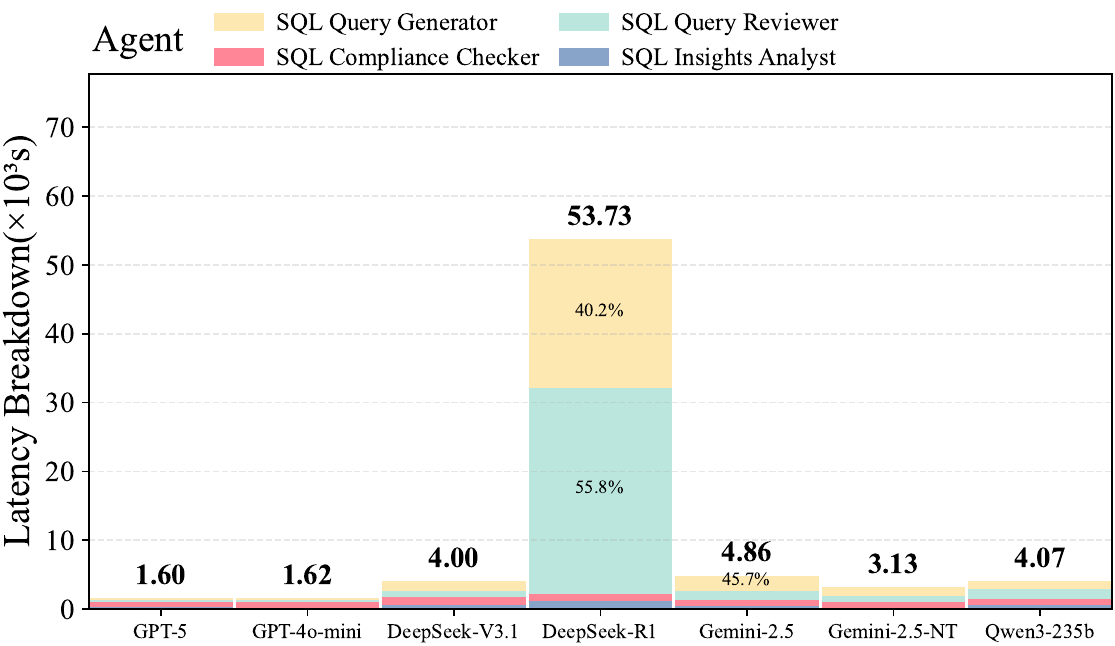}
    \end{minipage}
    \vspace{0.4em}

    \begin{minipage}{0.03\textwidth}
        \rotatebox{90}{\scriptsize \textbf{Recruitment}}
    \end{minipage}%
    \begin{minipage}{0.32\textwidth}
        \includegraphics[width=\linewidth]{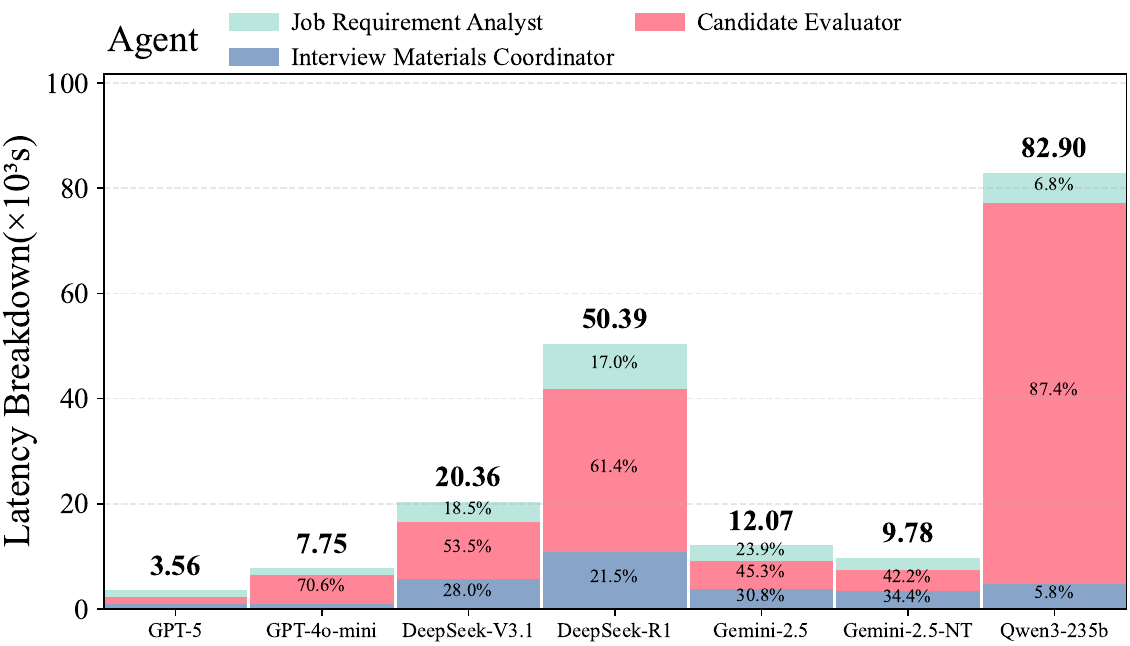}
    \end{minipage}%
    \begin{minipage}{0.32\textwidth}
        \includegraphics[width=\linewidth]{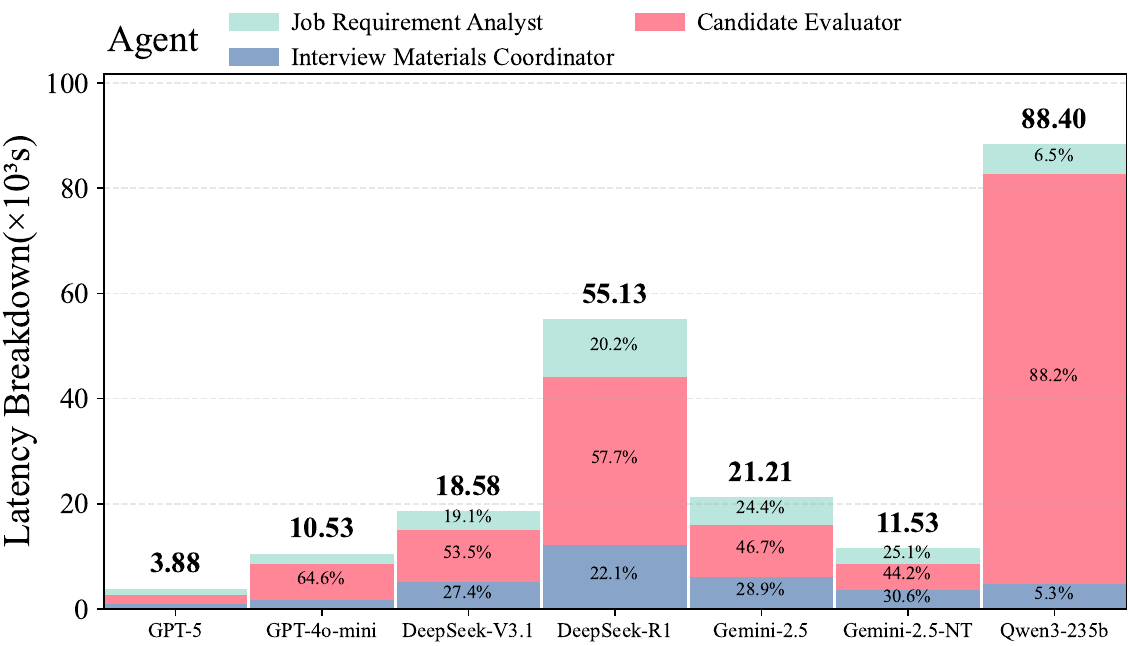}
    \end{minipage}%
    \begin{minipage}{0.32\textwidth}
        \includegraphics[width=\linewidth]{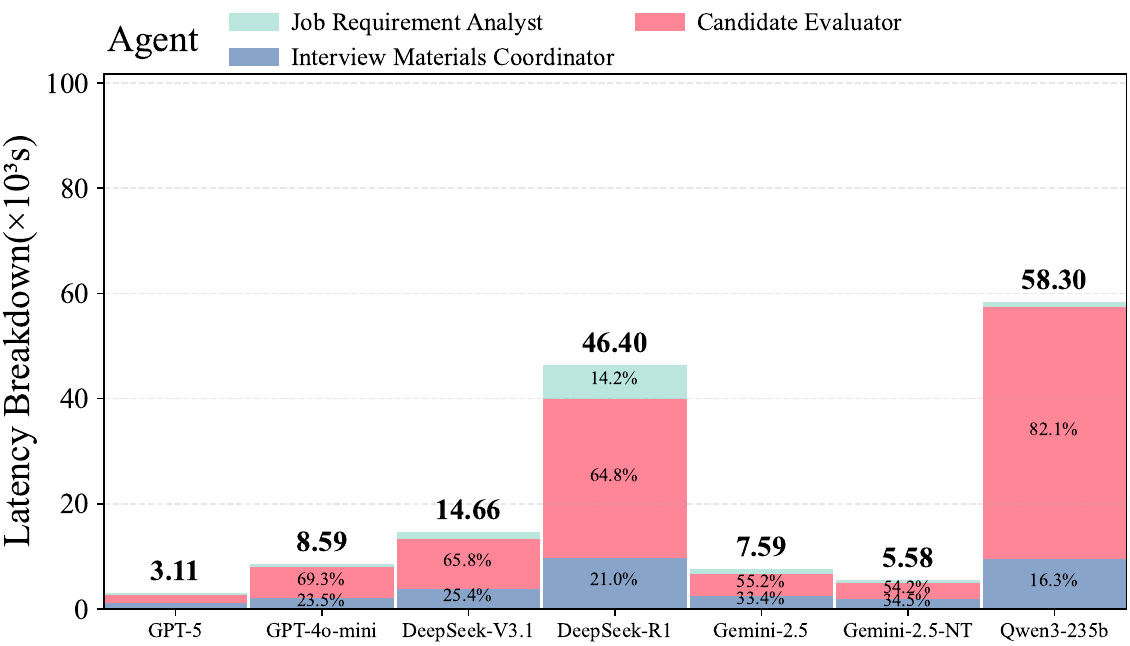}
    \end{minipage}
    \vspace{0.4em}
    
    \begin{minipage}{0.03\textwidth}
        \rotatebox{90}{\scriptsize \textbf{Landing Pg.}}
    \end{minipage}%
    \begin{minipage}{0.32\textwidth}
        \includegraphics[width=\linewidth]{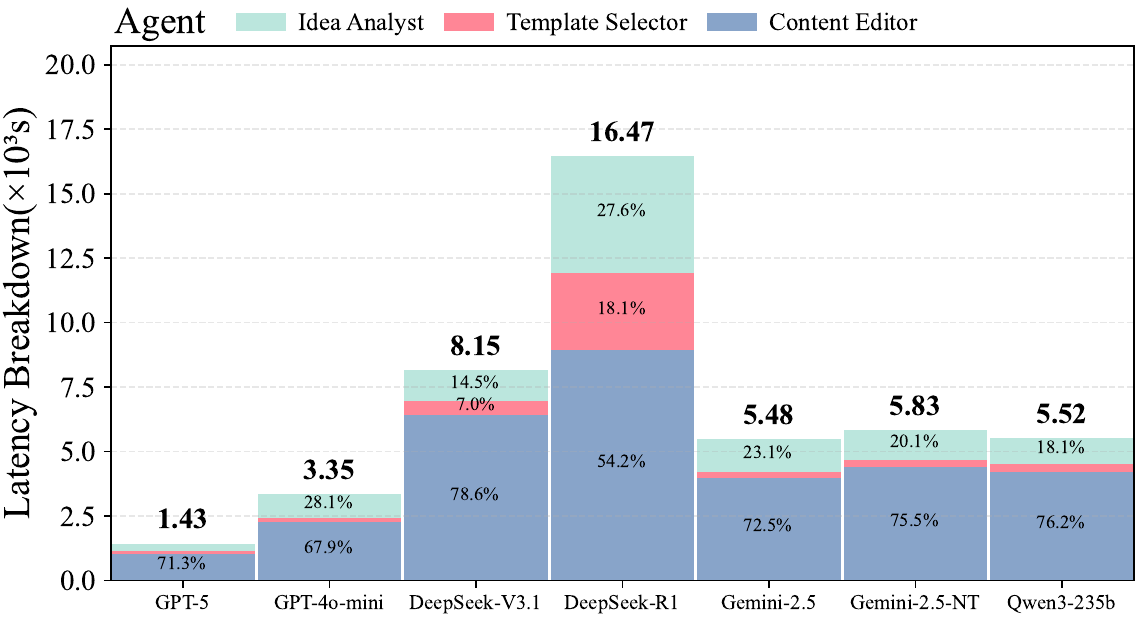}
    \end{minipage}%
    \begin{minipage}{0.32\textwidth}
        \includegraphics[width=\linewidth]{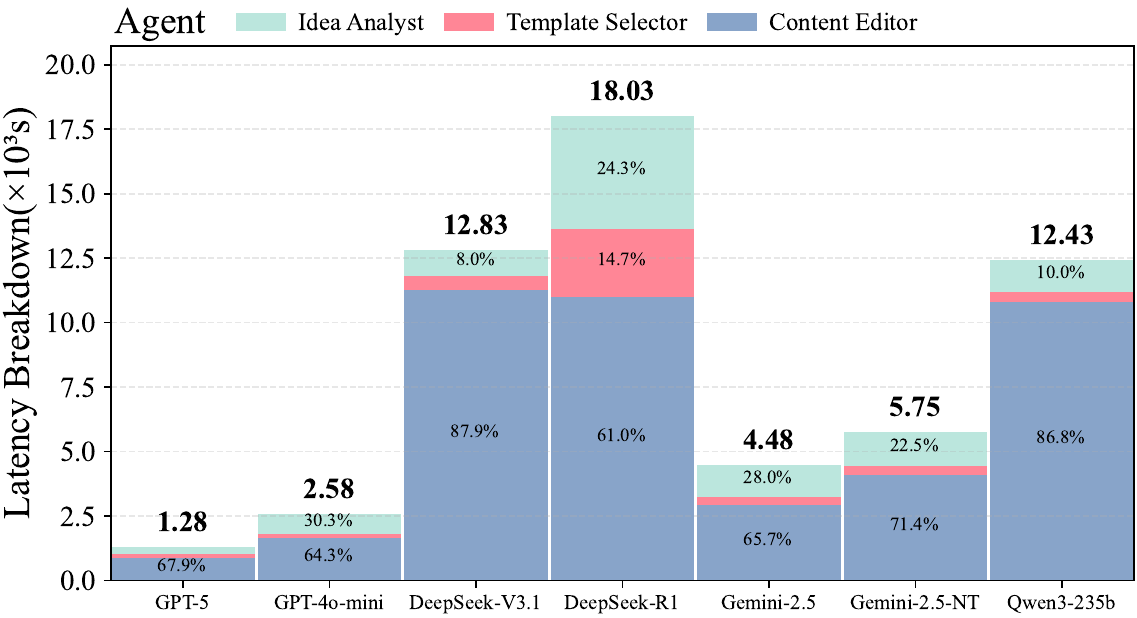}
    \end{minipage}%
    \begin{minipage}{0.32\textwidth}
        \includegraphics[width=\linewidth]{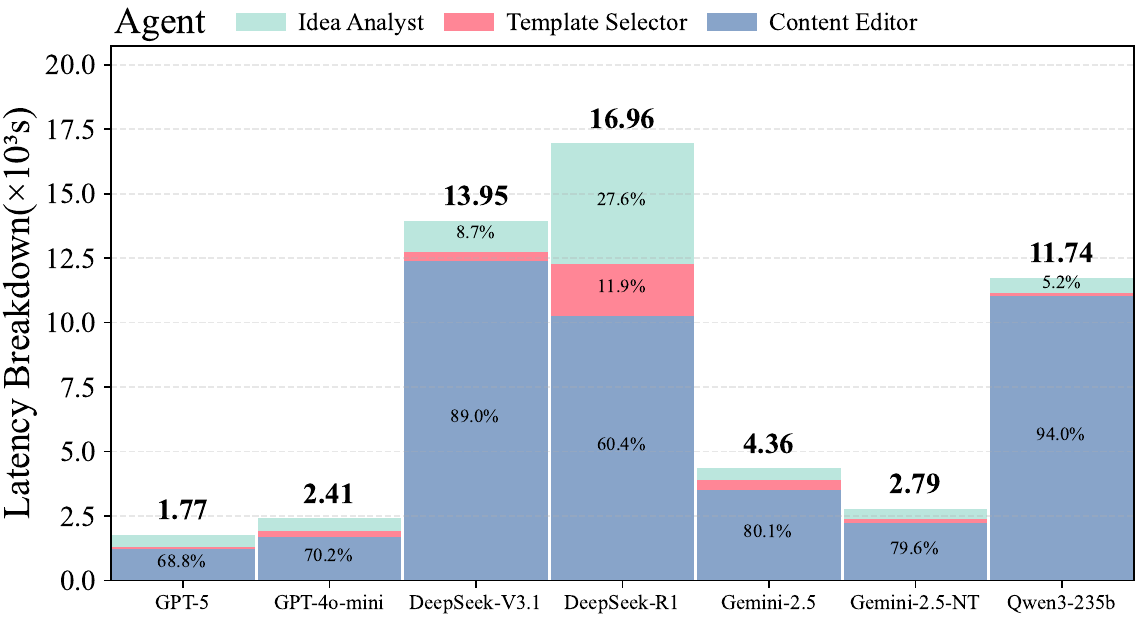}
    \end{minipage}
    \vspace{0.4em}

    \begin{minipage}{0.03\textwidth}
        \rotatebox{90}{\scriptsize \textbf{Social M. M.}}
    \end{minipage}%
    \begin{minipage}{0.32\textwidth}
        \includegraphics[width=\linewidth]{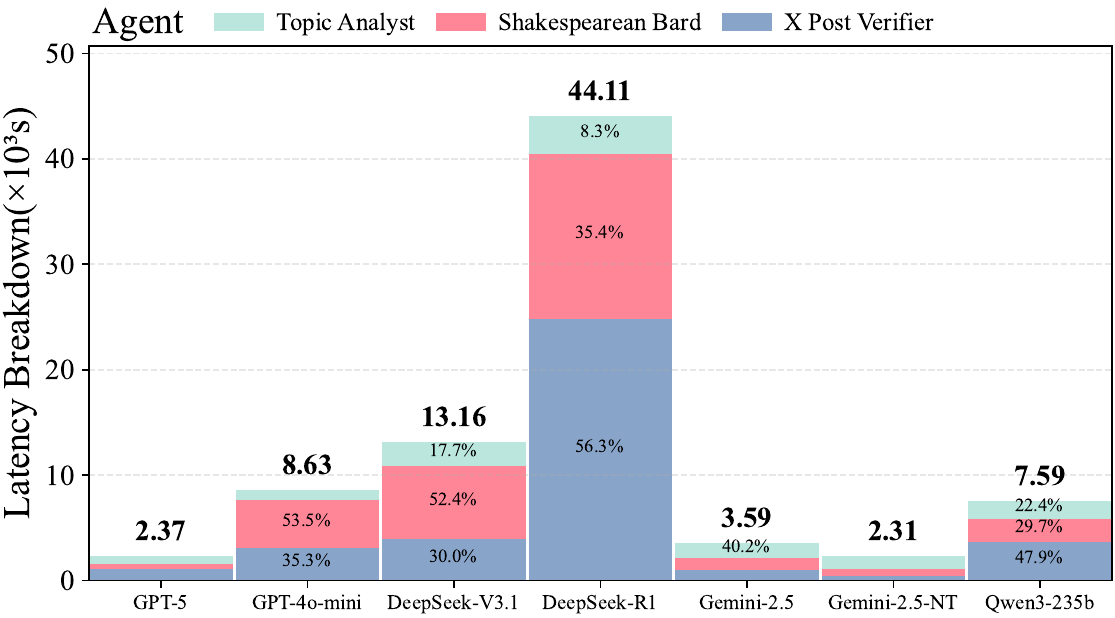}
    \end{minipage}%
    \begin{minipage}{0.32\textwidth}
        \includegraphics[width=\linewidth]{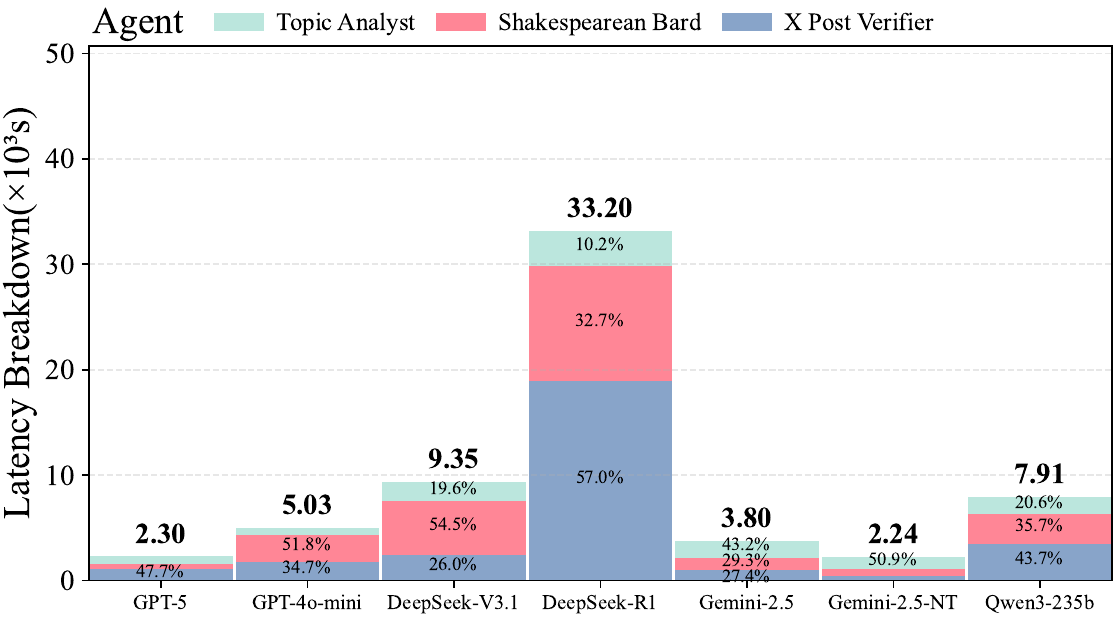}
    \end{minipage}%
    \begin{minipage}{0.32\textwidth}
        \includegraphics[width=\linewidth]{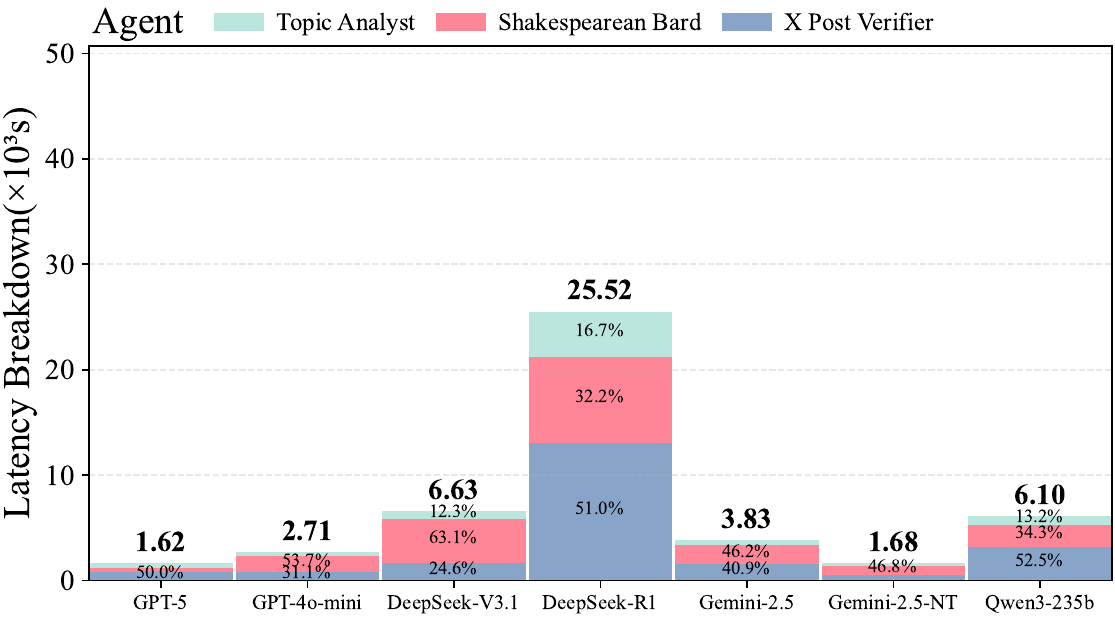}
    \end{minipage}
    \vspace{0.4em}

    \begin{minipage}{0.03\textwidth}
        \rotatebox{90}{\scriptsize \textbf{Book Writer}}
    \end{minipage}%
    \begin{minipage}{0.32\textwidth}
        \includegraphics[width=\linewidth]{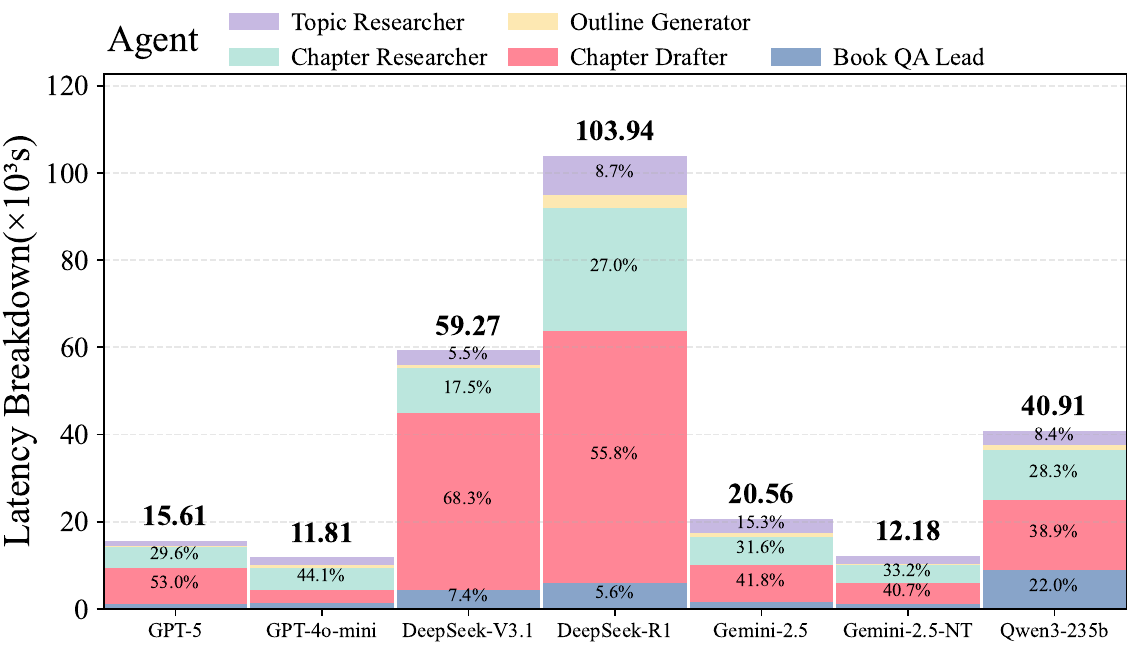}
    \end{minipage}%
    \begin{minipage}{0.32\textwidth}
        \includegraphics[width=\linewidth]{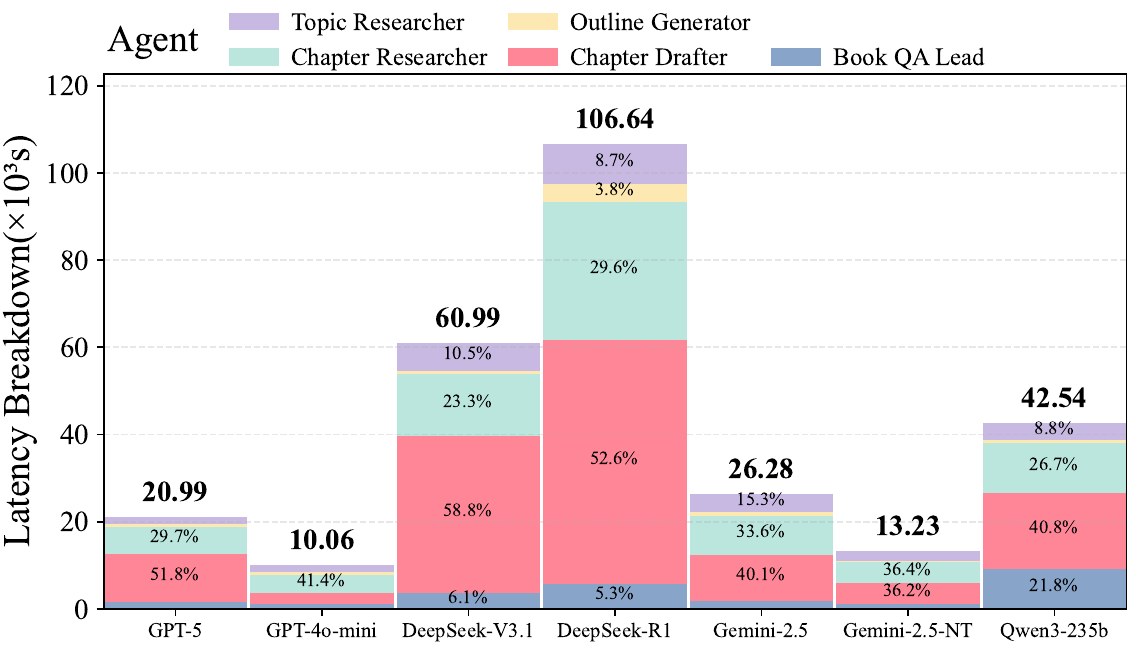}
    \end{minipage}%
    \begin{minipage}{0.32\textwidth}
        \includegraphics[width=\linewidth]{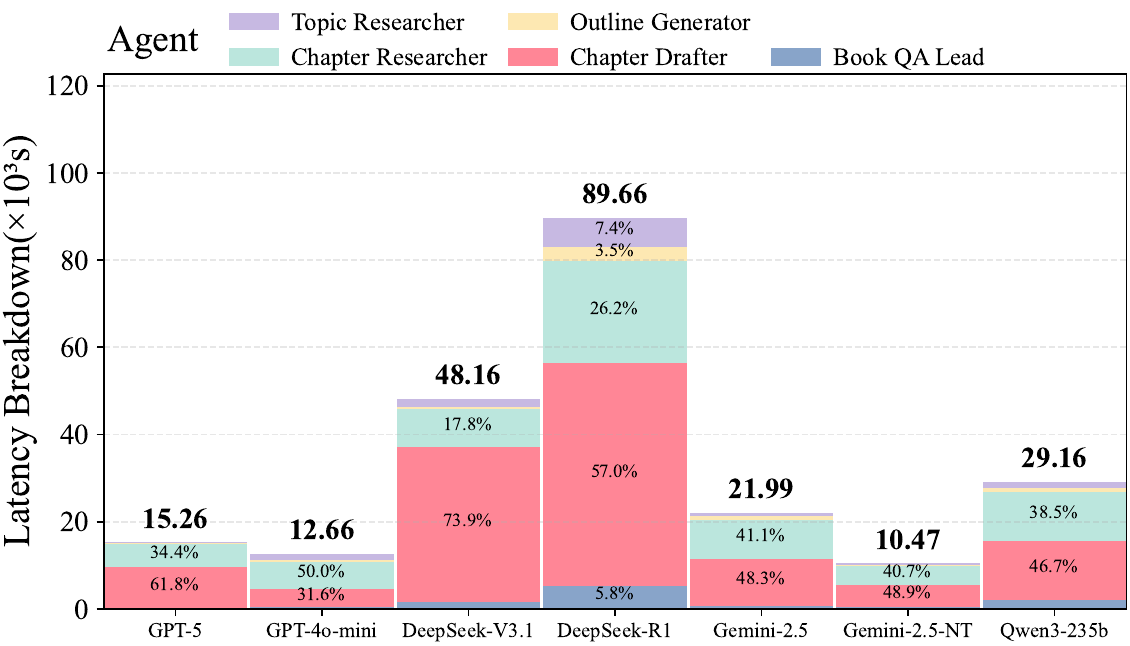}
    \end{minipage}

    \vspace{0.3em}
    \begin{minipage}{0.03\textwidth} \hfill \end{minipage}%
    \begin{minipage}{0.32\textwidth} \centering \textbf{MCP} \end{minipage}
    \begin{minipage}{0.32\textwidth} \centering \textbf{A2A} \end{minipage}
    \begin{minipage}{0.32\textwidth} \centering \textbf{H-A2A} \end{minipage}

    \caption{Comparison of agent latency breakdown for applications with three variants.}
    \label{fig:agent_time_share_3_cols}
    \vspace{-0.2in}
\end{figure*}

Moving beyond application-level aggregates, we dissect the internal latency distribution of individual agents to identify specific execution bottlenecks. Fig.~\ref{fig:agent_time_share_2_cols} and Fig.~\ref{fig:agent_time_share_3_cols} illustrate the latency contribution of each agent within the workflow.

\textbf{1. The Straggler Effect.}
The breakdown reveals an uneven latency distribution. Execution time is rarely uniform across agents. Instead, it is largely concentrated in a single ``straggler'' agent that processes the most complex sub-task, which dictating the overall system latency.

\begin{itemize}[leftmargin=*,label=\textbullet]
    \item \textbf{Dominant Nodes:} As shown in Fig.~\ref{fig:agent_time_share_3_cols}, in \texttt{Book Writer}, the \textit{Chapter Drafter} consistently accounting for an average of $52.7\%$ of the total duration, and peaking at $73.9\%$ in H-A2A variant with \texttt{DeepSeek-V3.1}. This pattern is even more pronounced in other applications. In \texttt{Recruitment Assistant} (A2A), the \textit{Candidate Evaluator} consumes $68.4\%$ of the latency budget, substantially exceeding the \textit{Job Requirement Analyst} ($15.1\%$) and \textit{Interview Materials Coordinator} ($16.4\%$). In extreme cases with \texttt{Qwen3-235b}, this disparity widens to $88.2\%$ versus $6.5\%$ and $5.3\%$, meaning that parallelizing non-dominant nodes offers limited benefits.
    
    \item \textbf{Reasoning Amplification:} Reasoning models increase this skew by extending the compute time of verification-heavy agents. A notable example is observed in \texttt{SQL Assistant} (MCP). While the \textit{SQL Query Reviewer} occupies a modest $20.4\%$ of the timeline under \texttt{GPT-5}, switching to \texttt{DeepSeek-R1} amplifies this share to $54.7\%$. This shift changes the critical path, transforming a balanced workflow into one bottlenecked exclusively by the reasoning model's verification loop.
\end{itemize}

Since the critical path is determined by the straggler, optimizing non-bottleneck agents yields minimal returns. In  \texttt{Landing Page Generator}, averaged across the three variants, the \textit{Content Editor} dominates nearly $74\%$ of the execution time, so cutting its latency by $50\%$ would substantially reduce end-to-end latency. In contrast, optimizing the faster \textit{Template Selector} (approx. $8\%$ share) would result in negligible improvements.
This requires targeted optimization. Developers should profile agent-level traces to identify the straggler and apply localized interventions. Strategies include swapping the bottleneck agent to a faster model (e.g., \texttt{GPT-4o-mini}) or pruning its specific prompt context. This approach is more cost-effective than uniformly upgrading the model tier for the entire system.

\finding{\textbf{Straggler Effect.} Latency in AI-Native systems skews toward a single straggler agent, consuming up to $88.2\%$ of end-to-end time. Performance engineering requires trace-based fault localization to target specific bottlenecks rather than global optimization strategies.}\label{fnd:agent_bottleneck}


\textbf{2. The Heterogeneity Paradox.}
Beyond individual agent bottlenecks, comparing architectural variants reveals a counter-intuitive phenomenon. It is often assumed that H-A2A introduces integration overhead, cross-framework semantic misalignment, and data format incompatibilities, potentially increasing latency. However, empirical data refutes this: H-A2A frequently outperforms the homogeneous A2A architecture in task completion speed. As evidenced by the aggregate results in Fig.~\ref{fig:agent_time_share_3_cols}, H-A2A reduces mean latency by $6.5\%$ to $31.3\%$ across projects. Notable examples include a mean reduction of $31.3\%$ ($-156.82$~s) in \texttt{Recruitment Assistant} (Overall) and $27.0\%$ ($-97.68$~s) in \texttt{Book Writer} (Overall). This suggests a heterogeneity paradox, driven by the dynamics of inter-agent interaction.

In homogeneous A2A systems, agents share the same prompt structures and error modes. When utilizing reasoning or highly verbose models (e.g., \texttt{Qwen3-235b} or \texttt{DeepSeek-R1}), this homogeneity creates a risk of synchronized reasoning loops. Specifically, if one agent initiates a redundant verification step, a peer agent with identical logic often validates rather than corrects this behavior. This results in a cycle where agents reinforce redundant verification. For instance, in \texttt{Recruitment Assistant} (Overall), \texttt{Qwen3-235b} suffers a massive mean latency penalty in the homogeneous setup ($1476.89$~s) but converges significantly faster in the heterogeneous setup ($973.11$~s), a mean reduction of over 500 seconds.

H-A2A can reduce synchronized reasoning loops by limiting how much execution context is shared across heterogeneous agents. Because agents run on different frameworks, the exchanged messages often contain less intermediate context, which makes it harder to sustain repeated verification. Specifically, cross-framework communication typically preserves the sender’s final outputs rather than its full intermediate reasoning chain, so the receiver is more likely to accept the input as a completed result and continue to the next step. As a result, the workflow tends to spend less time on iterative cross-checking and more time on progressing through the remaining steps.

However, this improvement is not universal. The paradox is most pronounced in complex workflows. In structured deterministic tasks like \texttt{Landing Page Generator} (Overall), the gain is small on average ($-6.5\%$). Furthermore, for frontier models~\cite{meinke2025frontiermodelscapableincontext} like \texttt{GPT-5}, heterogeneity can introduce a significant mean latency penalty, e.g., $+33.7\%$ in \texttt{SQL Assistant} (Overall) and $+30.3\%$ in \texttt{Landing Page Generator} (Overall), indicating that context fragmentation becomes an overhead when internal reasoning is already efficient.


\finding{\textbf{Heterogeneity Paradox}: H-A2A can accelerate workflow convergence by $6.5\%$ to $31.3\%$ in complex domains by reducing the likelihood of synchronized reasoning loops in homogeneous A2A. While this context stripping optimizes indecisive agents, it conversely incurs a latency penalty for frontier models in deterministic tasks, turning fragmentation from a feature into an overhead.}\label{fnd:heterogeneity_paradox}

\begin{figure}[t]
    \centering
    \begin{minipage}[t]{0.32\linewidth}
        \centering
        \includegraphics[width=\linewidth]{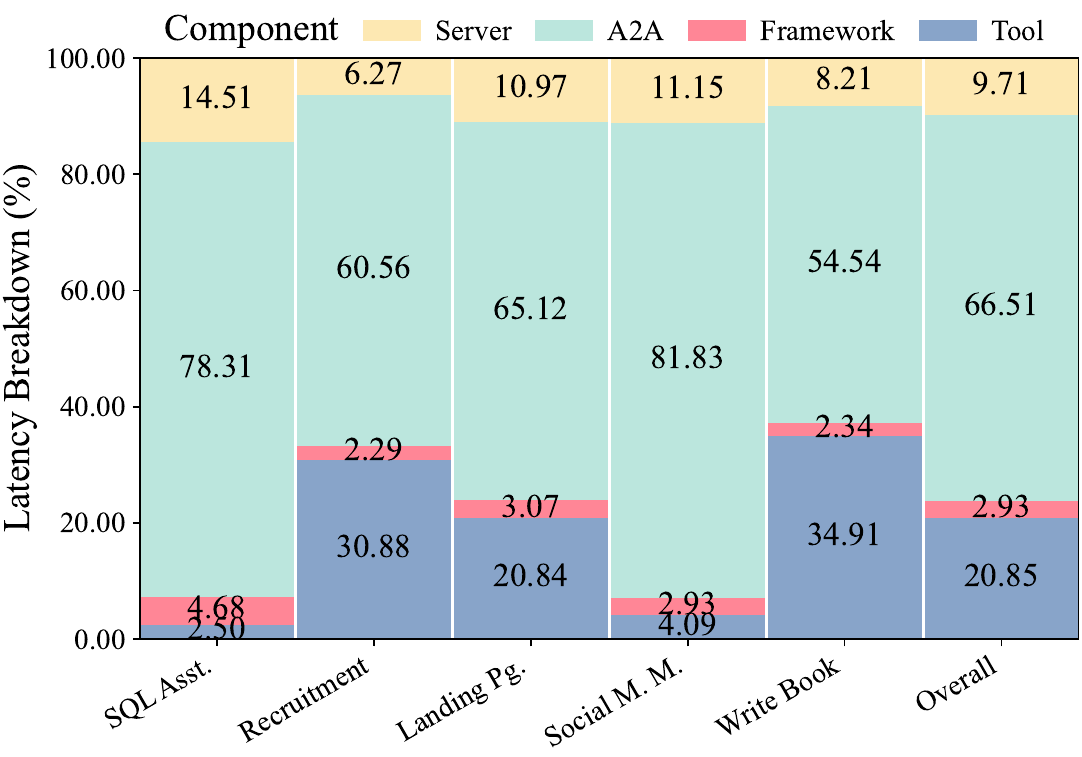}
        \caption{Normalized breakdown of non-LLM latency components for A2A projects.}
        \label{fig:model-overhead-a2a}
    \end{minipage}%
    \hfill
    \begin{minipage}[t]{0.32\linewidth}
        \centering
        \includegraphics[width=\linewidth]{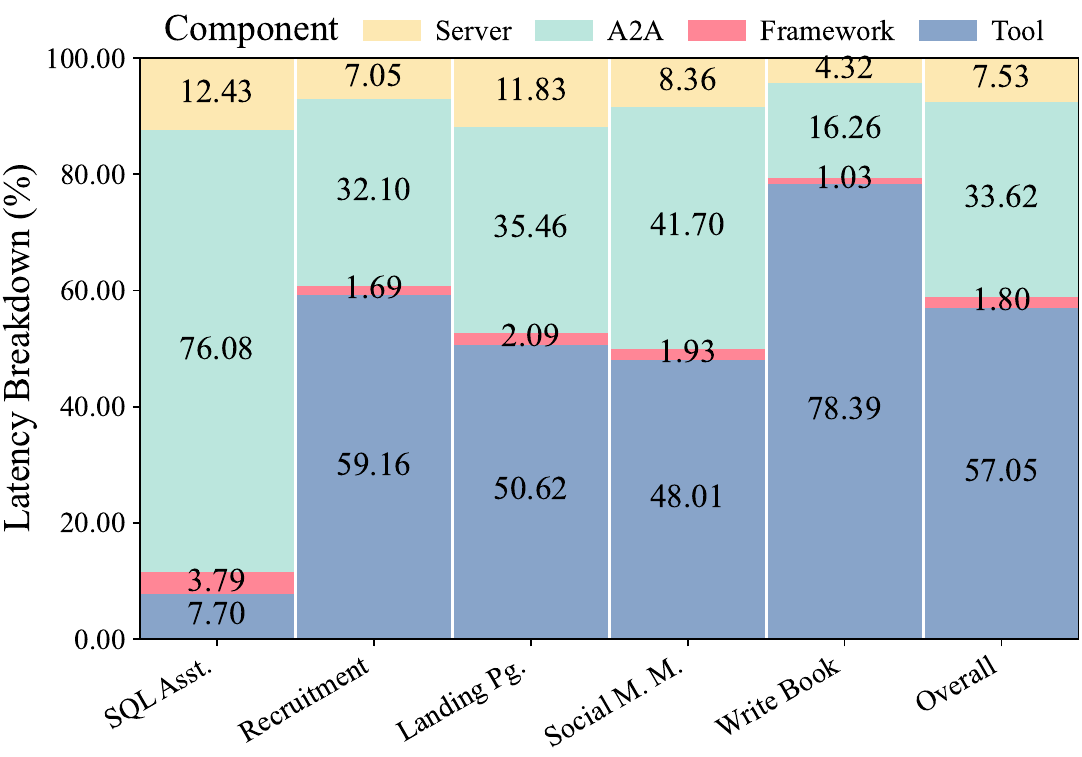}
        \caption{Normalized breakdown of non-LLM latency components for H-A2A projects.}
        \label{fig:model-overhead-mix}
    \end{minipage}%
    \hfill
    \begin{minipage}[t]{0.32\linewidth}
        \centering
        \includegraphics[width=\linewidth]{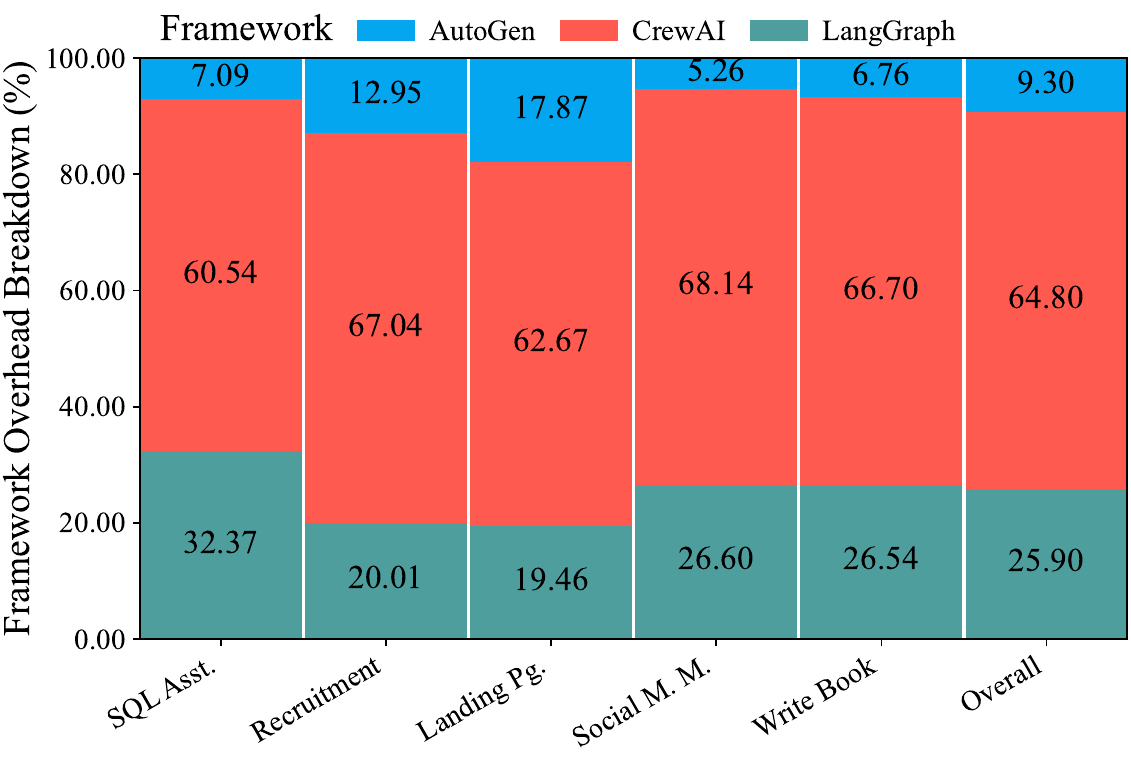}
        \caption{Framework overhead comparison across LangGraph, CrewAI, and AutoGen.}
        \label{fig:framework-overhead-ratio-mix-vs-a2a}
    \end{minipage}
    \vspace{-0.2in}
\end{figure}

\subsubsection{Infrastructure-Level Breakdown}

While LLM inference dominates the total duration, the remaining infrastructure latency offers valuable engineering insights. A granular decomposition of the non-LLM residual (averaging $\approx 1.1\%$ across all projects) uncovers a divergence between architectural styles.  To analyze this, Fig.~\ref{fig:model-overhead-a2a}, Fig.~\ref{fig:model-overhead-mix}, and Fig.~\ref{fig:framework-overhead-ratio-mix-vs-a2a} decompose the specific components, including tool execution, network protocols, and framework overhead.

\textbf{1. Protocol vs. Tool Dominance.}
The breakdown indicates that distributed overhead varies across configurations. In homogeneous A2A architectures (Fig.~\ref{fig:model-overhead-a2a}), protocol overhead is the primary factor. The A2A Protocol overhead (green bars) accounts for $\approx 66.51\%$ of the non-LLM latency on average. Conversely, in H-A2A architectures (Fig.~\ref{fig:model-overhead-mix}), tool execution becomes significant. In this setting, Tool Execution (blue bars) constitutes the majority ($\approx 57.05\%$) of the infrastructure time, exceeding the A2A protocol costs ($33.62\%$). This suggests that homogeneous protocols involve higher serialization overhead, while heterogeneous configurations shift the latency distribution, making tool computation the primary infrastructure component.

This inversion is driven by the interoperability barrier. Homogeneous setups often synchronize stateful contexts, such as full interaction history, which increases serialization costs. In contrast, heterogeneous interoperability forces a ``downgrade'' to strictly stateless JSON/text payloads. This strips away framework-specific bloat, reducing protocol overhead and exposing the raw computational cost of tool execution as the remaining bottleneck.

\textbf{2. Orchestration Overhead.}
Fig.~\ref{fig:framework-overhead-ratio-mix-vs-a2a} illustrates performance differences among agent frameworks. On average, \texttt{CrewAI}'s complex role-based orchestration incurs the highest overhead ($\approx 64.8\%$), substantially outpacing \texttt{LangGraph}'s graph-based state management ($25.9\%$) and \texttt{AutoGen}'s lean conversational architecture ($9.3\%$). 
This overhead imposes a hard latency floor. Although this cost is masked when using slow reasoning models (where inference dominates), it becomes relevant for high-frequency control loops utilizing lightweight models (e.g., \texttt{GPT-4o-mini}). In such scenarios, the heavy orchestration logic of frameworks like \texttt{CrewAI} might offset the speed benefits of the model, limiting the overall responsiveness of the system.


\finding{\textbf{Infrastructure Cost Inversion:} Architecture of AI-Native systems dictates infrastructure cost composition, shifting from protocol-dominant profiles in A2A to tool-dominant profiles in H-A2A. Furthermore, heavy framework orchestration imposes a rigid latency floor, effectively potentially the performance benefits of lightweight models.}\label{fnd:infra_overhead}


\subsection{RQ3: Dissecting Token Economics}\label{sec:rq3}
\label{sec:token-breakdown}

This section analyzes the economic efficiency of AI-Native systems by quantifying the cost of autonomy. Unlike traditional software where retries are cheap, agentic recovery incurs cumulative financial costs. We categorize execution outcomes into three states: \textbf{Direct Success} ($S_{d}$, success without retries), \textbf{Retry Success} ($S_{r}$, success with one or more retries), and \textbf{Failure} ($F$). Fig.~\ref{fig:token_2_cols} and Fig.~\ref{fig:token_3_cols} illustrate the token usage distribution. The data reveal that token consumption is driven not only by model selection, but also by the divergence between distributed and centralized architectures, variations in context transmission introduced by communication protocols, and the specific design of self-correction loops within the workflow.

\subsubsection{Cost of Self-Healing}

\begin{figure}[t!]
    \centering
    \includegraphics[width=\textwidth]{fig/Failure_Breakdown/Model_Legend_horizontal.pdf}
    \begin{minipage}{0.1\textwidth} \hfill \end{minipage}
    \begin{minipage}{0.05\textwidth}
        \rotatebox{90}{\scriptsize \textbf{Email Responder}}
    \end{minipage}%
    \begin{minipage}{0.35\textwidth}
        \includegraphics[width=\linewidth]{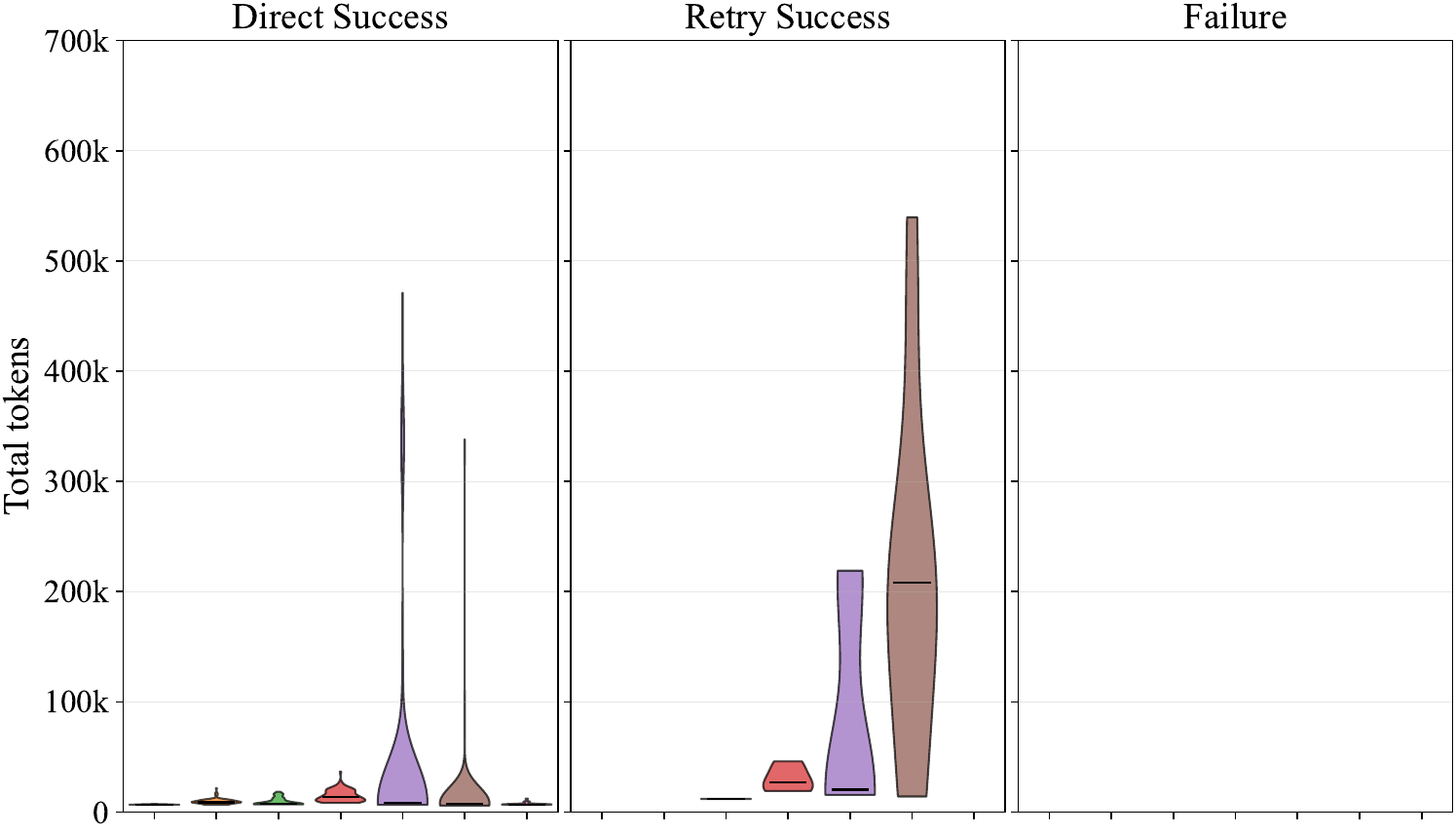}
    \end{minipage}%
    \hfill
    \begin{minipage}{0.35\textwidth}
        \includegraphics[width=\linewidth]{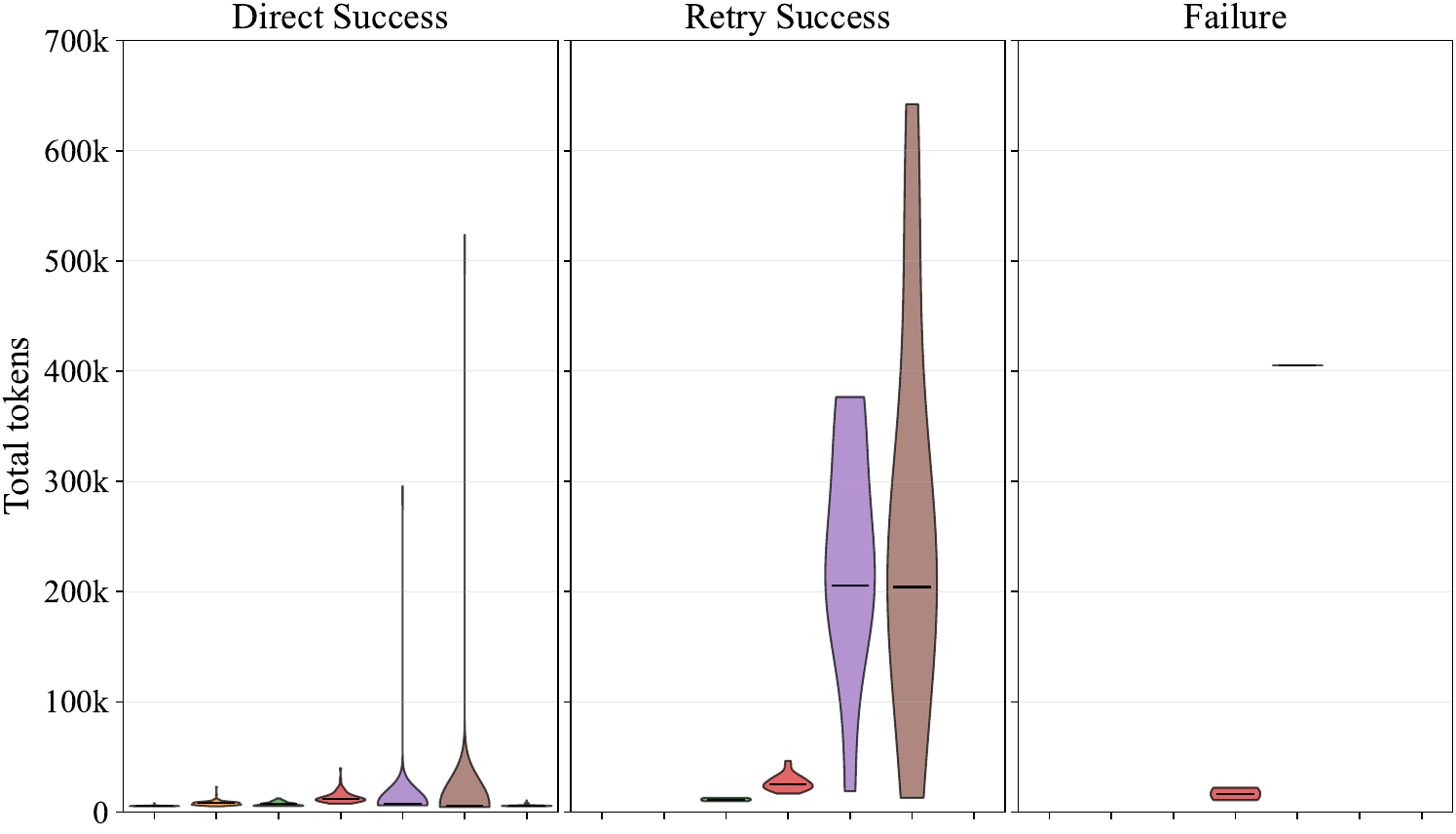}
    \end{minipage}
    \begin{minipage}{0.1\textwidth} \hfill \end{minipage}
    \vspace{0.4em}

    \begin{minipage}{0.1\textwidth} \hfill \end{minipage}
    \begin{minipage}{0.05\textwidth}
        \rotatebox{90}{\scriptsize \textbf{Game Builder}}
    \end{minipage}%
    \begin{minipage}{0.35\textwidth}
        \includegraphics[width=\linewidth]{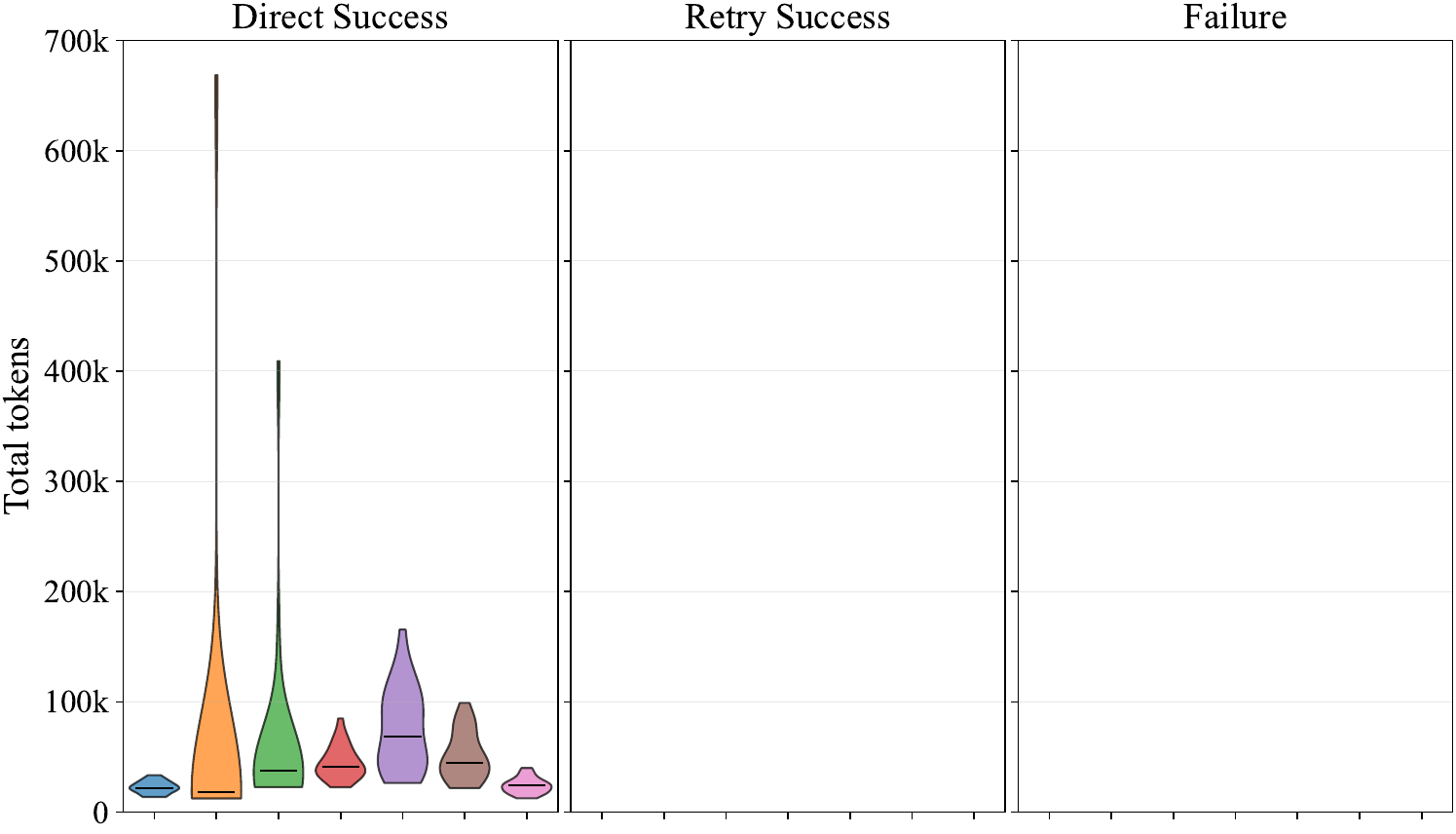}
    \end{minipage}%
    \hfill
    \begin{minipage}{0.35\textwidth}
        \includegraphics[width=\linewidth]{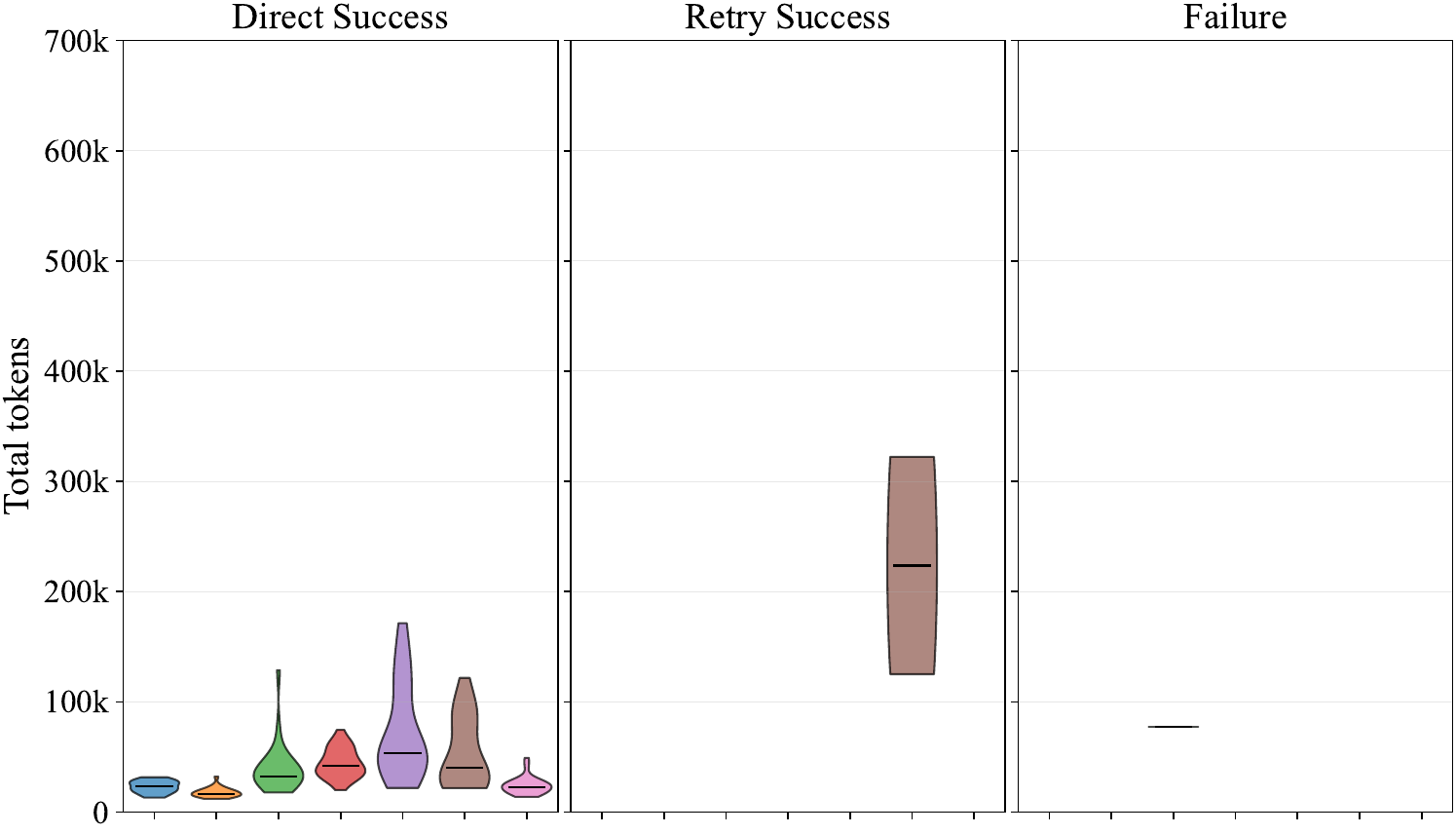}
    \end{minipage}
    \begin{minipage}{0.1\textwidth} \hfill \end{minipage}
    \vspace{0.4em}

    \begin{minipage}{0.1\textwidth} \hfill \end{minipage}
    \begin{minipage}{0.05\textwidth}
        \rotatebox{90}{\scriptsize \textbf{Markdown Val.}}
    \end{minipage}%
    \begin{minipage}{0.35\textwidth}
        \includegraphics[width=\linewidth]{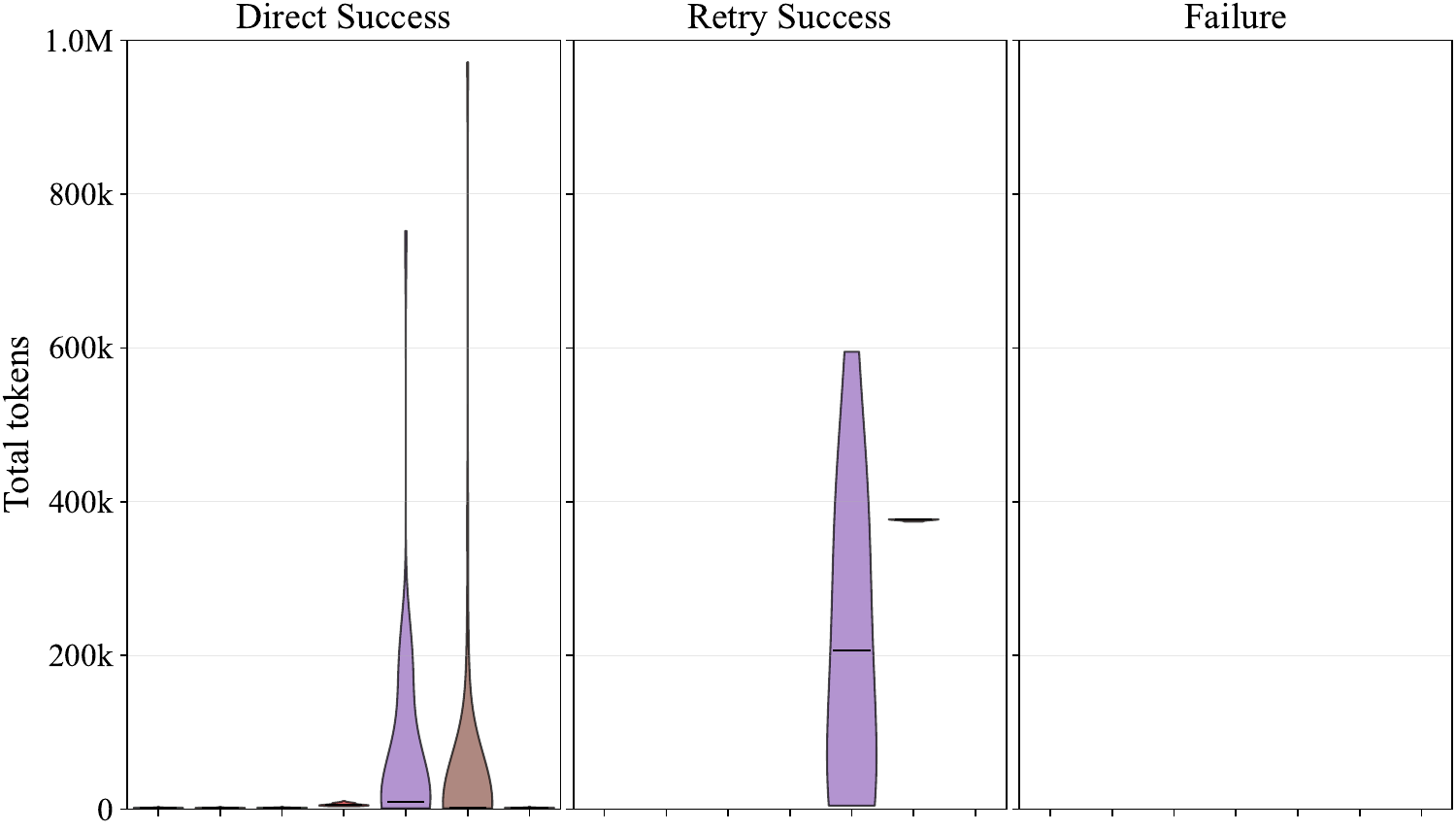}
    \end{minipage}%
    \hfill
    \begin{minipage}{0.35\textwidth}
        \includegraphics[width=\linewidth]{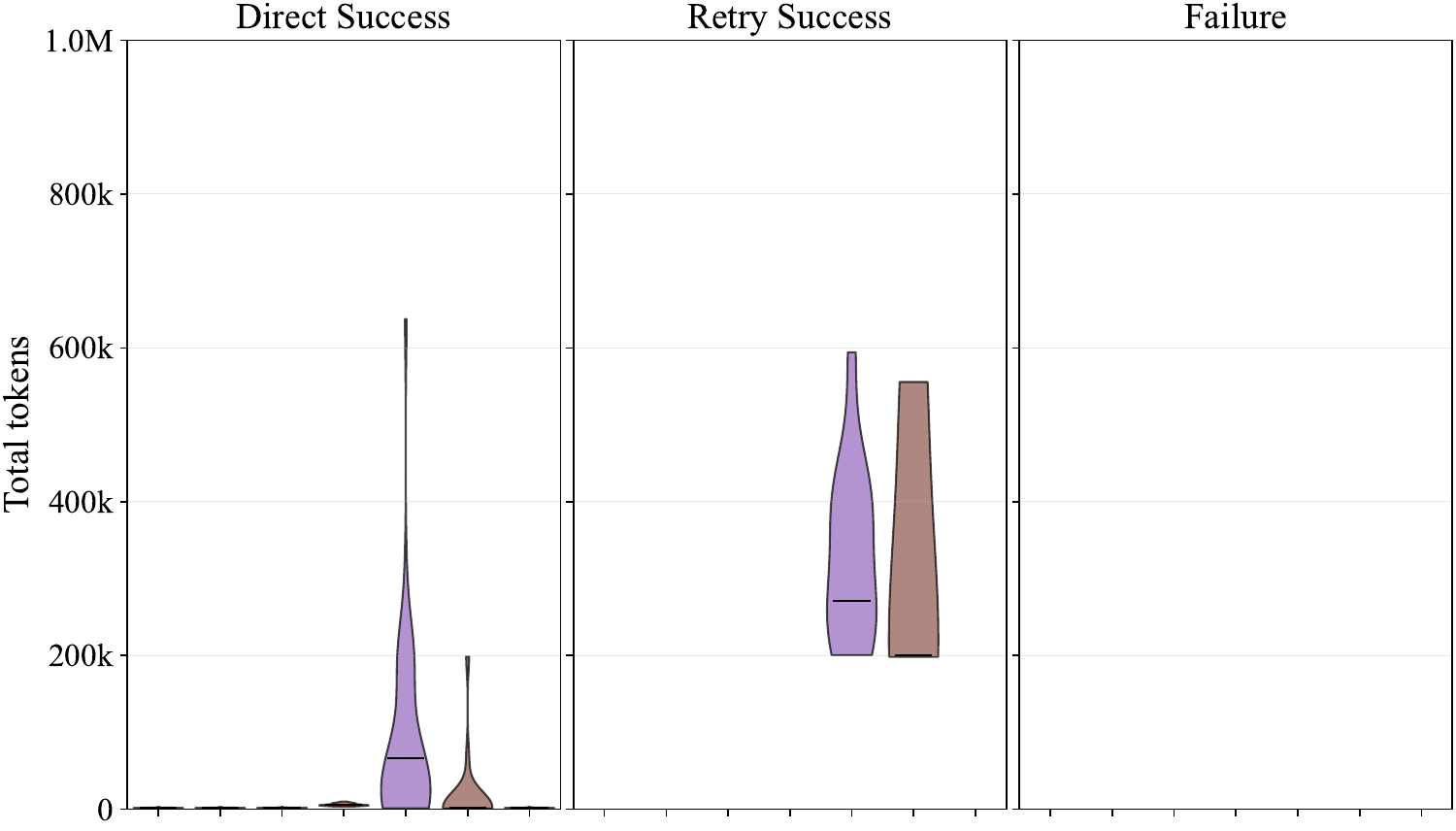}
    \end{minipage}
    \begin{minipage}{0.1\textwidth} \hfill \end{minipage}


    \vspace{0.3em}
    \begin{minipage}{0.15\textwidth} \hfill \end{minipage}
    \begin{minipage}{0.35\textwidth} \centering \textbf{\small{Pure CrewAI}} \end{minipage}
    \hfill
    \begin{minipage}{0.35\textwidth} \centering \textbf{\small{MCP}} \end{minipage}
    \begin{minipage}{0.1\textwidth} \hfill \end{minipage}

    \caption{Token usage breakdown across for applications with two variants.}
    \label{fig:token_2_cols}
\end{figure}

The data in Fig.~\ref{fig:token_2_cols} and Fig.~\ref{fig:token_3_cols} show that reliability often comes with a substantial token premium. Using our outcome categories, direct success $S_{d}$ averages approximately $48\text{k}$ tokens, while retry success $S_{r}$ rises to about $83\text{k}$ tokens. This represents an increase of approximately $70\%$. However, this difference varies across workloads and increases significantly when self-correction loops are triggered.

This trend is evident in complex multi-step workflows. For example, in \texttt{SQL Assistant} (H-A2A), \texttt{DeepSeek-R1} increases from $29\text{k}$ tokens in $S_{d}$ to about $89\text{k}$ in $S_{r}$ (approximately $3.1\times$). Similarly, under the MCP variant, token usage rises from $35\text{k}$ in $S_{d}$ to $93\text{k}$ in $S_{r}$  ($2.6\times$). In verification-intensive workloads, such as the \texttt{Social Media Manager} (A2A), \texttt{DeepSeek-R1} token usage increases from $46\text{k}$ to $165\text{k}$ ($3.6\times$).

Certain workloads demonstrate larger disparities, where retry successes are significantly more expensive than direct successes. In \texttt{Email Responder} (MCP), \texttt{Gemini-2.5-flash-nothinking} token usage jumps from $13\text{k}$ tokens in $S_{d}$ to approximately $200\text{k}$ in $S_{r}$ ($15.4\times$). Likewise, in \texttt{Markdown Validator} (Pure CrewAI), \texttt{Gemini-2.5-flash-nothinking} increases from $16\text{k}$ in $S_{d}$ to $318\text{k}$ in $S_{r}$ ($19.9\times$). These results correspond to the expanded distributions shown in Figures~\ref{fig:token_2_cols} and~\ref{fig:token_3_cols}, indicating that the retry steps consume a large portion of the token budget.


\finding{
\textbf{Reliability Cost}: Self-healing through retries induces non-linear cost increases. Retries elevate success rates but amplify token usage, particularly in reasoning models during recursive error correction, relative to direct successes.
}
\label{fnd:self_healing_cost}

\subsubsection{Expensive Failure Pattern}
In an ideal scenario, AI-Native systems would follow a ``Fail-Fast'' approach, ending execution promptly when a path is unviable to save resources. However, experimental data suggest that the economic cost of failure is closely associated with the workflow orchestration design, particularly the use of iterative self-correction loops. These contrasting patterns are consistently visible in Fig.~\ref{fig:token_3_cols}.


\textbf{1. Token Inflation in Recursive Workflows.} Applications with iterative self-correction suffer significant token inflation during failures. In particular, in \texttt{Social Media Manager}, failure tokens increase across all variants and models evaluated in Fig.~\ref{fig:token_3_cols}, with inflation spanning roughly $195\%$--$738\%$. Peak amplification occurs in the H-A2A variant, where \texttt{DeepSeek-R1} increases from approximately $40\text{k}$ to $337\text{k}$ tokens in $F$, i.e., from an increase of $738\%$. This pattern is also evident in other variants: in the A2A variant, \texttt{Qwen3-235b} consumes an average of $231\text{k}$ tokens in $F$, which represents an over $660\%$ increase compared to the runs in $S_{d}$ ($30\text{k}$ tokens). Similarly, \texttt{GPT-4o-mini} and \texttt{DeepSeek-R1} in the MCP variant of this task show token increases of $460.0\%$ and $450.3\%$ respectively. This suggests that agents in these workflows may continue to use the retry budget after already consuming substantial tokens, deviating from the intended trajectory, and still not guaranteeing recovery.

\begin{figure}[t]
  \centering
  \begin{minipage}{0.95\linewidth} 
    \centering
    
    \begin{minipage}[t]{0.49\linewidth}
      \vspace{0pt} 
      \centering
      \includegraphics[width=0.63\linewidth]{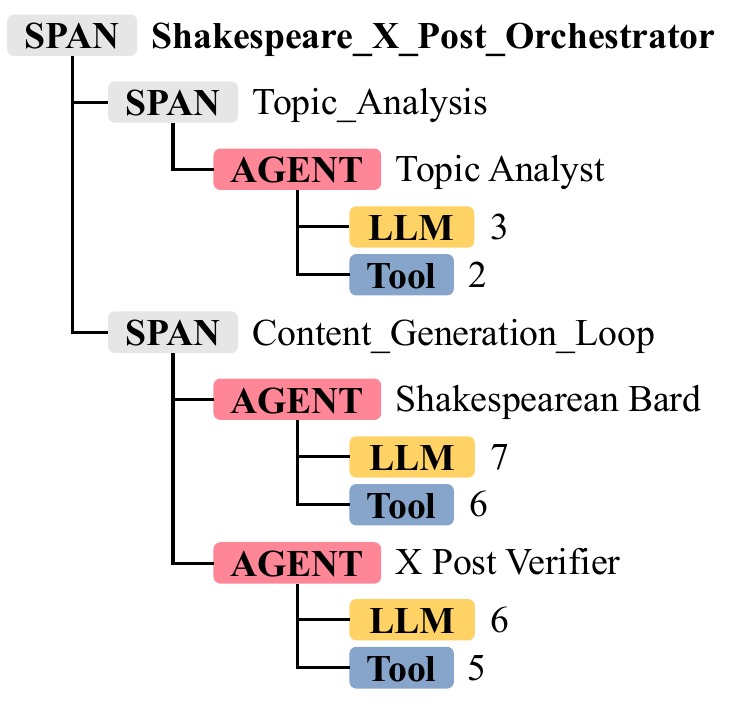}
      \centerline{\small (a) Golden Trace}
      \par\vspace{0.05cm} 
      \includegraphics[width=1\linewidth]{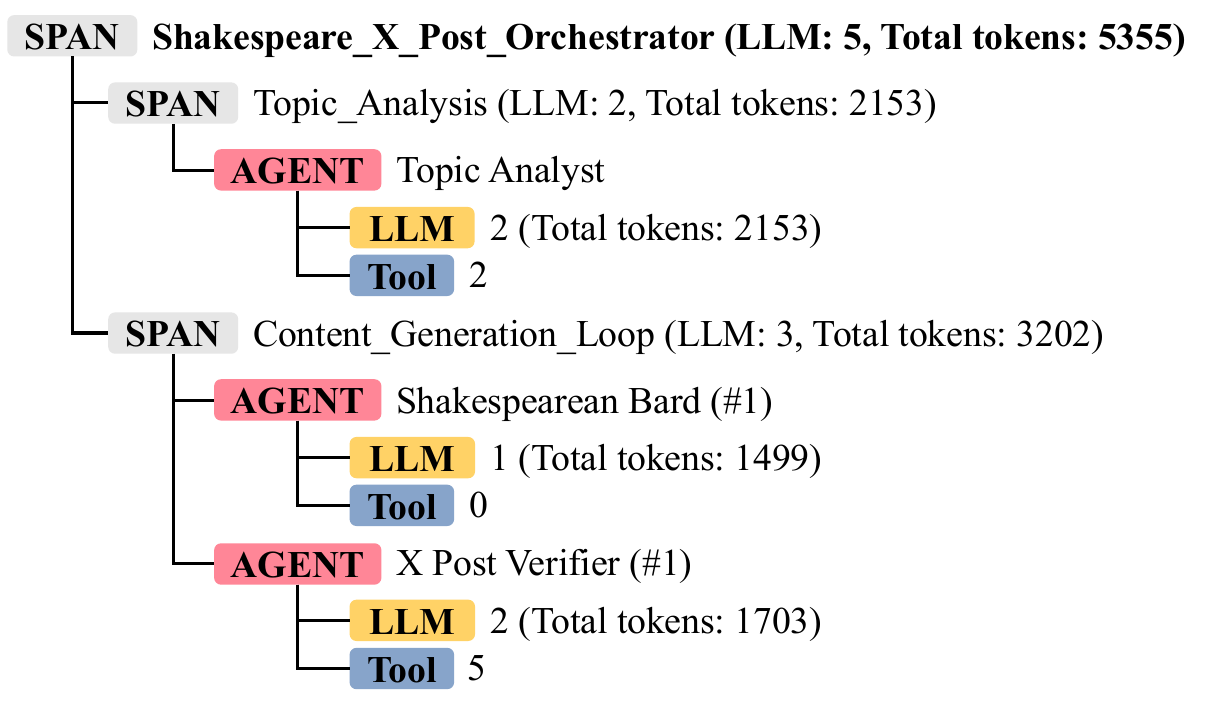}
      \centerline{\small (b) Direct Success Trace}
    \end{minipage}
    \hfill 
    \begin{minipage}[t]{0.40\linewidth}
      \vspace{0pt} 
      \centering
      \includegraphics[width=0.99\linewidth]{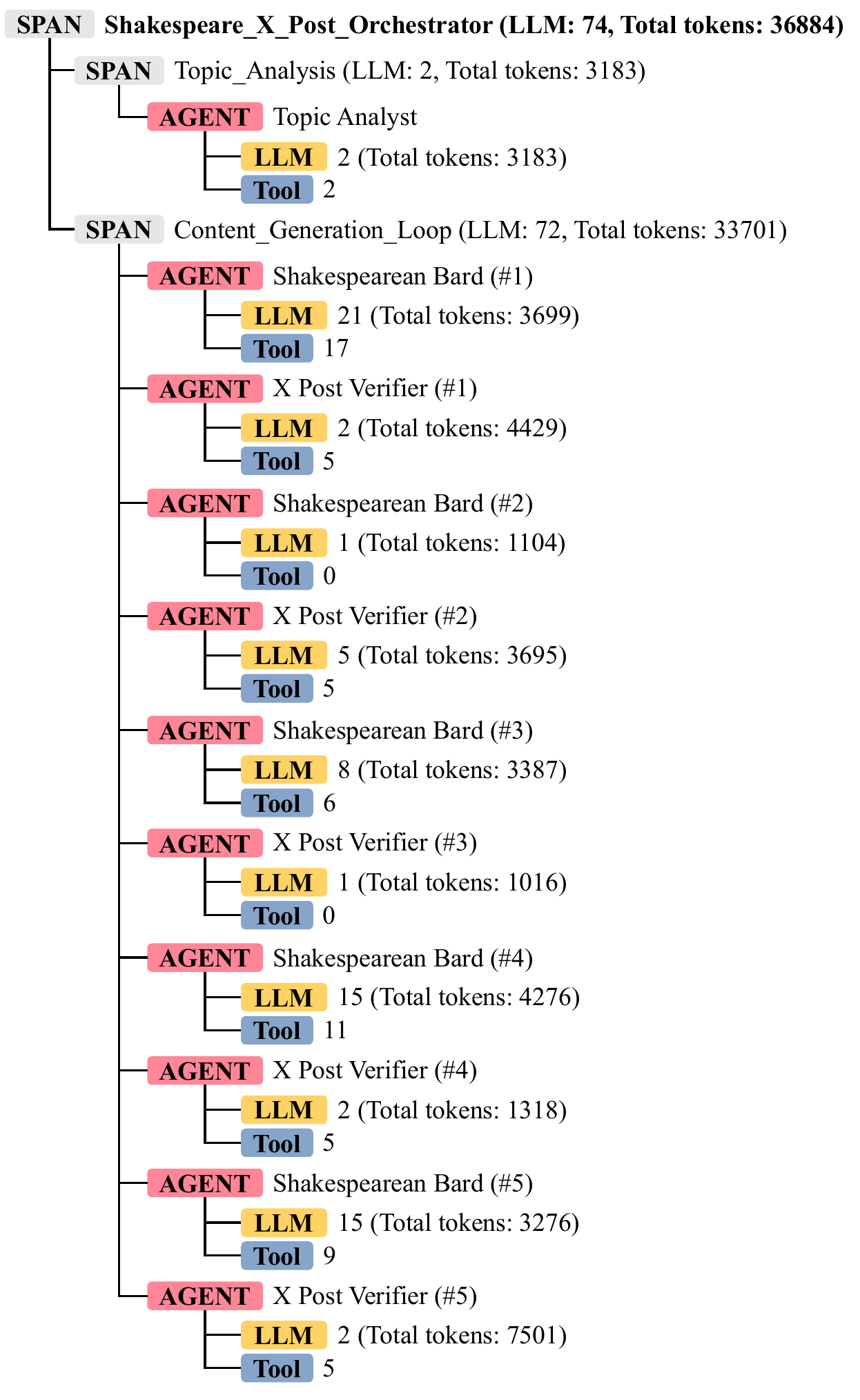}
      \centerline{\small (c) Failure Trace}
    \end{minipage}

    \caption{Comparison of token consumption on \texttt{Social Media Manager} across different execution states.}
    \label{fig:tokens-case}
    \vspace{-0.2in}
    
  \end{minipage}
\end{figure}

We can observe the mechanics of this inflation by examining the specific execution traces of the \texttt{Social Media Manager} shown in Fig.~\ref{fig:tokens-case}. As illustrated in the direct success trace, a functional run completes the \textit{Content \_Generation\_Loop} in a single iteration, incurring a cost of $5355$ tokens. In contrast, the failure trace shows that the system engages in repeated self-correction attempts. The \textit{X Post Verifier} rejects the \textit{Shakespearean Bard}'s output, triggering the system to use its full retry budget. The trace records five complete iterations (\#1 to \#5) of generation and critique. This process accumulates tokens at each step without producing a valid solution, resulting in a total consumption of $36884$ tokens. This represents a nearly $7\times$ increase, demonstrating how recursive architectures can accumulate costs during failure states.

\textbf{2. Fail-Fast Benefits in Linear Workflows.}
Workflows with limited local retries appear more consistent with the ``Fail-Fast'' principle. In \texttt{Landing Page Generator}, the majority of variants show that $F$ often use fewer tokens than $S_{d}$ (e.g., \texttt{GPT-4o-mini} in the A2A variant is the sole exception). Specifically, \texttt{DeepSeek-R1} in the A2A variant shows a $44.6\%$ lower cost in $F$ compared to $S_{d}$ ($26\text{k}$ vs.\ $47\text{k}$ tokens). This trend is also present in the MCP variant, where \texttt{Gemini-2.5-flash-nothinking} has a $F$ cost $85.2\%$ lower than its $S_{d}$ cost ($28\text{k}$ vs. $189\text{k}$ tokens). Since these projects use sequential hand-offs rather than intensive local self-healing, the system can terminate execution early when issues arise.

\finding{\textbf{Expensive Failure Pattern}: Current agentic systems invert the fail-fast principle, manifesting token inflation during failures.  Failed workflows frequently consume more token than successful ones as agents exhaust retry budgets on non-viable trajectories.}
\label{fnd:fail_slow}

\subsubsection{Protocol Cost of A2A}
\begin{figure}[t!] 
    \centering
    \includegraphics[width=\textwidth]{fig/Failure_Breakdown/Model_Legend_horizontal.pdf}
    \begin{minipage}{0.03\textwidth}
        \rotatebox{90}{\scriptsize \textbf{SQL Asst.}}
    \end{minipage}%
    \begin{minipage}{0.32\textwidth}
        \includegraphics[width=\linewidth]{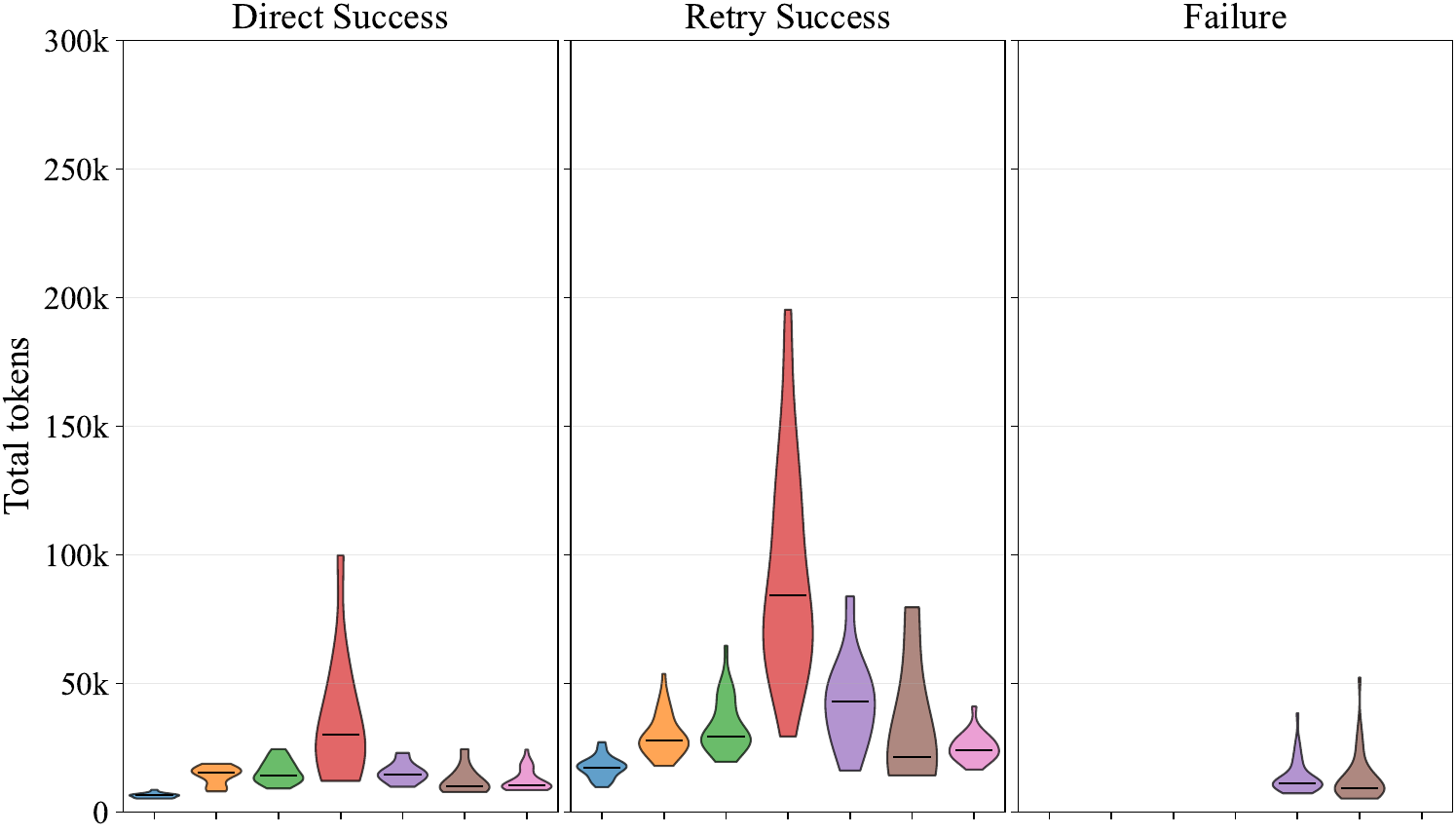}
    \end{minipage}%
    \begin{minipage}{0.32\textwidth}
        \includegraphics[width=\linewidth]{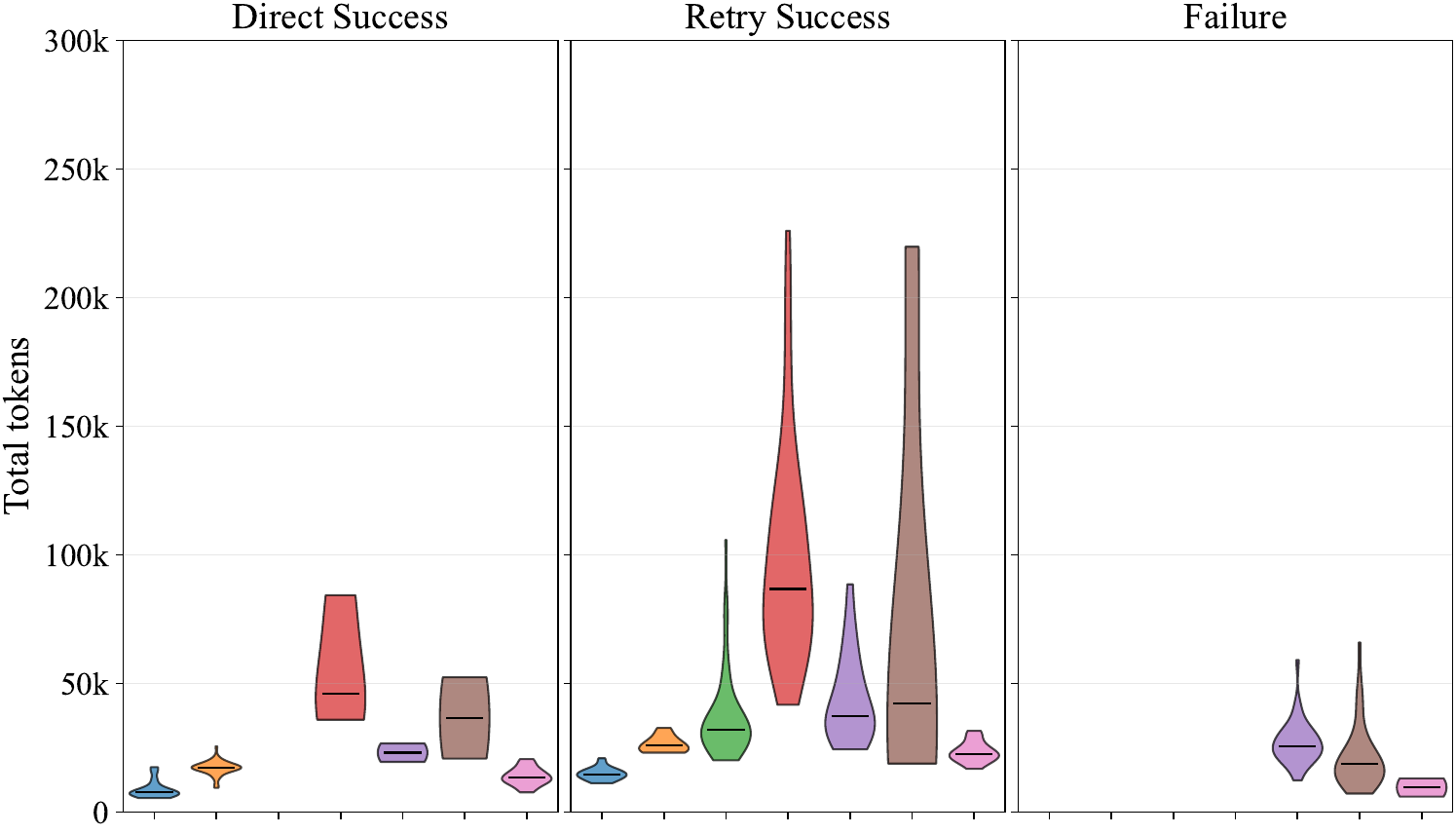}
    \end{minipage}%
    \begin{minipage}{0.32\textwidth}
        \includegraphics[width=\linewidth]{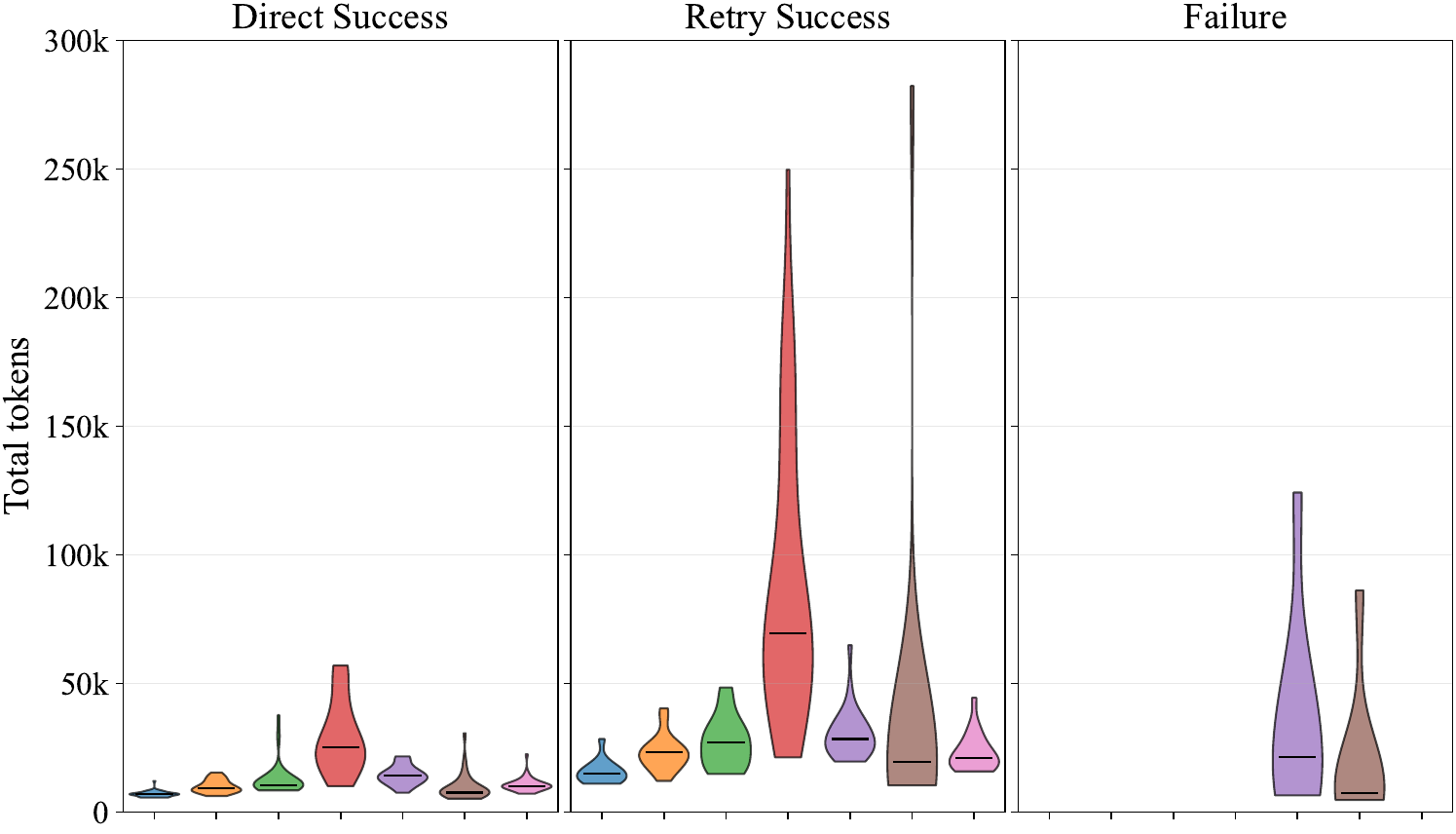}
    \end{minipage}
    \vspace{0.4em}

    \begin{minipage}{0.03\textwidth}
        \rotatebox{90}{\scriptsize \textbf{Recruitment}}
    \end{minipage}%
    \begin{minipage}{0.32\textwidth}
        \includegraphics[width=\linewidth]{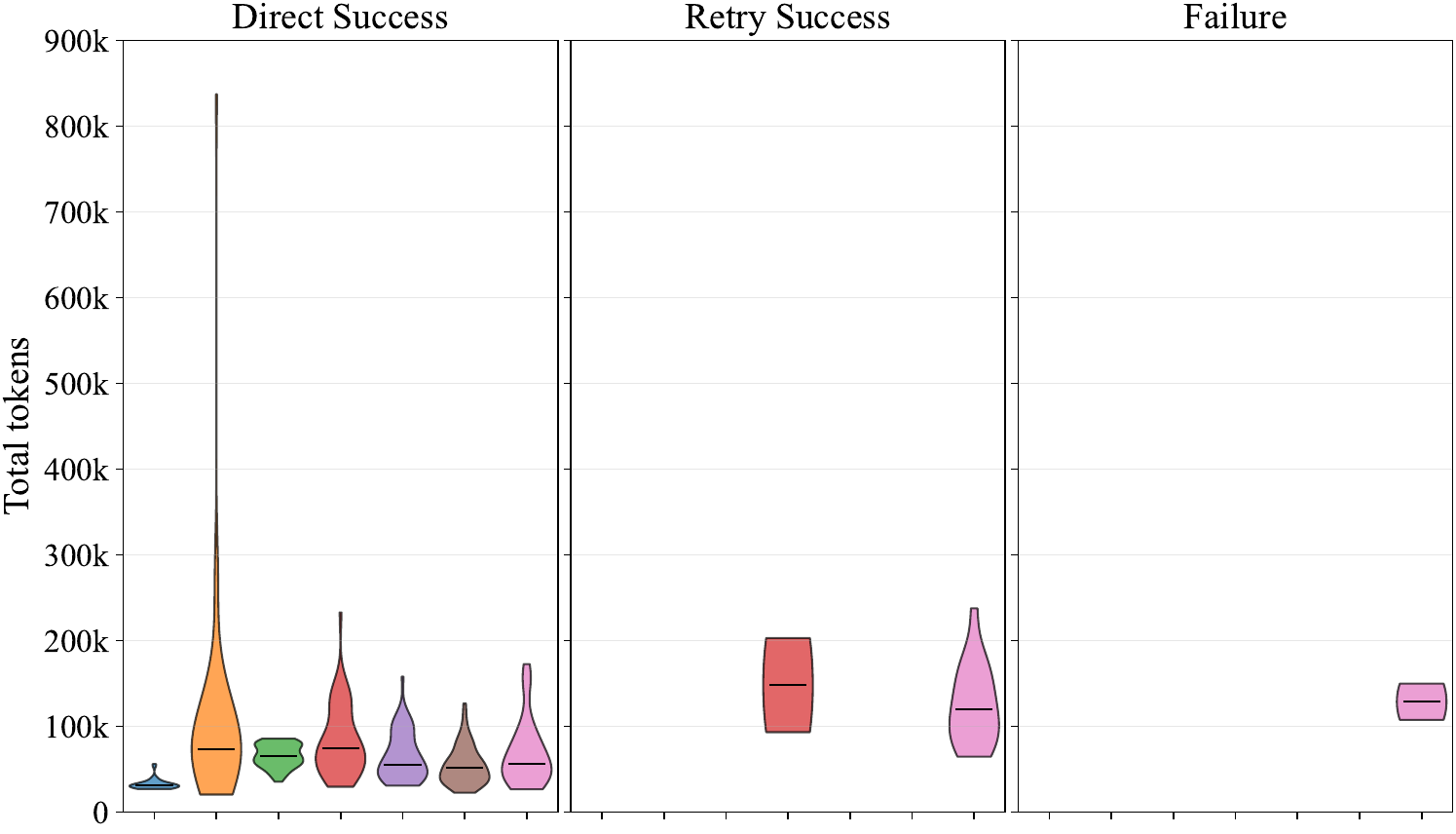}
    \end{minipage}%
    \begin{minipage}{0.32\textwidth}
        \includegraphics[width=\linewidth]{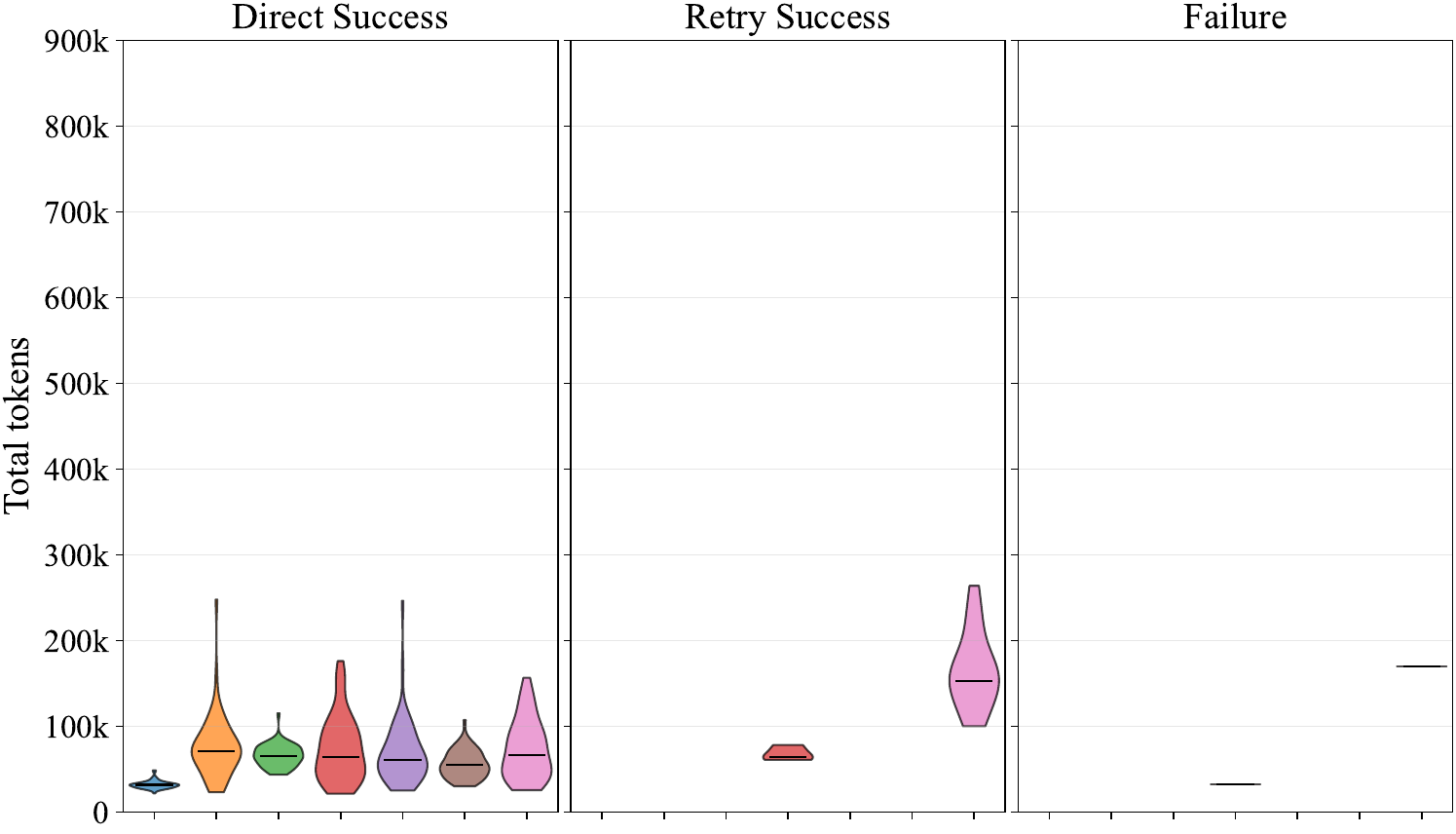}
    \end{minipage}%
    \begin{minipage}{0.32\textwidth}
        \includegraphics[width=\linewidth]{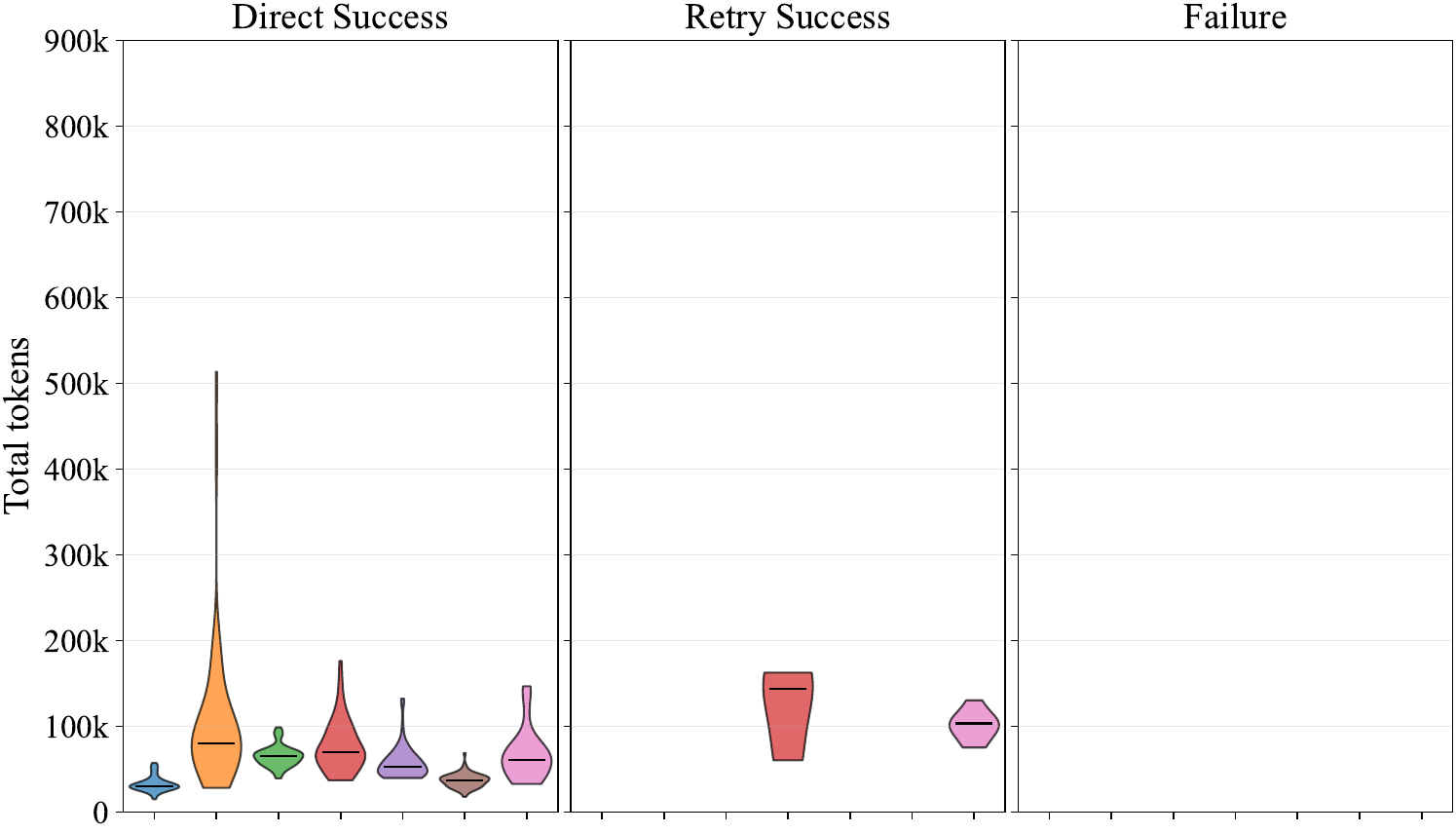}
    \end{minipage}
    \vspace{0.4em}
    
    \begin{minipage}{0.03\textwidth}
        \rotatebox{90}{\scriptsize \textbf{Landing Pg.}}
    \end{minipage}%
    \begin{minipage}{0.32\textwidth}
        \includegraphics[width=\linewidth]{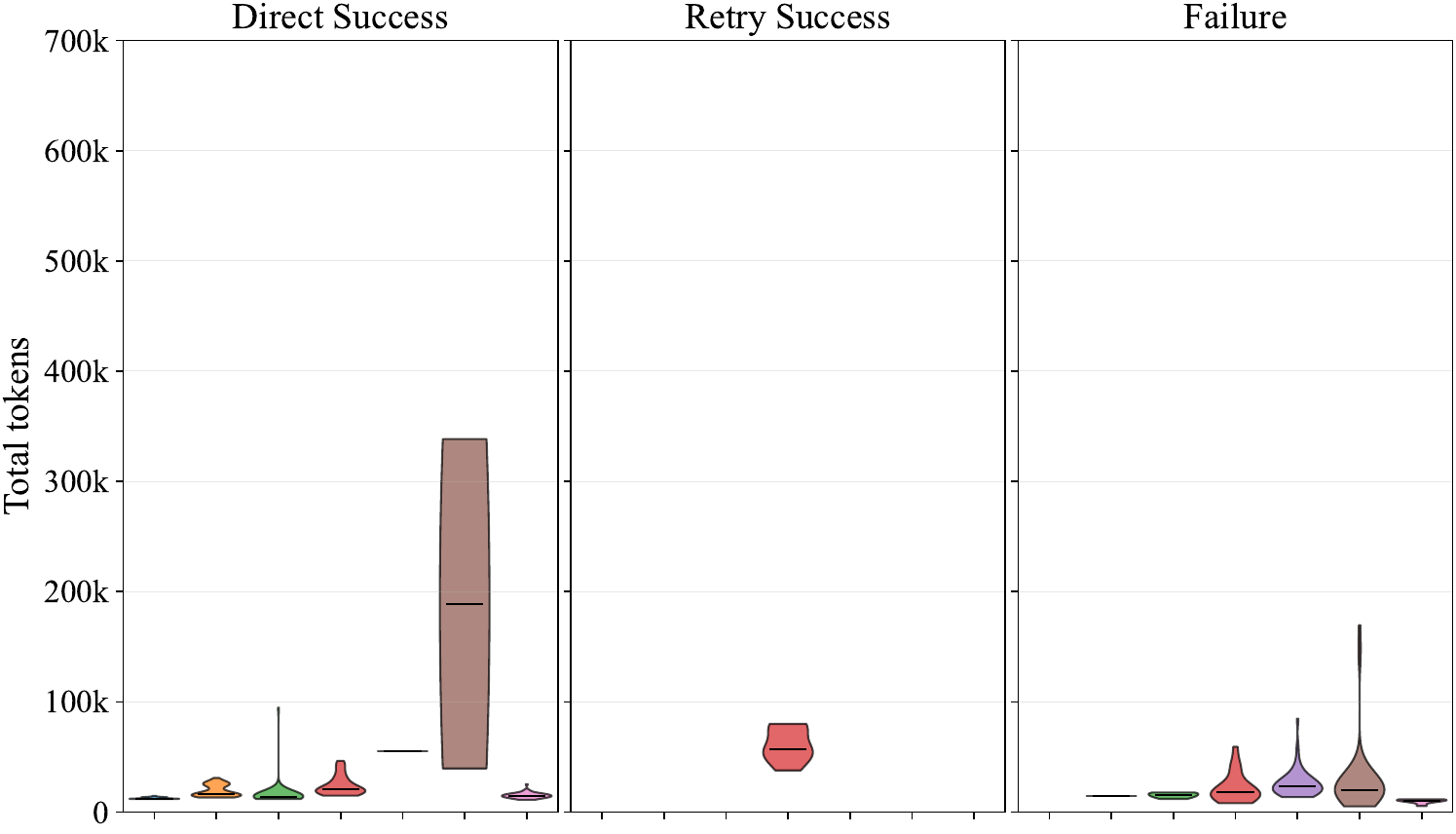}
    \end{minipage}%
    \begin{minipage}{0.32\textwidth}
        \includegraphics[width=\linewidth]{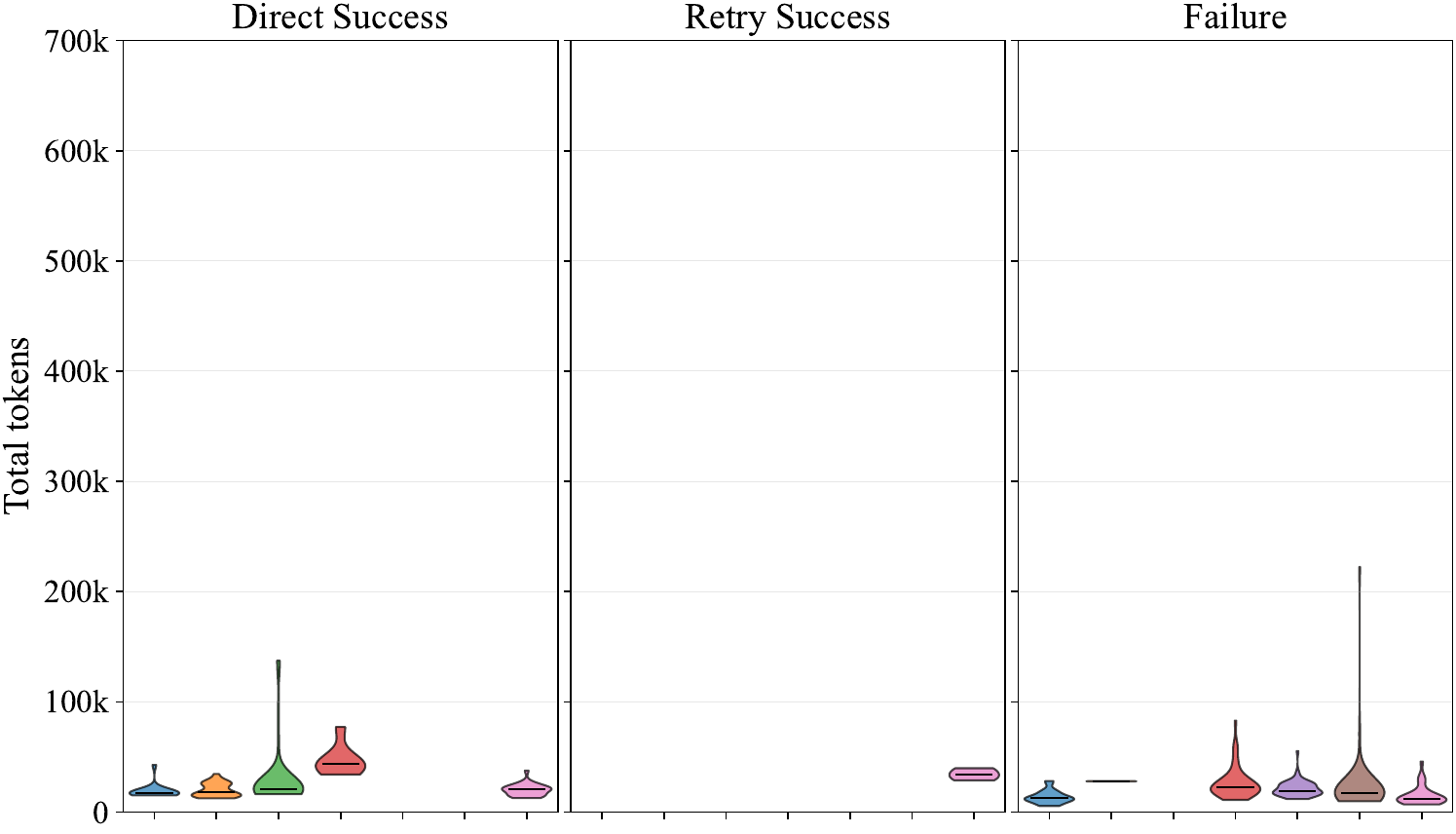}
    \end{minipage}%
    \begin{minipage}{0.32\textwidth}
        \includegraphics[width=\linewidth]{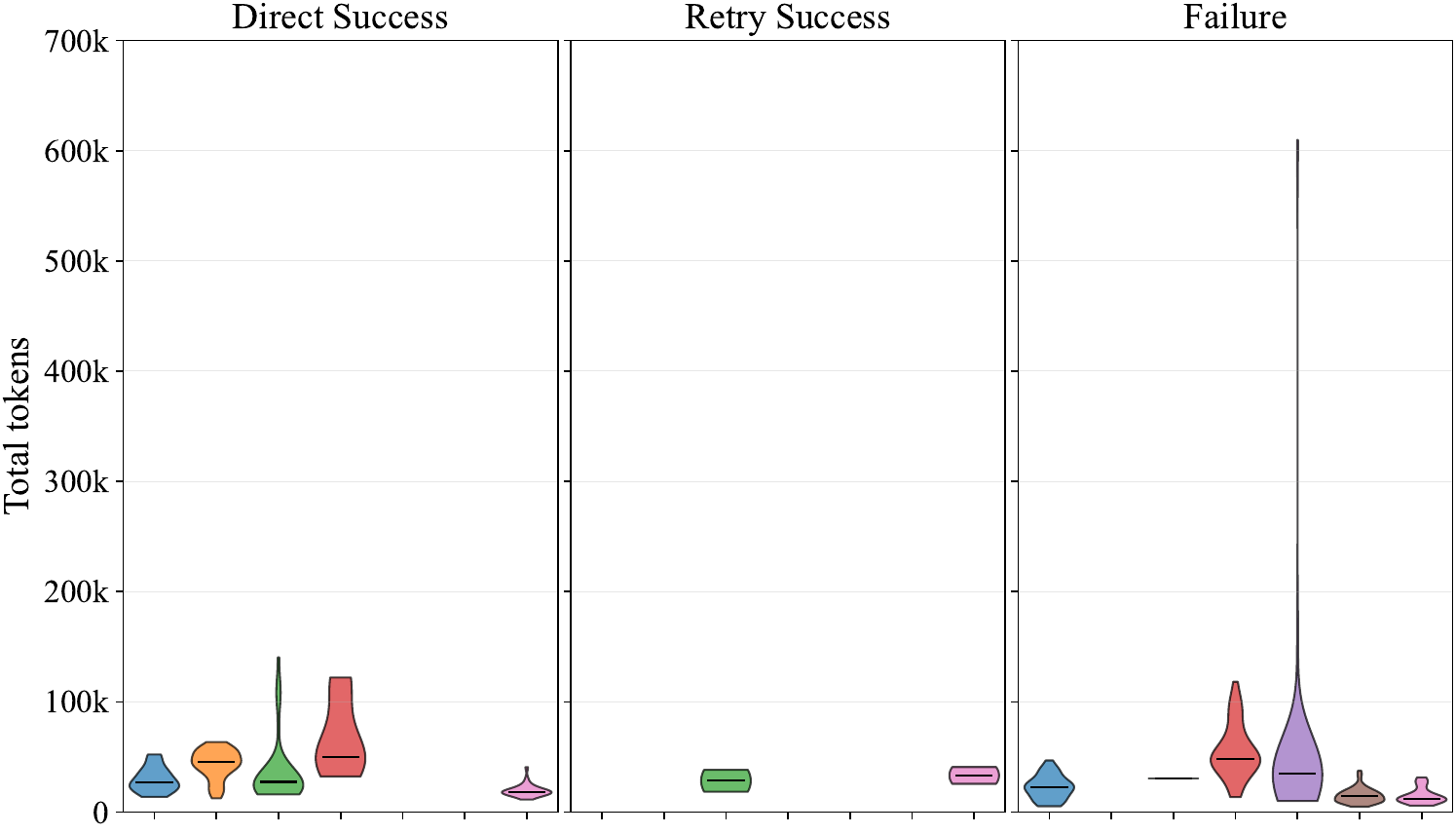}
    \end{minipage}
    \vspace{0.4em}

    \begin{minipage}{0.03\textwidth}
        \rotatebox{90}{\scriptsize \textbf{Social M. M.}}
    \end{minipage}%
    \begin{minipage}{0.32\textwidth}
        \includegraphics[width=\linewidth]{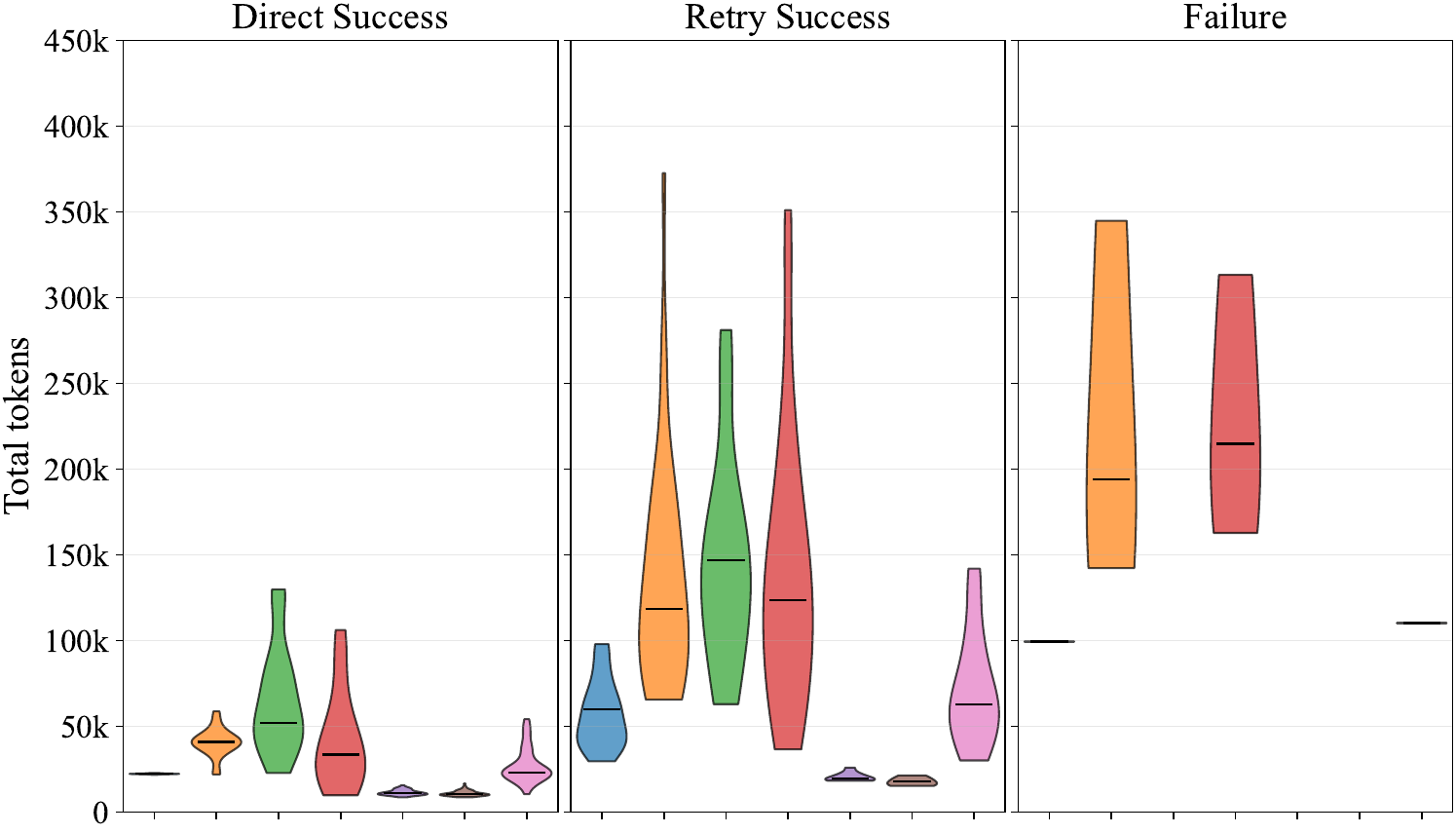}
    \end{minipage}%
    \begin{minipage}{0.32\textwidth}
        \includegraphics[width=\linewidth]{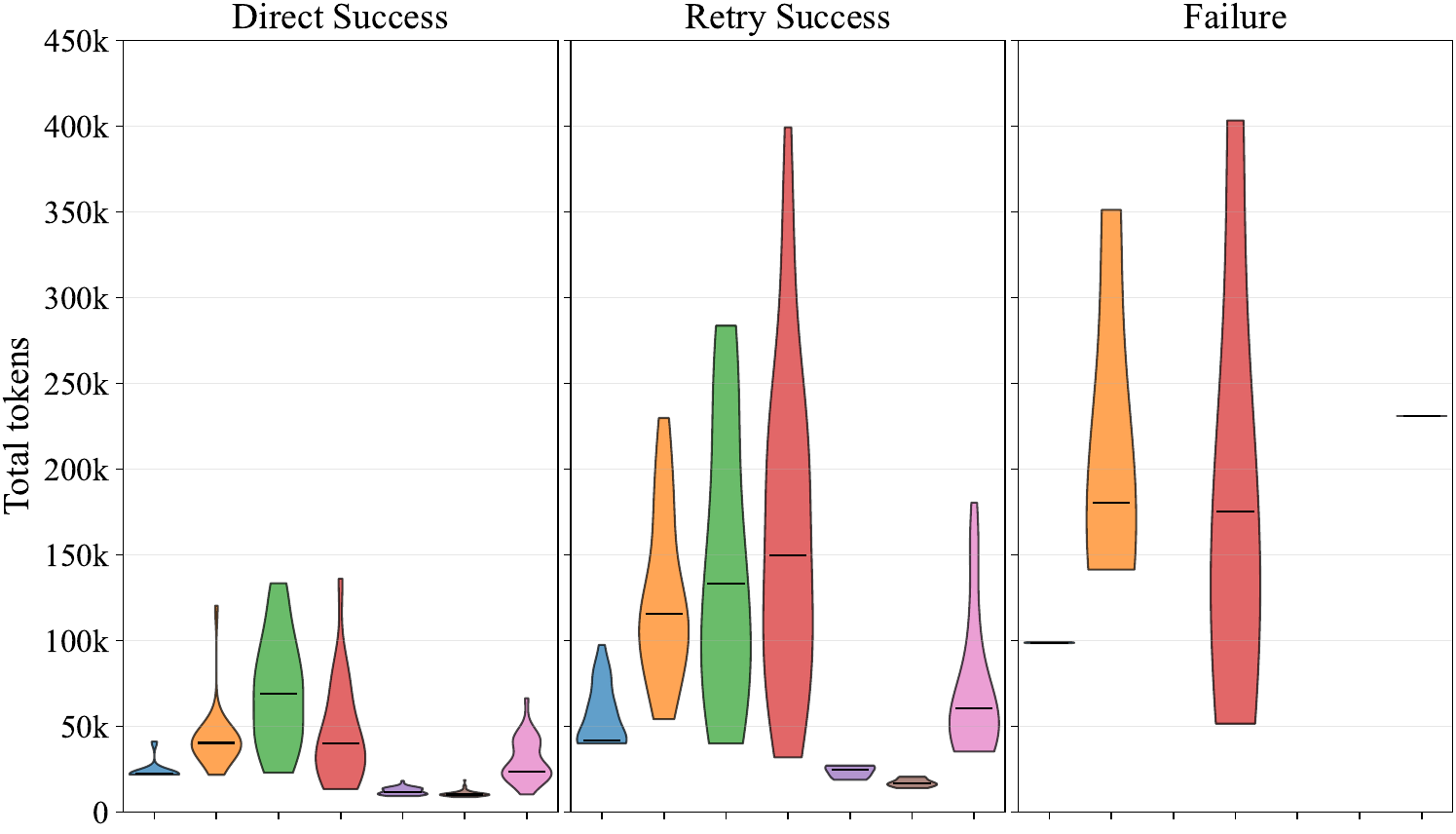}
    \end{minipage}%
    \begin{minipage}{0.32\textwidth}
        \includegraphics[width=\linewidth]{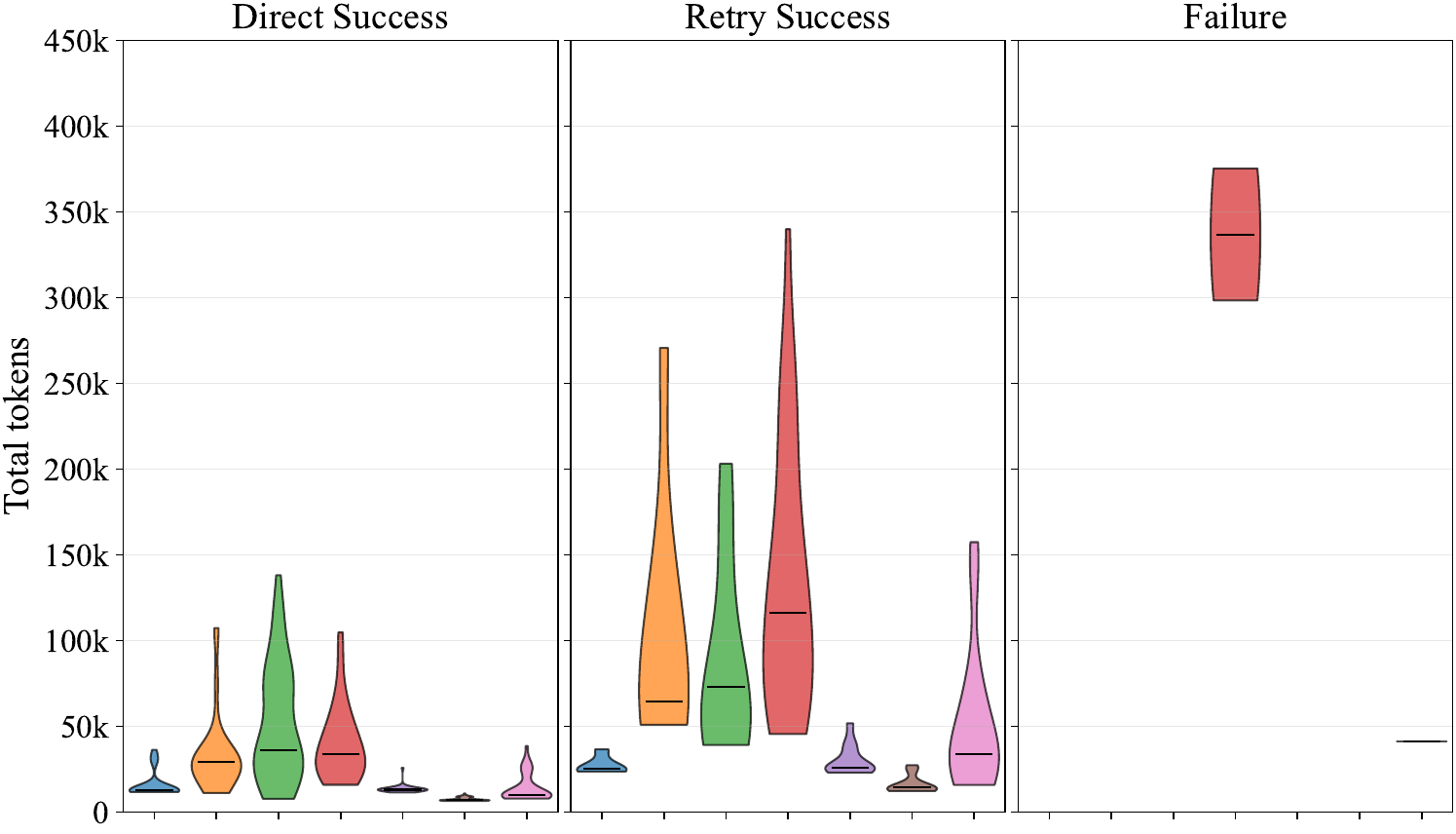}
    \end{minipage}
    \vspace{0.4em}

    \begin{minipage}{0.03\textwidth}
        \rotatebox{90}{\scriptsize \textbf{Book Writer}}
    \end{minipage}%
    \begin{minipage}{0.32\textwidth}
        \includegraphics[width=\linewidth]{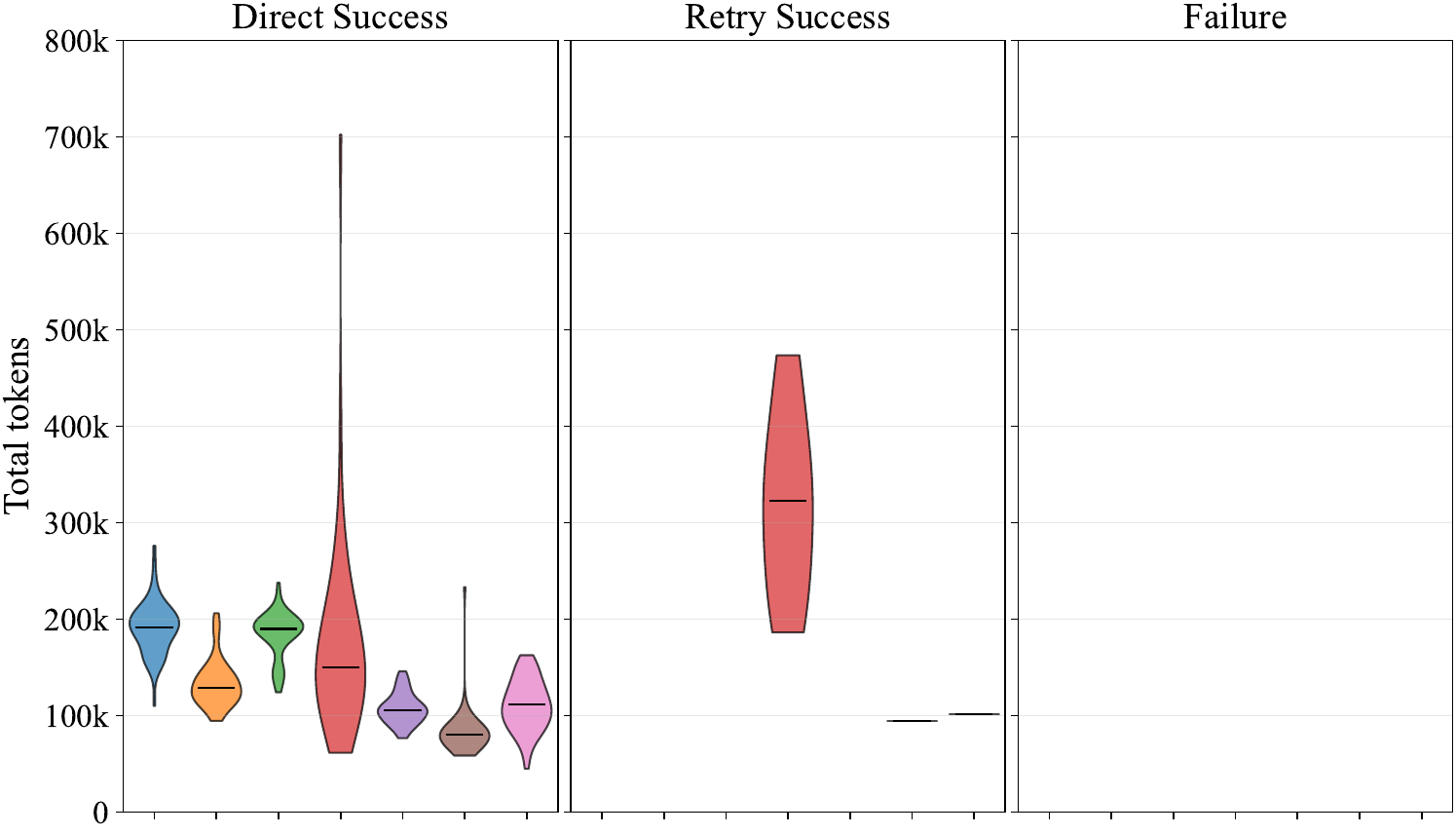}
    \end{minipage}%
    \begin{minipage}{0.32\textwidth}
        \includegraphics[width=\linewidth]{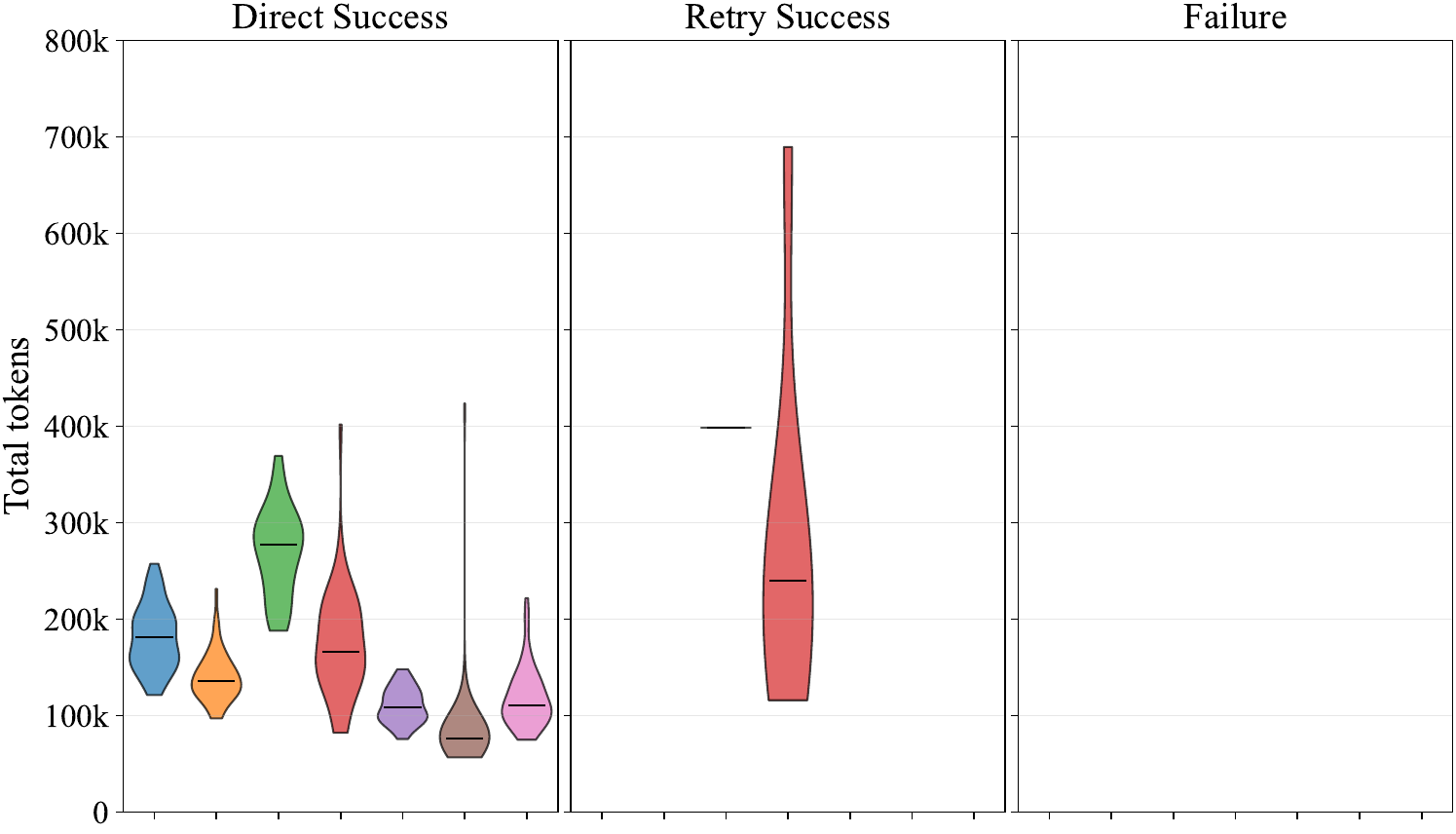}
    \end{minipage}%
    \begin{minipage}{0.32\textwidth}
        \includegraphics[width=\linewidth]{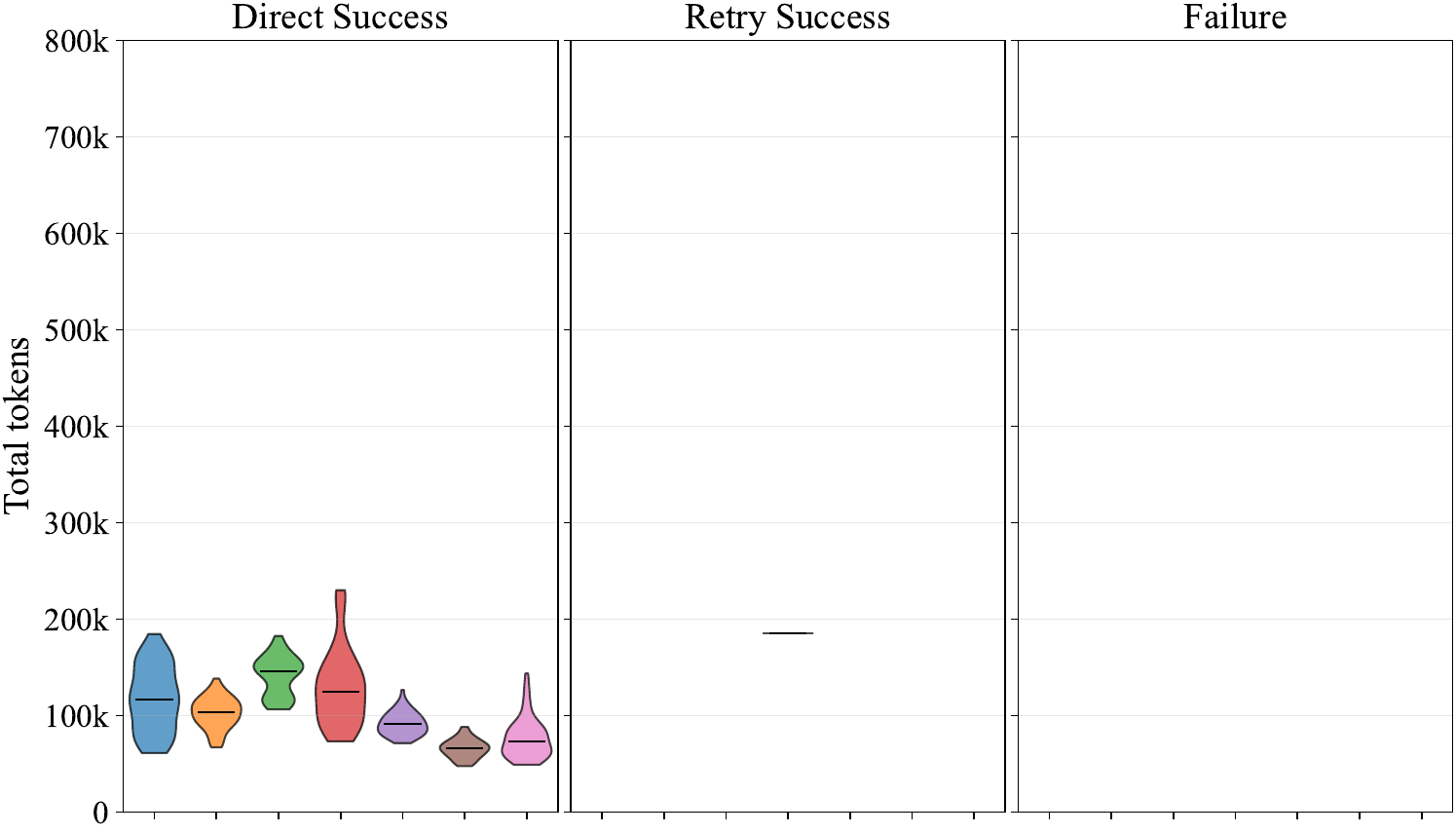}
    \end{minipage}

    \vspace{0.3em}
    \begin{minipage}{0.03\textwidth} \hfill \end{minipage}%
    \begin{minipage}{0.32\textwidth} \centering \textbf{MCP} \end{minipage}
    \begin{minipage}{0.32\textwidth} \centering \textbf{A2A} \end{minipage}
    \begin{minipage}{0.32\textwidth} \centering \textbf{H-A2A} \end{minipage}

    \caption{Token usage breakdown across for applications with threee variants.}
    \label{fig:token_3_cols}
\end{figure}

The analysis in RQ2 (\S~\ref{sec:rq2}) suggests that the infrastructure overhead introduced by the A2A protocol is limited. However, the economic analysis reveals a trade-off regarding token usage. While \texttt{Recruitment Assistant} (Overall) maintained comparable average token consumption ($-4.4\%$), all other projects (Fig.~\ref{fig:token_3_cols}) demonstrated an overall increase on average token volume when transitioning to distributed architectures utilizing the A2A protocol.

This upward trend is largely observed in specific model-task combinations. In \texttt{SQL Assistant} (Overall), the \texttt{Gemini-2.5-flash} series recorded average increases of approximately $70\%$, which can be attributed to the repeated transmission of complete schema definitions. Similarly, expressive models such as \texttt{DeepSeek-V3.1} showed marked increases, rising by $81.6\%$ on average in \texttt{Landing Page Generator} (Overall) and $47.7\%$ in \texttt{Book Writer} (Overall). These figures indicate that preserving semantic consistency across distributed agentic systems via the A2A protocol tends to involve increased token consumption.


\finding{\textbf{Context Redundancy Overhead:} While infrastructure latency remains minimal, A2A architectures necessitate context restatement to bridge isolated agents. This forces the system to recurrently feed the same state information into the LLM, resulting in significant token inflation (exceeding $80\%$) due to cumulative input redundancy.}\label{fnd:token_tax}

\section{Discussion}\label{sec:discussion}

\subsection{Implications}

\subsubsection{Implication 1: Capability-Aware Orchestration for AI-Native Systems}
\label{sec:implication-capability-orchestration}

RQ1 (\S~\ref{sec:rq1}) demonstrates that AI-Native systems cannot rely on static abstractions used in traditional software because they are characterized by probabilistic control flows. High-level categories like ``flagship'' or ``reasoning'' fail as reliable predictors for component behavior. Instead, reliability in AI-Native systems depends on architectural fit. our empiricl study identifies critical deviations where \texttt{GPT-5} prioritizes tool precision over process fidelity (Finding~\ref{fnd:parameter}) and \texttt{Gemini-2.5-flash} exhibits high completion rates with low adherence (Finding~\ref{fnd:reasoning}). We propose an architectural shift from static binding to adaptive orchestration.


\textbf{Hierarchical Hybrid Orchestration.}
Current monolithic designs force a trade-off between process fidelity and task precision. To resolve the parameter paradox (Finding~\ref{fnd:parameter}), we recommend a hybrid hierarchical architecture. System designers should decouple the \textit{Orchestration Plane} from the \textit{Execution Plane}. 
Adherence-strong models (e.g., \texttt{GPT-4o-mini}) should serve as the orchestration backbone, ensuring strict compliance with multi-step protocols and reducing context pollution. Conversely, capability-dense models (e.g., \texttt{GPT-5}, \texttt{DeepSeek-R1}) can be dynamically invoked only as leaf-node specialists for high-complexity sub-tasks. This hybrid approach leverages the comparative advantages of different model tiers, maximizing structural integrity without sacrificing decision quality~\cite{chen2025optimizingmodelselectioncompound,routerbench}.
Future work should explore automated learning frameworks that can dynamically discover optimal model-task pairings based on real-time feedback.

\textbf{Runtime Governance of Reasoning Mode.}
In AI-Native systems, reasoning mode in LLMs functions as a stochastic compute resource that introduces latency and cost volatility. Since reasoning mode can reverse performance (Finding~\ref{fnd:reasoning}) or significantly increase token consumption (Finding~\ref{fnd:self_healing_cost}), it requires a runtime governance policy. We recommend implementing a complexity classifier to activate reasoning only for sub-tasks where correctness is strictly predicated on the reasoning process. This justifies the operational risks of stochastic execution against the budget. Future research could investigate predictive complexity modeling, enabling the system to forecast the reasoning utility of a query before execution, thereby preventing expensive failures preemptively.

\textbf{Adaptive Protocol Selection Strategy.}
The communication layer determines the stability of the agent topology. To manage the trade-off identified in Finding~\ref{fnd:architecture}, architectures should move beyond rigid protocol binding. We recommend an adaptive selection strategy: defaulting to \texttt{MCP} for high-reliability, bounded workflows (to minimize hallucination), while restricting decentralized \texttt{A2A} protocols to specific autonomous contexts where creative exploration outweighs precision. This strategy also requires handling the bias in heterogeneous frameworks (Finding~\ref{fnd:heterogeneity_paradox}) to prevent completion-centric patterns from degrading structural integrity.
Ideally, future middleware should evolve towards ``Self-Negotiating Protocols,'' where agents can autonomously negotiate the optimal serialization format and context depth based on current network and cognitive loads~\cite{zhang2025agentorchestraor}.

\textbf{Structural Integrity Verification via Tracing.}
Finally, observability in AI-Native systems should transcend performance monitoring. The completion without faithfulness phenomenon confirms that aggregate scores can conceal procedural hallucinations. Consequently, distributed tracing should evolve into a behavioral auditing primitive. Systems should implement mandatory tool-use checks directly on the trace stream to invalidate runs that generate correct outputs through incorrect sequences. This repurposes observability to strictly enforce process fidelity in an inherently non-deterministic environment (Finding~\ref{fnd:architecture}). We envision the next generation of observability as proactive guardrails, where the tracing system moves from post-mortem analysis to runtime enforcement, intercepting and correcting deviant trajectories in real-time~\cite{rombaut2025watson}.

Overall, AI-Native system design requires a fundamental departure from static component assignment. The architecture should embrace probabilistic engineering where hybrid model hierarchies, reasoning policies, and protocol middleware are tuned dynamically to balance trajectory faithfulness, runtime robustness, and end-to-end completion.

\subsubsection{Implication 2: Straggler-Centric Performance Engineering}
\label{sec:implication-performance}

RQ2 (\S~\ref{sec:rq2}) challenges the conventional wisdom that network latency and serialization are the primary bottlenecks in distributed agent architectures. While network overhead remains a physical constraint, our data shows that in AI-Native systems, it is heavily masked by the inference monopoly (Finding~\ref{fnd:latency_dominance}). Since LLM computation currently consumes more than 99\% of execution time, the relative impact of protocol transport is diminished. Consequently, end-to-end latency is not governed by packet round-trips but by long-tail distributions driven by specific agent bottlenecks (Finding~\ref{fnd:agent_bottleneck}). We propose reorienting performance engineering from network optimization to computational profiling.

\textbf{Decoupled Micro-Agent Architectures.}
Engineers should move beyond the fear of distribution overhead. Although network latency is non-zero, it is overshadowed by inference costs in current LLM-based workflows (Finding~\ref{fnd:latency_dominance}). This allows systems to adopt decoupled Micro-Agent Architectures like A2A to maximize fault isolation and maintainability without incurring a prohibitive performance penalty. The primary constraint on distribution shifts from latency to the financial cost of context redundancy (Finding~\ref{fnd:token_tax}). Therefore, architectural boundaries should be defined by logical domains and token budgets rather than physical proximity. Future work should focus on semantic compression protocols in distributed protocol, which minimize the token tax of context redundancy by transmitting condensed vector representations rather than raw textual history between distributed agents~\cite{pan-etal-2024-llmlingua}.

\textbf{Critical Path Visualization via Tracing.}
Performance optimization in AI-Native systems follows a strict variation of Amdahl's Law where the slowest stochastic node dictates global latency. Here, distributed tracing should evolve from monitoring I/O blocking to visualizing the stochastic critical path. Observability platforms need to treat agentic spans as first-class citizens. They should highlight individual agent computation and tool times to pinpoint the straggler agents or tools (Finding~\ref{fnd:agent_bottleneck}) amidst concurrent executions. Without trace-level visibility into these stochastic bottlenecks, interventions like model swapping, context pruning, or parallelization remain speculative. We envision future observability platforms evolving into runtime optimization engines that can detect stragglers in real-time and dynamically trigger interventions~\cite{CRISP}, such as swapping to a faster model or parallelizing sub-tasks—to recover latency targets.

\textbf{Minimizing Orchestration Overhead.}
While network costs are masked, infrastructure overhead is not. The choice of orchestration framework acts as a hidden latency floor. Heavyweight frameworks like \texttt{CrewAI} introduce serialization delays that can significantly exceed raw network transport times (Finding~\ref{fnd:infra_overhead}). For latency-sensitive control loops, architects should prefer lightweight control planes such as \texttt{LangGraph} over thick abstractions. Furthermore, intentionally introducing structural heterogeneity can accelerate convergence. As the heterogeneity paradox (Finding~\ref{fnd:heterogeneity_paradox}) reveals, forcing agents to interact via rigid homogeneous protocols can induce synchronized reasoning loops that stall progress. Mixed-framework topologies can disrupt these loops and force faster system progression. Future research should explore agentic compilers that automatically transpile high-level workflow definitions into the most efficient runtime primitives~\cite{lin2025awarecompiler}, stripping away abstraction overhead based on the target deployment environment.

Overall, performance engineering in AI-Native systems should shift focus from the network layer to the compute layer. The goal is not merely to minimize round-trips but to leverage tracing to identify and accelerate the stragglers that dominate the critical path.

\subsubsection{Implication 3: Budget-Aware Governance for AI-Native Systems}
\label{sec:implication-budget}

RQ3 (\S~\ref{sec:rq3}) reveals a conflict between traditional resilience patterns and the economic reality of LLMs. While mechanisms like retries and circuit breakers are standard in distributed systems, our data shows that they can be counterproductive in AI-Native architectures. Specifically, the high cost of self-healing (Finding~\ref{fnd:self_healing_cost}) implies that indiscriminate recovery attempts often inflate costs without proportional accuracy gains. We propose redefining these classic patterns through the lens of budget-aware governance.

\textbf{From Mechanical Retries to Resource-Adaptive Planning.}
System architects should move beyond mechanical retries that simply repeat failed steps. Standard retries operate blindly and often deplete the token budget on invalid paths. To mitigate this, we propose resource-adaptive planning. This mechanism replaces the loop-until-success pattern with a dynamic planner that tracks the remaining budget state. When resources are plentiful, the agent utilizes a retry budget for broad exploration. As the budget dwindles, the planner forces a switch to exploitation strategies. This ensures that retries are treated as a finite investment rather than a default error-handling mechanism. Future work implies the development of economic controllers trained via Reinforcement Learning to learn the optimal stopping policy for retries based on task complexity and remaining funds~\cite{AAAI2024metareasoning}.

\textbf{Semantic Circuit Breakers.}
Traditional circuit breakers fail to protect AI-Native systems because they trigger on network errors rather than logic faults. An agent can successfully consume tokens while hallucinating, effectively bypassing standard breakers. We recommend implementing semantic circuit breakers powered by a distinct verification module. This module continuously checks intermediate outputs against logical constraints~\cite{zhang2025rvllm}. If a sub-task violates these constraints, the verifier trips the circuit immediately. This enforces a stop loss action that prunes the current branch. Unlike traditional breakers that stop all traffic, this semantic breaker redirects the agent to a new path before the budget is exhausted on a dead end. 

\textbf{Economic Fail-Fast.}
The principle of fail-fast should be redefined around token economics. Due to the cumulative nature of the token tax (Finding~\ref{fnd:token_tax}), prolonged recovery attempts often degrade performance as error traces pollute the context. Therefore, systems should enforce a policy of context bankruptcy. Instead of relying on timeout thresholds, the system should monitor the information gain of each step. If the entropy of the solution space fails to decrease after a set budget, the system declares bankruptcy and triggers a hard reset. It is more cost-effective to restart from a clean state than to pay for the continuous processing of a noisy and insolvent history. Future research should investigate entropy-based termination metrics, enabling systems to quantitatively measure the convergence rate of a reasoning chain and abort divergent trajectories mathematically.

In general, resilience in AI-Native systems requires upgrading standard primitives. Mechanical retries should evolve into adaptive planning, network circuit breakers into semantic verifiers, and simple timeouts into economic fail-fast policies backed by active state management.

\subsection{Threats to Validity}

\textbf{Internal Validity.}
The primary threat to internal validity stems from the inherent non-determinism of LLMs. Even with the temperature set to zero, commercial APIs like OpenAI and Google Gemini do not guarantee strict reproducibility due to sparse MoE architectures and non-deterministic CUDA operations. To mitigate this stochasticity, we repeated experiments multiple times to capture the variance in agent behavior rather than relying on single-shot results. Another threat is the opacity of model updates. Since we rely on closed-source APIs, silent updates or version deprecations during the data collection window could introduce inconsistencies. We mitigated this by locking specific model versions (e.g., \texttt{deepseek-v3.1-250821}) where possible and compressing the experimentation timeline to minimize drift.

\textbf{External Validity.}
A key concern is the generalizability of our findings to future models or different domains. The rapid evolution of LLMs means that absolute performance metrics may become obsolete quickly. However, our study focuses on architectural patterns (e.g., A2A vs. Centralized) and economic constraints (e.g., Token Tax) rather than leaderboard rankings. We argue that the identified phenomena, such as context pollution and the heterogeneity paradox, are structural challenges inherent to autoregressive models rather than artifacts of specific versions. While our benchmarks cover diverse tasks, they may not fully represent specialized domains like legal or medical counseling where domain-specific fine-tuning dominates general reasoning.

\textbf{Construct Validity.}
The measurement of ``correctness'' in agentic workflows is susceptible to metric bias. Reliance solely on final answer accuracy can miss subtle process errors or lucky guesses. We addressed this by implementing the multi-dimensional analysis discussed in RQ1, incorporating trace-level metrics and workflow adherence checks alongside standard pass rates. Additionally, prompt sensitivity poses a threat. It is possible that the performance gaps observed are partly due to prompts being inadvertently optimized for specific model families. To reduce this alignment bias, we utilized standardized prompt templates derived from community best practices (e.g., \texttt{CrewAI} defaults) and avoided model-specific prompt engineering to maintain a neutral evaluation baseline.

\section{Conclusion}\label{sec:conclusion}

The transition from Cloud-Native to AI-Native architectures represents a fundamental paradigm shift in software engineering. As applications evolve from deterministic microservices to probabilistic agentic services, traditional black-box evaluation methods have become insufficient. They measure what an agent achieves but fail to explain why it succeeds or fails. To address this observability crisis, we presented \bench, the first white-box benchmark suite grounded in distributed tracing and industry-standard protocols (MCP and A2A).

Our extensive empirical study re-examines several prevailing assumptions in the field. We observed a parameter paradox, where lightweight models frequently demonstrate adherence to engineering protocols that is comparable to their flagship counterparts. Furthermore, we identified a pervasive inference dominance in performance profiles, revealing that system latency is predominantly driven by LLM inference, rendering network overhead secondary. Economically, our data indicates that agentic systems are prone to an expensive failure pattern, where self-healing mechanisms can paradoxically risk acting as cost multipliers rather than safeguards.

Ultimately, this work advocates for a critical pivot in how we evaluate AI. The community must move beyond measuring raw reasoning capability to assessing system-level engineering characteristics. By treating agentic spans as first-class citizens within distributed traces, \bench provides the necessary visibility to transform stochastic agents into reliable, cost-effective, and observable software systems. We hope this open-source benchmark and dataset serves as a foundational platform for the next generation of AI-Native system research.


\bibliographystyle{ACM-Reference-Format}
\bibliography{ref}

@String{Computing = "Computing" }

@String{Springer = "Springer-Verlag" }

@ArtifactSoftware{R,
    title = {R: A Language and Environment for Statistical Computing},
    author = {{R Core Team}},
    organization = {R Foundation for Statistical Computing},
    address = {Vienna, Austria},
    year = {2019},
    url = {https://www.R-project.org/},
}

@inproceedings{AgentBench,
  author       = {Xiao Liu and
                  Hao Yu and
                  Hanchen Zhang and
                  Yifan Xu and
                  Xuanyu Lei and
                  Hanyu Lai and
                  Yu Gu and
                  Hangliang Ding and
                  Kaiwen Men and
                  Kejuan Yang and
                  Shudan Zhang and
                  Xiang Deng and
                  Aohan Zeng and
                  Zhengxiao Du and
                  Chenhui Zhang and
                  Sheng Shen and
                  Tianjun Zhang and
                  Yu Su and
                  Huan Sun and
                  Minlie Huang and
                  Yuxiao Dong and
                  Jie Tang},
  title        = {AgentBench: Evaluating LLMs as Agents},
  booktitle    = {{ICLR} 2024},
  publisher    = {OpenReview},
  year         = {2024},
  url          = {https://openreview.net/forum?id=zAdUB0aCTQ},
}

@inproceedings{MultiAgentBench,
  author       = {Kunlun Zhu and
                  Hongyi Du and
                  Zhaochen Hong and
                  Xiaocheng Yang and
                  Shuyi Guo and
                  Zhe Wang and
                  Zhenhailong Wang and
                  Cheng Qian and
                  Robert Tang and
                  Heng Ji and
                  Jiaxuan You},
  title        = {MultiAgentBench: Evaluating the Collaboration and Competition of {LLM} agents},
  booktitle    = {{ACL} 2025},
  pages        = {8580--8622},
  publisher    = {Association for Computational Linguistics},
  year         = {2025},
  url          = {https://aclanthology.org/2025.acl-long.421/},
}

@misc{MCP-AgentBench,
      title={MCP-AgentBench: Evaluating Real-World Language Agent Performance with MCP-Mediated Tools}, 
      author={Zikang Guo and Benfeng Xu and Chiwei Zhu and Wentao Hong and Xiaorui Wang and Zhendong Mao},
      year={2025},
      eprint={2509.09734},
      archivePrefix={arXiv},
      primaryClass={cs.CL},
      url={https://arxiv.org/abs/2509.09734}, 
}

@inproceedings{ALFWorld2021ICLR,
  author       = {Mohit Shridhar and
                  Xingdi Yuan and
                  Marc{-}Alexandre C{\^{o}}t{\'{e}} and
                  Yonatan Bisk and
                  Adam Trischler and
                  Matthew J. Hausknecht},
  title        = {ALFWorld: Aligning Text and Embodied Environments for Interactive
                  Learning},
  booktitle    = { {ICLR} 2021},
  publisher    = {OpenReview.net},
  year         = {2021},
  url          = {https://openreview.net/forum?id=0IOX0YcCdTn},
}

@inproceedings{KDDSurvey,
author = {Mohammadi, Mahmoud and Li, Yipeng and Lo, Jane and Yip, Wendy},
title = {Evaluation and Benchmarking of LLM Agents: A Survey},
year = {2025},
publisher = {ACM},
url = {https://doi.org/10.1145/3711896.3736570},
doi = {10.1145/3711896.3736570},
booktitle = {SIGKDD 2025},
pages = {6129–6139}
}

@article{HELM,
  author       = {Percy Liang and
                  Rishi Bommasani and
                  Tony Lee and
                  Dimitris Tsipras and
                  Dilara Soylu and
                  Michihiro Yasunaga and
                  Yian Zhang and
                  Deepak Narayanan and
                  Yuhuai Wu and
                  Ananya Kumar and
                  Benjamin Newman and
                  Binhang Yuan and
                  Bobby Yan and
                  Ce Zhang and
                  Christian Cosgrove and
                  Christopher D. Manning and
                  Christopher R{\'{e}} and
                  Diana Acosta{-}Navas and
                  Drew A. Hudson and
                  Eric Zelikman and
                  Esin Durmus and
                  Faisal Ladhak and
                  Frieda Rong and
                  Hongyu Ren and
                  Huaxiu Yao and
                  Jue Wang and
                  Keshav Santhanam and
                  Laurel J. Orr and
                  Lucia Zheng and
                  Mert Y{\"{u}}ksekg{\"{o}}n{\"{u}}l and
                  Mirac Suzgun and
                  Nathan Kim and
                  Neel Guha and
                  Niladri S. Chatterji and
                  Omar Khattab and
                  Peter Henderson and
                  Qian Huang and
                  Ryan Chi and
                  Sang Michael Xie and
                  Shibani Santurkar and
                  Surya Ganguli and
                  Tatsunori Hashimoto and
                  Thomas Icard and
                  Tianyi Zhang and
                  Vishrav Chaudhary and
                  William Wang and
                  Xuechen Li and
                  Yifan Mai and
                  Yuhui Zhang and
                  Yuta Koreeda},
  title        = {Holistic Evaluation of Language Models},
  journal      = {Trans. Mach. Learn. Res.},
  volume       = {2023},
  year         = {2023},
  url          = {https://openreview.net/forum?id=iO4LZibEqW},
}

@inproceedings{MINEDOJO,
author = {Fan, Linxi and Wang, Guanzhi and Jiang, Yunfan and Mandlekar, Ajay and Yang, Yuncong and Zhu, Haoyi and Tang, Andrew and Huang, De-An and Zhu, Yuke and Anandkumar, Anima},
title = {MINEDOJO: building open-ended embodied agents with internet-scale knowledge},
year = {2022},
publisher = {Curran Associates Inc.},
booktitle = {NeurIPS 2022},
articleno = {1333},
numpages = {20},
url       = {http://papers.nips.cc/paper\_files/paper/2022/hash/74a67268c5cc5910f64938cac4526a90-Abstract-Datasets\_and\_Benchmarks.html},
}

@inproceedings{WebShop,
  author       = {Shunyu Yao and
                  Howard Chen and
                  John Yang and
                  Karthik Narasimhan},
  title        = {WebShop: Towards Scalable Real-World Web Interaction with Grounded Language Agents},
  booktitle = {NeurIPS 2022},
  publisher = {Curran Associates Inc.},
  articleno = {1508},
  year         = {2022},
  url          = {http://papers.nips.cc/paper\_files/paper/2022/hash/82ad13ec01f9fe44c01cb91814fd7b8c-Abstract-Conference.html},
}

@inproceedings{Mind2Web,
  author       = {Xiang Deng and
                  Yu Gu and
                  Boyuan Zheng and
                  Shijie Chen and
                  Samual Stevens and
                  Boshi Wang and
                  Huan Sun and
                  Yu Su},
  title        = {Mind2Web: Towards a Generalist Agent for the Web},
  booktitle    = {NeurIPS 2023},
  publisher = {Curran Associates Inc.},
  year         = {2023},
  url          = {http://papers.nips.cc/paper\_files/paper/2023/hash/5950bf290a1570ea401bf98882128160-Abstract-Datasets\_and\_Benchmarks.html},
}

@inproceedings{Toolformer,
author = {Schick, Timo and Dwivedi-Yu, Jane and Dess\'{\i}, Roberto and Raileanu, Roberta and Lomeli, Maria and Hambro, Eric and Zettlemoyer, Luke and Cancedda, Nicola and Scialom, Thomas},
title = {Toolformer: language models can teach themselves to use tools},
year = {2023},
publisher = {Curran Associates Inc.},
booktitle    = {NeurIPS 2023},
url          = {http://papers.nips.cc/paper\_files/paper/2023/hash/d842425e4bf79ba039352da0f658a906-Abstract-Conference.html},
articleno = {2997},
numpages = {13},
series = {NIPS '23}
}

@inproceedings{ToolLLM,
  author       = {Yujia Qin and
                  Shihao Liang and
                  Yining Ye and
                  Kunlun Zhu and
                  Lan Yan and
                  Yaxi Lu and
                  Yankai Lin and
                  Xin Cong and
                  Xiangru Tang and
                  Bill Qian and
                  Sihan Zhao and
                  Lauren Hong and
                  Runchu Tian and
                  Ruobing Xie and
                  Jie Zhou and
                  Mark Gerstein and
                  Dahai Li and
                  Zhiyuan Liu and
                  Maosong Sun},
  title        = {ToolLLM: Facilitating Large Language Models to Master 16000+ Real-world
                  APIs},
  booktitle    = {{ICLR} 2024},
  publisher    = {OpenReview.net},
  year         = {2024},
  url          = {https://openreview.net/forum?id=dHng2O0Jjr},
}

@inproceedings{GAIA,
  author       = {Gr{\'{e}}goire Mialon and
                  Cl{\'{e}}mentine Fourrier and
                  Thomas Wolf and
                  Yann LeCun and
                  Thomas Scialom},
  title        = {{GAIA:} a benchmark for General {AI} Assistants},
  booktitle    = {{ICLR} 2024},
  publisher    = {OpenReview.net},
  year         = {2024},
  url          = {https://openreview.net/forum?id=fibxvahvs3},
}

@inproceedings{SWE-bench,
  author       = {Carlos E. Jimenez and
                  John Yang and
                  Alexander Wettig and
                  Shunyu Yao and
                  Kexin Pei and
                  Ofir Press and
                  Karthik R. Narasimhan},
  title        = {SWE-bench: Can Language Models Resolve Real-world Github Issues?},
  booktitle    = {{ICLR} 2024},
  publisher    = {OpenReview.net},
  year         = {2024},
  url          = {https://openreview.net/forum?id=VTF8yNQM66},
}

@inproceedings{MLAgentBench,
  author       = {Qian Huang and
                  Jian Vora and
                  Percy Liang and
                  Jure Leskovec},
  title        = {MLAgentBench: Evaluating Language Agents on Machine Learning Experimentation},
  booktitle    = {{ICML} 2024},
  publisher    = {OpenReview.net},
  year         = {2024},
  url          = {https://openreview.net/forum?id=1Fs1LvjYQW},
}

@article{MCP-Bench,
  author       = {Zhenting Wang and
                  Qi Chang and
                  Hemani Patel and
                  Shashank Biju and
                  Cheng{-}En Wu and
                  Quan Liu and
                  Aolin Ding and
                  Alireza Rezazadeh and
                  Ankit Shah and
                  Yujia Bao and
                  Eugene Siow},
  title        = {MCP-Bench: Benchmarking Tool-Using {LLM} Agents with Complex Real-World
                  Tasks via {MCP} Servers},
  journal      = {CoRR},
  volume       = {abs/2508.20453},
  year         = {2025},
  url          = {https://doi.org/10.48550/arXiv.2508.20453},
  doi          = {10.48550/ARXIV.2508.20453},
  eprinttype    = {arXiv},
  eprint       = {2508.20453},
}

@article{Microscaler,
  author       = {Guangba Yu and
                  Pengfei Chen and
                  Zibin Zheng},
  title        = {Microscaler: Cost-Effective Scaling for Microservice Applications
                  in the Cloud With an Online Learning Approach},
  journal      = {{IEEE} Trans. Cloud Comput.},
  volume       = {10},
  number       = {2},
  pages        = {1100--1116},
  year         = {2022},
  url          = {https://doi.org/10.1109/TCC.2020.2985352},
  doi          = {10.1109/TCC.2020.2985352},
}

@inproceedings{MicroRank,
  author       = {Guangba Yu and
                  Pengfei Chen and
                  Hongyang Chen and
                  Zijie Guan and
                  Zicheng Huang and
                  Linxiao Jing and
                  Tianjun Weng and
                  Xinmeng Sun and
                  Xiaoyun Li},
  title        = {MicroRank: End-to-End Latency Issue Localization with Extended Spectrum
                  Analysis in Microservice Environments},
  booktitle    = {{WWW} 2021},
  pages        = {3087--3098},
  publisher    = {{ACM}},
  year         = {2021},
  url          = {https://doi.org/10.1145/3442381.3449905},
  doi          = {10.1145/3442381.3449905},
}

@inproceedings{Nezha,
  author       = {Guangba Yu and
                  Pengfei Chen and
                  Yufeng Li and
                  Hongyang Chen and
                  Xiaoyun Li and
                  Zibin Zheng},
  title        = {Nezha: Interpretable Fine-Grained Root Causes Analysis for Microservices
                  on Multi-modal Observability Data},
  booktitle    = {{ESEC/FSE} 2023},
  pages        = {553--565},
  publisher    = {{ACM}},
  year         = {2023},
  url          = {https://doi.org/10.1145/3611643.3616249},
  doi          = {10.1145/3611643.3616249},
}

@inproceedings{Deathstar,
  author       = {Yu Gan and
                  Yanqi Zhang and
                  Dailun Cheng and
                  Ankitha Shetty and
                  Priyal Rathi and
                  Nayan Katarki and
                  Ariana Bruno and
                  Justin Hu and
                  Brian Ritchken and
                  Brendon Jackson and
                  Kelvin Hu and
                  Meghna Pancholi and
                  Yuan He and
                  Brett Clancy and
                  Chris Colen and
                  Fukang Wen and
                  Catherine Leung and
                  Siyuan Wang and
                  Leon Zaruvinsky and
                  Mateo Espinosa and
                  Rick Lin and
                  Zhongling Liu and
                  Jake Padilla and
                  Christina Delimitrou},
  title        = {An Open-Source Benchmark Suite for Microservices and Their Hardware-Software
                  Implications for Cloud {\&} Edge Systems},
  booktitle    = {2019},
  pages        = {3--18},
  publisher    = {{ACM}},
  year         = {2019},
  url          = {https://doi.org/10.1145/3297858.3304013},
  doi          = {10.1145/3297858.3304013},
}

@misc{Opentelemetry,
  author =       "Opentelemetry",
  year =         "2025",
  title =        "High-quality, ubiquitous, and portable telemetry to enable effective observability",
  lastaccessed = "Oct. 10, 2025",
  url =          "https://opentelemetry.io/",
}

@misc{Cursor,
  author =       "Cursor",
  year =         "2025",
  title =        "Cursor: AI Coding Assistant.",
  lastaccessed = "Oct. 10, 2025",
  url =          "https://www.cursor.com/",
}

@misc{Codex,
  author =       "OpenAI",
  year =         "2025",
  title =        "Openai codex.",
  lastaccessed = "Oct. 10, 2025",
  url =          "https://openai.c om/index/introducing-codex/,",
}

@misc{Windsurf,
  author =       "Windsurf",
  year =         "2025",
  title =        "Windsurf:Where developers are doing their best work.",
  lastaccessed = "Oct. 10, 2025",
  url =          "https://windsurf.com/,",
}

@misc{AIScientist,
      title={The AI Scientist: Towards Fully Automated Open-Ended Scientific Discovery}, 
      author={Chris Lu and Cong Lu and Robert Tjarko Lange and Jakob Foerster and Jeff Clune and David Ha},
      year={2024},
      eprint={2408.06292},
      archivePrefix={arXiv},
      primaryClass={cs.AI},
      url={https://arxiv.org/abs/2408.06292}, 
}

@misc{AIScientist1,
      title={Towards an AI co-scientist}, 
      author={Juraj Gottweis and Wei-Hung Weng and Alexander Daryin and Tao Tu and Anil Palepu and Petar Sirkovic and Artiom Myaskovsky and Felix Weissenberger and Keran Rong and Ryutaro Tanno and Khaled Saab and Dan Popovici and Jacob Blum and Fan Zhang and Katherine Chou and Avinatan Hassidim and Burak Gokturk and Amin Vahdat and Pushmeet Kohli and Yossi Matias and Andrew Carroll and Kavita Kulkarni and Nenad Tomasev and Yuan Guan and Vikram Dhillon and Eeshit Dhaval Vaishnav and Byron Lee and Tiago R D Costa and José R Penadés and Gary Peltz and Yunhan Xu and Annalisa Pawlosky and Alan Karthikesalingam and Vivek Natarajan},
      year={2025},
      eprint={2502.18864},
      archivePrefix={arXiv},
      primaryClass={cs.AI},
      url={https://arxiv.org/abs/2502.18864}, 
}

@misc{AIScientist2,
      title={The AI Scientist-v2: Workshop-Level Automated Scientific Discovery via Agentic Tree Search}, 
      author={Yutaro Yamada and Robert Tjarko Lange and Cong Lu and Shengran Hu and Chris Lu and Jakob Foerster and Jeff Clune and David Ha},
      year={2025},
      eprint={2504.08066},
      archivePrefix={arXiv},
      primaryClass={cs.AI},
      url={https://arxiv.org/abs/2504.08066}, 
}

@misc{crew,
  author =       "CrewAI",
  year =         "2025",
  title =        "Build a crew of AI Agents today, save countless hours forever",
  lastaccessed = "Oct. 10, 2025",
  url =          "https://www.crewai.com/",
}

@misc{finance1,
  author =       "IACPM and McKinsey",
  year =         "2025",
  title =        "Emerging generative ai use cases in credit: Research results",
  lastaccessed = "Oct. 10, 2025",
  url =          "https://iacpm.org/wp-content/uploads/2025/03/IACPM-McKinsey-Gen-AI-Webinar-2025.pdf",
}

@misc{finance2,
  author       = {Verhagen, Alexander and Luget, Angela and Conjeaud, Olivia and Stergiou, Vasiliki and Banerjee, Debanjan},
  title        = {How agentic {AI} can change the way banks fight financial crime},
  howpublished = {\url{https://www.mckinsey.com/capabilities/risk-and-resilience/our-insights/how-agentic-ai-can-change-the-way-banks-fight-financial-crime}},
  year         = {2025},
  lastaccessed = "Oct. 10, 2025",
}

@misc{finance3,
  author       = {Capgemini},
  title        = {Banks and insurers deploy ai agents to fight fraud and process applications, with plans for new roles to supervise the ai},
  howpublished = {\url{https://www.capgemini.com/us-en/news/press-releases/banks-and-insurers-deploy-ai-agents-to-fight-fraud-and-process-applications-with-plans-for-new-roles-to-supervise-the-ai/}},
  year         = {2025},
  lastaccessed = "Oct. 10, 2025",
}

@misc{openlit,
  author =       "Openlit",
  year =         "2025",
  title =        "Openlit: Open Source Platform for AI Engineering",
  lastaccessed = "Oct. 10, 2025",
  url =          "https://openlit.io/",
}

@misc{openinference,
  author =       "Arize",
  year =         "2025",
  title =        "What is openinference",
  lastaccessed = "Oct. 10, 2025",
  url =          "https://arize.com/docs/ax/observe/tracing-concepts/what-is-openinference",
}

@misc{CNCF,
  author =       "CNCF",
  year =         "2025",
  title =        "Cloud Native Computing Foundation (CNCF)",
  lastaccessed = "Oct. 10, 2025",
  url =          "https://github.com/CNCF",
}

@misc{autogen,
  author =       "autogen",
  year =         "2025",
  title =        "A programming framework for agentic AI",
  lastaccessed = "Oct. 10, 2025",
  url =          "https://github.com/microsoft/autogen",
}

@misc{langgraph,
  author =       "langgraph",
  year =         "2025",
  title =        "Balance agent control with agency",
  lastaccessed = "Oct. 10, 2025",
  url =          "https://www.langchain.com/langgraph",
}

@misc{mcp,
  author =       "MCP",
  year =         "2025",
  title =        "What is the Model Context Protocol (MCP)?",
  lastaccessed = "Oct. 10, 2025",
  url =          "https://modelcontextprotocol.io/docs/getting-started/intro",
}

@misc{a2a,
  author =       "Google",
  year =         "2025",
  title =        "What is A2A Protocol?",
  lastaccessed = "Oct. 10, 2025",
  url =          "https://a2a-protocol.org/",
}

@misc{langfuse,
  author =       "Langfuse",
  year =         "2025",
  title =        "Open Source LLM Engineering Platform",
  lastaccessed = "Oct. 10, 2025",
  url =          "https://langfuse.com/",
}

@article{TraStrainer,
author = {Huang, Haiyu and Zhang, Xiaoyu and Chen, Pengfei and He, Zilong and Chen, Zhiming and Yu, Guangba and Chen, Hongyang and Sun, Chen},
title = {TraStrainer: Adaptive Sampling for Distributed Traces with System Runtime State},
year = {2024},
publisher = {ACM},
volume = {1},
number = {FSE},
url = {https://doi.org/10.1145/3643748},
doi = {10.1145/3643748},
journal = {Proc. ACM Softw. Eng.},
articleno = {22},
}

@InProceedings{Microsketch,
author="Li, Yufeng
and Yu, Guangba
and Chen, Pengfei
and Zhang, Chuanfu
and Zheng, Zibin",
title="MicroSketch: Lightweight and Adaptive Sketch Based Performance Issue Detection and Localization in Microservice Systems",
booktitle="Service-Oriented Computing",
year="2022",
publisher="Springer",
pages="219--236",
}

@article{AINative,
  author       = {Yao Lu and
                  Song Bian and
                  Lequn Chen and
                  Yongjun He and
                  Yulong Hui and
                  Matthew Lentz and
                  Beibin Li and
                  Fei Liu and
                  Jialin Li and
                  Qi Liu and
                  Rui Liu and
                  Xiaoxuan Liu and
                  Lin Ma and
                  Kexin Rong and
                  Jianguo Wang and
                  Yingjun Wu and
                  Yongji Wu and
                  Huanchen Zhang and
                  Minjia Zhang and
                  Qizhen Zhang and
                  Tianyi Zhou and
                  Danyang Zhuo},
  title        = {Computing in the Era of Large Generative Models: From Cloud-Native
                  to AI-Native},
  journal      = {CoRR},
  volume       = {abs/2401.12230},
  year         = {2024},
  url          = {https://doi.org/10.48550/arXiv.2401.12230},
  doi          = {10.48550/ARXIV.2401.12230},
}

@article{AgenticService,
  author       = {Shuiguang Deng and
                  Hailiang Zhao and
                  Ziqi Wang and
                  Guanjie Cheng and
                  Peng Chen and
                  Wenzhuo Qian and
                  Zhiwei Ling and
                  Jianwei Yin and
                  Albert Y. Zomaya and
                  Schahram Dustdar},
  title        = {Agentic Services Computing},
  journal      = {CoRR},
  volume       = {abs/2509.24380},
  year         = {2025},
  url          = {https://doi.org/10.48550/arXiv.2509.24380},
  doi          = {10.48550/ARXIV.2509.24380},
  eprinttype    = {arXiv},
  eprint       = {2509.24380},
}

@article{ChaosSurvey,
author = {Yu, Guangba and Tan, Gou and Huang, Haojia and Zhang, Zhenyu and Chen, Pengfei and Natella, Roberto and Zheng, Zibin and Lyu, Michael R.},
title = {A Survey on Failure Analysis and Fault Injection in AI Systems},
year = {2025},
publisher = {Association for Computing Machinery},
url = {https://doi.org/10.1145/3732777},
doi = {10.1145/3732777},
journal = {ACM Trans. Softw. Eng. Methodol.},
}

@misc{zhu2025llmagentsfaillearn,
      title={Where LLM Agents Fail and How They can Learn From Failures}, 
      author={Kunlun Zhu and Zijia Liu and Bingxuan Li and Muxin Tian and Yingxuan Yang and Jiaxun Zhang and Pengrui Han and Qipeng Xie and Fuyang Cui and Weijia Zhang and Xiaoteng Ma and Xiaodong Yu and Gowtham Ramesh and Jialian Wu and Zicheng Liu and Pan Lu and James Zou and Jiaxuan You},
      year={2025},
      eprint={2509.25370},
      archivePrefix={arXiv},
      primaryClass={cs.AI},
      url={https://arxiv.org/abs/2509.25370}, 
}

@inproceedings{TS-InvarNet,
  author       = {Zijun Hu and
                  Pengfei Chen and
                  Guangba Yu and
                  Zilong He and
                  Xiaoyun Li},
  title        = {TS-InvarNet: Anomaly Detection and Localization based on Tempo-spatial
                  {KPI} Invariants in Distributed Services},
  booktitle    = {{ICWS} 2022},
  pages        = {109--119},
  publisher    = {{IEEE}},
  year         = {2022},
  url          = {https://doi.org/10.1109/ICWS55610.2022.00031},
  doi          = {10.1109/ICWS55610.2022.00031},
}

@misc{he2025trajectbench,
      title={TRAJECT-Bench:A Trajectory-Aware Benchmark for Evaluating Agentic Tool Use}, 
      author={Pengfei He and Zhenwei Dai and Bing He and Hui Liu and Xianfeng Tang and Hanqing Lu and Juanhui Li and Jiayuan Ding and Subhabrata Mukherjee and Suhang Wang and Yue Xing and Jiliang Tang and Benoit Dumoulin},
      year={2025},
      eprint={2510.04550},
      archivePrefix={arXiv},
      primaryClass={cs.AI},
      url={https://arxiv.org/abs/2510.04550}, 
}

@online{phishingEmailDataset,
  author = {Zefang Liu},
  title  = {phishing-email-dataset},
  year   = {2023},
  lastaccessed = "Oct. 10, 2025",
  url    = {https://huggingface.co/datasets/zefang-liu/phishing-email-dataset}
}

@online{resumeScoreDataset,
  author = {NETSOL Technologies Inc.},
  title  = {resume-score-details},
  year   = {2024},
  lastaccessed = "Oct. 10, 2025",
  url    = {https://huggingface.co/datasets/netsol/resume-score-details}
}

@online{markdownItDemo,
  author = {Alex Kocharin and Vitaly Puzrin},
  title  = {Markdown-it Demo Documents},
  year   = {2025},
  lastaccessed = "Oct. 10, 2025",
  url    = {https://markdown-it.github.io/}
}

@online{zsqlDataset,
  author = {Zerolink},
  title  = {zsql-sqlite-dpo},
  year   = {2023},
  lastaccessed = "Oct. 10, 2025",
  url    = {https://huggingface.co/datasets/zerolink/zsql-sqlite-dpo}
}

@article{hou2024bridging,
  author       = {Yupeng Hou and
                  Jiacheng Li and
                  Zhankui He and
                  An Yan and
                  Xiusi Chen and
                  Julian J. McAuley},
  title        = {Bridging Language and Items for Retrieval and Recommendation},
  journal      = {CoRR},
  volume       = {abs/2403.03952},
  year         = {2024},
  url          = {https://doi.org/10.48550/arXiv.2403.03952},
  doi          = {10.48550/ARXIV.2403.03952},
  eprinttype    = {arXiv},
  eprint       = {2403.03952},
}

@online{googlePlayDataset,
  author = {L. Gupta},
  title  = {Google Play Store Apps},
  year   = {2019},
  lastaccessed = "Oct. 10, 2025",
  url    = {https://www.kaggle.com/datasets/lava18/google-play-store-apps}
}

@online{reasoningStoryDataset,
  author = {rekrek},
  title  = {reasoning-engaging-story},
  year   = {2025},
  lastaccessed = "Oct. 10, 2025",
  url    = {https://huggingface.co/datasets/rekrek/reasoning-engaging-story}
}

@online{techKeywordsDataset,
  author = {Ida Silfverskiold},
  title  = {tech-keywords-topics-summary},
  year   = {2023},
  lastaccessed = "Oct. 10, 2025",
  url    = {https://huggingface.co/datasets/ilsilfverskiold/tech-keywords-topics-summary}
}

@article{chen2023frugalgpt,
  author       = {Lingjiao Chen and
                  Matei Zaharia and
                  James Zou},
  title        = {FrugalGPT: How to Use Large Language Models While Reducing Cost and
                  Improving Performance},
  journal      = {Trans. Mach. Learn. Res.},
  volume       = {2024},
  year         = {2024},
  url          = {https://openreview.net/forum?id=cSimKw5p6R},
}

@book{nygard2018release,
  title={Release It!: Design and Deploy Production-Ready Software},
  author={Nygard, Michael T.},
  edition={2nd},
  year={2018},
  publisher={Pragmatic Bookshelf}
}

@misc{pan2025measuringagentsproduction,
      title={Measuring Agents in Production}, 
      author={Melissa Z. Pan and Negar Arabzadeh and Riccardo Cogo and Yuxuan Zhu and Alexander Xiong and Lakshya A Agrawal and Huanzhi Mao and Emma Shen and Sid Pallerla and Liana Patel and Shu Liu and Tianneng Shi and Xiaoyuan Liu and Jared Quincy Davis and Emmanuele Lacavalla and Alessandro Basile and Shuyi Yang and Paul Castro and Daniel Kang and Joseph E. Gonzalez and Koushik Sen and Dawn Song and Ion Stoica and Matei Zaharia and Marquita Ellis},
      year={2025},
      eprint={2512.04123},
      archivePrefix={arXiv},
      primaryClass={cs.CY},
      url={https://arxiv.org/abs/2512.04123}, 
}

@misc{kim2025sciencescalingagentsystems,
      title={Towards a Science of Scaling Agent Systems}, 
      author={Yubin Kim and Ken Gu and Chanwoo Park and Chunjong Park and Samuel Schmidgall and A. Ali Heydari and Yao Yan and Zhihan Zhang and Yuchen Zhuang and Mark Malhotra and Paul Pu Liang and Hae Won Park and Yuzhe Yang and Xuhai Xu and Yilun Du and Shwetak Patel and Tim Althoff and Daniel McDuff and Xin Liu},
      year={2025},
      eprint={2512.08296},
      archivePrefix={arXiv},
      primaryClass={cs.AI},
      url={https://arxiv.org/abs/2512.08296}, 
}

@misc{gu2025surveyllmasajudge,
      title={A Survey on LLM-as-a-Judge}, 
      author={Jiawei Gu and Xuhui Jiang and Zhichao Shi and Hexiang Tan and Xuehao Zhai and Chengjin Xu and Wei Li and Yinghan Shen and Shengjie Ma and Honghao Liu and Saizhuo Wang and Kun Zhang and Yuanzhuo Wang and Wen Gao and Lionel Ni and Jian Guo},
      year={2025},
      eprint={2411.15594},
      archivePrefix={arXiv},
      primaryClass={cs.CL},
      url={https://arxiv.org/abs/2411.15594}, 
}

@misc{chen2025optimizingmodelselectioncompound,
      title={Optimizing Model Selection for Compound AI Systems}, 
      author={Lingjiao Chen and Jared Quincy Davis and Boris Hanin and Peter Bailis and Matei Zaharia and James Zou and Ion Stoica},
      year={2025},
      eprint={2502.14815},
      archivePrefix={arXiv},
      primaryClass={cs.AI},
      url={https://arxiv.org/abs/2502.14815}, 
}

@misc{meinke2025frontiermodelscapableincontext,
      title={Frontier Models are Capable of In-context Scheming}, 
      author={Alexander Meinke and Bronson Schoen and Jérémy Scheurer and Mikita Balesni and Rusheb Shah and Marius Hobbhahn},
      year={2025},
      eprint={2412.04984},
      archivePrefix={arXiv},
      primaryClass={cs.AI},
      url={https://arxiv.org/abs/2412.04984}, 
}

@misc{routerbench,
      title={RouterBench: A Benchmark for Multi-LLM Routing System}, 
      author={Qitian Jason Hu and Jacob Bieker and Xiuyu Li and Nan Jiang and Benjamin Keigwin and Gaurav Ranganath and Kurt Keutzer and Shriyash Kaustubh Upadhyay},
      year={2024},
      eprint={2403.12031},
      archivePrefix={arXiv},
      primaryClass={cs.LG},
      url={https://arxiv.org/abs/2403.12031}, 
}

@misc{rombaut2025watson,
      title={Watson: A Cognitive Observability Framework for the Reasoning of LLM-Powered Agents}, 
      author={Benjamin Rombaut and Sogol Masoumzadeh and Kirill Vasilevski and Dayi Lin and Ahmed E. Hassan},
      year={2025},
      eprint={2411.03455},
      archivePrefix={arXiv},
      primaryClass={cs.AI},
      url={https://arxiv.org/abs/2411.03455}, 
}

@inproceedings{pan-etal-2024-llmlingua,
    title = "{LLML}ingua-2: Data Distillation for Efficient and Faithful Task-Agnostic Prompt Compression",
    author = {Pan, Zhuoshi  and
      Wu, Qianhui  and
      Jiang, Huiqiang  and
      Xia, Menglin  and
      Luo, Xufang  and
      Zhang, Jue  and
      Lin, Qingwei  and
      R{\"u}hle, Victor  and
      Yang, Yuqing  and
      Lin, Chin-Yew  and
      Zhao, H. Vicky  and
      Qiu, Lili  and
      Zhang, Dongmei},
    booktitle = "ACL 2024 Finding",
    month = aug,
    year = "2024",
    address = "Bangkok, Thailand",
    publisher = "Association for Computational Linguistics",
    url = "https://aclanthology.org/2024.findings-acl.57/",
    doi = "10.18653/v1/2024.findings-acl.57",
    pages = "963--981",
}

@misc{zhang2025agentorchestraor,
      title={AgentOrchestra: Orchestrating Hierarchical Multi-Agent Intelligence with the Tool-Environment-Agent(TEA) Protocol}, 
      author={Wentao Zhang and Liang Zeng and Yuzhen Xiao and Yongcong Li and Ce Cui and Yilei Zhao and Rui Hu and Yang Liu and Yahui Zhou and Bo An},
      year={2025},
      eprint={2506.12508},
      archivePrefix={arXiv},
      primaryClass={cs.AI},
      url={https://arxiv.org/abs/2506.12508}, 
}

@inproceedings {CRISP,
author = {Zhizhou Zhang and Murali Krishna Ramanathan and Prithvi Raj and Abhishek Parwal and Timothy Sherwood and Milind Chabbi},
title = {{CRISP}: Critical Path Analysis of {Large-Scale} Microservice Architectures},
booktitle = {USENIX ATC 22},
year = {2022},
isbn = {978-1-939133-29-50},
address = {Carlsbad, CA},
pages = {655--672},
url = {https://www.usenix.org/conference/atc22/presentation/zhang-zhizhou},
publisher = {USENIX Association},
}

@misc{lin2025awarecompiler,
      title={AwareCompiler: Agentic Context-Aware Compiler Optimization via a Synergistic Knowledge-Data Driven Framework}, 
      author={Hongyu Lin and Haolin Pan and Haoran Luo and Yuchen Li and Kaichun Yao and Libo Zhang and Mingjie Xing and Yanjun Wu},
      year={2025},
      eprint={2510.11759},
      archivePrefix={arXiv},
      primaryClass={cs.PL},
      url={https://arxiv.org/abs/2510.11759}, 
}

@inproceedings{AAAI2024metareasoning,
author = {Budd, Matthew and Lacerda, Bruno and Hawes, Nick},
title = {Stop! planner time: metareasoning for probabilistic planning using learned performance profiles},
year = {2024},
isbn = {978-1-57735-887-9},
publisher = {AAAI Press},
url = {https://doi.org/10.1609/aaai.v38i18.29983},
doi = {10.1609/aaai.v38i18.29983},
articleno = {2235},
numpages = {8},
booktitle    = {{AAAI} 2024},
}

@misc{zhang2025rvllm,
      title={RvLLM: LLM Runtime Verification with Domain Knowledge}, 
      author={Yedi Zhang and Sun Yi Emma and Annabelle Lee Jia En and Jin Song Dong},
      year={2025},
      eprint={2505.18585},
      archivePrefix={arXiv},
      primaryClass={cs.AI},
      url={https://arxiv.org/abs/2505.18585}, 
}

\end{document}